\magnification=\magstep1 \overfullrule=0pt 
\advance\hoffset by -0.35truecm   
\font\tenmsb=msbm10       \font\sevenmsb=msbm7
\font\fivemsb=msbm5       \newfam\msbfam
\textfont\msbfam=\tenmsb  \scriptfont\msbfam=\sevenmsb
\scriptscriptfont\msbfam=\fivemsb
\def\Bbb#1{{\fam\msbfam\relax#1}}
\def\R{{\Bbb R}}\def\Z{{\Bbb Z}}
\def\C{{\Bbb C}}\def\N{{\Bbb N}}\def\T{{\Bbb T}}

\font\gross=cmr10 scaled \magstep2

\font\kap=cmcsc10
\font\tpwrt=cmtt8 scaled \magstep2
\def\qed{{\vrule height4pt width4pt depth1pt}}
\def\lb{\lbrack}\def\rb{\rbrack}  \def\q#1{$\lb{\sl #1}\,\rb$}
\def\bn{\bigskip\noindent} \def\mn{\medskip\smallskip\noindent}
\def\sn{\smallskip\noindent} 
\def\lra{\longrightarrow} 
\def\lla{\longleftarrow}  
\def\a{\hbox{$\cal A$}}   \def\b{\hbox{$\cal B$}}
\def\m{\hbox{$\cal M$}}   
\def\h{\hbox{$\cal H$}}   \def\e{\hbox{$\cal E$}}
   \def\S{{\cal S}}
\def\cald{{\cal D}}       \def\calbd{\overline{{\cal D}}}
\def\one{{\bf 1}}          
\def\tmc{T^*M^{\C}}       
\def\End{{\rm End\,}}     \def\Lim{\mathop{\rm Lim}}
\def\lie{{\bf g}}         \def\cinfty{C^{\infty}(M)}
\def\Asy{\hbox{{\tpwrt A}}}\def\Tot{\hbox{$T_{\rm tot}$}}
\def\d{\hbox{$\cal D$}}   \def\bard{\overline{\hbox{$\cal D$}}}
\def\ttd{\hbox{{\tt d}}}  \def\ttdst{\hbox{${\tt d}^*$}}
\def\td{\hbox{$\widetilde{{\tt d}}$}}
\def\tdst{\hbox{$\widetilde{\tt d}{}^*$}}
\def\dd{{\tt d}}
\def\Spv{\hbox{${\rm Spin}(V)$}}   
\def\Spn{\hbox{${\rm Spin}(n)$}}
\def\Sprn{\hbox{${\rm Spin}(\R^n)$}}
\def\Spcv{\hbox{${\rm Spin}^c(V)$}}
\def\spcv{\hbox{${\rm spin}^c(V)$}}
\def\spv{\hbox{${\rm spin}(V)$}}
\def\barint{{\int \kern-9pt {\rm -} \ }}
\def\Barint{{\int \kern-10pt - \ }}
\def\barintbeta{{\int_{{\scriptscriptstyle \beta}} 
      \kern-11pt {\rm -} \ }}
\def\Barintbeta{{\int_{\beta} \kern-12pt - \ }}
\def\Neqone{\hbox{$N = 1$} \kern -31.48pt \hbox{$N = 1$}\ }
\def\Noneone{\hbox{$N = (1,1)$} \kern -48.25pt \hbox{$N = (1,1)$}\ }
\def\Nonezero{\hbox{$N = 1$} \kern -31.70pt \hbox{$N = 1$}\ }
\def\Ntwotwo{\hbox{$N = (2,2)$} \kern -48.25pt \hbox{$N = (2,2)$}\ }
\def\Nfourfour{\hbox{$N = (4,4)$} \kern -48.25pt \hbox{$N = (4,4)$}\ }
\def\MM{\hbox{$M$} \kern -14.5pt \hbox{$M$}\ }
\def\CC{\hbox{${}^c$} \kern -7.5pt \hbox{${}^c$}}
\def\CCVV{\hbox{$^c(V)$} \kern -24pt \hbox{$^c(V)$}}
\def\oddots{\mathinner{\mkern1mu\raise1pt\vbox{\kern7pt\hbox{.}}
  \mkern2mu\raise4pt\hbox{.}\mkern2mu\raise7pt\hbox{.}\mkern1mu}}
\def\bigwedge{{\Lambda}}
\def\llcorner{\raise1.7pt\vbox{
     \hbox{\vrule height1.9pt width.3pt $\underline{\phantom{n}}\,$}}}
\setbox1=\hbox{$\lra$}
\def\frarrow{\raise0.36pt\hbox{$\lra$}\hskip-\wd1
\raise0.18pt\hbox{$\lra$}\hskip-\wd1\hbox{$\lra$}\hskip-\wd1
\raise-0.18pt\hbox{$\lra$}\hskip-\wd1\raise-0.35pt\hbox{$\lra$}}
\def\fuarrow{\hbox{$\uparrow$}\kern-4.55pt\hbox{$\uparrow$} 
\kern-4.56pt\hbox{$\uparrow$}\kern-4.5pt\raise.17pt\hbox{$|$}}
\def\frtlarrow{\raise-2.5pt\hbox{$\frarrow$}
\hskip-\wd1{\raise2.5pt\hbox{$\lla$}}}
\def\futdarrow{\fuarrow\!\downarrow}
\def\sqr#1#2{{\vcenter{\vbox{\hrule height.#2pt
       \hbox{\vrule width.#2pt height #1pt \kern#1pt 
         \vrule width.#2pt}\hrule  height.#2pt}}}}
\def\square{\mathchoice\sqr{6.5}{38}\sqr{6.5}{38}\sqr{2.1}3\sqr{1.5}3}
\def\bigwedge{{\Lambda}}
\def\Omti{\hbox{$\widetilde{\Omega}^k({\cal A})$} 
  \kern -32pt\hbox{$\widetilde{\Omega}^k({\cal A})$}\ }
\def\Omtipar{\hbox{$\widetilde{\Omega}^{\ 1}_{\partial,\overline{
 \partial}}({\cal A})$} \kern -38.7pt\hbox{$\widetilde{\Omega}^{\ 1}_{
 \partial,\overline{\partial}}  ({\cal A})$}\ }
\def\cinfty{C^{\infty}(M)}
\def\oddots{\mathinner{\mkern1mu\raise1pt\vbox{\kern7pt\hbox{.}}
  \mkern2mu\raise4pt\hbox{.}\mkern2mu\raise7pt\hbox{.}\mkern1mu}}
\def\ttR{\hbox{{\tpwrt R}}}\def\ttT{\hbox{{\tpwrt T}}}
\def\ttRic{\hbox{{\tpwrt Ric}}}\def\ttr{\hbox{{\tpwrt r}}}
\def\omdt{{\raise-1.65pt\hbox{$\widetilde{\phantom{\widetilde{\Omega}
  }}$} \kern - 7.4pt\hbox{$\widetilde{\Omega}$}}}
\def\nadt{{\raise-1.65pt\hbox{$\widetilde{\phantom{\widetilde{\nabla}
  }}$} \kern - 8.5pt\hbox{$\widetilde{\nabla}$}}}
\def\pabt{{\raise-1.2pt\hbox{$\widetilde{\phantom{\overline{\partial}
  }}$} \kern - 6.2pt\hbox{$\overline{\partial}$}}}
\def\scpbp{{\scriptscriptstyle \!\partial\!,\!\bar\partial}}
\def\c{\hbox{{\tpwrt c}}}
\def\cedille#1{\setbox0=\hbox{#1}\ifdim\ht0=1ex \accent'30 #1%
 \else{\ooalign{\hidewidth\char'30\hidewidth\crcr\unbox0}}\fi}
\def\gaw{Gaw\cedille edzki}\def\qdq{\quad\dotfill\quad}
{\nopagenumbers
\rightline{ETH-TH/96-45}
\sn
\bn\bn\bn
\centerline{{\gross 
Supersymmetric Quantum Theory}}
\bn
\centerline{{\gross and}}
\bn
\centerline{{\gross (Non-Commutative)}}
\bn
\centerline{{\gross Differential Geometry}}
\bn\bn
\vfill \centerline{by}\bn\bn \vfill
\centerline{{ J.\ Fr\"ohlich, O.\ Grandjean${}^*$ and 
   A.\ Recknagel${}^{**}$}}
\bn\bn
\centerline{Institut f\"ur Theoretische Physik, ETH-H\"onggerberg}
\centerline{CH-8093 Z\"urich, Switzerland}
\bn\bn\vfill\vfill\vfill\bn
\vfill\vfill\vfill
\smallskip
${}^*\ $ work supported in part by the Swiss National 
Foundation \hfill\break
\indent ${}^{**}$ work supported in part by a European Union 
HCM Fellowship 
\medskip 
e-mail addresses: juerg, grandj and anderl@itp.phys.ethz.ch 
\eject
\vfill $$\phantom{Leerseite}$$\vfill
\eject}
\pageno=-1\parindent=32pt
\leftline{\bf Contents}
\bn\mn
1. Introduction and summary of results\qdq \ 1
\medskip
\item{1.1} Geometry and quantum physics \qdq \ 2
\item{1.2} Pauli's electron \qdq \ 3
\item{1.3} Supersymmetric quantum field theory and geometry put 
   into  perspective\qdq 14
\bn
2. Canonical anti-commutation relations, Clifford algebras, and 
   Spin${}^c$-groups \qdq 17
\medskip
\item{2.1} Clifford algebras \qdq 17
\item{2.2} The group Spin${}^c(V)$ \qdq 18
\item{2.3} Exterior algebra and spin representations \qdq 20
\bn
3. Algebraic formulation of differential geometry\qdq 22
\medskip
\item{3.1} The $N=1$ formulation of Riemannian geometry\qdq 22
\item{3.2} The $N=(1,1)$ formulation of Riemannian geometry\qdq31
\item{3.3} Lie groups\qdq 41
\item{3.4} Algebraic formulation of complex geometry\qdq 42
\item{3.5} Hyperk\"ahler geometry\qdq 53
\item{3.6} Symplectic geometry\qdq 60
\bn
4. Supersymmetry classification of geometries\qdq 65
\bn
5. Non-commutative geometry\qdq 77
\medskip
\item{5.1} The $N=1$ formulation of non-commutative geometry\qdq 77
\item{5.2} The $N=(1,1)$ formulation of non-commutative
    geometry\qdq 93 
\item{5.3} Hermitian and K\"ahler non-commutative geometry\qdq 103
\item{5.4} The $N=(4,4)$ spectral data\qdq 111 
\item{5.5} Symplectic non-commutative geometry\qdq 111
\bn
6. Directions for future work\qdq 114
\bn
References \qdq 119
\vfill\eject 
{\nopagenumbers\phantom{xxx}\vfill\eject}
\pageno=1\parindent=25pt  
\leftline{\bf 1. Introduction and summary of results}
\bn
In this paper we describe an approach to differential topology and 
geometry rooted in supersymmetric quantum theory. We show how 
the basic concepts and notions of differential geometry emerge {}from 
concepts and notions of the quantum theory of non-relativistic 
particles with spin, and how the classification of different types 
of differential geometry follows the classification of 
supersymmetries.
\sn
Historically, the development of geometry has been closely linked 
to that of {\sl classical} physical theory. Geometry was born, in 
antique Greece, out of concrete problems in geodesy. Much later, 
the invention and development of differential geometry took place 
in parallel to the development of classical mechanics and 
electromagnetism. A basic notion in differential geometry is that of 
parameterized curves on a manifold. This notion is intimately 
related to the one of trajectories of point particles central in  
classical mechanics: Tangent vectors correspond to velocities, vector 
fields to force laws. There are many further examples of such 
interrelations. To mention one, the work of de Rham and Chern was 
inspired and influenced by Maxwell's theory of electromagnetism, a  
classical field theory.
\sn
In this century, the foundations of physics have changed radically 
with the discovery of quantum mechanics. Quantum theory is more 
fundamental than classical physics. It is therefore natural to 
ask how differential geometry can be rediscovered -- starting {}from the 
basic notions of quantum theory. In this work, we outline a plausible 
answer to this question. As a payoff, we find natural generalizations 
of differential geometry to {\sl non-commutative} differential 
geometry, in the sense of Connes. 
\sn
Connes' theory naturally leads to a far-reaching generalization of 
geometry which makes it possible to study highly singular and 
genuinely non-commutative spaces as 
{\sl geometric} spaces. Such generalizations are called for by deep 
problems in quantum mechanics, quantum field theory and quantum 
gravity. Quantum mechanics entails that phase space must be deformed 
to a non-commutative space; moreover, attempts to reconcile quantum 
theory with general relativity naturally lead to the idea that 
space-time is non-commutative. 
\sn
The idea to reconsider and generalize algebraic topology and 
differential geometry {}from the point of view of operator theory and 
quantum theory is a guiding principle in Connes' work on 
non-commutative geometry, the development and results of which are 
summarized in detail in \q{Co1\!-\!4} and the references therein. 
Connes' work is basic for the approach presented in this 
paper. \hfill\break
\noindent  That supersymmetry and supersymmetric quantum field theory 
provide natural formulations of problems in differential topology and 
geometry and a variety of novel tools was recognized by Witten, 
see e.g.\ \q{Wi1,2}. \hfill\break 
\noindent  In this paper, we attempt to combine the two threads 
coming {}from the work of Connes and of Witten: Notions and concepts {}from 
quantum mechanics and operator theory and {}from supersymmetry play 
fundamental roles in the approach to classical and non-commutative 
geometry described here. 
\sn    
A key example of supersymmetric quantum mechanics is Pauli's theory 
of non-relativistic spinning electrons, which will be described later 
in this introduction. It forms the basis in an analysis of classical
differential geometry starting {}from notions of quantum theory. Although 
we shall not solve any difficult problem in algebraic topology or 
differential geometry in this paper, we expect that the perspective of 
topology and geometry offered by supersymmetric quantum mechanics and 
the generalizations it suggests will turn out to be useful and 
productive. 
\mn
In this first section, we give a short sketch of 
our approach and of the main results. A detailed account of the 
algebraic formulation of classical geometry is contained in 
section 3, with some preparatory material collected in 
section 2. The fourth section is reserved for an overview 
of the various types of differential geometry treated before, 
but now classified according to the supersymmetry  
encoded in the algebraic data. This provides us with 
a better understanding of the hierarchy of geometries, 
and further facilitates the task to generalize our classical notions 
to the non-commutative case. This problem is dealt with in section 5. 
In the concluding section, we describe some important open problems 
arising in our work, and sketch a number of further generalizations. 
\bn\bn
\leftline{\bf 1.1 Geometry and quantum physics}
\bn
In quantum mechanics, one is used to identify {\sl pure states} of the  
physical system under consideration with {\sl unit rays} in a separable 
Hilbert space. This Hilbert space carries a representation of an 
``observable algebra'', in the following denoted by  ${\cal F}_{\hbar}$. 
Imagine that the system consists of  
a quantum mechanical particle propagating on a manifold $M$. Then 
${\cal F}_{\hbar}$ can be identified with a deformation of the 
algebra of functions over the phase space $T^*M$. The 
non-commutativity of ${\cal F}_{\hbar}$ can be traced back to the 
Heisenberg uncertainty relations 
$$
{\scriptstyle \triangle}p_j {\scriptstyle \triangle}q^j 
\geq{\hbar\over2} \eqno(1.1)$$
for each $j=1,\ldots, {\rm dim}M$, 
where $p_j, q^j$ are Darboux coordinates of $T^*M$, and $\hbar$ 
is Planck's constant divided by $2\pi$. In the Schr\"odinger picture, 
one could identify ${\cal F}_{\hbar}$ with the algebra of 
pseudo-differential operators. We will always choose it in such a 
way that it contains the algebra \a\ of smooth functions over the 
configuration space $M$ as a maximal abelian sub-algebra. 
\sn Starting {}from the algebra ${\cal F}_{\hbar}$ of observables, 
viewed as an abstract ${}^*$-algebra, one is thus interested in 
constructing ${}^*$-representations of ${\cal F}_{\hbar}$ on separable 
Hilbert spaces.  Traditionally, this is accomplished as follows: Given 
$({\cal F}_{\hbar}, \a)$, or equivalently $({\cal F}_{\hbar},M)$, and 
assuming e.g.\ that $M$ is smooth, one chooses a {\sl Riemannian metric} 
$g$ on $M$. If the quantum mechanical particle is a ``{\sl neutral 
scalar particle}'' one would take as a Hilbert space 
$$
\h_{\rm sc} = L^2(M, dvol_g)\ ,
\eqno(1.2)$$
where integration is defined with the {\sl Riemannian volume form}
$dvol_g$  on $M$. The algebra ${\cal F}_{\hbar}$ 
has a natural ${}^*$-representation on $\h_{\rm sc}$. 
\sn
The generator of the quantum mechanical time evolution, i.e.\ 
the {\sl Hamiltonian}, of a system describing a neutral scalar 
particle is chosen to be 
$$
H = -{\hbar^2\over2m} \triangle_g + v\ ,
\eqno(1.3)$$
where $\triangle_g$ is the Laplace-Beltrami operator on smooth 
functions on $M$ (in physics defined to be negative definite), and 
the ``potential'' $v$ is some function on $M$, i.e,\ $v\in\a$. 
Finally, $m$ is the mass of the particle. 
\sn
Using ideas of Connes, see e.g.\ \q{Co1}, it has been verified in 
\q{FG} -- and is recalled in section 3.1.3 -- that the manifold $M$ 
and the {\sl geodesic distance} on $M$ can 
be reconstructed    {}from the the abstract ``{\sl spectral triple}'' 
$(\a, \h_{\rm sc}, H)$. Hence the Riemannian geometry of the 
configuration space $M$ is encoded into the following data of 
quantum mechanics: 
\smallskip
\item{$i$)} the abelian algebra \a\ of position measurements; 
\smallskip
\item{$ii$)} the Hilbert space $\h_{\rm sc}$ of pure state vectors; 
\smallskip
\item{$iii$)} the Hamiltonian $H$ generating the time evolution of the 
system. 
\mn 
Of course, as physicists, we would prefer exploring the Riemannian 
geometry of $M$ starting {}from the data $({\cal F}_{\hbar}, 
\h_{\rm sc}, H)$. This poses some interesting mathematical
questions which are only partially answered. Since, ultimately, 
momentum measurements are reduced to position measurements, it is 
tolerable to take the spectral triple $(\a, \h_{\rm sc}, H)$ as 
a starting point. But even then the reconstruction of the differential 
topology and geometry of $M$ {}from the quantum mechanics of a scalar 
particle, as encoded in the latter triple, is quite cumbersome and 
unnatural. Fortunately, nature has invented particles with {\sl spin} 
which are much better suited for an exploration of the geometry of $M$.  
\bn\bn
\leftline{\bf 1.2 Pauli's electron}
\bn
Low-energy electrons can be treated as non-relativistic quantum 
mechanical point particles with spin. Here is a sketch of {\sl Pauli's 
quantum mechanics} of non-relativistic electrons and positrons. 
\sn
Physical space is described in terms of a smooth, orientable 
Riemannian (spin${}^c$) manifold $(M,g)$ of dimension $n$. Let $T^*M$ 
be the cotangent bundle of $M$ and denote by $\Lambda^{\bullet}M$
the bundle of completely anti-symmetric covariant tensors over $M$. 
Let $\Omega^{\bullet}(M)$ be the space of smooth sections of 
$\Lambda^{\bullet}M$, i.e., of smooth differential forms on $M$, 
and $\Omega_{\C}^{\bullet}(M) =\Omega^{\bullet}(M) \otimes \C$ 
its complexification.  
Since we are given a Riemannian metric on $M$, $\Lambda^{\bullet}M$
is equipped with a Hermitian structure which, together with the 
Riemannian volume element $dvol_g$, determines a scalar product 
$(\cdot,\cdot)_g$ on $\Omega_{\C}^{\bullet}(M)$. Let $\h_{\rm e-p}$
denote the Hilbert space completion of $\Omega_{\C}^{\bullet}(M)$  
in the norm given by the scalar product $(\cdot,\cdot)_g$. Hence 
$\h_{\rm e-p}$ is the space of complex-valued square-integrable 
differential forms on $M$. The meaning of the subscript ``e-p'' 
will become clear soon. 
\sn
Given a 1-form $\xi\in\Omega^1(M)$, let $X$ be the 
vector field corresponding to $\xi$ by the equation 
$$
\xi(Y) = g(X,Y)
$$
for any smooth vector field $Y$. For every $\xi\in\Omega^1(M)
\otimes\C$, we define two operators on $\h_{\rm e-p}$: 
$$
a^*(\xi) \psi = \xi\wedge\psi 
\eqno(1.4)$$
and
$$
a(\xi)\psi = X \llcorner\, \psi 
\eqno(1.5)$$
for all $\psi\in\h_{\rm e-p}$. In (1.5), $\llcorner$ denotes interior 
multiplication. Thus  $a^*(\xi)$ is the adjoint of 
the operator $a(\overline{\xi})$ on $\h_{\rm e-p}$. 
One verifies that, for arbitrary $\xi, \eta \in 
\Omega^1(M)\otimes\C$ (corresponding to vector fields $X, Y$ via $g$), 
$$
\lbrace\,a(\xi), a(\eta)\,\rbrace = 
\lbrace\,a^*(\xi), a^*(\eta)\,\rbrace = 0 
\eqno(1.6)$$
and 
$$
\lbrace\,a(\xi), a^*(\eta)\,\rbrace = g(X,Y)\cdot\one \equiv 
g(\xi,\eta)\cdot\one\ ;
\eqno(1.7)$$
here $\lbrace\,A,B\,\rbrace := AB+BA$ denotes the anti-commutator of 
$A$ and $B$, and we have used the same symbol for the metric on 
vector fields and 1-forms. Eqs.\ (1.6) and (1.7) are called {\sl 
canonical anti-commutation relations} and are basic in the 
description of fermions in physics.
\sn
Next, for every {\sl real} $\xi\in\Omega^1(M)$, we define two 
anti-commuting anti-selfadjoint operators $\Gamma(\xi)$ and 
$\overline{\Gamma}(\xi)$ on $\h_{\rm e-p}$ by 
$$\eqalignno{
\Gamma(\xi) &= a^*(\xi) - a(\xi)\ , &(1.8)\cr
\overline{\Gamma}(\xi) &= i\bigl(a^*(\xi) + a(\xi)\bigr)\ . 
&(1.9)\cr}
$$
One checks that 
$$\eqalignno{
\lbrace\,\Gamma(\xi),\Gamma(\eta)\,\rbrace &= 
\lbrace\,\overline{\Gamma}(\xi),\overline{\Gamma}(\eta)\,\rbrace 
= -2g(\xi,\eta)\ , &(1.10)\cr
\lbrace\,\Gamma(\xi),\overline{\Gamma}(\eta)\,\rbrace &= 0\ , 
&(1.11)\cr}$$
for arbitrary $\xi$ and $\eta$ in $\Omega^1(M)$. Thus $\Gamma(\xi)$ 
and $\overline{\Gamma}(\xi)$, $\xi\in\Omega^1(M)$, are anti-commuting
sections of two isomorphic {\sl Clifford bundles} $Cl(M)$ over $M$. 
\sn
An $n$-dimensional Riemannian manifold $(M,g)$ is a spin${}^c$ manifold 
if and only if $M$ is oriented and there exists a complex Hermitian 
vector bundle $S$ of rank $2^{\lb{n\over2}\rb}$ over $M$, where 
$\lb k\rb$ denotes the integer part of $k$, together with a bundle 
homomorphism $c\,:\ T^*M \lra {\rm End}\,(S)$ such that 
$$\eqalign{
&c(\xi)+c(\xi)^* = 0 \cr
&c(\xi)^*c(\xi) = g(\xi,\xi)\cdot\one\cr}
\eqno(1.12)$$
for all cotangent vectors $\xi\in T^*M$. Above, the adjoint is defined 
with respect to the Hermitian structure $(\cdot,\cdot)_S$ on $S$. 
The completion of the space of sections $\Gamma(S)$ in the norm 
induced by $(\cdot,\cdot)_S$ is a 
Hilbert space which we denote by $\h_{\rm e}$, the Hilbert space of 
{\sl square-integrable Pauli-Dirac spinors}. The homomorphism $c$ 
extends uniquely to an {\sl irreducible unitary Clifford action} of 
$T_x^*M$ on $S_x$ for all $x\in M$.  
\sn
If $M$ is an even-dimensional, orientable Riemannian manifold then 
there is an element $\gamma$ in the Clifford bundle generated by the 
operators $\Gamma(\xi)$, $\xi\in\Omega^1(M)$, which anti-commutes with 
every $\Gamma(\xi)$ and satisfies $\gamma^2=\one$; $\gamma$ corresponds 
to the Riemannian volume form on $M$. We conclude that there exists an 
isomorphism 
$$
i\,:\ \Omega^{\bullet}_{\C}(M)\lra\overline{S} \otimes_{\cal A} S 
\eqno(1.13)$$
where $\a=C^{\infty}(M)$ if $M$ is smooth (and $\a=C(M)$ for topological 
manifolds), and where $\overline{S}$ is the ``charge-conjugate'' bundle 
of $S$, constructed {}from $S$ by complex conjugation of the transition 
functions, see section 3.2.2. Upon ``transporting'' $c$ to 
$\overline{S}$, this bundle receives a natural Clifford action 
$\overline{c}$, and and the map $i$ is an intertwiner satisfying   
$$\eqalign{
i\circ\Gamma(\xi) &= (\one\otimes c(\xi)) \circ i\ ,\cr 
i\circ\overline{\Gamma}(\xi) &= 
(\overline{c}(\xi)\otimes\gamma)\circ i\ \cr}
\eqno(1.14)$$
for all $\xi\in\Omega^1(M)$. The ``volume element'' $\gamma$ has been 
inserted so as to ensure that the Clifford action $\one\otimes c$ on 
$\overline{S}\otimes_{\cal A} S$ anti-commutes with the second action 
$\overline{c}\otimes\gamma$.  
\sn
If $M$ is an odd-dimensional spin${}^c$ manifold then $\gamma$ is 
central and we use the Pauli matrices $\tau_1=\pmatrix{0&1\cr1&0\cr}$ 
and  $\tau_3 = \pmatrix{1&0\cr0&-1\cr}$ 
in order to obtain anti-commuting Clifford actions 
$\one\otimes c \otimes \tau_3$ and $\overline{c}\otimes\one\otimes\tau_1$  
on the bundle $\overline{S} \otimes_{\cal A} S \otimes \C^2$; as before, 
there is an isomorphism 
$$
i\,:\ \Omega^{\bullet}_{\C}(M) \lra \overline{S} \otimes_{\cal A} 
S \otimes \C^2           \eqno(1.15)$$
which intertwines the Clifford actions: 
$$\eqalign{
i\circ\Gamma(\xi) &=            
(\one\otimes c(\xi) \otimes \tau_3) \circ i\ ,\quad\cr
i\circ\overline{\Gamma}(\xi)&= 
(\overline{c}(\xi)\otimes\one\otimes\tau_1)\circ i\ .\cr}
\eqno(1.16)$$
\sn
A connection $\nabla^S$ on $S$ is called a {\sl spin${}^c$ connection} 
iff it satisfies
$$
\nabla^S_{\!X}\bigl(c(\eta)\psi) = c(\nabla_{\!X} \eta) \psi 
+ c(\eta)\nabla^S_{\!X}\psi
\eqno(1.17)$$
for any vector field $X$, any 1-form $\eta$ and any section 
$\psi\in\Gamma(S)$, where $\nabla$ is a connection on $T^*M$. We say 
that $\nabla^S$ is compatible with the Levi-Civita connection iff, 
in (1.17), $\nabla=\nabla^{{\rm L.C.}}$. 
\sn
If $\nabla^S_1$ and $\nabla^S_2$ are two spin${}^c$ connections 
compatible with $\nabla^{{\rm L.C.}}$ then 
$$
\bigl(\nabla^S_1 - \nabla^S_2\bigr) \psi = i\,\alpha \otimes \psi 
\eqno(1.18)$$
for some real 1-form $\alpha\in\Omega^1(M)$. In physics, $\alpha$ is the 
difference of two {\sl electromagnetic vector potentials}. If 
$R_{\nabla^S}$ denotes the curvature of a spin${}^c$ connection 
$\nabla^S$ then
$$
2^{-\lb{n\over2}\rb} {\rm tr}\,\bigl(R_{\nabla}^S(X,Y)\bigr) = F_{2A}(X,Y)
\eqno(1.19)$$
for arbitrary vector fields $X,Y$, where $F_{2A}$ is the curvature 
(``the electromagnetic field tensor'') of a U(1)-connection $2A$ 
(``electromagnetic vector potential'') on a line bundle canonically 
associated to $S$; see sections 2 and 3.2.2. 
\sn
The {\sl (Pauli-)Dirac operator} associated with a spin${}^c$ 
connection $\nabla^S$ is defined by 
$$
D_A = c \circ \nabla^S\ .
\eqno(1.20)$$
\sn
We are now prepared to say what is meant by {\sl Pauli's quantum 
mechanics of non-relativistic electrons}. As a Hilbert space of pure 
state vectors one chooses $\h_{\rm e}$, the space of square-integrable 
Pauli-Dirac spinors. The dynamics of an electron (with the gyromagnetic 
factor $g$, measuring the strength of the magnetic moment of the electron, 
set equal to 2) is generated by the Hamiltonian 
$$
H_A ={\hbar^2\over2m} D_A^2 + v ={\hbar^2\over2m}\bigl(-\triangle^S_A 
+ {r\over4} + c(F_A) \bigr) + v 
\eqno(1.21)$$
where the electrostatic potential $v$ is a function on $M$ and $r$ 
denotes the scalar curvature. There are well-known sufficient 
conditions on $M$ and $v$ which guarantee that $H_A$ is self-adjoint 
and bounded {}from below on $\h_{\rm e}$ -- and there are less well-known 
ones, see \q{FLL,LL,LY}. As an algebra of ``observables'' associated 
with a quantum mechanical electron, one chooses the algebra 
$\a=C^{\infty}(M)$, possibly enlarged to $\a=C(M)$. 
\sn
One may ask in how far the geometry of $M$ is encoded in the spectral 
triple $(\a, \h_{\rm e}, D_A)$ associated with a non-relativistic 
electron. The answer given by Connes, see \q{Co1}, is that  
$(\a, \h_{\rm e}, D_A)$ encodes the differential topology and 
Riemannian geometry of $M$ {\sl completely}. This story is told in 
detail in section 3, so we are not anticipating it here. 
\mn
{}From a physics point of view, it is more natural to work with an 
algebra ${\cal F}_{\hbar}$ of pseudo-differential operators that acts 
on $\h_{\rm e}$ and is invariant under the Heisenberg picture dynamics 
generated by $H_A$. Thus one should consider the triple $({\cal F}_{
\hbar}, \h_{\rm e}, D_A)$. When ${\cal F}_{\hbar}$, $\h_{\rm e}$ and 
$D_A$ are given as {\sl abstract data}, some interesting mathematical 
problems connected with the reconstruction of $M$ remain to be solved.  
\mn
If $v=0$, then 
$$
H_A = \Biggl(\sqrt{ {\hbar^2\over2m}}\, D_A  \Biggr)^2 
\eqno(1.22)$$
i.e., $H_A$ is the square of a ``{\sl supercharge}'' 
$Q_A =\sqrt{{\hbar^2\over2m}}\,D_A$. If $M$ is even-dimensional then, 
as discussed above, the Riemannian volume form determines a section 
$\gamma$ of the Clifford bundle which 
is a unitary involution with the property that 
$$
\lb\,\gamma,a\,\rb = 0\quad\hbox{for all}\ a\in\a\quad
(\hbox{or for all}\  a\in{\cal F}_{\hbar})
$$
but 
$$
\{\,\gamma,Q_A\,\} = 0\ .
\eqno(1.23)$$
Thus $\gamma$ defines a $\Z_2$-grading. The data $(\a, \h_{\rm e}, Q_A, 
\gamma)$ yield an example of what physicists call $N=1$ 
{\sl supersymmetric quantum mechanics}.
\mn
In order to describe the twin of Pauli's electron, the 
{\sl non-relativistic positron}, we have to pass {}from $S$ to the 
charge-conjugate spinor bundle $\overline{S}$. The latter inherits 
a spin${}^c$ connection $\nabla^{\overline{S}}$ {}from $S$, which can 
be defined, {\sl locally}, by using the (local) isomorphism 
$S\cong\overline{S}$ and setting 
$\bigl( \nabla^S - \nabla^{\overline{S}} \bigr) \psi =2i\,A\otimes\psi$ 
for $\psi\in S\cong\overline{S}\,$; precise formulas are given in 
section 3.2.2. The space of square integrable  sections $\h_{\rm p}$ 
of $\overline{S}$ is, however, canonically isomorphic to $\h_{\rm e}$, 
and thus the description of the positron involves the same algebra of
observables, $\a$ or ${\cal F}_{\hbar}$, and the same Hilbert space,
now denoted by $\h_{\rm p}$ ($=\h_{\rm e}$); we only replace the 
operator $D_A$ by 
$$
\overline{D}_A = \overline{c} \circ \nabla^{\overline{S}}
\eqno(1.24)$$ 
and set 
$$
\overline{H}_A = {\hbar^2\over2m} \overline{D}_A^2 - v \ .
\eqno(1.25)$$
The physical interpretation of these changes is simply that we have 
reversed the sign of the electric charge of the particle, keeping 
everything else, such as its mass $m$, unchanged. 
\mn
The third character of the play is the (non-relativistic) 
{\sl positronium}, the ground state of a bound pair of an electron 
and a positron. Here, ``ground state'' means that we ignore the 
relative motion of electron and positron.  As an algebra
of ``observables'', we continue to use \a. The Hilbert space of 
pure state vectors of the positronium ground state is
$$\eqalignno{
\h_{\rm e-p} &= \h_{\rm p}\otimes_{\cal A} \h_{\rm e}\quad\ 
\hbox{if dim$M$ is even}\,,
&(1.26)\cr
\h_{\rm e-p} &= \bigl(\h_{\rm p}\otimes_{\cal A} \h_{\rm e}\bigr)_+
\oplus \bigl(\h_{\rm p}\otimes_{\cal A} \h_{\rm e}\bigr)_- 
\cr
&\cong \bigl(\h_{\rm p}\otimes_{\cal A} \h_{\rm e}\bigr) \otimes \C^2
\quad\ \hbox{if dim$M$ is odd}\ .
&(1.27)\cr}$$
Elements in $\bigl(\h_{\rm p}\otimes_{\cal A}\h_{\rm e}\bigr)_+$ 
are even, elements in $\bigl(\h_{\rm p}\otimes_{\cal A} 
\h_{\rm e}\bigr)_-$ are odd under reversing the orientation of $M$, 
i.e.\ under space reflection.   \hfill\break
\noindent We can define a connection $\widetilde{\nabla}$ on 
$\h_{\rm e-p}$ as follows: If $\phi\in\h_{\rm e-p}$ is given by 
$\phi=\psi_1 \otimes \psi_2 (\otimes u)$, $\psi_1\in\h_{\rm p}$, 
$\psi_2\in\h_{\rm e}$, ($u\in\C^2$), we set 
$$
\widetilde{\nabla}\phi = \bigl(\nabla^{\overline{S}}\psi_1\bigr)
\otimes \psi_2 (\otimes u) + \psi_1 \otimes \nabla^S\psi_2 
(\otimes u)\ .
\eqno(1.28)$$
Given $\nabla^S$, this defines $\widetilde{\nabla}$ uniquely, and 
using the intertwiners (1.13,15) it turns out that in fact 
$\widetilde{\nabla} = \nabla^{\rm L.C.}$, see Lemma 3.13 below. 
Observe that $\widetilde{\nabla}$ is independent of the 
virtual U(1)-connection $A$ -- which, physically, is related to 
the fact that the electric charge of positronium is zero. 
\sn
We can now introduce two first order differential operators on 
$\h_{\rm e-p}$,
$$
\d = \Gamma\circ\widetilde{\nabla}\ ,\quad
\bard = \overline{\Gamma}\circ\widetilde{\nabla}\ ,
\eqno(1.29)$$
with $\Gamma$ and $\overline{\Gamma}$ defined as in (1.8,9), 
see also (1.14,16). The details of this construction are explained in 
section 3.2.2; see also section 5.2.5 for an extension to the 
non-commutative case. 
\sn
If $\nabla^S$ is {\sl compatible with the Levi-Civita connection} 
then $\d$ and $\bard$ satisfy the algebra  
$$
\{\,\d,\bard\,\} = 0\ ,\quad \d^2 = \bard{}^2 
\eqno(1.30)$$
and are (formally) self-adjoint on $\h_{\rm e-p}$. 
\sn
The quantum theory of positronium is formulated in terms of the algebra 
of ``observables'' \a, the Hilbert space $\h_{\rm e-p}$ and the 
Hamiltonian 
$$
H = {\hbar^2\over2\mu} \d^2 = {\hbar^2\over2\mu} \bard{}^2\ ,
\eqno(1.31)$$
assuming (1.30), i.e.\ that $\nabla^S$ is compatible with the 
Levi-Civita connection; $\mu=2m$ is the mass of the positronium. The 
Weitzenb\"ock formula says that 
$$
H={\hbar^2\over2\mu}\, \bigl( -\triangle +{r\over4}-{1\over8}R_{ijkl}
\overline{\Gamma}{}^i\overline{\Gamma}{}^j\Gamma^k\Gamma^l \bigr)
\eqno(1.32)$$
where $\triangle = g^{ij}(\nabla_i\nabla_j - \Gamma_{ij}^k\nabla_k)$  
in terms of the Christoffel symbols $\Gamma_{ij}^k$ of the Levi-Civita 
connection, where $r$ is the scalar curvature  and $R_{ijkl}$ are 
the components of the Riemann curvature tensor in local coordinates 
$q^j$; finally, $\Gamma^j=\Gamma(dq^j)$, and analogously for 
$\overline{\Gamma}{}^j$. In (1.32), the summation convention has 
has been used. \hfill\break
\noindent Of course, we have seen in (1.13-16) that $\h_{\rm e-p}$ is 
simply the Hilbert space of square-integrable (complexified) 
differential forms; hence it is no surprise that the Hamiltonian $H$ 
of positronium is proportional to the usual Laplacian on differential 
forms. 
\sn
It is convenient to introduce operators \ttd\ and \ttdst\ given by 
$$
\ttd = {1\over2}\,\bigl(\d-i\bard\bigr)\ ,\quad 
\ttdst = {1\over2}\,\bigl(\d+i\bard\bigr)\ .
\eqno(1.33)$$
Then the relations (1.30) show that 
$$
\ttd^2 = \bigl(\ttdst\bigr){}^2 = 0 
\eqno(1.34)$$ 
and 
$$
{\hbar^2\over2\mu}\,(\ttd\ttdst + \ttdst\ttd) = H \ .
\eqno(1.35)$$
Using (1.8,9,29,33), one sees that 
$$
\ttd =a^*\circ \widetilde{\nabla} = \Asy\circ\widetilde{\nabla}\ ,
\eqno(1.36)$$
where $a^*$ is defined in (1.4) and \Asy\ denotes anti-symmetrization. 
In local coordinates, 
$$
\ttd = a^*(dq^j) \widetilde{\nabla}_j \ .
$$
Since the {\sl torsion} $T(\widetilde{\nabla})$ of a connection 
$\widetilde{\nabla}$ on $\Omega^{\bullet}(M)$ is defined by 
$$
T(\widetilde{\nabla}) = d -  \Asy\circ\widetilde{\nabla} 
\eqno(1.37)$$
where $d$ denotes {\sl exterior differentiation}, we conclude that 
$$
\ttd = d\quad\Longleftrightarrow\quad T(\widetilde{\nabla}) = 0
\quad\Longrightarrow\quad\hbox{relations (1.30) hold} 
$$
and that vice versa eqs.\ (1.30) imply $d=\ttd$  if we additionally 
assume that $\widetilde{\nabla}$ is a metric connection, which 
guarantees that $\d$ and $\bard$ are symmetric operators on 
$\h_{\rm e-p}$. Thus, $\ttd=d$\ is exterior differentiation precisely 
if $\widetilde{\nabla}$ is the Levi-Civita connection 
$\nabla^{\rm L.C.}$ on $\h_{\rm e-p}$. It follows that the quantum 
mechanics of positronium can be formulated on general orientable 
Riemannian manifolds $(M,g)$ which {\sl need not} be spin${}^c$. 
\sn
One easily verifies that, no matter whether the dimension of $M$ is 
even or odd, there always exists a $\Z_2$-grading $\gamma$ on 
$\h_{\rm e-p}$ such that 
$$
\{\,\gamma,\ttd\,\} = \{\,\gamma,\ttdst\,\} = 0\ ,\quad 
\lb\,\gamma,a\,\rb = 0 
\eqno(1.38)$$
for all $a\in\a$. The operator $\gamma$ has eigenvalue $+1$ on
even-degree and $-1$ on odd-degree differential forms. 
\sn
The spectral data $(\a, \h_{\rm e-p}, \ttd, \ttdst, \gamma)$ define an 
example of what physicists call $N=(1,1)$ {\sl supersymmetric quantum 
mechanics}: There are two supercharges \ttd\ and \ttdst\ (or $\d$ 
and $\bard$) satisfying the algebra (1.34,35) (or (1.30), respectively). 
\sn
When $\ttd=d$ (exterior differentiation) the $\Z_2$-grading $\gamma$ 
can be replaced by a {\sl $\Z$-grading}\ $T_{\rm tot}$ counting the 
total degree of differential forms, and one can add to the spectral 
data described so far a {\sl unitary Hodge operator} $*$ such that 
$\lb\,*,a\,\rb=0$ for all $a\in\a$ and $*\ttd = \zeta \ttdst *\,$, where 
$\zeta$ is a phase factor.
\sn
The spectral data 
$$
(\a, \h_{\rm e-p}, \ttd, \ttdst, T_{\rm tot}, *)
\eqno(1.39)$$          are said to define a 
model of $N=2$ {\sl supersymmetric quantum mechanics}. 
\sn
It is important to distinguish $N=(1,1)$ {}from $N=2$ supersymmetry: 
Every $N=2$ supersymmetry is an $N=(1,1)$ supersymmetry, but the 
converse does not usually hold, even in the context of classical 
geometry. This is seen by considering geometry with torsion: Let 
$\nabla^S$ be a spin connection with torsion, i.e.\ the connection 
$\nabla$ on $T^*M$ determined by $\nabla^S$ as in eq.\ (1.17) has 
non-vanishing torsion $T(\nabla)$. We then redefine what we mean by the 
(charge-conjugate) connection $\nabla^{\overline{S}}$, namely (locally) 
$$
\bigl( \nabla^S - \nabla^{\overline{S}}\bigr) \psi = 2iA \otimes \psi 
+ c(T(\nabla)) \otimes \psi 
$$
for all $\psi\in\h_{\rm e}$; in local coordinates, 
$$
c(T(\nabla))\otimes\psi = {1\over2}\,dq^i\otimes 
T_{ijk}c(dq^j)c(dq^k)\psi\ .
$$
The connection $\widetilde{\nabla}$ on $\h_{\rm e-p}$ is defined in 
terms of $\nabla^S$ and $\nabla^{\overline{S}}$ as in (1.28). We assume 
that $d\,T(\nabla)=0$ and introduce two Dirac operators $\d$ and $\bard$ 
by 
$$\eqalignno{
\d &= \Gamma\circ \widetilde{\nabla}- {1\over6}\, \Gamma(T(\nabla)) 
&(1.40)\cr
\noalign{\noindent{\rm and}\hfill} 
\bard &= \overline{\Gamma}\circ \widetilde{\nabla}
+ {1\over6}\, \overline{\Gamma}(T(\nabla)) \ .
&(1.41)\cr}$$
Then the $N=(1,1)$ algebra (1.30) holds, but there is no natural 
$\Z$-grading $T_{\rm tot}$ with the property that 
$\lb\,T_{\rm tot},\ttd\,\rm = \ttd$ and 
$\lb\,T_{\rm tot},\ttdst\,\rb =\ttdst$ for \ttd\ 
and \ttdst\  as in (1.33). One can again derive a Weitzenb\"ock 
formula; it  can be used to show that, in various examples, the 
Hamiltonian $H={\hbar\over2\mu} \d^2$ is strictly positive, in which 
case one says that the supersymmetry is broken \q{Wi1} and that the 
indices of $\d$ and $\bard$ vanish, which has implications for the 
topology of the manifold $M$. The present example is discussed in some 
detail in \q{FG}. 
\sn
One can verify that the de Rham-Hodge theory and the differential 
geometry of a Riemannian manifold $(M,g)$ are completely encoded in the 
$N=2$ set of spectral data $(\a, \h_{\rm e-p}, \ttd, \ttdst, 
T_{\rm tot}, *)$. This theme is developed in great detail in 
section 3.2. \hfill\break
\noindent For purposes of physics (in particular, in analyzing the 
geometry of quantum field theory and string theory), it would be
desirable to replace \a\  by a suitable algebra ${\cal F}_{\hbar}$ of
pseudo-differential operators on $M$. The resulting change in 
perspective will be discussed in future work. 
\mn
Mathematicians not interested in quantum physics may ask what one gains 
by reformulating differential topology and geometry in terms of spectral 
data, such as those provided by $N=1$ (electron) or $N=2$ (positronium)
supersymmetric quantum mechanics, beyond a slick algebraic 
reformulation. The answer -- as emphasized by Connes -- is:\ {\sl 
generality}$\,$! Supersymmetric quantum mechanics enables us to study 
{\sl highly singular spaces} or {\sl discrete objects}, like graphs, 
lattices and aperiodic tilings (see e.g.\ \q{Co1}),  and also  
{\sl non-commutative spaces}, like quantum groups, as {\sl geometric 
spaces}, and to extend standard constructions and tools of algebraic 
topology or of differential geometry to this more general context, so 
as to yield non-trivial results. The general theory is developed in 
section 5. Below, we shall argue that quantum physics actually {\sl 
forces} us to generalize the basic notions and concepts of geometry.  
\mn
The principle that the time evolution of a quantum mechanical system is 
a one-parameter {\sl unitary group} on a {\sl Hilbert} space, whose 
generator is the Hamiltonian of the system (a {\sl self-adjoint} 
operator), entails that the study of supersymmetric quantum mechanics 
is the study of {\sl metric} geometry. Let us ask then how we would 
study manifolds like {\sl symplectic} manifolds that are, a priori, 
{\sl not} endowed with a metric. The example of symplectic manifolds is 
instructive, so we sketch what one does (see also sections 3.6 and 5.5). 
\sn
Let $(M,\omega)$ be a symplectic manifold. The symplectic form $\omega$ 
is a globally defined closed 2-form. It is known that every symplectic 
manifold can be equipped with an almost complex structure $J$ such that 
the tensor $g$ defined by 
$$
g(X,Y) = -\omega(JX,Y) \ ,
\eqno(1.42)$$ 
for all vector fields $X,\,Y$, is a Riemannian metric on $M$. Thus, we 
can study the Riemannian manifold $(M,g)$, with $g$ {}from (1.42), by 
exploring the quantum mechanical propagation of e.g.\ positronium on 
$M$, using the spectral data $(\a, \h_{\rm e-p}, \ttd, \ttdst, 
T_{\rm tot}, *)$ of $N=2$ supersymmetric quantum mechanics, with 
$\a=C^{\infty}(M)$ -- or $\a=C(M)$, etc., depending on the smoothness 
of $M$. We must ask how these data ``know'' that $M$ is symplectic. 
The answer, developed in sections 3.6 and 5.5, is as follows: We can 
view the $\Z$-grading $T_{\rm tot}$ as the generator of a 
U(1)-symmetry (a ``global gauge symmetry'') of the system. It may 
happen that this symmetry can be enlarged to an SU(2)-symmetry, with 
generators $L^1, L^2, L^3$ acting on $\h_{\rm e-p}$ such that they 
commute with all elements of \a\ and have the 
following additional properties:
\smallskip
\item{$i$)} $\quad L^3 = T_{\rm tot} - {n\over2}\quad{\rm with}\quad
  n={\rm dim}\,M$. 
\sn
Defining $L^{\pm}= L^1 \pm i L^2$, the structure equations of
su(2)$\,=\,$Lie(SU(2)) imply that
\smallskip 
\item{$ii$)} $\quad \lb\,L^3,L^{\pm}\,\rb= \pm 2 L^{\pm}\
,\quad\lb\,L^+,L^{-}\,\rb=L^3\ $, 
\sn
and, since in quantum mechanics symmetries are represented unitarily,
\smallskip
\item{$iii$)} $\quad\bigl(L^3\bigr)^*= L^3\ ,\quad\bigl(L^{\pm}
\bigr)^* = L^{\mp}\ .$ 
\sn
We also assume that 
\smallskip
\item{$iv$)} $\quad \lb\,L^+,\ttd\,\rb=0\ $,  
\sn
hence $L^-$ commutes with \ttdst\ by property $iii$). Next we define 
an operator $\tdst$ by 
$$
\tdst = \lb\,L^-,\ttd\,\rb\ ; 
\eqno(1.43)$$
it satisfies $\lb\,L^+,\tdst\,\rb= \ttd$ because of $ii$) and $iii$), 
and also 
$$
\{\,\tdst,\ttd\,\} = 0 
$$
since \ttd\ is nilpotent. Assuming, moreover, that 
\smallskip\item{$v$)} $\quad \lb\,L^-,\tdst\,\rb = 0$ 
\sn 
we find that $\bigl( \ttd, \tdst \bigr)$ transforms as a doublet under 
the adjoint action of $L^3, L^+, L^-$  and that $\tdst$ is 
{\sl nilpotent}. Thus,  $\bigl( \td, -\ttd^* \bigr)$ with $\td = 
(\tdst)^*$ is an SU(2)-doublet, too, and $\td{}^2=0$. 
\sn
The theorem is that the spectral data 
$$
(\a,\ \h_{\rm e-p},\ \ttd,\ \ttd^*,\ \{L^3,L^+,L^-\},\ *) 
\eqno(1.44)$$
with properties $i)-v)$ assumed to be valid, encode the geometry of a 
{\sl symplectic} manifold $(M,\omega)$ equipped with the metric $g$ 
defined in (1.42). The identifications are as follows: 
$$\eqalign{
&L^3 = T_{\rm tot}-{n\over2}\ ,\quad L^+ = \omega\wedge\ = {1\over2} 
\omega_{ij}a^*(dq^i)a^*(dq^j)\ ,\cr
&L^- = (L^+)^*={1\over2}(\omega^{-1})^{ij}a(\partial_i)a(\partial_j)\ .
\cr}$$
Assumption $iv)$ is equivalent to $d\omega=0$. Further details can be 
found in section 3.6. 
\sn
We say that the spectral data (1.44) define $N=4^s$ supersymmetric 
quantum mechanics, because there are four ``supersymmetry generators'' 
$\ttd, \tdst, \td, \ttd^*$; the superscript $s$ stands for 
``symplectic''. 
\sn
Note that we are {\sl not} claiming that 
$$
\{\,\ttd,\td\,\}=0
\eqno(1.45)$$
because this equation does, in general, not hold. However, if it holds 
then $(M,\omega)$ is in fact a {\sl K\"ahler manifold}, 
with the $J$ {}from eq.\ (1.42) as its complex structure and $\omega$ as 
its K\"ahler form. Defining 
$$
\partial = {1\over2}\,(\ttd - i \td)\ ,\quad
\overline{\partial} = {1\over2}\,(\ttd + i \td)\ ,
\eqno(1.46)$$
one finds that, thanks to eqs.\ (1.43, 45) and because $\ttd$ and 
$\td$ are nilpotent, 
$$
\partial^2 = \overline{\partial}{}^2 = 0\ ,\quad 
\{\,\partial,\partial^*\,\}= \{\,\overline{\partial},\overline{\partial}
{}^*\,\} \ .
\eqno(1.47)$$
There is a useful alternative way of saying what it is that identifies 
a symplectic manifold $(M,\omega)$ as a K\"ahler manifold: Eq.\ (1.45) 
is a consequence of the assumption that an $N=4^s$ supersymmetric 
quantum mechanical model has an {\sl additional} U(1)-symmetry -- 
which, in physics jargon, one is tempted to call a ``global chiral 
U(1)-gauge symmetry'': We define 
$$\eqalign{
\ttd_{\theta} &= \ \ \cos\theta\, \ttd + \sin\theta\, \td\ ,\cr
\td_{\theta}  &= -\sin\theta\, \ttd + \cos\theta\, \td\ ,\cr}
\eqno(1.48)$$
and assume that  $\bigl(\ttd_{\theta},\td^*_{\theta} \bigr)$ and  
$\bigl( \td_{\theta}, -\ttd_{\theta}^* \bigr)$ are again SU(2)-doublets 
with the same properties as  $\bigl( \ttd, \tdst \bigr)$ and 
$\bigl(\td, -\ttd^*\bigr)$, for all real angles $\theta$. Then the 
nilpotency of \ttd, \td\ and of $\td_{\theta}$ for all $\theta$ implies 
eq.\ (1.45). Furthermore 
$$
\partial_{\theta} =  {1\over2}\,(\ttd_{\theta} - i \td_{\theta})= 
 e^{i\theta}\, \partial\ ,\quad
\overline{\partial}_{\theta} = {1\over2}\,(\ttd_{\theta} + i 
\td_{\theta})=  e^{-i\theta}\, \overline{\partial}\ .
$$
Assuming that the symmetry (1.48) is implemented by a one-parameter 
unitary group on $\h_{\rm e-p}$ with an infinitesimal generator 
denoted by $J_0$, we find that 
$$
\lb\,J_0, {\tt d}\,\rb = -i\,\td\ ,\quad\quad
\lb\,J_0, \td\,\rb = i\,{\tt d}\ .
\eqno(1.49)$$
Geometrically, $J_0$ can be expressed in terms of the {\sl complex 
structure} $J$ on a K\"ahler manifold -- it is bilinear in $a^*$  
and $a$ with coefficients given by $J$. Defining 
$$ T :=  {1\over2}\,(L^3 + J_0)\ ,\quad\quad\, 
\overline{T} := {1\over2}\,(L^3 - J_0)\ ,
\eqno(1.50)$$
one checks that 
$$\eqalign{
&\lb\,T,\partial\,\rb = \partial\ ,\quad\quad
\lb\,T,\overline{\partial}\,\rb =0\ ,  \cr
&\lb\,\overline{T},\partial\,\rb = 0\ ,\quad\quad
\lb\,\overline{T},\overline{\partial}\,\rb =\overline{\partial}\ .\cr}
\eqno(1.51)$$
Thus $T$ is the holomorphic and $\overline{T}$ the anti-holomorphic 
$\Z$-grading of complex differential forms, 
see sections 3.4, 5.3 and also 5.6.  The spectral data 
$$
(\a,\ \h_{\rm e-p},\ \ttd,\ \ttd{}^*,\ \{\,L^3,L^+,L^-\,\},\ J_0,\ *)
\eqno(1.52)$$
belong to  $N=4^+$ supersymmetric mechanics. We have seen that 
they contain the spectral data
$$
(\a,\ \h_{\rm e-p},\ \partial,\ \partial^*,\ \overline{\partial},\ 
\overline{\partial}{}^*,\ T,\ \overline{T},\ *)
\eqno(1.53)$$
characterizing {\sl K\"ahler manifolds}. We say that these define 
$N=(2,2)$ supersymmetric quantum mechanics. If one drops the 
requirement that $\partial$ anti-commutes with 
$\overline{\partial}{}^*$ (amounting to the {\sl breaking} of the SU(2) 
symmetry generated by $L^3,L^+,L^-\,$) the data (1.53) characterize 
{\sl complex Hermitian manifolds}, see section 3.4.2.   
\sn
Alternatively, complex Hermitian manifolds can be described by $N=2$ 
spectral data as in eq.\ (1.39) with an additional U(1) symmetry 
generated by a self-adjoint operator $J_0$ with the property that 
$\td := i\,\lb\,J_0,\ttd\,\rb$ is nilpotent, and different {}from $\ttd$. 
Then $\td$ and $\ttd$ anti-commute, and one may define $\partial$ and
$\overline{\partial}$ through eqs.\ (1.46). One verifies that 
$$
\partial^2 = \overline{\partial}{}^2 = 0\ \quad{\rm and}\quad\ 
\{\,\partial,\overline{\partial}\,\}=0\ .
$$
\sn
Having proceeded thus far, one might think that on certain K\"ahler 
manifolds with special properties the U(1) symmetries generated by $T$ 
and $\overline{T}$ are embedded into SU(2) symmetries with generators 
$T^3=T,\ T^+,\ T^-$ (analogously for the anti-holomorphic generators) 
which satisfy  properties $i)$ through $v)$ {}from above, with $\ttd$ 
and $\ttd^*$ replaced by $\partial$ and $\partial^*$, and such that 
$\widetilde{\partial}{}^* = \lb\,T^-,\partial\,\rb\,$ -- as well as 
analogous relations for the anti-holomorphic generators.  \hfill\break
Alternatively, one might assume that, besides the SU(2) symmetry 
generated by $L^3, L^+, L^-$ there are actually {\sl two} ``chiral'' 
U(1) symmetries with generators $I_0$ and $J_0$, enlarging the original 
U(1) symmetry. 
\sn
Indeed, this kind of symmetry enhancement can happen, and what one 
finds are spectral data characterizing {\sl Hyperk\"ahler manifolds}, 
see sections 3.5 and 5.4. The two ways of enlarging the 
SU(2)$\,\times\,$U(1) symmetry of K\"ahler manifolds to larger symmetry 
groups characteristic of Hyperk\"ahler manifolds are {\sl equivalent} 
by a theorem of Beauville, see \q{Bes} and section 3.5. The resulting 
spectral data define what is called $N=(4,4)$ supersymmetric quantum 
mechanics, having two sets of four supercharges, 
$\{\,\partial,\,\widetilde{\partial}{}^*,\,
\overline{\partial}{}^*,\,\pabt\,\}$ and $\{\,\widetilde{\partial},\,
\partial^*,\,\pabt{}^*,\,\overline{\partial}\,\}$, with the property 
that each set transforms in the fundamental representation 
of Sp(4) -- see section 4 for the details. \hfill\break 
\noindent This yields the data of $N=8$ supersymmetric quantum 
mechanics -- {}from which we can climb on to $N=(8,8)$ or $N=16$ 
supersymmetric quantum mechanics and enter the realm of very rigid 
geometries of symmetric spaces with special holonomy groups \q{Joy,Bes}. 
\sn
Of course, the operators 
$$
I := \exp(-i\pi\,I_0)\ ,\quad\quad J := \exp(-i\pi\,J_0)\ ,
\quad\quad K := IJ   
\eqno(1.54)$$
in the group of ``chiral symmetries'' of the spectral data of 
$N=(4,4)$ supersymmetric quantum mechanics correspond to the three 
complex structures of Hyperk\"ahler geometry. One may then try to go 
ahead and enlarge these ``chiral'' symmetries by adding further complex 
structures, see e.g.\ \q{WTN} and references therein for some formal 
considerations in this direction. 
\mn
We could now do our journey through the land of geometry and 
supersymmetric quantum mechanics in reverse and pass {}from special 
(rigid) geometries, i.e., supersymmetric quantum mechanics with high 
symmetry, to more general ones by reducing the supersymmetry algebra. 
The passage {}from special to more general geometries then appears in 
the form of {\sl supersymmetry breaking} in supersymmetric quantum 
mechanics (in a way that is apparent {}from our previous discussion). 
The symmetry generators in the formulation of geometry as 
supersymmetric quantum mechanics are bilinear expressions in the 
creation and annihilation operators $a^*$ and $a$ {}from eqs.\ (1.4,5), 
with coefficients that are tensors of rank two. It is quite 
straightforward to find conditions that guarantee that such tensors 
generate symmetries and hence to understand what kind of {\sl 
deformations} of geometry {\sl preserve} or {\sl break} the symmetries. 
Furthermore, the general transformation theory of quantum mechanics 
enables us to describe the deformation theory of the supersymmetry
generators $(D_A;\ \d,\bard;\ {\rm or}\ \ttd, \ttd^*)$ including 
isospectral deformations (as unitary transformations). Deformations of 
\ttd\ and $\ttd^*$ played an important role in Witten's proof of the 
Morse inequalities \q{Wi2} and in exploring geometries involving 
anti-symmetric tensor fields such as torsion -- see the example 
described in eqs.\ (1.40,41) -- which are important in string theory. 
\bn
We hope we have made our main point clear: {\sl Pauli's quantum 
mechanics of the non-relativistic electron and of positronium on a 
general manifold (along with its internal symmetries) neatly encodes 
and classifies all types of differential geometry.} 
\bn\bn
\leftline{\bf 1.3 Supersymmetric quantum field theory and 
geometry put into perspective}
\bn
A non-linear $\sigma$-model is a field theory of maps {}from a parameter 
``space-time'' $\Sigma$ to a target space $M$. Under suitable 
conditions on $M$, a non-linear $\sigma$-model can be extended to 
a supersymmetric theory. (See e.g.\ \q{FG} for a brief introduction 
to this subject.) One tends to imagine that such models can be 
quantized. When $\Sigma$ is the real line $\R$, this is indeed 
possible, and one recovers supersymmetric quantum mechanics in 
the sense explained in the previous subsection. 
When $\Sigma=S^1\times\R$, there is hope that quantization is 
possible, and one obtains an analytic tool to explore the 
infinite-dimensional geometry of loop space $M^{S^1}$. When 
$\Sigma=L\times\R$ with dim$\,L\geq2$, the situation is far less 
clear, but a supersymmetric non-linear $\sigma$-model with 
parameter space-time $L\times\R$ could be used to explore the 
geometry of $M^L$ -- i.e.\ of poorly understood infinite-dimensional 
manifolds. 
\sn
A (quantized) supersymmetric $\sigma$-model with $n$ global 
supersymmetries and with parameter space
$\Sigma=\T^d\times\R$, where $\T^d$ is a $d$-dimensional torus,
formally determines a model of 
supersymmetric quantum mechanics by {\sl dimensional reduction}: 
The supersymmetry algebra of the non-linear $\sigma$-model with 
this $\Sigma$ contains the algebra of infinitesimal translations 
on $\Sigma$; by restricting the theory to the 
Hilbert sub-space that carries the {\sl trivial} representation of 
the group of translations of $\T^d$ one obtains a model of 
supersymmetric quantum mechanics. The supersymmetry algebra can be 
reduced to this ``zero-momentum'' subspace, and the restricted 
algebra is of the form discussed in section 1.2 (see also section 4). 
Starting {}from $\hat n$ supersymmetries and a $d+1\,$-dimensional 
parameter space-time $\Sigma$, one ends up with a model of $N=(n,n)$
supersymmetric quantum mechanics where $n=\hat n$ for $d=1,2$, 
$n=2\hat n$ for $d=3,4$, $n=4\hat n$ for $d=5,6$ and $n=8\hat n$ for 
$d=7,8$ (see e.g.\ \q{Au}). The resulting supersymmetric quantum 
mechanics (when restricted to an even smaller subspace of ``zero 
modes'') is expected to encode the geometry of the target space $M$ 
-- in, roughly speaking, the sense outlined in section 1.2. {}From what 
we have learned there, it follows at once that {\sl target spaces} of 
$\sigma$-models with {\sl many supersymmetries} or {\sl with a 
high-dimensional parameter space-time} must have {\sl very special 
geometries}.  
\sn
This insight is not new. It has been gained in a number of papers, 
starting in the early eighties with work of Alvarez-Gaum\'e and 
Freedman \q{AGF}, see also \q{BSN,Ni,WN}. Supersymmetric quantum 
mechanics is a rather old idea, too, beginning {}from papers by Witten 
\q{Wi1,Wi2} -- which, as is well known, 
had a lot of impact on mathematics. In later works,   
``supersymmetry proofs'' of the index theorem were given 
\q{AG,FW,Ge}. The reader may find a few comments on the history of 
global supersymmetry in chapter 4; for many more details see e.g.\ 
\q{WeB,Wes}.
\mn
Our discussion in section 1.2 has made it 
clear that, in the context of finite-dimensional classical 
manifolds, (globally) supersymmetric quantum mechanics -- as it 
emerges {}from the study of non-relativistic electrons and positrons -- 
is just another name for classical differential topology and 
geometry. Actually, this is a general fact: {\sl Global supersymmetry, 
whether in quantum mechanics or in quantum field theory, is just 
another name for the differential topology and geometry of (certain) 
spaces.}
\mn
In the following, we indicate why globally supersymmetric {\sl quantum 
field theory}, too, is nothing more than geometry of 
infinite-dimensional spaces. However, once we pass {}from supersymmetric 
quantum mechanics to quantum field theory, there are surprises.
\sn
A (quantum) field theory of Bose fields can always be thought of as 
a {\sl $\sigma$-model} (linear or non-linear), i.e., as a theory 
of maps {}from a parameter ``space-time'' $\Sigma=L\times\R$ to a target 
space $M$. At the level of classical field theory, we may attempt to 
render such a model globally supersymmetric, and then to 
quantize it. As we have discussed above, the resulting quantum field 
theory -- if it exists -- provides us with the {\sl spectral data} to 
explore the geometry of what one might conjecture to be some version 
of the formal infinite-dimensional manifold $M^L$. It may happen 
that the quantum field theory exhibits some form of invariance under 
re-parametrizations of parameter space-time $\Sigma$ (though such an 
invariance is often destroyed by {\sl anomalies}, even if present at 
the classical level). However, when $\Sigma=\R$, re-parametrization 
invariance can be imposed; when e.g.\ $\Sigma=S^1\times\R$, it leads 
us to the tree-level formulation of first-quantized string theory. 
Let us be content with ``conformal invariance'' and study 
(supersymmetric) {\sl conformal} $\sigma$-models of maps {}from parameter 
space-time $\Sigma=S^1\times\R$ to a target space $M$ which, for 
concreteness, we choose to be a smooth, compact Riemannian manifold 
without boundary, at the classical level. An example would be $M=G$, 
some compact (simply laced) Lie group. The corresponding field theory 
is the supersymmetric Wess-Zumino-Witten model \q{Wi3}, which is 
surprisingly well understood. Thanks to the theory of Kac-Moody 
algebras and centrally extended loop groups, see e.g.\ \q{GeW,FGK,PS}, 
its quantum theory is under fairly complete mathematical control and 
provides us with the spectral data of some $N=(1,1)$ supersymmetric 
quantum theory (related to the example discussed in eqs.\ (1.42,43) 
with ``broken supersymmetry''; see e.g.\ \q{FG}). We might expect that 
these spectral data encode the geometry of the loop space $G^{S^1}$ 
over $G$. The {\sl surprise} is, though, that they encode the geometry 
of loop space over a {\sl quantum deformation} of $G$ (where the 
deformation parameter depends on the {\sl level} $k$ of the 
Wess-Zumino-Witten model in such a way that, formally, 
$k\rightarrow\infty$ corresponds to the classical limit). We expect 
that this is an example of a {\sl general phenomenon}: Quantized 
supersymmetric $\sigma$-models with parameter space-time $\Sigma$ of 
dimension $d\geq2$ -- assuming that they exist -- tend to provide us 
with the spectral data of a supersymmetric quantum theory which 
encodes the geometry of a ``quantum deformation'' of the target 
space of the underlying classical $\sigma$-model.  
\mn                This is one reason why 
quantum physics forces us to go beyond classical differential 
geometry. A second, more fundamental subject which calls 
for ``quantum geometry'' is the outstanding problem of unifying the 
quantum theory of matter with the theory of gravity within a 
theory of {\sl quantum gravity}. It was argued in \q{DFR} and \q{F}, 
and undoubtedly in other places before, 
that in such a theory space-time coordinates should no longer 
commute with each other. In \q{DFR}, the main question was how 
quantum field theory on a ``non-commutative Minkowski space-time'' 
can be formulated technically, in particular how to construct 
representations of the basic commutation relations, while the 
uncertainty relations for the coordinates were assumed as an input. 
In the second reference, the emphasis was put on showing that it is 
the presence of quantum mechanical matter in physical space-time 
which leads to those uncertainties and, therefore, to a 
non-commutative structure of the underlying space-time. The main idea 
is to combine the Heisenberg uncertainty relations with the horizon 
radius of a black hole, which leads to lower bounds on the size of an 
observable event -- irrespective of whether black holes are really 
black or radiate. Since the final theory of quantum gravity is 
unknown as yet, the arguments given in \q{DFR,F} are necessarily 
heuristic. Nevertheless, they strongly indicate that in order 
to formulate a fundamental theory of ``space-time-matter'' it is 
necessary to go beyond classical geometry. These remarks motivate 
the attempts presented in section 5. 
\vfill
\sn{\bf Acknowledgements}
\bn 
We thank  A.\ Alekseev,  O.\ Augenstein, J.-B.\ Bourgignon, 
A.\ Chamseddine, G.\ Felder, C.\ Voisin and M.\ Walze for 
interesting discussions, comments and collaborations. We are 
particularly grateful to A.\ Connes and K.\ \gaw\ for their 
invaluable help  and inspiration and to A.\ Connes for advance 
copies of some of his works. J.F.\ thanks the I.H.\'E.S.\ for 
hospitality during various stages of work presented in this paper.  
\eject
\pageno=17  
\leftline{\bf 2. Canonical anti-commutation relations, Clifford algebras, 
and Spin\CC-groups}
\bn 
In this section, we review some standard material
about Clifford algebras and the groups Spin$^c$ which
will be needed later on. In subsection 2.2, we follow 
\q{Sa}, where the reader may 
find detailed proofs and further references. 
\bn\bn
\leftline{\bf 2.1 Clifford algebras}
\bn 
Let ${\cal F}_k$ be the unital complex algebra generated 
by elements $b^i$ and their adjoints $b^{i*}$, 
$i=1,\ldots,k$, satisfying the 
{\sl canonical anti-commutation relations} (CAR)
$$
\{ b^i,b^j \} = \{ b^{i*},b^{j*} \}=0\ , 
\ \ \{ b^i,b^{j*} \} = \delta^{ij}\ . 
\eqno(2.1)$$
Note that here we regard the $b^i$ as abstract generators
of a ${}^*$-algebra; we can alternatively identify them  
with basis elements of a real Euclidian vector space. Then 
the Kronecker symbol $\delta^{ij}$ would have to be replaced 
with the inner product of $b^i$ and $b^j$. A representation 
$$
\pi\,:\ {\cal F}_k \lra \End(W)
$$
of ${\cal F}_k$ on a Hermitian vector space $W$ is 
called unitary if $\pi(b^{i*})=\pi(b^i)^*$. 
In fact, the algebra ${\cal F}_k$ has a unique 
irreducible unitary representation which is 
explicitly realized on $(\C^2)^{\otimes k}$ by
setting
$$\eqalignno{
b^j &= \tau_3\otimes\ldots\otimes\tau_3\otimes
  \tau_-\otimes\one\otimes\ldots\otimes\one
&(2.2)\cr
b^{j*} &= \tau_3\otimes\ldots\otimes\tau_3\otimes
  \tau_+\otimes\one\otimes\ldots\otimes\one
&(2.3)\cr}$$
with $\tau_{\pm}:= {1\over2}\,(\tau_1\pm i\tau_2)$  
in the $j\,$th position; $\tau_a$ denote the 
Pauli matrices 
$$ 
\tau_1=\pmatrix{0&1\cr 1&0\cr}\ ,\ \ 
\tau_2=\pmatrix{0&-i\cr i&0\cr}\ ,\ \ 
\tau_3=\pmatrix{1&0\cr 0&-1\cr}\ . 
$$
This representation is faithful, and therefore
${\cal F}_k$ is isomorphic to the full matrix 
algebra $M(2^k,\C)$. 
\sn
Let $V$ be a {\sl real} $n$-dimensional vector space 
with scalar product $(\cdot,\cdot)$. The 
{\sl complexified Clifford algebra} $Cl(V)$ 
associated to $V$ is the unital complex algebra 
generated by vectors $v,w\in V$ 
subject to the
relations 
$$
\{ v,w \} = -2(v,w)   \ .
\eqno(2.4)
$$ 
If $e^1,\ldots,e^n$ is an orthonormal basis of 
$V$, the relations (2.4) are equivalent to 
$$
\{ e^i,e^j \} = -2\delta^{ij} \ .
\eqno(2.5)$$
A ${}^*$-operation in $Cl(V)$ is induced by anti-linear 
and involutive extension of $v^*:= -v$ for all 
$v\in V$, and, accordingly, a representation 
$$
\pi\,:\ Cl(V) \lra W
$$
of $Cl(V)$ on a Hermitian vector space $W$ is 
called unitary if $\pi(v)^*=-\pi(v)$, for all
$v\in V$. 
\sn
Let $n=2k+p$ where $p=0,1$ is the parity of
$n$. For $j=1,\ldots,k$, consider the following
generators of the Clifford algebra:
$$\eqalignno{
b^j &= {1\over2}\bigl(e^{2j-1}- i e^{2j}\bigr)
&(2.6)\cr  
b^{j*} &= -{1\over2}\bigl(e^{2j-1}+ i e^{2j}\bigr)
&(2.7)\cr}$$
and, for $p=1$, 
$$
\nu = i^{k+1} e^1e^2\cdots e^n \ .
\eqno(2.8)$$  
It is easy to check that the elements $b^j,b^{j*}$ 
satisfy the CAR, and therefore, the Clifford 
algebra is isomorphic to the full matrix algebra
$M(2^k,\C)$ for $p=0$. If $p=1$, the 
relations
$$
\lb\, b^j,\nu \,\rb = \lb\, b^{j*},\nu \,\rb = 0\ ,
\ \ \nu^2=1
\eqno(2.9)$$
show that $Cl(V)$ decomposes into a direct sum 
$$
Cl(V) =  {1\over2}(1-\nu)\cdot Cl(V)
\oplus {1\over2}(1+\nu)\cdot Cl(V)
\eqno(2.10)$$
and each factor is again isomorphic to $M(2^k,\C)$.
Thus, for an odd-dimensional vector space $V$, 
the Clifford algebra $Cl(V)$ has two unitary 
irreducible representations, related to each other 
by space reflection. Note also that the trace on 
$Cl(V)$, inherited {}from the isomorphism to the 
matrix algebra, after appropriate normalization 
defines a scalar product on $Cl(V)$, 
$$
(a,b) = 2^{-\lb{n+1\over2}\rb}{\rm tr}\,(a^*b)\ ,
$$
$a,b\in CL(V)$, which extends the scalar product on $V$. 
\bn\bn
\leftline{\bf 2.2 The group Spin\CCVV}
\bn
We now assume that the real vector space $V$ is oriented,
and as before we denote by $x\mapsto x^*$, $x\in Cl(V)$
the unique anti-linear anti-automorphism obtained by 
extending $v\mapsto -v$, $v\in V$. Then we can 
define the group \Spv\ as 
$$
\Spv :=  \{ x\in Cl_{\R}^{{\rm ev}}(V)\,|\,
xx^* =1\,,\ xVx^* \subset V  \}
\eqno(2.11)$$
where $Cl_{\R}^{{\rm ev}}(V)$ denotes the real
sub-algebra of $Cl(V)$ generated by products of 
an even number of elements of $V$. We will 
sometimes use the notation\ \Spn\  as an 
abbreviation for\ \Sprn. The group\ \Spcv\ 
is then given by 
$$
\Spcv := \{ e^{i\alpha}x\,|\, 
\alpha\in\R\,,\ x\in\Spv\} \ .
\eqno(2.12)$$  
For each $x\in\Spcv$, let 
$$
Ad(x)v := xvx^*
$$
denote the adjoint action on $v\in V$. 
There is a short exact sequence -- see \q{Sa} --
$$
1\lra {\rm U}(1) \lra \Spcv\, {\buildrel Ad\over\lra}\,
{\rm SO}(V) \lra 1  \ .
\eqno(2.13)$$
\mn
{\bf Definition 2.1} \q{Sa}\quad A {\sl spin$^c$-structure} 
on $V$ is a pair $(W,\Gamma)$ of a $2^k$-dimensional 
Hermitian vector space $W$ and a linear map 
$\Gamma: V\lra\End(W)$ such that 
$$\eqalignno{
&\Gamma(v)^* + \Gamma(v) = 0\ , 
&(2.14)\cr
&\Gamma(v)^*\Gamma(v) = (v,v)\cdot 1 \ .
&(2.15)\cr}$$
A {\sl spin$^c$-isomorphism} {}from $(V_0,W_0,\Gamma_0)$
to $(V_1,W_1,\Gamma_1)$ is a pair $(A,\Phi)$ where 
$A$ is an orientation preserving orthogonal map {}from 
$V_0$ to $V_1$ and $\Phi$ is a unitary operator {}from
$W_0$ to $W_1$ intertwining $\Gamma_0$ and $\Gamma_1$
``up to $A$'', more precisely
$$
\Phi \Gamma_0(v) \Phi^{-1} = \Gamma_1(Av)
\eqno(2.16)$$
for all $v\in V_0$. The set of all such 
spin$^c$-isomorphisms {}from $W_0$ to $W_1$ is denoted by 
Hom$^{{\rm spin}^c}(W_0,W_1)$. For $W_0=W_1$ this is 
a group; in fact: 
\mn
{\bf Proposition 2.2} \q{Sa}\quad Let $(W,\Gamma)$ be a 
spin$^c$-structure on $V$; then the map 
$$\Xi_{\Gamma}: \cases {&$\!\!\Spcv \lra\ {\rm Hom}^{{\rm spin}^c}(W,W)$\cr
           &$\vphantom{\sum}\quad\quad x\quad\longmapsto\
(Ad(x),\Gamma(x))$\cr}\ , 
$$
where $\Gamma$ now denotes the extension to $\Spcv$\ of $\Gamma$ in 
Definition 2.1, is an isomorphism.
\mn
Finally, we describe the Lie algebra, \spcv, of \Spcv. {}From 
the definitions of \Spv\  and \Spcv, it follows immediately
that
$$
\spv = \{ \xi\in Cl^{\rm ev}(V)\,|\, \xi+\xi^*=0, 
    \lb\,\xi,V\,\rb \subset V \} 
$$
and 
$$
\spcv = \spv \oplus {\rm u}(1) \ .
\eqno(2.17)$$  
Furthermore, a direct computation proves that 
$$
\spv = Cl_2(V) := \bigl\{ \sum_{i,j}a_{ij}e^ie^j\,|\, a_{ij} =
-a_{ji} \in\R \bigr\}  
\eqno(2.18)$$
as expected: Spin$(V)$ is locally isomorphic to SO$(V)$. 
\sn
For the description of connections on Spin$^c$-manifolds in 
subsection 3.2.2, 
we will need some further properties of spin$^c$-isomorphisms. 
Let $(A_t,\Phi_t)$, $t\in\lb-1,1\rb$, be a smooth 1-parameter
family in Hom$^{{\rm spin}^c}(W,W)$ with $(A_0,\Phi_0)=(\one,\one)$,
and set $dA= {d\over dt}A_t|_{t=0}$ and 
$d\Phi= {d\over dt}\Phi_t|_{t=0}$. Then, differentiating 
(2.16) at $t=0$ yields 
$$
\lb\, d\Phi,\Gamma(e^i)\,\rb = \Gamma(dA\,e^i) = 
dA^i_{\phantom{i}j}\Gamma(e^j)\ .
\eqno(2.19)$$
Since $d\Phi$ is an element of the Lie algebra spin$^c(V)$ 
acting on $W$, it is of the form -- see (2.17,18) --
$$
d\Phi = \sum_{i,j}a_{ij}\Gamma(e^i)\Gamma(e^j)
+ i\,\delta A
\eqno(2.20)$$
for some anti-symmetric matrix $a_{ij}$ and some
$\delta A\in\R$. It follows that 
$$
\lb\, d\Phi,\Gamma(e^i)\,\rb = 4a_{ij}\Gamma(e^j)\ ,
$$
and with (2.19) and (2.20) we get 
$$
d\Phi = {1\over4} dA^i_{\phantom{i}j}\Gamma(e^i)
\Gamma(e^j)+ {i}\delta A\ .
\eqno(2.21)$$
Let $e^{i\alpha_t}x_t$, $x_t\in\,$Spin$(V)$, be 
the path in Spin$^c(V)$ corresponding to 
$(A_t,\Phi_t)$ by the isomorphism of Proposition
2.2. Then the scalar term in (2.21) is given by
$$
 \delta A = {d\over dt}\alpha_t|_{t=0}\ .
\eqno(2.22)$$
\bn\bn
\leftline{\bf 2.3 Exterior algebra and spin 
representations}
\bn
Let $V^{\C}$ be the complexification of $V$, 
with a $\C$-bilinear inner product induced 
by the scalar product on $V$,  and let 
$\bigwedge{}^{\bullet}V^{\C}$ be the associated
exterior algebra. On the latter space, there 
exists a canonical scalar product with respect 
to which the set of elements 
$$
e^{i_1}\wedge \ldots \wedge e^{i_{\nu}}\ ,\quad 
1\leq i_1< \ldots < i_{\nu}\leq n 
$$
is an orthonormal basis. For each $v\in V$, 
we define 
$$
a^*(v)\,:\ \bigwedge{}^{\bullet}V^{\C}\lra
 \bigwedge{}^{\bullet}V^{\C}\ ,\quad
\omega\longmapsto a^*(v)\omega := v\wedge\omega\ ,
$$
and we denote by $a(v)$ the adjoint of $a^*(v)$. 
This operator $a(v)= v\,\llcorner$ is contraction 
with $v$, more explicitly  
$$
a(v)\,e^{i_1}\wedge \ldots \wedge e^{i_{\nu}} = 
\sum_{\alpha=1}^{\nu} (-1)^{\alpha+1} 
(e^{i_{\alpha}},v) e^{i_1}\wedge \ldots 
\wedge \widehat{e^{i_{\alpha}}}\wedge\ldots 
\wedge e^{i_{\nu}}\ ,
\eqno(2.23)$$
where $\ \widehat{\cdot}\ $ denotes omission of the term. 
$a^*(v)$ and $a(v)$ are extended to $v\in V^{\C}$ by
$\C$-linearity; note that, for $w\in V^{\C}$, 
taking adjoints involves complex conjugation, i.e.,  
$(a^*(w))^* = a(\bar w)$. 
The operators $a^{i*}:= a^*(e^i)$ and $a^i:=a(e^i)$
satisfy the CAR (2.1), and thus 
$$
\Gamma(v) := a^*(v)-a(v)\ ,\quad 
\overline{\Gamma}(v) := i(a^*(v)+a(v)) 
\eqno(2.24)$$
define two anti-commuting unitary representations 
of the Clifford algebra on $\bigwedge{}^{\bullet}V^{\C}$. 
We study the even and the odd dimensional cases 
separately:
\sn
Let $n={\rm dim}V$ be even ($p=0$). We introduce the 
element 
$$\gamma = i^k \Gamma(e_1)\Gamma(e_2)\ldots
   \Gamma(e_n) 
\eqno(2.25)$$
which anti-commutes with all $\Gamma(v)$ 
and satisfies $\gamma^2=1$. Since the Clifford 
algebra has only one irreducible representation,
say $c$ on $S$, we can write
$$
\bigwedge^{\bullet}V^\C\cong S\otimes \overline{S}\ ,
\quad \Gamma(v) \cong c(v)\otimes \one
\eqno(2.26)$$
where, at this point, $\overline{S}$ is just some 
multiplicity space. Counting dimensions of course 
gives $\overline{S}\cong S$ as vector spaces. 
Furthermore, the equation $\lb\,\Gamma(v),\gamma
\overline{\Gamma}(w)\,\rb=0$ for all $v,w\in V$ 
implies 
$$
\overline{\Gamma}(v) = \gamma\otimes\overline{c}(v)
\eqno(2.27)$$
so $\overline{S}$ is also isomorphic to the unique
irreducible representation of $Cl(V)$. (The $\gamma$ 
in the first tensor factor of (2.27) ensures  
that the two Clifford actions anti-commute.) 
\sn
Let $n={\rm dim}V$ be odd ($p=1$). We now define 
$$
\gamma = i^{k+1} \Gamma(e_1)\Gamma(e_2)\ldots
   \Gamma(e_n) = \Gamma(\nu)\ ,
\eqno(2.28)$$
see (2.8); this element commutes with all 
$\Gamma(v)$ and satisfies $\gamma^2=1$. The 
Clifford algebra now has two irreducible 
representations corresponding to $\gamma=\pm1$,
see (2.10), and one can show that both 
eigenvalues of $\gamma$ come with multiplicity
$2^{n-1}$. Thus we have
$$
\bigwedge{}^{\bullet}V^\C\cong S\otimes\C^2\otimes\overline{S}\ .
\eqno(2.29)$$
Explicit expressions for $\Gamma(v)$ and $\overline{\Gamma}(v)$
are given by 
$$
\Gamma(v) = c(v)\otimes\tau_3\otimes\one\ ,\quad
\overline{\Gamma}(v) = \one\otimes\tau_1\otimes
\overline{c}(v)\ .
\eqno(2.30)$$
Note that eq.\ (2.30) implies anti-commutativity of $\Gamma$
and $\overline{\Gamma}$, as well as $\{\overline{\Gamma}(v),
\gamma\}=0$. Any other expression for $\Gamma$ and 
$\overline{\Gamma}$ is related to (2.30) by a unitary 
conjugation. 
\bn\bn
\vfill\eject
\leftline{\bf 3. Algebraic formulation of differential 
   geometry}
\bn 
Historically, the first step towards an algebraic 
formulation of geometry was Gelfand's Theorem 
showing that every unital abelian $C^*$-algebra
\a\ is isomorphic to the algebra $C(X)$ of complex-valued
continuous functions over some compact Hausdorff space. The 
points of $X$ can be identified with the prime ideals of $C(X)$. 
\sn
Swan's Lemma, stating that every finite dimensional vector 
bundle $E$ over a compact topological space  $X$ can equivalently 
be viewed as a finitely generated projective module
over $C(X)$, provided a further link between the theory 
of operator algebras and topology of spaces.
\sn
Connes' work made it possible to treat geometrical 
aspects by means of operator algebraic techniques and, 
more importantly, led to proposals of how to describe  (compact) 
non-commutative spaces and their geometry by means
of (unital) non-commutative \hbox{*-algebras}. 
\sn
In this chapter, we shall give algebraic formulations 
of classical Riemannian, symplectic, Hermitian, K\"ahler, 
and Hyper-K\"ahler geometry which are tailor-made for 
generalization to the non-commutative setting. 
Compact Lie groups are discussed as a special example 
in section 3.3.
\bn\bn
\leftline{\bf 3.1 The \Neqone formulation of 
Riemannian geometry}
\bn
In this section, we explain how to encode a Riemannian
manifold into what will be called $N=1$ spectral data.
The special case of spin manifolds, which is more
familiar to physicists, will be mentioned in section 3.1.2. 
Following Connes \q{Co1}, sections 3.1.3 - 6 describe 
how to reconstruct the manifold {}from the spectral data. 
In the last section, we will also show that the Riemannian
aspects of the manifold may be recovered. 
\bn\bn
\leftline{\bf 3.1.1 The \Neqone spectral data associated 
to a Riemannian manifold}
\bn
Let $M$ be a smooth compact real manifold of dimension $n$
with Riemannian metric $g$. (Throughout this paper, we 
restrict ourselves to compact manifolds, even when not 
mentioned explicitly.) The complexified cotangent bundle 
$\tmc$ carries a symmetric $\C$-bilinear form induced by $g$,
so we can write $Cl_p(M)$ as a shorthand for the 
{\sl Clifford algebra} $Cl(T_p^*M) = Cl_{\R}(T_p^*M)\otimes\C$ 
at $p\in M$, compare 
section 2.1, where now the Riemannian metric enters the
relations: 
$$
\lbrace\, \omega,\eta \,\} := \omega\eta+\eta\omega = 
  -2g(\omega,\eta)
\eqno(3.1)$$
The {\sl Clifford bundle} $Cl(M)$ is the bundle whose fiber at $p$ 
is $Cl_p(M)$. The space of smooth sections $\Gamma(Cl(M))$ 
carries an algebra structure given by pointwise Clifford 
multiplication. The Levi-Civita connection 
$\nabla$ on $\tmc$ induces a connection on $Cl(M)$, also 
denoted by $\nabla$, such that 
$$
\nabla_X(\sigma_1\sigma_2) = (\nabla_X\sigma_1)\sigma_2 
  + \sigma_1\nabla_X\sigma_2 
\eqno(3.2)$$ 
for all $\sigma_1, \sigma_2 \in \Gamma(Cl(M))$ and $X\in \Gamma(TM)$.  
\hfill\break
Let $C^{\infty}(M)$ denote the algebra of smooth complex valued
functions on $M$. A complex vector bundle $S$ over $M$ with 
Hermitian structure $\langle\cdot,\cdot\rangle_S$ is 
said to carry a {\sl unitary Clifford action} if there is 
a $C^{\infty}(M)$-linear map
$$
c: \Gamma(Cl(M)) \lra \Gamma(\End S)
$$
such that 
$$
\{\,c(\sigma),c(\sigma')\,\}=-2g(\sigma,\sigma') 
\cdot {\rm id}_S
$$ 
and
$$
\langle c(\sigma)s_1,s_2\rangle_S + 
\langle s_1,c(\sigma)s_2\rangle_S =0
\eqno(3.3)$$
for all $s_1,s_2 \in \Gamma(S)$ and all $\sigma,\sigma'\in\Gamma(T^*M)$. 
Then, a connection $\nabla^S$ on $S$ is called a 
{\sl Clifford connection} iff it is Hermitian and 
$$
\nabla^S(c(\sigma)s) = c(\nabla\sigma)s + c(\sigma)\nabla^S s
\eqno(3.4)$$
for all $\sigma\in\Gamma(Cl(M)), s\in\Gamma(S)$, where $\nabla$ 
denotes the Levi-Civita connection. 
\mn
{\bf Definition 3.1}\quad A {\sl Dirac bundle} is a Hermitian
vector bundle $S$ together with a unitary Clifford action and
a Clifford connection. 
\mn
If $S$ is a Dirac bundle then the {\sl Dirac operator} on $\Gamma(S)$ 
is defined by the composition of maps 
$$
D: \Gamma(S)\, {\buildrel \nabla^S \over \lra}\, \Gamma(\tmc\otimes S)
\,{\buildrel c \over \lra}\, \Gamma(S)   \ .
\eqno(3.5)$$
With local coordinates $x^{\mu}$ on $M$, and setting   
$$
\gamma^{\mu} := c(dx^{\mu}) \ ,
\eqno(3.6)$$ 
the Dirac operator reads as usual 
$$
D = \gamma^{\mu} \nabla^S_{\mu}\ .
\eqno(3.7)$$
Since the manifold $M$ is Riemannian it carries a canonical 
1-density locally given by $\sqrt{g}d^nx$, and functions
can be integrated over $M$ with this 1-density. Therefore, 
the space $\Gamma(S)$ is supplied with a canonical scalar
product 
$$
(s_1,s_2) = \int_M \langle s_1,s_2 \rangle_S \sqrt{g}d^nx\ ,
\eqno(3.8)$$
and we denote the Hilbert space completion of $\Gamma(S)$
with respect to this scalar product by $L^2(S)$. The 
Dirac operator on $\Gamma(S)$ extends to a self-adjoint 
operator on $L^2(S)$ \q{Su,LaM}. 
\mn  
{\bf Proposition 3.2}\quad On a Riemannian manifold $(M,g)$,
the complexified bundle of differential forms $\bigwedge^{\bullet}M^{\C} 
:= \bigwedge^{\bullet}T^*M \otimes\C$
carries a canonical Dirac bundle structure. 
\sn
{\kap Sketch of proof}: Let $p\in M$ and $\{\theta^i\}_{i=1,\ldots,n}$
be an orthonormal basis of $T^*_pM$. Then by declaring
the set $\{\,\theta^{i_1}\wedge\ldots\wedge\theta^{i_k} 
| 1\leq k \leq n\,;\ i_1<i_2< \ldots i_k \,\}$ to be an 
orthonormal basis of $\bigwedge{}^{\bullet}T^*_pM^{\C}$, we obtain 
a scalar product that does not depend on the choice of the
basis $\{\theta^i\}$, and this defines a Hermitian 
structure on $\bigwedge{}^{\bullet}M^{\C}$. For any $\omega\in\Gamma(T^*M)$,
let $\epsilon(\omega)$ be the wedging operator by $\omega$, 
and $\iota(\omega)$ the contraction operator by $\omega$, see (2.23), 
$$\eqalign{
\epsilon(\omega)\, \eta &:= \omega \wedge \eta , 
\cr
\iota(\omega)\,\eta    &:= \omega\,\llcorner \eta   \ , 
\cr}\eqno(3.9)$$
for all $\eta\in\Gamma(\bigwedge{}^{\bullet}M^{\C})$.
Then the map 
$$
c\,:\ \cases{ &$\Gamma(T^*M) \lra \Gamma(\End\bigwedge{}^{\bullet}M^{\C})$
\cr
&$\quad\quad\quad\omega \longmapsto c(\omega) := 
\epsilon(\omega)-\iota(\omega)$
\cr}\eqno(3.10)$$
extends to a unitary Clifford action on $\bigwedge{}^{\bullet}M^{\C}$
and the Levi-Civita connection on $\bigwedge{}^{\bullet}M^{\C}$ is a
Clifford connection.    \hfill\qed
\mn
Notice that the canonical Dirac operator on 
$\bigwedge{}^{\bullet}M^{\C}$ is given by the signature 
operator $D=d+d^*$. 
\mn
The proposition shows that Dirac bundles always exist; thus,
given a compact Riemannian manifold $M$ -- which need not be 
spin --, we can choose a Dirac bundle $S$ over $M$ and obtain 
what we will call an {\sl $N=1$ spectral 
triple} $(C^{\infty}(M), L^2(S), D)$. In the following 
sections we shall prove that the Riemannian manifold $M$
can be recovered {}from the spectral triple associated to 
any Dirac bundle. 
\sn
For detailed proofs and a survey on applications of
the theory of Dirac operators see \q{BGV, LaM}.    
\bn\bn
\leftline{\bf 3.1.2 The special case of spin manifolds}
\bn
Let $M$ be a spin manifold, i.e.\  the first and the second 
Stiefel-Whitney class of $M$ vanish. Then, in particular, $M$ is  
oriented, and there is a canonical volume form determined
by the Riemannian metric, which allows to define integration
(of functions) over $M$. Since $M$ is spin, it furthermore carries
a Hermitian bundle $S$ of irreducible representations of the
Clifford algebra, i.e.\ of Dirac spinors. We associate to 
the manifold $M$ the $N=1$ spectral triple 
$(C^{\infty}(M), L^2(S), D)$, where $C^{\infty}(M)$ is 
the algebra of smooth complex valued functions on $M$, 
$L^2(S)$ is the Hilbert space of square integrable spinors, 
and $D$ is {\sl the} Dirac operator associated to the 
spin-connection on $S$. The same data can be used if the manifold 
is only spin$^c$, i.e., on $M$ one can define spinors coupled to 
a U(1) gauge field. (The Dirac operator $D_A$ then depends on 
this gauge field $A$, but we will see later that $D_A$  
enters e.g.\ the differential forms only through commutators, 
therefore $A$ drops out again.)  In cohomological terms, 
a manifold is  spin$^c$ if its first Stiefel-Whitney class 
vanishes, and its second one is the mod-2 reduction of an integral
class. One can show that, in particular, every smooth compact 
oriented 4-manifold has a spin$^c$-structure, see e.g.\ \q{Sa}. 
\sn
For the remainder of section 3.1, we return to the case 
of a general Dirac operator on a Riemannian manifold $(M,g)$.
\bn\bn
\leftline{\bf 3.1.3 The manifold \MM as a metric space}
\bn
The algebra $C^{\infty}(M)$ is a ${}^*$-algebra with involution
given by complex conjugation, and it acts on the Hilbert 
space $L^2(S)$ by bounded (multiplication) operators. Thus, 
$C^{\infty}(M)$ is equipped with a $C^*$-norm, and taking 
its $C^*$-closure yields the algebra of continuous functions 
on $M$. Thus, with the help of Gelfand's Theorem, we can
reconstruct $M$ as a compact 
Hausdorff space {}from the spectral triple. 
\sn
Beyond that, we can recover the geodesic distance on $M$. To 
this end, note that, in local coordinates,  a Dirac operator can be 
written as 
$$
D= \gamma^{\mu}(\partial_{\mu}+\omega_{\mu})
\eqno(3.11)$$
where the $\omega_{\mu}$ are local sections of $\End S$.
It follows immediately that for any function $f\in C^{\infty}(M)$
we have 
$$
\lb\, D,f \,\rb = \gamma^{\mu}\partial_{\mu}f = c(df)\ ,
\eqno(3.12)$$
i.e.\ the commutator $\lb\, D,f \,\rb$ acts on $L^2(S)$ simply
as Clifford multiplication by $df$. 
\mn
{\bf Proposition 3.3} \q{Co1}\quad The geodesic distance 
between two points $p,q\in M$ is given by 
$$
d(p,q) = \sup_f \bigl\{\, |f(p)-f(q)|\ :\ \Vert\, \lb\, D,f \,\rb\,
\Vert_{L^2(S)}  \leq 1  \,\bigr\}  \ .
$$
\sn
{\kap Sketch of proof}: Using the Clifford algebra relations (3.1) 
and the unitarity of the Clifford action (3.3), one verifies
that 
$$
 \Vert\, \lb\, D,f \,\rb\,
\Vert^2_{L^2(S)}\ =\  \Vert c(df) \Vert^2_{L^2(S)}\  
=\ \sup_{p\in M} g(df,df)(p)\ =\ \Vert f \Vert^2_{{\rm Lip}}
$$
where the Lipschitz norm $\Vert\cdot\Vert_{{\rm Lip}}$
is defined by 
$$
\Vert f \Vert_{{\rm Lip}} = \sup_{p,q\in M}
{|f(p)-f(q)|\over d(p,q)}\ . 
$$
This proves that the right hand side in the proposition 
is bounded by the geodesic distance. 
Conversely, it is easy to verify that 
the function $f(x)=d(p,x)$ realizes the supremum. \hfill\qed
\mn
We would like to mention that the geodesic distance of Proposition 
3.3 can also be recovered {}from the algebra $C^{\infty}(M)$ and the 
Laplace operator $\triangle =\triangle_g$ associated with a Riemannian
metric $g$ on $M$, see \q{FG}. More precisely, we have 
$$
d(p,q) = \sup_f \bigl\{\,|f(p)-f(q)|\ :\ \Vert\,{1\over2}\,( \triangle f^2
+ f^2 \triangle) - f \triangle f\, \Vert_{L^2(S)}  \leq 1 \, \bigr\}  \ .
$$ 
In this way, a great deal of classical differential geometry can also be 
extracted {}from a triple $(C^{\infty}(M), L^2\bigl(\Lambda^{\bullet}M^{\C}\bigr), 
\triangle)$, the technicalities are, however, much more involved than 
with the Dirac operator. 
\bn\bn
\leftline{\bf 3.1.4 Reconstruction of differential forms} 
\bn
We begin with some preliminary remarks on the structure 
of the Clifford bundle and its relation to differential forms.
We define
$$
\Gamma(Cl^{(k)}(M)) := {\rm span}\{ \omega_1\omega_2\cdots
\omega_k \,\vert\, \omega_i \in \Gamma(\tmc) \}\ ,
\eqno(3.13)$$
where the product is given by pointwise Clifford
multiplication. Using the defining relations (3.1) we obtain
the inclusions
$$
\Gamma(Cl^{(k)}(M))\supset\Gamma(Cl^{(k-2)}(M))
\supset\Gamma(Cl^{(k-4)}(M))\supset \ldots \ .
\eqno(3.14)$$
The {\sl symbol map} $\sigma_k : \Gamma(Cl^{(k)}(M)) \lra
\Gamma(\bigwedge{}^kM^{\C})$ is the $\C$-linear map 
$$
\sigma_k(\omega_1\cdots \omega_k) := \omega_1\wedge\ldots
\wedge \omega_k\ ,
\eqno(3.15)$$
and its kernel is given by 
$$
{\rm ker}\,\sigma_k = \Gamma(Cl^{(k-2)}(M)) 
\eqno(3.16)$$
so that there is an isomorphism 
$$
\Gamma(\bigwedge{}^kM^{\C}) \cong \Gamma(Cl^{(k)}(M)) / 
\Gamma(Cl^{(k-2)}(M))\ .
\eqno(3.17)$$
With this result we shall reconstruct the space of differential
forms on $M$ {}from the spectral triple $(\a, L^2(S), D)$, where
we have set $\a:=C^{\infty}(M)$ to simplify notations.
\mn
We first introduce an abstract object, the {\sl algebra
of universal forms} $\Omega^{\bullet}(\a)$, see \q{CoK}, 
which also plays an important role in the non-commutative setting: 
$\Omega^{\bullet}(\a)$ is the unital complex algebra generated 
by elements $f$ and $\delta f$, $f\in\a$, subject to the 
relations 
$$\eqalignno{
\delta(\lambda f+g) &= \lambda \delta f + \delta g 
&(3.18)\cr
\delta(fg) &= (\delta f)g + f\delta g
\quad{\rm (Leibniz\ rule)}  
&(3.19)\cr}$$
where $\lambda\in\C$ and $f,g\in\a$. Note that even in the context 
of classical manifold theory, $\Omega^{\bullet}(\a)$ is {\sl not} 
(graded) commutative. 
\sn
Using the Leibniz rule, one easily sees that $\Omega^{\bullet}(\a)$
is $\Z$-graded, 
$$\eqalignno{
\Omega^{\bullet}(\a) &= \bigoplus_{k=0}^{\infty}\Omega^{k}(\a)\ ,
&(3.20)\cr
\Omega^{k}(\a) &= \bigl\lbrace\, \sum_{i=1}^N f_0^i \delta f_1^i \ldots
\delta f_k^i\ \vert N\in\N\,, f_j^i \in \a \,\bigr\rbrace \ .
&(3.21)\cr}$$
This algebra is also endowed with an involution given by
$$
f^* = \bar f \ ,\quad (\delta f)^* = - \delta \bar f 
\eqno(3.22)$$
for each $f\in\a$. 
\sn
The ``symbol'' $\delta$ can in fact be viewed as a nilpotent 
derivation of degree 1 by setting 
$$
\delta(f_0 \delta f_1 \ldots \delta f_k) = 1\cdot 
\delta f_0 \delta f_1 \ldots \delta f_k\ ;
\eqno(3.23)$$
nilpotency follows {}from $\delta 1 = 0$ which in turn is a 
consequence of the Leibniz rule (3.19). There is a 
${}^*$-representation $\pi$ of $\Omega^{\bullet}(\a)$ on $L^2(S)$ 
by bounded operators defined via 
$$
\pi(f) = f\ ,\quad \pi(\delta f) = \lb\, D,f \,\rb = c(df)\ . 
\eqno(3.24)$$
\mn
{\bf Lemma 3.4}\quad The representation $\pi$ maps homogeneous
elements of degree $k$ onto sections of $Cl^{(k)}(M)$, i.e.\ 
$$
\pi(\Omega^k(\a)) = \Gamma(Cl^{(k)}(M))\ .
$$ 
\sn
{\kap Proof}: The inclusion ``$\subset$'' is clear by (3.24). For the 
other direction, let $x^{\mu}$ be a coordinate system in 
$U\subset M$; then the space $\Gamma(Cl^{(k)}(M))\big\vert_U$ is
linearly generated over $C^{\infty}(U)$ by elements of the form 
$$
\gamma^{\mu_1}\cdots \gamma^{\mu_k} =  
c(dx^{\mu_1})\cdots c(dx^{\mu_k}) = 
[D,x^{\mu_1}]\cdots [D,x^{\mu_k}]\ ,
$$ 
and this proves the reverse inclusion using a partition 
of unity. \hfill\qed
\mn
Note that in the last step we implicitly used the fact that the 
kernel of the multiplicative Clifford action does not intersect 
$Cl^{(k)}(M))$. This is certainly true if the dimension $n$ of 
$M$ is even since all representations of the Clifford algebra
are faithful in that situation. If $n$ is odd, consider the  
``chirality element'' $\gamma$ in the center of the Clifford
algebra: 
All elements in the kernel of $c$ are annihilated by one of 
the projections ${1\over2}(1\pm\gamma)$; but since $n$ is 
odd, such elements are not homogeneous for the $\Z_2$-grading
of $Cl(M)$ and it follows that $c$ is indeed faithful on 
$Cl^{(k)}(M)$ for all $k$. 
\sn
For the reconstruction of differential forms it remains to 
identify those elements of $\Omega^k(\a)$ whose images are
in $\Gamma(Cl^{(k-2)}(M))$. We define 
$$
J^k := {\rm ker}\,\pi |_{\Omega^k({\cal A})}\ ; 
\eqno(3.25)$$
$\bigoplus_k J^k$ is {\sl not} a differential ideal in 
$\Omega^{\bullet}(\a)$, since there are 
$\omega\in\Omega^k(\a)$ such 
that $\pi(\omega)=0$ but $\pi(\delta\omega)\neq0$.  
Typical elements with this property are (see \q{VGB})
$$
\omega=f\delta f -{1\over2}\delta(f^2)
$$
for all $f\in\a$ with $\pi(\delta f)\neq 0$; these $\omega$
satisfy
$$\eqalign{
\pi(\omega) &=  fc(df) - {1\over2}c(d(f^2)) = 0 \ ,
\cr
\pi(\delta\omega) &= \pi(\delta f\delta f) = 
c(df)c(df) = -g(df,df) \neq 0\ .
\cr}$$  
\mn
{\bf Lemma 3.5}\quad The representation $\pi$ maps the elements
of $\delta J^k$ onto sections of $Cl^{(k-1)}(M)$, i.e.\ 
$$
\pi(\delta J^k) = \Gamma(Cl^{(k-1)}(M))\ .
$$
\sn 
{\kap Proof}: Writing an arbitrary element $\omega \in \Omega^k(\a)$
as a linear combination of elements $f_0\delta f_1\ldots\delta f_k$,
it is easy to verify that $\sigma_{k+1}(\pi(\delta\omega)) = 
d\sigma_k(\pi(\omega))$, see (3.15). Therefore, $\omega\in J^k$ implies
$\pi(\delta\omega)\in {\rm ker}\,\sigma_{k+1}=\Gamma(Cl^{(k-1)}(M))$,
cf.\ (3.16), which proves that the left hand side of the 
equation above is contained in the right hand side. 
In order to show the reverse inclusion, we notice that elements 
of the form $\tau = -g(df_0,df_0)c(df_1)\ldots c(df_{k-1})$, 
$f_i\in\a$, generate $\Gamma(Cl^{(k-1)}(M))$ and that 
$\tau = \pi(\delta\omega)$ with $\omega=(f_0\delta f_0-{1\over2}\delta(f_0^2))
\delta f_1\ldots \delta f_{k-1} \in J^k$.  \hfill\qed 
\mn
The following proposition concludes our construction of 
differential forms.
\mn
{\bf Proposition 3.6} \q{Co1}\quad For each $k\in\Z_+$, 
there is an isomorphism 
$$
\Omega_D^k(\a) := \Omega^k(\a)/(J^k + \delta J^{k-1}) 
\ \cong\  \Gamma(\bigwedge{}^kM^{\C}) \ .
$$
{\kap Proof}: Combining Lemmata 3.4 and 3.5 with equation (3.17),
we get 
$$\eqalignno{ 
\Omega_D^k(\a) &= \Omega^k(\a)/(J^k + \delta J^{k-1}) 
\cong \pi(\Omega^k(\a))/\pi(\delta J^{k-1}) 
&\cr 
&\cong \Gamma(Cl^{(k)}(M))/\Gamma(Cl^{(k-2)}(M))
\cong \Gamma(\bigwedge{}^kM^{\C})\ .
&\qed\cr}$$    
\mn
It is easy to verify that $\Omega^{\bullet}_D(\a)= 
\bigoplus_{k=0}^{\infty}\Omega_D^k(\a)$ is a graded
algebra isomorphic to $\Gamma(\bigwedge{}^{\bullet}M^{\C})$ 
and that the derivation $\delta$ descends to a differential
$d$ on $\Omega^{\bullet}_D(\a)$ which corresponds to 
exterior differentiation on the space of differential
forms. Thus, we see that the spectral data 
$(C^{\infty}(M), L^2(S), D)$ indeed encode the 
differential algebra of forms, and therefore in 
particular the de Rham cohomology ring. 
\bn\bn  
\leftline{\bf 3.1.5 Integration and Hilbert space structure 
on differential forms}
\bn
The integration theory on $M$ may be recovered {}from 
the $N=1$ spectral data in two different ways, 
namely   \hfill\break
1) using the Dixmier trace ${\rm Tr}_{\omega}(\cdot|D|^{-n})$
and Connes' trace theorem, see \q{Co3}, or  \hfill\break
2) through the heat operator $\exp(-\varepsilon D^2)$ and the
heat kernel expansion.
\sn
The first approach is described in \q{Co1,3}; we shall follow
the second one. In both cases, mild restrictions have to be 
placed on the class of admissible Dirac bundles.
\mn
{\bf Definition 3.7}\quad A $\Z_2$-{\sl graded Dirac bundle}
is a Dirac bundle together with a section $\gamma\in
\Gamma(\End S)$ such that $\gamma^2=1$, $\{\gamma,c(\omega)\}
=0$ for each $\omega\in\Gamma(T^*M^{\C})$, and 
$\langle \gamma s_1,\gamma s_2\rangle = 
\langle s_1, s_2 \rangle$ for all $s_1,s_2\in\Gamma(S)$. 
\mn
Note that the spin bundle of an even dimensional spin
manifold is $\Z_2$-graded by the chirality element $\gamma$.
\sn
In the following, we shall only consider $\Z_2$-graded 
Dirac bundles -- which is in fact no restriction since 
any Dirac bundle $S$ can be made into a $Z_2$-graded one
by introducing $S' := S \oplus S \cong S \otimes \C^2$ 
with grading $\gamma := 1 \otimes \tau_1$ and 
Clifford action $c'(\omega) := c(\omega)\otimes\tau_3$
for all $\omega\in\tmc$; here, the $\tau_i$ are the Pauli
matrices. 
\sn
The Dirac operator is a self-adjoint first order differential
operator with symbol $\sigma_{\xi}(p,D) = i c(\xi)$, 
where $p\in M$ and $\xi\in T^*_pM^{\C}$. It follows that 
$D^2$ is a self-adjoint second  order differential
operator with symbol $\sigma_{\xi}(p,D^2) = g(\xi,\xi)$, 
i.e.\ it is a (generalized) Laplacian. The heat operator
$\exp(-\varepsilon D^2)$, $\varepsilon>0$, therefore has a 
smooth kernel $k_{\varepsilon}(x,y)$ such that 
$$
\bigl(e^{-\varepsilon D^2}s\bigr)(x) = 
\int_M k_{\varepsilon}(x,y) s(y) \sqrt{g} d^nx  
\eqno(3.26)$$
for all sections $s\in\Gamma(S)$. 
The operator $k_{\varepsilon}(x,y)$ is a linear map {}from
$S_y$ to $S_x$ and it has the asymptotic expansion \q{BGV}
$$
k_{\varepsilon}(x,y) = {1\over(4\pi \varepsilon)^{n/2}} 
e^{-d(x,y)^2/(4\varepsilon)} \psi(d(x,y)) \tau(x,y) 
+ {\rm O}(\varepsilon^{1-n/2})
\eqno(3.27)$$
where $d(x,y)$ is the geodesic distance between $x$ and $y$, 
$\psi$ is a positive cutoff function equal to 1 near 0 and 
vanishing outside a neighborhood of 0, and $\tau(x,y)$ denotes 
parallel transport {}from $S_y$ to $S_x$ along the unique
geodesic lying in supp$\psi$. 
\sn
Let $\sigma\in\Gamma(Cl(M))$, then the operator 
$c(\sigma)\exp(-\varepsilon D^2)$ has a kernel 
$c(\sigma)(x)k_{\varepsilon}(x,y)$,  
and its trace is given by 
$$\eqalign{
{\rm Tr}\bigl(c(\sigma)e^{-\varepsilon D^2}\bigr) &=
\int_M {\rm tr}(c(\sigma)(x)k_{\varepsilon}(x,x))
\sqrt{g} d^nx  \cr
&=
{1\over(4\pi t)^{n/2}} \int_M {\rm tr}(c(\sigma)(x)) 
\sqrt{g} d^nx + {\rm O}(t^{1-n/2})
\cr}\eqno(3.28)$$
where we have used (3.27) in the second equation. 
\sn
Let $\theta^i$ be a local orthonormal basis of $\Gamma(T^*M)$
over $U\subset M$, then any section $\sigma\in\Gamma(Cl(M))$
can locally be written as 
$$
\sigma = \sigma_0 + \sigma_i\theta^i + \sigma_{i_1i_2}
\theta^{i_1}\theta^{i_2} + \ldots + \sigma_{i_1\ldots i_n}
\theta^{i_1}\cdots\theta^{i_n}
\eqno(3.29)$$
with totally anti-symmetric functions $\sigma_{i_1\ldots i_k}$.
Using the $\Z_2$-grading of the bundle $S$, one finds that
$$
{\rm tr}(c(\sigma)(x)) = {\rm Rk}(S)\cdot\sigma_0(x)
\eqno(3.30)$$
where Rk$(S)$ denotes the rank of $S$. Putting together 
(3.28) and (3.30), we obtain 
$$
\Barint c(\sigma) := \lim_{\varepsilon\to0^+}
{{\rm Tr}\bigl(c(\sigma)e^{-\varepsilon D^2}\bigr)
\over {\rm Tr}\bigl(e^{-\varepsilon D^2}\bigr)} 
= {1\over{\rm Vol}(M)} \int_M \sigma_0(x)\sqrt{g}d^nx \ .
\eqno(3.31)$$
The integral $\barint$ induces a scalar product on the 
space $\Gamma(Cl(M))$, 
$$
(c(\sigma_1),c(\sigma_2)) := \Barint c(\sigma_1)c(\sigma_2)^*
\eqno(3.32)$$
and it is easy to verify that the projection of a section 
$\phi\in\Gamma(Cl^{(k)}(M))$ -- see (3.13) -- onto the 
orthogonal complement of $\Gamma(Cl^{(k-2)}(M))$ is the
differential form $\sigma_k(\phi)$ -- see (3.15) -- 
corresponding to the equivalence class 
$[\phi] \in\Gamma(Cl^{(k)}(M))/ \Gamma(Cl^{(k-2)}(M))
= \Omega_D^k(\a)$. 
Furthermore, the induced scalar product on differential 
forms is the usual one, i.e.\ 
$$
(\sigma_k(\phi),\sigma_k(\psi)) = {1\over{\rm Vol}(M)}
\int_M \phi_{i_1\ldots i_k}\psi^{i_1\ldots i_k}\sqrt{g}d^nx \ .
\eqno(3.33)$$
In summary, we see that the $N=1$ spectral triple 
also encodes the Hilbert space structure on the space
of differential forms. 
\bn\bn
\leftline{\bf 3.1.6 Vector bundles and Hermitian structures}
\bn
Let $E$ be a complex vector bundle over $M$. Its space of 
sections $\e=\Gamma(E)$ is a left- (and right-) module over 
$C^{\infty}(M)$, and using a partition of unity, we see that $\e$ is 
finitely generated. Furthermore, there exists a vector 
bundle $E'$ such that $E\oplus E'$ is a trivial bundle, 
see \q{Sw}, which means that the corresponding module 
$\e\oplus\e'$ is free. In other words, $\e$ is a 
finitely generated projective left module over $C^{\infty}(M)$. 
It turns out that the converse statement is also true, i.e.\ 
every module $\e$ of the above type is isomorphic to the
space of sections of a complex vector bundle $E$ over $M$,
which is unique up to isomorphism 
\q{Sw}. Thus we have a characterization of complex vector
bundles over $M$ in terms of the algebra $C^{\infty}(M)$
alone. A Hermitian structure on a bundle (a module) $\e$ 
is then simply defined as a non-degenerate positive 
$C^{\infty}(M)$-sesqui-linear map {}from $\e\times\e$ to 
$C^{\infty}(M)$. 
\sn
The extension of the Riemannian metric to a Hermitian
structure on differential forms is recovered by defining
$g(\phi,\bar\psi)$, $\phi,\psi\in\Gamma(\bigwedge{}^kM^{\C})$, 
to be the unique function on $M$ such that 
$$
\int_M g(\phi,\bar\psi) f = \Barint c(\phi)c(\psi)^* f
$$
for all $f\in C^{\infty}(M)$. 
\sn
At this stage we notice that more or less any global definition 
of connection, metric connection, curvature, torsion, etc.\ 
will fit into the present algebraic formalism. This completes 
our reconstruction of the Riemannian manifold $M$ {}from an
$N=1$ spectral triple. 
\bn\bn
\vfil\eject
\leftline{\bf 3.2 The \Noneone formulation of Riemannian 
geometry}
\bn
In this section, we describe an algebraic formulation of 
Riemannian geometry that apparently requires 
slightly more structure on the manifold than the 
$N=1$ setting. The advantage is 
that the reconstruction procedure and the non-commutative
generalization are easier {}from the $N=(1,1)$ point of 
view, see subsection 3.2.4. In addition, in 3.2.2, we will in 
fact see that every Riemannian manifold provides this 
larger set of data.  
\bn\bn
\leftline{\bf 3.2.1 The \Noneone spectral data} 
\bn
{\bf Definition 3.8}\quad Let $M$ be a Riemannian manifold. 
An $N=(1,1)$ {\sl Dirac bundle over $M$} is a $\Z_2$-graded
Dirac bundle $\S$ with Clifford action $\c$ and connection 
$\nabla^S$, see Definition 3.7, together with a second Clifford action 
$\overline{\c}$ such that   
\smallskip
\item{1)} $(\S,\overline{\c}, \nabla^S)$ is a  $\Z_2$-graded 
Dirac bundle; 
\smallskip
\item{2)} $\{ \c(\sigma_1),\overline{\c}(\sigma_2)\} =0$ for all 
$\sigma_1,\sigma_2\in\Gamma(\tmc)$; 
\smallskip
\item{3)} the Dirac operators $\d=\c\circ\nabla^S$ and 
$\bard=\overline{\c}\circ\nabla^S$ satisfy the relations
$$ 
\{\d,\bard\} =0\ ,\quad \d^2=\bard^2\ .
$$
\mn
Given an $N=(1,1)$ Dirac bundle $\S$ over $M$ we can 
introduce an operator {\tt d} on $\Gamma(\S)$  and its adjoint by  
$$
{\tt d}\,:= {1\over2}(\d - i\bard) \ ,\quad 
{\tt d}^*\,:={1\over2}(\d + i\bard)\ ;  
\eqno(3.34)$$ 
both of these operators are nilpotent: ${\tt d}^2=0$. 
\bn\bn
\leftline{\bf 3.2.2 Spin\CC-manifolds and the canonical
\Noneone Dirac bundle} 
\bn
In this section we prove that spin$^c$-manifolds carry a 
canonical $N=(1,1)$ Dirac bundle which can be constructed 
in terms of the spinor bundle $S$. Moreover, it turns out that 
the Dirac bundle is isomorphic to the bundle of differential 
forms, which means that in fact to any compact Riemannian manifold 
one can associate an $N=(1,1)$ Dirac bundle, as mentioned at 
the beginning of section 3.2. In the present subsection, we 
systematically avoid the language of principal bundles 
since it is difficult to generalize to the non-commutative 
setting. Instead, we make use of the material of section 
2 and of some basic knowledge of vector bundles to convert 
those ``local'' notions into global ones.    
\sn
The definition of a spin$^c$-manifold we use is a direct 
generalization of a spin$^c$-structure on an oriented 
vector space introduced in Definition 2.1: 
\mn
{\bf Definition 3.9} \q{Sa}\quad A {\sl spin$^c$-manifold} 
is an oriented Riemannian manifold $(M,g)$ of dimension $n$ 
together with a complex Hermitian vector bundle $S$ of 
rank $2^{\lb{n\over2}\rb}$, where $\lb k\rb$ denotes 
the integer part of $k$, and a bundle homomorphism 
$c\,:\ T^*M \lra \End(S)$ such that 
$$\eqalign{
&c(\omega)+c(\omega)^* =0 
\cr
&c(\omega)^*c(\omega) = g(\omega,\omega) 
\cr}\eqno(3.35)$$
for all $\omega\in T^*M$. 
\mn
Let $M$ be a compact spin$^c$-manifold. Note that $c$ 
extends uniquely to an irreducible unitary Clifford 
action of $Cl(M)$ on $S$. Recall that a Clifford connection 
$\nabla^S$ on $S$ is a Hermitian connection compatible 
with the Levi-Civita covariant derivative $\nabla$ 
on $T^*M$, see (3.4). Here, we follow the terminology of 
\q{BGV}; in the context of Spin${}^c$-geometry and 
particularly in the study of Seiberg-Witten invariants, 
what we call a Clifford connection on the 
Spin${}^c$-bundle is referred to as a Spin${}^c$-connection 
compatible with the Levi-Civita connection.  
\mn
{\bf Proposition 3.10} \q{Sa}\quad If $\nabla^S$ and 
$\widetilde{\nabla}^S$ are Clifford connections on $S$, 
then their difference is a 1-form with values in 
the purely imaginary numbers, i.e.\ 
$$
\nabla^S - \widetilde{\nabla}^S \in \Gamma(T^*M\otimes i\R)\ .
$$
\sn
{\kap Sketch of Proof}: The compatibility with $\nabla$ implies 
that 
$$
(\nabla^S_X - \widetilde{\nabla}^S_X) c(\omega) =
c(\omega) (\nabla^S_X - \widetilde{\nabla}^S_X)
$$
for all $\omega\in\Gamma(Cl(M))$ and $X\in \Gamma(TM)$. 
Since the commutant of $\Gamma(Cl(M))$ in $\End S$ equals 
$C^{\infty}(M)$, the element $\nabla^S_X - \widetilde{\nabla}^S_X$
is just a function over the manifold. Because of the  
Hermiticity condition, it takes purely imaginary values.   \hfill\qed
\mn
Note that the 1-form taking purely imaginary values may 
be interpreted {}from the physical point of view as an 
electromagnetic vector potential; thus, it is charged 
spinning particles that can propagate on a spin$^c$-manifold.
\mn
By virtue of Proposition 3.10, we can describe all 
Clifford connections on $S$ if we only present a 
construction for one of them, which will be done 
in the following. The main idea is to glue together 
the local data introduced in section 2 with the 
help of suitably chosen local trivializations of $T^*M$
and $S$: 
\sn
Fix a real oriented $n$-dimensional vector space $V$
with scalar product $(\cdot,\cdot)$ and a spin$^c$-structure
$(W,\Gamma)$ on $V$. Let $U\subset M$ be a 
contractible open set and 
$$
h\,:\ S\big\vert_U \lra U\times W
$$
be a local trivialization of $S$ with the property   
$$
\langle \sigma_1,\sigma_2\rangle_S=
\langle h(\sigma_1),h(\sigma_2)\rangle_W\ . 
\eqno(3.36)$$
We define a local trivialization of $T^*M$ 
$$
\varphi\,:\ T^*M\big\vert_U \lra U\times V
$$
by requiring 
$$
\bigl(\Gamma(\varphi(\omega))\bigr)w = h\bigl(c(\omega)h^{-1}(w)\bigr)
\eqno(3.37)$$
for all $\omega\in T^*M\big\vert_U$ and $w\in W$. (Such a $\varphi$ 
exists because $c$ and $\Gamma$ both implement irreducible 
representations of algebras that are isomorphic to $Cl(M)\big\vert_U$.)
The map $\varphi$ satisfies 
$$
(\varphi(\omega),\varphi(\eta)) = g(\omega,\eta)
$$
for all $\omega,\eta\in T^*M\big\vert_U$. A triple $(U,\varphi,h)$ 
as above is called an {\sl admissible trivialization}
of $T^*M$ and $S$ with respect to $(V,W,\Gamma)$. 
\mn
{\bf Lemma 3.11}\quad The structure group of $S$ can be 
reduced to Spin$^c(V)$. 
\sn
{\kap Sketch of Proof}: Choose a finite covering $\{U_{\alpha}\}$     
of $M$ by contractible open sets such that 
$(U_{\alpha}, \varphi_{\alpha}, h_{\alpha})$ are admissible 
trivializations of $T^*M$ and $S$. Then the joint 
transition functions 
$$
(A_{\alpha\beta},\Phi_{\alpha\beta}) :=   
(\varphi_{\alpha}\circ\varphi^{-1}_{\beta},
h_{\alpha}\circ h^{-1}_{\beta})
$$
of $T^*M$ and $S$ are easily seen to take values in 
Hom$^{{\rm spin}^c}(W,W) \cong \Spcv$ by Proposition 2.2. \hfill\qed
\sn
With the help of this lemma, one may also show that Definition 3.9 of  
a spin$^c$-manifold is equivalent to the usual definition of 
Spin$^c$-structures via principal bundles.  
\mn
Let $\delta\,:\ {\rm Spin}^c(n) \lra {\rm U}(1)$ be 
the group homomorphism defined by 
$$
\delta (e^{i\alpha}x) = e^{2i\alpha}\ ,
\eqno(3.38)
$$ 
$x\in {\rm Spin}(n)$, and $(U_{\alpha}, \varphi_{\alpha}, 
h_{\alpha})$ be a set of admissible trivializations 
of $T^*M$ and $S$. The U$(1)$-valued functions
$\delta\circ((A_{\alpha\beta},\Phi_{\alpha\beta}))$
-- see the proof of Lemma 3.11 -- define a Hermitian
line bundle $L$. The bundle $S$ can be written locally
as $S_{\R}\otimes L^{1/2}$ where $S_{\R}$ is a bundle of 
irreducible representations of the real Clifford algebra
and $L^{1/2}$ is the square root of $L$. It may happen 
that neither $S_{\R}$ nor $L^{1/2}$ are globally defined.
The ``bundle'' $L^{1/2}$ is therefore called virtual 
bundle. Notice that an admissible trivialization $(U,\varphi,h)$ 
of $T^*M$ and $S$ also determines a trivialization of $L$. 
Let $\nabla^L$ be a Hermitian connection on $L$ and $2A$ 
its connection 1-form in the trivialization  $(U,\varphi,h)$. 
We choose orthonormal bases $e^i$ and $\epsilon^p$ of $V$ 
and $W$, respectively; then $\theta^i=\varphi^{-1}(e^i)$ 
and $\sigma^p=h^{-1}(\epsilon^p)$ are orthonormal frames
of $\Gamma(T^*M\big\vert_U)$ and $\Gamma(S\big\vert_U)$, respectively, 
and we denote by $\vartheta_i$ the frame of $\Gamma(TM\big\vert_U)$
dual to $\theta^i$. 
\mn
{\bf Lemma 3.12}\quad Let $\omega^j_{\phantom{j}k}(\vartheta_i)$
denote the coefficients of the Levi-Civita connection in the basis
$\vartheta_i$ of $\Gamma(TM\big\vert_U)$, and let $2A$ be the 
connection 1-form of a Hermitian connection on $L$; then the formula
$$
\nabla^S\sigma^p = \theta^i \otimes 
\bigl( -{1\over4} \omega^j_{\phantom{j}k}(\vartheta_i)
c(\theta^j)c(\theta^k)+{i} A(\vartheta_i) 
\bigr) \sigma^p 
$$ 
defines a Clifford connection on $S$. 
\sn
{\kap Sketch of Proof}: The expression for $\nabla^S\sigma^p$ does not
depend on the choice of bases $e^i$ and $\epsilon^p$ since 
any two such choices are related by a global transformation 
over $U$. If $(\tilde U, \tilde\varphi, \tilde h)$ is 
another admissible local trivialization of $T^*M$ and $S$ with the 
same $V$ and $(W,\Gamma)$ , there 
is a Hom$^{{\rm spin}^c}(W,W)$-valued transition function 
$(A,\Phi)_x$, $x\in U\cap\tilde U$, such 
that $\tilde\sigma^p = \Phi^p_{\phantom{p}q}\sigma^q$ and
$\tilde\theta^i = A^i_{\phantom{i}j}\theta^j$; then we have
$$\eqalign{
\nabla^S\tilde\sigma^p &= d\Phi^p_{\phantom{p}q}\bigl(\Phi^{-1}
\bigr)^q_{\phantom{q}r} \otimes \tilde\sigma^r + 
\Phi^p_{\phantom{p}q}\nabla^S\sigma^q 
\cr &=  d\Phi^p_{\phantom{p}q}\bigl(\Phi^{-1}
\bigr)^q_{\phantom{q}r} \otimes \tilde\sigma^r + 
\theta^i \otimes 
\bigl( -{1\over4} \omega^j_{\phantom{j}k}(\vartheta_i)
c(\theta^j)c(\theta^k)+{i} A(\vartheta_i) 
\bigr) \tilde\sigma^p\ . 
\cr}$$
According to Proposition 2.2, we may define  
$$
e^{i\alpha_x}\xi_x = \Xi^{-1}_{\Gamma} \bigl((A^i_{\phantom{i}j},
\Phi^p_{\phantom{p}q})_x\bigr) \in C^{\infty}(U\cap\tilde U,
{\rm Spin}^c(V))\ 
$$
where $\xi_x\in C^{\infty}(U\cap\tilde U,{\rm Spin}(V))$, 
and {}from (2.21,22) and the compatibility properties of 
admissible trivializations we obtain 
$$
d\Phi^p_{\phantom{p}q}\bigl(\Phi^{-1}
\bigr)^q_{\phantom{q}r} \otimes \tilde\sigma^r 
= {1\over4} dA^j_{\phantom{j}l}(A^{-1})^l_{\phantom{l}k}
\otimes c(\tilde\theta^j) c(\tilde\theta^k)
\tilde\sigma^p + i  d\alpha \ .
$$
The transformation formula for the Levi-Civita connection
is as usual: 
$$
{1\over4} \omega^j_{\phantom{j}k}(\vartheta_i)
c(\theta^j)c(\theta^k) = 
{1\over4} \tilde\omega^j_{\phantom{j}k}(\vartheta_i)
c(\tilde\theta^j)c(\tilde\theta^k)
+{1\over4} dA^j_{\phantom{j}l}(A^{-1})^l_{\phantom{l}k}
\otimes c(\tilde\theta^j) c(\tilde\theta^k)
$$
Finally, using the transformation of the U(1)-field, 
$$
\tilde A = A + d\alpha \ ,
$$
and putting all the terms together, we see that $\nabla^S$
is indeed well-defined. A direct computation shows that it  
also satisfies the properties of a Clifford connection.  \hfill\qed
\sn
Note that ${i}A$ transforms like a connection on 
$L^{1/2}$. Therefore it is called virtual connection. 
\mn
We will now use the results of Section 2.3 to establish 
a relation to differential forms. Let $\{\varphi_{\alpha
\beta}\}$ be a family of Spin$^c(n)$-valued transition 
functions for $S$. Using the automorphism 
$$
 e^{i\alpha} \xi \mapsto \overline{e^{i\alpha} \xi} =  
 e^{-i\alpha} \xi
\eqno(3.39)$$
of Spin$^c(n)$, where $\xi\in$ Spin$(n)$, we define $\overline{S}$
to be the vector bundle of rank $2^{\lb{n\over2}\rb}$
defined by the transition functions $\{\overline{\varphi_{
\alpha\beta}}\}$. Then $\overline{S}$ is also a Dirac bundle 
with Clifford action $\bar c$ -- which is just $c$ transferred
to the ``new'' bundle -- 
and its Clifford connection $\nabla^{\bar S}$ reads 
$$
\nabla^{\bar S}\bar\sigma^p = \theta^i \otimes 
\bigl( -{1\over4} \omega^j_{\phantom{j}k}(\vartheta_i)
\overline{c}(\theta^j)\overline{c}(\theta^k)-{i} A(\vartheta_i) 
\bigr) \bar\sigma^p 
\eqno(3.40)$$ 
in an admissible trivialization. The vector bundle $S\otimes\overline{S}$ 
has SO$(n)$ as its structure group, and there is a vector
bundle isomorphism 
$$
\psi\,:\ \bigwedge{}^{\bullet}M^{\C} \lra \S
\eqno(3.41)$$
with $\S=S\otimes\overline{S}$ if $n$ is even and 
$\S=S\otimes\C^2\otimes\overline{S}$ if $n$ is odd.  
Furthermore, this isomorphism can be chosen in such a 
way that the correspondences (2.24-30) hold, and therefore
we can in particular use $\psi$ to push the two anti-commuting
Clifford actions $\Gamma$ and $\overline{\Gamma}$ (2.24) 
to $\S$, where we call them $\c$ and $\overline{\c}$, respectively. 
(Of course, we could have immediately identified $\c$ with 
$c\otimes\one$ and $\overline{\c}$ with $\gamma\otimes\overline{c}$.)
\sn 
Let $\nabla^{\C^2}$ be the canonical flat connection on 
$M\times \C^2$ and $\nabla^{\rm tot}$ be the tensor
product connection on $\S$, i.e. 
$$
\nabla^{\rm tot}= \nabla^{S}+\nabla^{\C^2}+\nabla^{\bar S}
\eqno(3.42)$$
where of course the middle term only appears if $n$ is odd.
\mn
{\bf Lemma 3.13}\quad The isomorphism $\psi$ in (3.41) 
maps the Levi-Civita connection to $\nabla^{\rm tot}$.
\sn
{\kap Proof}: Using Lemma 3.12 and eq.\ (3.40), we see that in 
an admissible trivialization the U(1)-field $2A$ does not 
contribute to the connection coefficients of $\nabla^{\rm tot}$,
i.e.\  
$$
\nabla^{\rm tot} = \theta^i \otimes 
\bigl( -{1\over4} \omega^j_{\phantom{j}k}(\vartheta_i)
(c(\theta^j)c(\theta^k)+\overline{c}(\theta^j)\overline{c}(\theta^k))
\bigr)\ .
$$
By (2.24-30) we get 
$$
\psi^{-1}\circ\nabla^{\rm tot}\circ\psi = 
 \theta^i \otimes (-\omega^j_{\phantom{j}k}(\vartheta_i)
a^{k*}a^j)
$$
where $a^{k*}=\theta^k\wedge$ and $a^k$ is its adjoint. 
But the last expression is just the Levi-Civita connection 
in the basis $\theta^i$.   \hfill\qed
\mn
We use the isomorphism $\psi$ of eq.\ (3.41) to define two Dirac 
operators on $\S$, 
$$
\d =  \c\circ \nabla^{\rm tot}\ ,\quad
\bard = \overline{\c}\circ \nabla^{\rm tot}\ .
\eqno(3.43)$$
\sn
{\bf Lemma 3.14}\quad The operators $\d$ and $\bard$ satisfy
$$\eqalign{
\psi^{-1}\d\psi &= d + d^* \ ,
\cr
\psi^{-1}\bard\psi &= i(d - d^*)\ . 
\cr}\eqno(3.44)$$
\sn
{\kap Proof}: This follows {}from (2.24) and the equations  
$$
d=a^{i*}\circ\nabla_{\vartheta_i}\ ,\quad
d^*=-a^{i}\circ\nabla_{\vartheta_i}\ ,
$$
see e.g.\ \q{BGV}; $\nabla$ denotes the (torsion-free)
Levi-Civita connection. \hfill\qed
\mn
With Lemma 3.14 it is easy to see that $\d$ and $\bard$ 
fulfill the relations 
$$\{\d,\bard\}=0\quad {\rm and}\quad \d^2=\bard^2\ .
$$ 
of Definition 3.8. Thus, the bundle $\S = S\otimes\overline{S}$ -- 
or $\S=S\otimes\C^2\otimes\overline{S}$ for $n$ odd --  
is an $N=(1,1)$ Dirac bundle over
$M$. $\S$ is furthermore isomorphic to the bundle of 
differential forms, and this proves that any 
Riemannian manifold carries an $N=(1,1)$ structure. 
\bn\bn
\leftline{\bf 3.2.3 Structure theorem for \Noneone Dirac 
bundles} 
\bn
In the previous section we have seen that any Riemannian
manifold carries a canonical $N=(1,1)$ Dirac bundle, see
Definitions 3.1, 3.7, 3.8, namely  
the bundle of differential forms together with the Levi-Civita
connection and the Clifford actions $\Gamma$ and $\overline{\Gamma}$
as in (2.24). In this section, we prove that any $N=(1,1)$
Dirac bundle is obtained by tensoring this canonical one 
with a flat Hermitian vector bundle. 
\sn
Let $\S$ be an $N=(1,1)$ Dirac bundle with Clifford actions 
$\c$ and $\overline{\c}$ and Clifford connection $\nabla^S$. 
We define 
$$
E := {\rm Hom}_{Cl}(\bigwedge{}^{\bullet}M^{\C},\S)
\eqno(3.45)$$
to be the vector bundle of bundle maps 
$$
\varphi: \bigwedge{}^{\bullet}M^{\C} \lra \S 
$$
which intertwine the Clifford actions on $\bigwedge^{\bullet}M^{\C}$
and $\S$, i.e.\ 
$$
\varphi \circ \Gamma(\omega) = \c(\omega)\circ\varphi\ ,\quad 
\varphi \circ \overline{\Gamma}(\omega) = \overline{\c}(\omega)\circ\varphi\ , 
$$
for all $\omega\in\bigwedge^{\bullet}M^{\C}$.
\mn
{\bf Lemma 3.15}\quad The bundle map 
$$
\psi\,:\ \cases{ &$\bigwedge{}^{\bullet}M^{\C}\otimes E \lra \S$\cr
&\quad\quad\quad\ $\vphantom{\int}\omega\otimes\varphi \longmapsto 
\varphi(\omega)$\cr}
$$
is an isomorphism, and the Clifford algebra acts 
trivially on $E$, i.e.\ for all $\omega\in Cl(M)$ we have
$$\eqalign{
&\c(\omega) \circ \psi = 
\psi\circ(\Gamma(\omega)\otimes\one)\ ,
\cr 
&\overline{\c}(\omega) \circ \psi= 
\psi \circ (\overline{\Gamma}(\omega)\otimes\one)\ . 
\cr}$$ 
\sn
{\kap Proof}: For each $x\in M$, the two anti-commuting 
unitary representations $\c$ and $\overline{\c}$ of 
$Cl(M)|_x$ on $\S_x$ may be viewed as one unitary 
representation of $Cl(M\times M)|_{(x,x)}\,$. Since
$M\times M$ is even-dimensional, $\S_x$ is a 
multiple of the unique irreducible unitary representation
of $Cl(M\times M)|_{(x,x)}$ which is precisely 
$\bigwedge^{\bullet}T^*_xM^{\C}$ with actions $\Gamma$ and 
$\overline{\Gamma}$. This proves that $\psi$ is an isomorphism, 
and since $E$ is the ``multiplicity space'' it is clear
that the Clifford algebra acts trivially on it. \hfill\qed
\mn
In the following we will identify the bundles 
$\bigwedge^{\bullet}M^{\C}\otimes E$ and $\S$ with the help 
of $\psi$. 
\mn  
{\bf Lemma 3.16}\quad The sesqui-linear map 
$$
\langle\cdot,\cdot\rangle_E\, :\ \cases{&$\Gamma(E)\times\Gamma(E)
\lra C^{\infty}(M)$
\cr 
&\quad\quad\quad$\vphantom{\int^T}(\varphi,\varphi') \longmapsto 
\langle\varphi(1),\varphi'(1)\rangle_{\S}\ ,$
\cr}$$
where $1\in\Gamma(\bigwedge^{\bullet}M^{\C})$ denotes the 
identity function, defines a Hermitian structure on $E$. 
Furthermore, the Hermitian structure on $\S$ factorizes 
as 
$$
\langle\cdot,\cdot\rangle_{\S} = 
\langle\cdot,\cdot\rangle\otimes\langle\cdot,\cdot\rangle_E
$$
where $\langle\cdot,\cdot\rangle$ denotes the canonical 
Hermitian structure on $\bigwedge^{\bullet}M^{\C}$ given 
by the Riemannian metric. 
\sn
{\kap Proof}: Let $\theta^i$ be an orthonormal basis of 
$\Gamma(\bigwedge^{\bullet}M^{\C})$ over a coordinate 
chart $U\subset M$ and $\varphi,\varphi'\in\Gamma(E\big\vert_U)$. 
We denote by $a^{i*}=a^*(\theta^i)$ the wedging operator 
by $\theta^i$ on $\Gamma(\bigwedge^{\bullet}M^{\C}\big\vert_U)$.
Since $a^{i*}$ is a linear combination of $\Gamma(\theta^i)$
and $\overline{\Gamma}(\theta^i)$, it commutes with $\varphi$. 
If $\varphi(1)=0$ then we have 
$$\varphi(\theta^{i_1}\wedge\ldots\wedge\theta^{i_p})
= \varphi(a^{i_1*}\ldots a^{i_p*}\cdot 1)
= a^{i_1*}\ldots a^{i_p*}\cdot\varphi(1) =0
$$
for all $1\leq i_1 <\ldots <i_p \leq n= {\rm dim}\,M$, and 
therefore $\varphi\equiv 0$; this  
proves that $\langle\cdot,\cdot\rangle_E$ is 
a Hermitian structure on $E$. \hfill\break
\noindent The contraction operators
$a^i=a(\theta^i)$ also commute with $\varphi$ and $\varphi'$,  
and it follows that $a^i\varphi(1)=a^i\varphi'(1)=0$ for 
all $i$. Using the CAR we get 
$$\eqalign{
\langle \varphi&(\theta^{i_1}\wedge\ldots\wedge\theta^{i_p}),
\varphi'(\theta^{j_1}\wedge\ldots\wedge\theta^{j_q})\rangle_{\S}
= \langle a^{i_1*}\ldots\, a^{i_p*}\cdot\varphi(1), 
 a^{j_1*}\ldots\, a^{j_q*}\cdot\varphi'(1) \rangle_{\S} 
\cr &= \langle \varphi(1), a^{i_p}\ldots\, a^{i_1}a^{j_1*}\ldots   
\,a^{j_q*}\cdot\varphi'(1) \rangle_{\S} 
= \delta_{p,q}\,{\rm sgn}\,{j_1\ldots j_q \choose i_1\ldots i_p} 
\langle\varphi(1),\varphi'(1)\rangle_{\S} 
\cr &= \langle \theta^{i_1}\wedge\ldots\wedge\theta^{i_p},
\theta^{j_1}\wedge\ldots\wedge\theta^{j_q}\rangle
\langle\varphi,\varphi'\rangle_E 
\cr}
$$
where in the second equality we have used the unitarity of 
the Clifford actions. The result follows by linearity.  \hfill\qed
\mn
{\bf Lemma 3.17}\quad The connection $\nabla^S$ on $\S$ 
is a tensor product connection, i.e. 
$$
\nabla^S = \nabla\otimes\one + \one\otimes\nabla^E
$$
for some covariant derivative $\nabla^E$ on $E$; as usual, 
$\nabla$ denotes the Levi-Civita connection. 
\sn
{\kap Proof}: We define 
$$
(\nabla^E\varphi)(\omega) = \nabla^S(\varphi(\omega))
-\varphi(\nabla\omega)
$$
for all $\varphi\in\Gamma(E)$ and $\omega\in\Gamma
(\bigwedge^{\bullet}M^{\C})$. It is readily seen that $\nabla^E\varphi$ 
is tensorial, i.e.\ $C^{\infty}(M)$-linear, in $\omega$.   
Furthermore, since $\nabla^S$ and $\nabla$ are Clifford connections
compatible with the Levi-Civita connection, we have 
$$
(\nabla^E\varphi)(\Gamma(\eta)\omega) = \c(\eta)(\nabla^E\varphi)
(\omega)
$$
for all $\omega,\eta\in\Gamma(\bigwedge^{\bullet}M^{\C})$
and $\varphi\in\Gamma(E)$, and the analogous relation with 
the ``barred'' Clifford actions $\overline{\Gamma}$
and $\overline{\c}$ holds as well. This proves that $\nabla^E$ maps 
$\Gamma(E)$ into $\Gamma(T^*M\otimes E)$,
in other words, $\nabla^E$ is a connection on $E$. Finally, 
the equation 
$$
\nabla^S(\omega\otimes\varphi) = \nabla\omega\otimes\varphi
+ \omega\otimes\nabla^E\varphi
$$
is obvious {}from the definition of $\nabla^E$.    \hfill\qed
\mn
{\bf Lemma 3.18}\quad Let $\d=\c\circ\nabla^S$ and
$\bard = \overline{\c}\circ \nabla^S$ be the two Dirac operators 
on $\S$ as in Definition 3.8, and let $R(\nabla^E)$ be the 
curvature tensor of $\nabla^E$. Then the following statements
are equivalent:
\smallskip
\item {1)} $\{\d, \bard\} = 0$ 
\smallskip  
\item {2)} $\d^2 = \bard^2 $ 
\smallskip  
\item {3)} $R(\nabla^E)= 0$ 
\sn 
{\kap Proof}: First of all, we use Lemma 3.17 and the isomorphism $\psi$ of 
Lemma 3.15 to write  $\d\,(\omega\otimes\xi) = \Gamma(dx^{\mu})
(\nabla_{\mu}\,\omega\otimes\xi+\omega\otimes\nabla^E_{\mu}\xi)$. {}From 
the previous section, we know that the 
operators $\d_{\wedge}:= \Gamma(dx^{\mu})\nabla_{\mu}$ 
and $\bard_{\wedge}:= \overline{\Gamma}(dx^{\mu})\nabla_{\mu}$   
on the space $\Gamma(\bigwedge^{\bullet}M^{\C})$ satisfy the first
two conditions. In addition, we exploit the fact that 
$\nabla$ is a Clifford connection, i.e. 
$$
\nabla_{\mu}\Gamma(dx^{\nu})\,\omega = 
-\Gamma_{\mu\sigma}^{\nu}\Gamma(dx^{\sigma})\,\omega  
+ \Gamma(dx^{\nu})\nabla_{\mu}\omega 
$$ 
-- and similarly with $\overline{\Gamma}$ instead of $\Gamma$ --
for all $\omega\in \Gamma(\bigwedge^{\bullet}M^{\C})$; 
here $\Gamma_{\mu\sigma}^{\nu}$ are the Christoffel
symbols. With this, it is straightforward, if slightly 
lengthy, to calculate 
$$\eqalign{
\{\d,\bard\}\,&(\omega\otimes\xi) =  
\{\d_{\wedge},\bard_{\wedge}\}\,(\omega\otimes\xi) 
\cr &+ \{\Gamma(dx^{\mu}),\overline{\Gamma}(dx^{\nu})\} 
\nabla_{\nu}\omega\otimes\nabla^E_{\mu}\xi + 
 \{\Gamma(dx^{\mu}),\overline{\Gamma}(dx^{\nu})\} 
\nabla_{\mu}\omega\otimes\nabla^E_{\nu}\xi 
\cr &- \{\Gamma(dx^{\mu}),\overline{\Gamma}(dx^{\nu})\} 
\Gamma_{\mu\nu}^{\lambda}\omega\otimes\nabla^E_{\lambda}\xi
+ \Gamma(dx^{\mu})\overline{\Gamma}(dx^{\nu})\,
\omega\otimes\lb\,\nabla^E_{\mu},\nabla^E_{\nu}\,\rb\xi 
\cr &=  \Gamma(dx^{\mu})\overline{\Gamma}(dx^{\nu})\, 
\omega\otimes R^E(\partial_{\mu},\partial_{\nu})\xi\ ,
\cr}$$
which shows that condition 3 implies 1. On the other 
hand, if 1 holds then, taking the commutator of the 
last expression with $\overline{\Gamma}(dx^{\sigma})$
and then the anti-commutator with $\Gamma(dx^{\lambda})$, 
we obtain 3. Next, a similar computation as above gives 
$$\eqalign{
\d^2\, &(\omega\otimes\xi) = \d_{\wedge}^2\omega\otimes\xi
- 2g^{\mu\nu}\nabla_{\mu}\omega\otimes\nabla^E_{\nu}\xi
+ g^{\mu\nu}\Gamma_{\mu\nu}^{\lambda}\omega\otimes
\nabla^E_{\lambda}\xi 
\cr &- g^{\mu\nu} \omega\otimes\nabla^E_{\mu}\nabla^E_{\nu}\xi 
+ {1\over2} \Gamma(dx^{\mu})\Gamma(dx^{\nu}) \omega\otimes
R^E(\partial_{\mu},\partial_{\nu})\xi 
\cr}$$
where the Riemannian metric arises {}from the basic relation (3.1) 
of the Clifford algebra; for $\bard^2$, one obtains the same 
formula with $\Gamma$ replaced by $\overline{\Gamma}$. Therefore, 
the difference of the squares of the Dirac operators is
$$
\d^2-\bard^2 = {1\over2}\bigl(\Gamma(dx^{\mu})\Gamma(dx^{\nu})   
-\overline{\Gamma}(dx^{\mu})\overline{\Gamma}(dx^{\nu})\bigr)\,
\omega\otimes R^E(\partial_{\mu},\partial_{\nu})\xi \ ,
$$
which proves the equivalence of the last two statements
of the lemma.  \hfill\qed
\mn
Altogether, Lemmata 3.15 - 3.18 imply the following 
structure theorem for $N=(1,1)$ Dirac bundles:
\mn
\eject
\noindent{\bf Theorem 3.19}\quad Any $N=(1,1)$ Dirac bundle 
$\S$ on $M$, see Definition 3.8, 
is of the form $\S=\bigwedge^{\bullet}M^{\C}\otimes E$
where $E$ is a Hermitian vector bundle over $M$ endowed with 
a flat connection. Conversely, any flat Hermitian vector 
bundle defines an $N=(1,1)$ Dirac bundle.   
\mn
In view of the ideas on the quantum mechanical picture of differential 
geometry outlined in the introduction, it is natural to interpret the 
typical fiber of the bundle $E$ as a space of internal degrees of 
freedom, on which some gauge group may act. If this gauge group is 
of the second kind, dynamical gauge fields with an associated field
strength appear. For this situation, Theorem 3.19 states that the 
algebraic structure of the $N=(1,1)$ Dirac bundle is spoiled by the  
presence of gauge fields with non-vanishing field strength tensors. 
\bn\bn
\leftline{\bf 3.2.4 Reconstruction of differential forms} 
\bn
Let $\S$ be an $N=(1,1)$ Dirac bundle over a Riemannian 
manifold $M$. We associate to $M$ the spectral data
$(C^{\infty}(M), L^2(\S), \d, \bard, \gamma)$, where $\gamma$ 
denotes the $\Z_2$-grading operator on $\S$. As in Lemma
3.4, we see that linear combinations of operators 
$f_0 [\d,f_1] \ldots [\d,f_k] = f_0 \c(df_1)\ldots \c(df_k)
\in \c(\Gamma(Cl^{(k)}(M)))$, with $k\in\N$\ and 
$f_i\in C^{\infty}(M)$, generate the algebra 
$\c(\Gamma(Cl(M)))$ -- and that $\overline{\c}(\Gamma(Cl(M)))$ 
is obtained in the same way using $\bard$. Thus, 
we recover the two anti-commuting Clifford actions
$\c$ and $\overline{\c}$ on $\S$, which will turn out to simplify 
the reconstruction of differential forms {}from the $N=(1,1)$ 
spectral data considerably. First of all, the ideals 
$J^k$ introduced in eq.\ (3.25) become differential ideals 
in the $N=(1,1)$ case, even in the non-commutative generalization, 
see section 5, such that working with the ``reduced'' 
algebra of forms $\Omega^{\bullet}_{{\rm d}}(\a)$ technically 
becomes much easier -- the Dirac operator $D$ {}from the $N=1$ framework 
is now replaced by a nilpotent operator {\tt d}, see eq.\ (3.34). 
Moreover, in the classical case we may even disregard the algebra 
of universal forms completely and define $\Omega^{\bullet}_{{\rm d}}(\a)$
directly without worrying about quotients. In order to show this, 
we first state the following representation theoretic  
\mn
{\bf Lemma 3.20}\quad Let $Cl(V)$ be the (complexified) Clifford
algebra over a Euclidean vector space $V$ of dimension $n$, 
and $\c,\,\overline{\c}$ be two anti-commuting unitary representations 
of $Cl(V)$ on a vector space $W$. Then the map 
$$
\pi(x) := {1\over2}\bigl(\c(x)-i\overline{\c}(x)\bigr)\ \in \End W\ ,
\quad x\in V^{\C}
$$ 
defines a faithful representation of the exterior algebra 
$\bigwedge{}^{\bullet}V^{\C}$ on $W$ by 
$$
\pi(x_1\wedge\ldots\wedge x_k)) := \pi(x_1)\cdots\pi(x_k)\ ,
\quad x_i\in V^{\C}\ .
$$ 
\sn
{\kap Proof}: Let $e_i$ be an orthonormal basis of $V$ and 
define $a^*_i := \pi(e_i)$ as well as $a_i := \pi(e_i)^*$. 
It is easy to verify that the operators $a_i$ satisfy the 
canonical anti-commutation relations (CAR). The result 
follows since a representation of CAR is necessarily 
faithful and the algebra generated by the $a_i^*$ is 
isomorphic to $\bigwedge{}^{\bullet}V^{\C}$.   \hfill\qed
\mn
The result of this lemma trivially extends to coordinate 
neighborhoods on $M$ and, using a partition of unity, to 
$M$ itself. Recalling the form of the operator {\tt d} 
defined in eq.\ (3.34), we see that the algebra of 
differential forms over $M$ is linearly generated by products 
of commutators of functions with {\tt d}, i.e., 
$$
\Omega^k_{\rm d}(\a) = \bigl\lbrace\,  
\sum_{i=1}^N f_0^i \lb\,{\tt d},f_1^i\,\rb \ldots \lb\,{\tt d},f_k^i\,\rb 
\, | \,  f_j^i \in C^{\infty}(M)\,\bigr\rbrace\ .
$$ 
Note that forms of even degree commute with $\gamma$, whereas forms 
of odd degree anti-commute with $\gamma$. Since the operator {\tt d} 
is nilpotent, the exterior derivative of a form $\omega$ 
is directly given by 
$$
   d\,\omega = \lb\,{\tt d},\omega\,\rb_g
\eqno(3.46)$$ 
where $\lb\cdot,\cdot\rb_g$ is the graded commutator. 
\sn
Let us mention here that, in the classical case, the counting of 
the degree of differential forms provides a $\Z$-grading $T$ on 
the set of spectral data $(C^{\infty}(M), L^2({\cal S}), 
{\tt d}, {\tt d}^*, \gamma)$ -- where we have passed {}from $\d,\ 
\bard$ to nilpotent differentials using (3.34); thus we automatically 
obtain a description of classical Riemannian manifolds by a 
{\sl set of  $N=2$ spectral data} $(C^{\infty}(M), L^2({\cal S}), 
{\tt d}, {\tt d}^*, T)$ as introduced in section 1.2. 
\bn\bn
\leftline{\bf 3.2.5 Integration}
\bn
We define the integral and the scalar product on 
differential forms as in the $N=1$ setting:
$$
\Barint \omega :=  \lim_{\varepsilon\to0^+}
{{\rm Tr}\bigl(\omega e^{-\varepsilon \d^2}\bigr)
\over {\rm Tr}\bigl(e^{-\varepsilon \d^2}\bigr)} 
= {1\over{\rm Rk}(S){\rm Vol}(M)} \int_M 
{\rm tr}(\omega(x)) \sqrt{g}d^nx \ .
\eqno(3.47)$$
and 
$$
(\omega,\eta) := \Barint \omega\eta^*
\eqno(3.48)$$
for all differential forms $\omega$ and $\eta$ acting 
on $\S$. The fact that this gives the correct result 
follows {}from  
\mn
{\bf Lemma 3.21} Let $\theta^i$ be and orthonormal basis of 
$T^*_pM$ and let $a_i^*$ denote the action of $\theta^i$ 
on $\S$. Then 
$$
{\rm tr}(a^*_{i_1}\ldots a^*_{i_k}a_{j_1}\ldots a_{j_l})
= {{\rm Rk}(S)\over2^k}\,{\rm sgn}{j_1\ldots j_l \choose i_k\ldots i_1}
$$ 
where $ {\rm sgn}{j_1\ldots j_l \choose i_k\ldots i_1}$ denotes
the sign of the permutation ${j_1\ldots j_l \choose i_k\ldots i_1}$.
\bn\bn
\eject
\leftline{\bf 3.3 Lie groups}
\bn
Before we turn to the various types of complex geometries,
we briefly recall some facts on this important special
case of real Riemannian manifolds. Here, all the geometrical
objects like the metric or the Levi-Civita connection 
can be expressed through Lie algebraic data. 
\sn 
Let $G$ be a compact connected semi-simple Lie group. 
For each $g\in G$, we denote by $L_g$ the natural 
left action of $g$ 
on $G$. The Lie algebra \lie\ of $G$ is 
the space of left-invariant vector fields, i.e.\  
$$
\lie = \bigl\{ X\in \Gamma(TG)\,\big\vert\, L_{g*}X=X \circ L_g
\ \forall g\in G\bigr\}\ ,
$$
which is canonically isomorphic to the tangent space 
$T_eG$ at the unit of $G$: To $X\in T_eG$ associate
the left-invariant vector field $\tilde X$ with
values $\tilde X(h)=L_{h*}X$ for all $h\in G$. 
\sn
The Lie algebra \lie\ acts on itself by the adjoint representation
$$
ad_X(Y) = \lb\, X,Y \,\rb \ ,\quad X,Y\in\lie\ ,
$$ 
and we can introduce 
the symmetric $\lie$-invariant Killing form on \lie, 
$$
\langle X,Y \rangle = {\rm Tr}(ad_X\circ ad_Y)\ ,\ \ X,Y\in\lie
$$ 
which is non-degenerate (since $G$ is semi-simple) and
negative definite. The Killing form provides a 
Riemannian metric on all of $TG$ by putting 
$$
g(X,Y) = - \langle L_{h^{-1}*}X,L_{h^{-1}*}Y \rangle\ ,\ \ 
X,Y \in T_hG \ . 
$$
Thus, $G$ carries a canonical Riemannian structure and the 
general $N=1$ or $N=(1,1)$ formalisms can be applied. 
But let us instead give more concrete expressions for the 
metric and, in particular, for the Levi-Civita connection
as well as the exterior derivative which are special to 
Lie groups with their enhanced structure. 
\sn
To this end, we choose a basis $\{\vartheta_i\}$, $i=1,\ldots,n$, 
of left-invariant vector fields with dual basis $\{\theta_i\}$ 
of $\Gamma(T^*G)$. The structure constants $f^k_{ij}$ of $G$ with
respect to this basis are defined by 
$$
\lb\, \vartheta_i,\vartheta_j\,\rb = f^k_{ij} \vartheta_k\ .
$$
First note that the metric is given in terms of the 
structure constants as
$$
g(\vartheta_i,\vartheta_j) = -f_{il}^k f_{jk}^l\ .
$$
Next, using the \lie-invariance of the Killing form, 
$$
\langle \lb\, X,Y\,\rb,Z\rangle +\langle Y,\lb\, X,Z\,\rb\rangle =0  
$$
for all $X,Y,Z \in\lie$, the conditions of metricity and 
vanishing torsion immediately yield the following 
formula for the 
Levi-Civita connection on $G\,$: 
$$
\nabla\vartheta_i 
= {1\over2} f^j_{ki}\, \theta^k\otimes\vartheta_j\ .
$$
If we define, as previously, two sets of operators on 
$\bigwedge{}^{\bullet}T^*G$  
$$
a^{i*} = \theta^i\wedge\ ,\ \ a_i = \vartheta_i\,\llcorner\ ,
\quad i=1,\ldots,n\ ,
$$
-- which satisfy the CAR --  then we can write the Levi-Civita 
covariant derivative on differential forms as 
$$
\nabla = \theta^i \otimes (\vartheta_i-{1\over2} f^k_{ij}
a^{j*}a_k)\ ;
\eqno(3.49)$$
here $\vartheta_i$ acts as a derivation of the 
coefficient functions: 
$$
\vartheta_i(\omega_{j_1\ldots j_k}\theta^{j_1}\wedge
\ldots\wedge\theta^{j_k}) 
= \vartheta_i(\omega_{j_1\ldots j_k})
\theta^{j_1}\wedge\ldots\wedge\theta^{j_k}\ .
$$
Since $\nabla$ has no torsion, we find a simple expression
for the exterior derivative on $G$ in terms of the 
fermionic operators above and the structure constants 
of the Lie algebra:
$$
d = a^{i*}\nabla_i = a^{i*}\vartheta_i - 
{1\over2} f^k_{ij}\,a^{i*}a^{j*}a_k \ .
\eqno(3.50)$$
Actually, this formula holds for all finite-dimensional
Lie groups independently of the additional requirements 
listed at the beginning of this subsection. 
The operator on the rhs of (3.50) is known
to mathematicians as the coboundary operator in Lie 
algebra cohomology, and to physicists as the BRST charge.
Also infinite-dimensional generalizations of (3.50)
play an important role in string theory. 
\mn
For a compact connected semi-simple Lie group $G$, the 
following identities hold for the lowest cohomology
groups -- see e.g.\ \q{CE} --
$$
H^1(G)=H^2(G)=H^4(G)=0\ ,\quad H^3(G)\neq0\ .  
$$
\bn\bn
\leftline{\bf 3.4 Algebraic formulation of complex geometry}
\bn
In this section, we will discuss the spectral data associated to 
complex, Hermitian and K\"ahler manifolds. Since these data always
contain an $N=(1,1)$ substructure, the reconstruction of the 
Riemannian aspects proceeds along the lines of the previous 
sections, and therefore we only show how the additional features  
are recovered. First, algebraic conditions that tell if  
a Riemannian manifold admits a complex structure are introduced.  
Afterwards, we will turn to Hermitian manifolds and to the 
special case of complex geometry which is the most interesting 
one {}from the string point of view, namely K\"ahler manifolds.
This will lead to what we will call $N=(2,2)$ data. 
In addition, we present an algebraic characterization of 
holomorphic vector bundles and connections in 
section 3.4.4.       \hfill\break
\noindent Although they are also complex manifolds, we will discuss 
Hyperk\"ahler manifolds only in section 3.5, since their 
definition in terms of  algebraic relations displays a 
higher type of supersymmetry called $N=(4,4)$.  \hfill\break 
\noindent  For more definitions and further details on 
complex differential geometry, the reader is referred to 
the standard literature, e.g. \q{Ko, KN, Wel}.  
\bn\bn
\leftline{\bf 3.4.1 Complex manifolds}
\bn
Let $(C^{\infty}(M), L^2(\S), \d, \bard, \gamma)$ be $N=(1,1)$
spectral data associated to a compact Riemannian manifold
$M$. In section 3.2 we have seen that using the operator 
${\tt d}={1\over2}(\d-i\bard)$ we can recover the graded algebra
of differential forms on $M$ as an operator algebra 
$\Omega_{\rm d}^{\bullet}(M)$ on $L^2(\S)$. 
\mn
{\bf Definition 3.22}\quad The $N=(1,1)$ spectral data 
$(C^{\infty}(M), L^2(\S), \d, \bard, \gamma)$ are
called {\sl complex} if there is an operator $T$ on 
$L^2(\S)$ with the following properties:
\smallskip
\item {a)} $G:= \lb\, T,{\tt d}\, \rb$ satisfies $G^2 = 0$; \hfill (3.51)
\smallskip  
\item {b)} $\lb\, T,G \,\rb = G $;  \hfill (3.52)
\smallskip  
\item {c)} $\lb\, T,f \,\rb = 0$ for all $f\in  C^{\infty}(M)$;  \hfill (3.53) 
\smallskip
\item {d)} $\lb\, T,\omega \,\rb \in \Omega^1(M)$ for all 1-forms 
   $\omega\in\Omega^1(M)$ acting on $L^2(\S)$; \hfill (3.54)
\smallskip
\item {e)} put $\overline{G} := {\tt d} - G $ and define 
$$\eqalign{
\Omega^{(1,0)}(M) &= \bigl\{ \sum_{i=1}^N f_0^i\lb\, G,
f^i_1\,\rb\,|\, f^i_j \in C^{\infty}(M),\ N\in\N \bigr\}\ , 
\cr
\Omega^{(0,1)}(M) &= \bigl\{\sum_{i=1}^N f_0^i\lb\,\overline{G},
f^i_1\,\rb\,|\, f^i_j \in C^{\infty}(M),\ N\in\N \bigr\}\ , 
\cr}$$
then the anti-linear map 
$$
\natural\,:\ \cases{&$\Omega^{(1,0)}(M)\lra\Omega^{(0,1)}(M)$ 
\cr
&\ \ $\vphantom{\int}f_0\lb\, G,f_1\,\rb \longmapsto \bar f_0\lb\, \overline{G},
\bar f_1\,\rb$
\cr}\eqno(3.55)$$
is a well-defined isomorphism; here, complex conjugation of functions 
is the ${}^*$-opera\-tion in $C^{\infty}(M)$. 
\mn    
Comparing with the usual  notions of 
complex geometry, one recognizes that $T$ simply counts the
holomorphic degree $p$ of a $(p,q)$ form in the Dolbeault
complex. 
On the whole, Definition 3.22 of a ``complex manifold'' looks 
very different {}from the ones given usually which in particular 
involve the existence of an almost complex structure on 
the tangent bundle. However, the latter notion cannot 
be used conveniently in non-commutative geometry so that we 
are forced to replace it by an algebraic characterization 
which is natural {}from a physical point of view and  
coincides with the usual definition in the classical case  
as shown below. It is not clear whether 
all the five conditions listed above are independent {}from
each other, though there are examples of Riemannian manifolds 
without complex structure that satisfy a) through d) but not e). 
\mn
\eject
\noindent {\bf Lemma 3.23}\quad The following identities hold 
for complex spectral data:
\smallskip
\item {1)} $\lb\, T,\overline{G}\,\rb =0$
\smallskip
\item {2)} $\{\, G,{\tt d}\,\} =0$
\smallskip
\item {3)} $\overline{G}{}^2 =0$
\smallskip
\item {4)} $\{\, G,\overline{G}\,\} =0$
\sn
{\kap Proof}: 1) $\lb\, T,\overline{G}\,\rb =\lb\, T,{\tt d}-G \,\rb = 0$ 
by (3.51) and (3.52).  \hfill\break
\noindent 2) $\{ G,{\tt d}\} = \{\lb\, T,{\tt d} \,\rb,{\tt d}\} = 0$ by the
Jacobi identity and since ${\tt d}$ is nilpotent. \hfill\break
\noindent 3)  $\overline{G}^2 = ({\tt d}-G)^2=0$ by (3.51) and 
part 2) of the lemma.    \hfill\break
\noindent 4) $\{ G,\overline{G}\} = \{G,{\tt d}-G\} =0$ by (3.51) 
and 2).    \hfill\qed
\mn
For each $f\in C^{\infty}(M)$ we define 
$$
D^{(1,0)}f = \lb\, G,f\,\rb\ ,\quad 
D^{(0,1)}f = \lb\, \overline{G},f\,\rb\ ,\quad
D^{(1,1)}f = \{G,\lb\, \overline{G},f\,\rb\}\ ,\quad
\eqno(3.56)$$
and for $r,s\in\Z_+$ we set 
$$
\Omega^{(r,s)}(M) = \bigl\{ \sum_{i=1}^N f_0^i D^{(\alpha_1,\beta_1)}f^i_1
\ldots D^{(\alpha_k,\beta_k)}f^i_k\,\big|\, f^i_j \in C^{\infty}(M),\ N\in\N,\ 
{\scriptstyle \sum_n\alpha_n = r\,,\ \sum_n\beta_n = s}\,\bigr\}\ . 
\eqno(3.57)$$
Note that this definition of the vector space of differential forms
is symmetric in $G$ and $\overline{G}$ in spite of the choice made  
in defining $D^{(1,1)}$ above, since the Jacobi identity yields 
$$
\{G,\lb\, \overline{G},f\,\rb\} =- \{\overline{G},\lb\, G,f\,\rb\}
$$
for all $f\in C^{\infty}(M)$.    
\mn
{\bf Proposition 3.24}\quad The space of $p$-forms decomposes
into a direct sum 
$$
\Omega^p(M) = \bigoplus_{r+s=p} \Omega^{(r,s)}(M)\ ,
\eqno(3.58)$$
and the induced bi-grading on $\Omega^{\bullet}(M)$ is
compatible with the operators $G$ and $\overline{G}$, i.e.
$$\eqalignno{
&\lb\, G,\omega\,\rb_g \in \Omega^{(r+1,s)}(M)\ ,
&(3.59)\cr
&\lb\, \overline{G},\omega\,\rb_g \in \Omega^{(r,s+1)}(M)\ ,
&(3.60)\cr}$$
for any $\omega\in\Omega^{(r,s)}(M)$; as before, $\lb\cdot,
\cdot\,\rb_g$ denotes the graded commutator. 
\sn
{\kap Proof}: The conditions (3.52,53) on the operator $T$ 
and part 1) of Lemma 3.23 imply that 
$\lb\, T,D^{(\alpha,\beta)}f \,\rb = \alpha D^{(\alpha,\beta)}f$, 
and therefore we have  $\lb\, T,\omega\,\rb = r \omega$ for
any $\omega\in\Omega^{(r,s)}(M)$, which proves that the summands 
in the rhs of (3.58) have trivial intersection.  
Obviously, ${\tt d}=G+\overline{G}$ implies the inclusion 
$$
\Omega^p(M) \subset \bigoplus_{r+s=p} \Omega^{(r,s)}(M)\ ;
$$
moreover, in the case $p=1$ the conditions (3.53) and (3.54) give, 
for any function $f\in C^{\infty}(M)$, 
$$
\lb\, G,f \,\rb = \bigl\lb\lb\, T,{\tt d}\,\rb,f\,\bigr\rb =  
\bigl\lb\, T,\lb\, {\tt d},f\,\rb\bigr\rb \in \Omega^1(M)
$$
and
$$
\lb\, \overline{G},f \,\rb = \lb\, {\tt d},f\,\rb - \lb\, G,f\,\rb \in \Omega^1(M)
$$
which shows that $\Omega^1(M)=\Omega^{(1,0)}(M)\oplus\Omega^{(0,1)}(M)$. 
For an arbitrary $\omega\in\Omega^{(0,1)}(M)$ we have 
$$
\lb\, G,\omega\,\rb_g = \bigl\lb\lb\, T,{\tt d}\,\rb,\omega\,\bigr\rb_g = 
-\bigl\lb\lb\, {\tt d},\omega\,\rb_g,T\,\bigr\rb + \bigl\lb\lb\,\omega,T\,\rb,
{\tt d}\,\bigr\rb_g \in \Omega^{2}(M) 
$$
and this gives, in particular, that $\bigl\{G,\lb\,\overline{G},f\,\rb\bigr\} 
\in \Omega^2(M)$ for all $f\in C^{\infty}(M)$. Since $\Omega^{(1,0)}(M)$, 
$\Omega^{(0,1)}(M)$ and $\Omega^{(1,1)}(M)$ generate all of 
$\bigoplus_{r,s}\Omega^{(r,s)}(M)$, equality (3.58) follows. The next
statement, eq.\ (3.59), is established by the computation    
$$
\bigl\lb\, T,\lb\, G,\omega\,\rb_g\,\bigr\rb  = 
-\bigl\lb\, G,\lb\,\omega,T\,\rb\bigr\rb_g + (-1)^{r+s+1} \bigl\lb\,\omega,\lb\, T,
G\,\rb\bigr\rb_g = (r+1) \lb\, G,\omega\,\rb_g\ .
$$
Finally, (3.60) follows {}from 
$$
\lb\, \overline{G},\omega\,\rb_g = \lb\, {\tt d},\omega\,\rb_g - \lb\, G,\omega\,\rb_g 
\in \Omega^{r+s+1}(M)
$$ 
and 
$$ 
\bigl\lb\, T,\lb\,\overline{G},\omega\,\rb_g\,\bigr\rb = r \lb\, \overline{G},
\omega\,\rb_g 
$$ 
for $\omega\in\Omega^{(r,s)}(M)$.   \hfill\qed
\mn
To make contact with the usual definition of complex manifolds, 
let us define a $\C$-linear operator $J$ on $\Omega^1(M)$ by 
$$
J|_{\Omega^{(1,0)}(M)} = i\ ,\quad J|_{\Omega^{(0,1)}(M)} = - i\ .
$$
\mn
{\bf Lemma 3.25}\quad The operator $J$ on $\Omega^1(M)$ is an 
almost complex structure on $M$, i.e.\  it is a real operator 
with square $-\one\,$:
\smallskip
\item {1)} $(J\omega)^{\natural}  = J(\omega^{\natural})$
\smallskip
\item {2)} $ J^2 = -\one$
\sn
{\kap Proof}: Property 2) is trivial. Let $\omega\in\Omega^{1}(M)$, then
by Proposition 3.24 we can write 
$$
\omega = \sum_{i=1}^N f_0^i \lb\, G,f_1^i\,\rb + 
\sum_{j=1}^{N'} g_0^j \lb\, \overline{G},g_1^j\,\rb 
$$
for some functions $f^i_k, g^j_k \in C^{\infty}(M)$; now we simply 
insert the definition (3.55) of the anti-linear map $\natural$: 
$$
(J\omega)^{\natural} = \bigl(i \sum_{i=1}^N f_0^i \lb\, G,f_1^i\,\rb -i 
\sum_{j=1}^{N'} g_0^j \lb\, \overline{G},g_1^j\,\rb \bigr)^{\natural}
=  -i \sum_{i=1}^N \bar f_0^i \lb\, \overline{G},\bar f_1^i\,\rb + i 
\sum_{j=1}^{N'} \bar g_0^j \lb\, G,\bar g_1^j\,\rb 
= J(\omega^{\natural})
$$\hfill\qed
\eject
\noindent{\bf Theorem 3.26}\quad The operator $J$ on $\Omega^1(M)$ 
is a complex structure, i.e.\ $M$ is a complex manifold.
\sn
{\kap Proof}: It only remains to show that the almost complex 
structure $J$ is integrable: Since ${\tt d}=G+\overline{G}$, we 
have 
$$
\lb\, {\tt d},\omega\,\rb_g \in \Omega^{(r+1,s)}(M) \oplus \Omega^{(r,s+1)}(M)
$$
for $\omega \in \Omega^{(r,s)}(M)$, and this is equivalent to 
the required property of $J$; see e.g.\ \q{KN}.  \hfill\qed
\mn
Theorem 3.26 shows that if complex $N=(1,1)$ spectral data as 
defined in 3.22 are associated to a compact Riemannian 
manifold $M$, then the latter is indeed a complex manifold. 
\bn
Let us add as a remark that we could have given an alternative algebraic 
definition of complex manifolds by postulating the existence of two 
commuting operators $T$ and $\overline{T}$ on the space of forms 
which satisfy conditions c) and d) of Definition 3.22 and such that 
for $G := \lb\,T, {\tt d}\,\rb$ and $\overline{G}:= \lb\,\overline{T}, 
{\tt d}\,\rb$ the conditions a) and b) hold. The fifth requirement 
of Definition 3.22 can be replaced by the condition that $T$ and 
$-\overline{T}$ are unitarily equivalent by the Hodge operator. 
For the non-commutative generalization of complex geometry, see 
section 5.3, we will in fact work with such sets of data. 
\bn\bn
{\bf 3.4.2 Hermitian manifolds}
\bn
Here, we describe the additional structure needed on a 
set of complex $N=(1,1)$ spectral data in order to make
the underlying manifold -- in the classical case -- into 
a Hermitian manifold. Not too surprisingly, the relevant 
condition is one of Hermiticity. 
\mn
{\bf Definition 3.27}\quad The complex $N=(1,1)$ 
spectral data $(\cinfty,L^2(\S),G,\overline{G},T,\gamma)$
are called {\sl Hermitian} if the 
operator $T$ is self-adjoint. 
\mn
In the following, we shall prove that a manifold giving 
rise to Hermitian spectral data is indeed Hermitian. 
Recall that by Theorem 3.19, any $N=(1,1)$ bundle $\S$
is of the form 
$$ 
\S = \bigwedge{}^{\bullet}M^{\C}\otimes E
$$
where $E$ is a Hermitian vector bundle equipped with 
a flat Hermitian connection $\nabla^E$; the Clifford 
actions $\c$ and $\overline{\c}$ act trivially on $E$. 
\mn
{\bf Lemma 3.28}\quad The operator $T$ on $\Gamma(\S)$ 
has degree zero with respect to the natural grading 
on $\S$, i.e.\ 
$$
T\,:\ \Gamma(\bigwedge{}^{p}M^{\C}\otimes E) \lra 
\Gamma(\bigwedge{}^{p}M^{\C}\otimes E)  \ .
$$
\sn
{\kap Proof}: For any $\omega\in\Gamma(T^*M^{\C})$, denote by $a^*(\omega)$
the wedging operator by $\omega$, and set $a(\omega) := 
\bigl(a^*(\overline{\omega})\bigr)^*$, i.e.\ $a(\omega)$ acts
as contraction by $\omega$. Since $a^*(\omega)\in\Omega^1(M)$
regarded as the space of 1-forms acting on $\Gamma(\S)$, 
condition (3.54) yields
$$
\lb\, T, a^*(\omega) \,\rb = a^*(T\omega)
$$
for some 1-form $T\omega\in\Gamma(T^*M^{\C})$, and since by definition 
$T$ is self-adjoint, we get 
$$
\lb\, T, a(\overline{\omega}) \,\rb = - a(\overline{T\omega})\ .
$$
This is already sufficient to prove the lemma for $p=0$, since 
it implies that for any $\omega\in\Gamma(T^*M)$ and 
$\xi\in\Gamma(E)$ -- 1 is the constant function on $M$ --  
$$
a(\omega)T(1\otimes\xi) = \lb\, a(\omega),T\,\rb (1\otimes\xi) 
+ T (a(\omega)1 \otimes \xi) = a(\overline{T\overline{\omega}}) 1
\otimes\xi = 0\ ,
$$
i.e.\ $T(1\otimes\xi)$ is annihilated by all contraction operators 
with 1-forms and therefore 
$$
T (1\otimes\xi)  \in \Gamma(\bigwedge{}^{0}M^{\C}\otimes E) \ .
$$
Now take $\omega\in\Gamma(\bigwedge{}^{p}M^{\C})$ and $\xi\in\Gamma(E)$, 
then 
$$
T (\omega\otimes\xi) = \lb\,T, a^*(\omega)\,\rb\, (1\otimes\xi) 
+ a^*(\omega)T(1\otimes\xi) \ , 
$$
and since $\lb\, T,a^*(\omega)\,\rb \in \Omega^p(M)$ by (3.54), we 
obtain the desired result 
$$
T (\omega\otimes\xi)\in  \Gamma(\bigwedge{}^{p}M^{\C}\otimes E)\ .
$$
\hfill\qed
\mn
{\bf Theorem 3.29}\quad A compact Riemannian manifold $M$
corresponding to Hermitian spectral data is Hermitian.
\sn
{\kap Proof}: Since Hermitian spectral data are in particular complex 
$N=(1,1)$ data, the mani\-fold is complex by the results of the 
last subsection. Hermiticity of $M$ is equivalent to the fact 
that the spaces $\bigwedge{}^{(1,0)}M^{\C}$ 
and $\bigwedge{}^{(0,1)}M^{\C}$ are orthogonal to each other with 
respect to the sesqui-linear form induced by the Riemannian metric. 
By (3.53), the operator $T$ acts on the fibers, and since it is 
self-adjoint we can choose an eigenvector 
$\xi\in\Gamma(\bigwedge{}^{\bullet}M^{\C}\otimes E)_x$ of $T$ at 
$x\in M$ with eigenvalue $\lambda\in\R$, i.e.\ 
$$ 
T\xi=\lambda\xi\ .
$$ 
Let $\omega\in \bigwedge{}^{(1,0)}T_x^*M$ and $\eta\in
\bigwedge{}^{(0,1)}T_x^*M$; then (3.52) and Lemma 3.23, statement 1),
imply that   
$$\eqalign{
T(\omega\otimes\xi) &= (\lambda+1)\,\omega\otimes\xi\ ,
\cr
T(\eta\otimes\xi) &= \lambda\,\eta\otimes\xi\ .
\cr}$$
Since $T$ is self-adjoint, $\omega\otimes\xi$ and $\eta\otimes\xi$
are orthogonal to each other, and with Lemma 3.16 we obtain 
$\langle \omega,\eta\rangle =0$.    \hfill\qed
\mn
As was remarked at the end of section 3.4.1, we could also require 
the existence of two commuting and, in the Hermitian case, self-adjoint 
operators $T$ and $\overline{T}$ on \h. Equivalently, the approach of 
section 1.2 could have been chosen, involving U(1) generators 
$T_{\rm tot}$ and $J_0$, the second operator being directly related 
to the complex structure $J$. 
\bn\bn
{\bf 3.4.3 K\"ahler manifolds and Dolbeault cohomology} 
\bn
Let $M$ be a Hermitian manifold with Riemannian metric $g$ and 
complex structure $J$. We define the {\sl fundamental 2-form}
$\Omega$ by 
$$
\Omega(X,Y) = g(JX,Y)
$$
for all $X,Y \in TM$. It is readily verified that $\Omega$ 
is a (1,1)-form on $M$. 
\mn
{\bf Definition 3.30}\quad A Hermitian manifold $M$ is 
a {\sl K\"ahler manifold} if its fundamental 2-form 
is closed, i.e.\ 
$$ 
d\Omega =0\ .
$$
In this case the fundamental form is called {\sl K\"ahler 
form}. 
\mn
The following alternative characterization of K\"ahler 
manifolds is very useful for computations (for a proof 
see e.g.\ \q{KN,Wel}).
\mn
{\bf Theorem 3.31}\quad A Hermitian manifold is K\"ahler 
if and only if for each $p\in M$ there exists a system 
$\{z^{\mu}\}$ of holomorphic 
coordinates around $p$, also called 
{\sl complex geodesic system}, such that 
$$
g_{\mu\bar\nu}(p)=\delta_{\mu\bar\nu}\ ,\quad  
\partial_{\lambda}g_{\mu\bar\nu}(p)= 0 \ ,\quad  
\partial_{\bar\lambda} g_{\mu\bar\nu}(p)=  0\ .
$$
\mn
To a compact Hermitian manifold of complex dimension $n$ we can 
associate the canonical Hermitian spectral data 
$\S=\bigwedge{}^{\bullet}M^{\C}$, $G=\partial$, $\overline{G}
=\bar\partial$, and the operator $T$ acting on 
$\bigwedge{}^{(p,q)}M^{\C}$ by multiplication with $p-{n\over2}$. 
Whether such a manifold -- i.e. such a set of spectral data --
is K\"ahler, can also be formulated in an algebraic fashion: 
\mn
{\bf Theorem and Definition 3.32}\quad The following statements are 
equivalent: 
\smallskip
\item {1)} The manifold $M$ is K\"ahler.
\smallskip
\item {2)} $\{\partial,\bar\partial^*\}=0$.
\smallskip
\item {3)} The holomorphic and anti-holomorphic Laplacians 
$\square\, = \{\partial,\partial^*\}$ resp.\ $\overline{\square}=
\{\bar\partial,\bar\partial^*\}$ coincide: 
$$
\square\, = \overline{\square}\ .
$$
\sn
Hermitian spectral data with differential operators $G\equiv \partial$ 
and $\overline{G}\equiv\bar\partial$ that satisfy conditions
2) and 3) will be called $N=(2,2)$ or {\sl K\"ahler spectral data}. 
\sn 
{\kap Proof}: It is a well-known fact that 1) implies 2) and 3) -- 
see e.g.\ \q{Wel}. In order to show the converse, we choose local 
coordinates $z^{\mu},\ \bar z^{\mu}$, and use the operators 
$a^{\mu\,*} = dz^{\mu}\wedge\,,\ a^{\bar\mu\,*} = d\bar z^{\mu}\wedge$
as before; together with their adjoints, they satisfy the relations 
$$
\{\,a^{\mu\,*}, a_{\nu}\,\} = \{\,a^{\bar\mu\,*}, a_{\bar\nu}\,\} = 
\delta^{\mu}_{\nu}\ ,
$$
while all other anti-commutators vanish; of course, $a_{\mu} = 
g_{\mu\bar\nu}a^{\bar\nu}$ etc.    \hfill\break 
\noindent    {}From the decomposition of ${\tt d}^* = -a^{\mu}\nabla_{\mu}
- a^{\bar\mu}\nabla_{\bar\mu}$ -- here, $\nabla$ is the Levi-Civita 
connection, therefore no torsion terms appear in this formula -- 
into ${\tt d}^* = \partial^* + \overline{\partial}{}^*$ we obtain 
the expressions
$$\eqalign{
\partial^* &= -a^{\bar\mu}\partial_{\bar\mu} 
+ \Gamma^{\bar\sigma}_{\bar\mu\bar\nu} g^{\bar\mu\rho} 
a_{\rho}a^{\bar\nu\,*}a_{\bar\sigma} 
+ \Gamma^{\sigma}_{\mu\bar\nu} g^{\mu\bar\rho} 
a_{\bar\rho}a^{\bar\nu\,*}a_{\sigma} 
+ \Gamma^{\sigma}_{\bar\mu\nu} g^{\bar\mu\rho} 
a_{\rho}a^{\nu\,*}a_{\sigma} \ ,
\cr
\overline{\partial}{}^* &= -a^{\mu}\partial_{\mu} 
+ \Gamma^{\sigma}_{\mu\nu} g^{\mu\bar\rho} 
a_{\bar\rho}a^{\nu\,*}a_{\sigma} 
+ \Gamma^{\bar\sigma}_{\bar\mu\nu} g^{\bar\mu\rho} 
a_{\rho}a^{\nu\,*}a_{\bar\sigma} 
+ \Gamma^{\bar\sigma}_{\mu\bar\nu} g^{\mu\bar\rho} 
a_{\bar\rho}a^{\bar\nu\,*}a_{\bar\sigma} \ .
\cr}$$
The $\Gamma$'s are the Christoffel symbols of the Levi-Civita 
connection; since $g_{\mu\nu}=g_{\bar\mu\bar\nu}=0$, they satisfy 
$\Gamma^{\bar\sigma}_{\mu\nu} = \Gamma^{\sigma}_{\bar\mu\bar\nu} 
= 0$. \hfill\break
\noindent Now let us consider the anti-commutator $\{\,\partial,
\overline{\partial}{}^*\,\}\,$, which, locally, is a sum of 
differential operators of order 0 and 1. If condition 2) of the 
theorem holds, these must vanish separately, and this leads to equations
on the Christoffel symbols. With the help of the operators $a^{\mu\,*}$
etc., we compute for the first order part (with $\partial=a^{\mu\,*}
\partial_{\mu}\,$)
$$
\{\,\partial,\overline{\partial}{}^*\,\} = 
\bigl(- \partial_{\mu}g^{\sigma\bar\nu} 
- g^{\rho\bar\nu}\Gamma^{\sigma}_{\rho\mu} 
+ g^{\bar\rho\sigma}\Gamma^{\bar\nu}_{\bar\rho\mu} \bigr)\,
a^{\mu\,*}a_{\bar\nu}\partial_{\sigma}\,+\ldots\ ,
$$
where the dots substitute for the zeroth order terms that were omitted. 
Thus, we arrive at the implication 
$$
\{\,\partial,\overline{\partial}{}^*\,\} = 0\ \Longrightarrow\ 
\partial_{\mu}g^{\sigma\bar\nu} = - g^{\rho\bar\nu}
\Gamma^{\sigma}_{\rho\mu} + g^{\bar\rho\sigma}
\Gamma^{\bar\nu}_{\bar\rho\mu}\ .
$$
But since the Levi-Civita connection is metric, i.e.\ 
$$
\nabla g = 0\ \Longleftrightarrow\ 
\partial_{\mu}g^{\sigma\bar\nu} = - g^{\rho\bar\nu}
\Gamma^{\sigma}_{\rho\mu} -  g^{\bar\rho\sigma}
\Gamma^{\bar\nu}_{\bar\rho\mu} \ ,
$$
we further conclude that 
$$
\Gamma^{\bar\nu}_{\mu\bar\rho} = 
\overline{\Gamma^{\nu}_{\bar\mu\rho}} = 0
$$
so that the only non-vanishing Christoffel symbols are 
$\Gamma^{\sigma}_{\mu\nu}$ and $\Gamma^{\bar\sigma}_{\bar\mu\bar\nu}$.
This is precisely the condition for the complex structure to be 
covariantly constant, thus the manifold $M$ is K\"ahler. \hfill\break
\noindent Let us now show that also condition 3) leads to this consequence. 
In the same fashion as before, we calculate, locally,  the first order parts 
of the operators $\square\,$ and $\overline{\square}\,$. We find
$$\eqalign{
\{\,\partial,\partial^*\,\} &=  
-\partial_{\mu}g^{\bar\sigma\nu}a^{\mu\,*}a_{\nu}\partial_{\bar\sigma} 
+\bigl( g^{\bar\rho\sigma}\Gamma^{\bar\nu}_{\bar\rho\bar\mu} - g^{\rho\bar\nu}
\Gamma^{\sigma}_{\rho\bar\mu} \bigr)\, a^{\bar\mu\,*}a_{\bar\nu}\partial_{\sigma}
+ \bigl( g^{\bar\rho\sigma}\Gamma^{\nu}_{\bar\rho\mu} - g^{\bar\rho\nu}
\Gamma^{\sigma}_{\bar\rho\mu} \bigr)\, a^{\mu\,*}a_{\nu}\partial_{\sigma}
\cr
&\quad+ 2 g^{\mu\bar\nu}\Gamma^{\sigma}_{\mu\bar\nu}\partial_{\sigma}\,+\ldots\ ,
\cr}$$
where the dots indicate zeroth and second order differential operators which
we do not need to know explicitly. The expression for $\overline{\square}\,$ is 
simply obtained by complex conjugation of the above.   \hfill\break
\noindent Assume that condition 3) holds. Then, $\square
-\overline{\square}=0$, and, in particular, the term in the difference which 
is proportional to the operator 
$a^{\bar\mu\,*}a_{\bar\nu}\partial_{\sigma}$ has to vanish:
$$
\bigl( g^{\bar\rho\sigma}\Gamma^{\bar\nu}_{\bar\rho\bar\mu} -  g^{\rho\bar\nu}
\Gamma^{\sigma}_{\rho\bar\mu} + \partial_{\bar\mu} g^{\sigma\bar\nu} \bigr)\,
a^{\bar\mu\,*}a_{\bar\nu}\partial_{\sigma} = 0 \ .
$$
But this coefficient is just the complex conjugate of the first order part 
of $\{\,\partial,\overline{\partial}{}^*\,\}$ computed above, so we may in the 
same way conclude that $M$ is K\"ahler if $\square =\overline{\square}\,$. \hfill\qed 
\mn
For a Hermitian manifold, we can study not only ordinary 
de Rham theory of differential forms, but also the 
cohomology of the complex 
$$
\Gamma(\bigwedge{}^{(p,0)}M^{\C}) 
{\buildrel \bar\partial \over \lra} 
\Gamma(\bigwedge{}^{(p,1)}M^{\C}) 
{\buildrel \bar\partial \over \lra} 
\ldots
{\buildrel \bar\partial \over \lra} 
\Gamma(\bigwedge{}^{(p,q)}M^{\C}) 
{\buildrel \bar\partial \over \lra} 
\Gamma(\bigwedge{}^{(p,q+1)}M^{\C}) 
{\buildrel \bar\partial \over \lra} 
\ldots 
$$
which is called {\sl Dolbeault cohomology} and denoted by 
$H^{(p,\bullet)}(M)$; the dimensions $h^{p,q}$ 
of the spaces $H^{(p,q)}(M)$ are called {\sl Hodge numbers}. 
It is customary to present the Hodge numbers of a Hermitian 
manifold of complex dimension $n$ in the form of 
the so-called {\sl Hodge diamond}:
\def\phan{\phantom{xxxxxxxxxx}}\def\pha{\phantom{xxxx}}
\def\phant{\phantom{xxxxxxxxxxx}}\def\ph{\phantom{xxx}}
$$\eqalign{
&\phan\ h^{n,n}\vphantom{\sum} \cr
&\pha\ \ \  h^{n,n-1}\ph h^{n-1,1}\cr 
\noalign{\vskip -.2truecm}
&\ph\oddots\phant\ph\ddots\cr 
&h^{n,0}\ \ \ \ h^{n-1,1}\ \ \cdots\ \ \   
 h^{1,n-1}\ \ \ \ h^{0,n}\cr
\noalign{\vskip -.2truecm}
&\ph\ddots\phant\ph\oddots\cr
&\pha\ \ \  h^{1,0}\ph\ \ \ \ \ h^{0,1}\cr  
&\phan\ h^{0,0}\vphantom{\sum}\cr
}$$
If the manifold is K\"ahler there are several 
equations between these numbers as well as the 
{\sl Betti numbers} $b^p = {\rm dim}\,H_{\rm dR}^p(M)$:
\mn
{\bf Theorem 3.33}\quad Let $M$ be a K\"ahler manifold 
of complex dimension $n$; then
\smallskip 
\item {1)} $h^{p,q} = h^{q,p}\,$; 
\smallskip
\item {2)} $h^{p,q} = h^{n-p,n-q}\,$;
\smallskip
\item {3)} $b^p = \sum_{r+s=p} h^{r,s}\,$;
\smallskip
\item {4)} $b^{2p-1}$ is even;
\smallskip 
\item {5)} $b^{2p}\geq 1$ for all $1\leq p \leq n\,$. 
\sn
{\kap Proof}: {}From the Hodge decomposition theorem, see e.g.\ \q{Wel},  
$$
\Gamma(\bigwedge{}^{(p,q)}M^{\C}) = {\cal H}^{(p,q)}(M^{\C}) 
\oplus \bar\partial\Gamma(\bigwedge{}^{(p,q-1)}M^{\C}) 
\oplus \bar\partial^*\Gamma(\bigwedge{}^{(p,q+1)}M^{\C})\ ,
$$
and the fact that the Laplace operators satisfy
$$
\square = \overline{\square} = {1\over2}\triangle
$$ 
we obtain statement 1) using complex conjugation, 
statement 2) with the help of the Hodge $*$-operator, and
statement 3) using $\overline{\square}={1\over2}\triangle$;  
fact 4) is a consequence of 1) and 3), and finally 5) 
follows {}from the existence of 
the K\"ahler form $\Omega$ along with its powers up to
$\Omega^{\wedge n}$,  which are closed but not 
exact since the latter is the volume form on $M$. \hfill\qed
\mn
A subclass of K\"ahler manifolds which is of particular 
interest for superstring theory is given by the so-called 
Calabi-Yau manifolds.
\sn
{\bf Definition 3.34}\quad A {\sl Calabi-Yau manifold} is a
compact K\"ahler manifold whose first Chern class vanishes.
\sn
It is clear that a Ricci flat K\"ahler manifold has vanishing 
first Chern class; that the converse is also true was 
conjectured by Calabi \q{Cal} and proven by Yau \q{Y}: 
\mn
{\bf Theorem 3.35}\quad Let $M$ be a compact K\"ahler 
manifold with vanishing first Chern class and $\Omega$ 
be its K\"ahler form. Then there exists a unique 
Ricci flat metric whose K\"ahler form is in the same
cohomology class as $\Omega$. 
\mn
Calabi-Yau manifolds have a number of striking properties. 
We list two of them in the following theorem, the proof 
of which can be found e.g.\ in \q{Can}.
\sn
{\bf Theorem 3.36}\quad Let $M$ be a Calabi-Yau manifold 
of complex dimension $n$ such that the Euler characteristic 
$\chi(M)\neq 0$.     
Then the first Betti number of $M$ vanishes, 
$$
b_1 = 2h^{1,0} = 0\ .
$$
Furthermore, there exists a harmonic $(n,0)$-form on $M$
which is covariantly constant with respect to the Ricci 
flat metric. In particular, for $n=3$, one has $h^{3,0}=1$.
\mn
The existence of a covariantly constant $(n,0)$-form 
supplies a further symmetry of the Hodge numbers, 
$$
h^{p,0} = h^{n-p,0}\ ,
$$
which -- after some additional work -- permits to fix 
the Hodge diamond of a  Calabi-Yau three-fold with 
$\chi(M)\neq0$ up to two numbers:
\def\phn{\phantom{xxxxxxxxx}}\def\pha{\phantom{xxxx}}
\def\ph{\phantom{xxx}}
$$\eqalign{
&\phn 1 \cr
&\pha\ \ \  0\ph\ph 0\cr 
&\ \ \ \ \ 0\ \pha  h^{1,1}\ \ \ \ \ \ \ 0\cr
&1\ \ \ \ \ \ \  h^{2,1}\ph\ h^{2,1}\ \ \ \ \ \ \ 1\cr  
&\ \ \ \ \ 0\ \pha  h^{1,1}\ \ \ \ \ \ \ 0\cr
&\pha\ \ \ 0 \ph\ph 0\cr 
&\phn 1 \cr
}$$  
It can be shown that, for an arbitrary K\"ahler manifold 
$M$, deformations of the complex and  K\"ahler structures 
are parameterized by elements of $H^{(2,1)}(M)$ and 
$H^{(1,1)}(M)$, respectively, which gives a geometrical
meaning to the two free Hodge numbers above. But let us remark 
that the latter also have a deep interpretation within string 
theory, see e.g. \q{GSW}. Investigations of Calabi-Yau 
manifolds within this context have led to the conjecture 
of a new symmetry among Calabi-Yau manifolds which 
has become important for purely mathematical considerations,
too: The idea of {\sl mirror symmetry} suggests  
that to an $n$-dimensional 
Calabi-Yau manifold $M$ one can associate another 
Calabi-Yau $n$-fold $\widetilde{M}$ such that in 
particular the Hodge numbers satisfy 
$$
h_M^{p,q} = h_{\widetilde{M}}^{n-p,q}\ ;
$$
for $n=3$, this means e.g.\ that deformations of the complex
structure on $M$ correspond to deformations of the 
K\"ahler structure on $\widetilde{M}$ and vice versa. {}From the 
point of view of superconformal quantum field theory, it is almost 
trivial to predict the existence of ``mirrors'' of 
Calabi-Yau manifolds; their explicit construction is already 
much more involved, even within this physical context.
The general investigation of mirror symmetry 
in mathematically rigorous terms is far {}from being complete. 
\bn\bn 
\leftline{\bf 3.4.4 Holomorphic vector bundles and connections}
\bn
Let $(\cinfty,L^2(\S),G,\overline{G},T,\gamma)$
be the canonical spectral data associated to a Hermitian 
manifold $M$, see the paragraph preceding Theorem 3.32. We have 
seen in the previous sections how
to recover the complex geometry {}from these. Now we will 
characterize holomorphic bundles over $M$ as well as 
holomorphic connections on such bundles. 
\sn
The bundle of 1-forms on $M$ decomposes into the 
direct sum
$$
T^*M^{\C} = \bigwedge{}^{(1,0)}M^{\C} 
\oplus \bigwedge{}^{(0,1)}M^{\C} 
$$
of the bundles of holomorphic and anti-holomorphic 1-forms. 
Let $E$ be a complex vector bundle over $M$ with a 
connection $\nabla$, 
$$
\nabla\,:\ \Gamma(E) \lra \Gamma(T^*M^{\C}\otimes E)\ .
$$
We can decompose $\nabla$ according to its range, i.e.\ 
we write $\nabla=\nabla^{(1,0)}+\nabla^{(0,1)}$ with 
$$
\nabla^{(\alpha,\beta)}\,:\ \Gamma(E) \lra 
\Gamma(\bigwedge{}^{(\alpha,\beta)}M^{\C}\otimes E)\ ,
$$
$\alpha,\beta=0,1$. The Leibniz rule for $\nabla$ is 
refined as follows:
$$\eqalign{
\nabla^{(1,0)}(f\xi) &= \partial f\otimes\xi + f\nabla^{(1,0)}\xi
\cr 
\nabla^{(0,1)}(f\xi) &= \bar\partial f\otimes\xi + f\nabla^{(0,1)}\xi
\cr}$$
for all $f\in\cinfty$ and $\xi\in\Gamma(E)$. The characterization 
of holomorphic vector bundles over $M$ is based on the following 
theorem, the proof of which can be found e.g.\ in \q{Ko}: 
\mn
{\bf Theorem 3.37}\quad Let $E$ be a complex vector bundle over 
a complex manifold $M$ and $\nabla$ be a connection on $E$ such 
that the map 
$$
\nabla^{(0,1)}\circ\nabla^{(0,1)}\,:\ \Gamma(E) \lra 
\Gamma(\bigwedge{}^{(0,2)}M^{\C}\otimes E)
$$
vanishes. Then there is a unique holomorphic vector bundle 
structure on $E$ such that $\nabla^{(0,1)}\xi=0$ for any 
local holomorphic section $\xi$ of $E$. 
\mn
Recall that complex vector bundles over $M$ are in one-to-one
correspondence with finitely generated projective modules 
over $\cinfty$. Altogether, this leads us to the following 
definitions:
\mn
{\bf Definition 3.38}\quad A {\sl complex structure} on a finitely 
generated projective left $\cinfty$-module \e\ is a connection 
$\nabla$ on \e\ such that $\nabla^{(0,1)}\circ\nabla^{(0,1)}=0$. 
The pair $(\e,\nabla)$ is then called a {\sl holomorphic vector
bundle}. Any other connection $\widetilde{\nabla}$ on $(\e,\nabla)$ is 
a {\sl holomorphic connection} if 
$$
\nabla - \widetilde{\nabla} \in {\rm Hom}_{\cinfty}\bigl(\e,\Omega^{(1,0)}(M)
\otimes_{\cinfty}\e\bigr)\ .
\eqno(3.61)$$
The complex structures $\nabla$ and $\widetilde{\nabla}$ are called
{\sl equivalent} if (3.61) holds. 
\bn\bn
{\bf 3.5 Hyperk\"ahler manifolds}
\bn 
After having discussed the case of K\"ahler manifolds at some 
length, we will now focus on an even more special type of 
complex geometry whose algebraic characterization will 
involve an $N=(4,4)$ supersymmetry algebra. Hyperk\"ahler 
manifolds are interesting {}from the mathematical point 
of view since they admit a metric with vanishing 
Einstein tensor. They also were discussed in the 
physical literature in the context of $N=4$ supersymmetric 
non-linear sigma models \q{AG,HKLR} on a classical target, 
but it seems that the possibility of a direct 
algebraic interpretation of the Hyperk\"ahler axioms has 
been overlooked by now. This is also the reason why we will 
be more explicit in the proofs contained in this section. 
\bn\bn
\leftline{\bf 3.5.1 Definition and basic properties of 
Hyperk\"ahler manifolds}
\bn
{\bf Definition 3.39}\quad A Riemannian manifold $(M,g)$ 
is a {\sl Hyperk\"ahler manifold} if it carries three 
complex structures $I, J$ and $K$ satisfying the 
quaternion algebra 
$$
IJ=-JI=K\ ,\quad JK=-KJ=I\ ,\quad KI=-IK=J\ ,
\eqno(3.62)$$
and such that $g$ is a K\"ahler metric for $I, J$ and $K$. 
\mn
{}From the representation theory of the quaternion algebra
we conclude that a Hyperk\"ahler manifold 
must satisfy 
$$
{\rm dim}_{\R}M \equiv 0\ {\rm mod}\,4\ .
\eqno(3.63)$$ 
Let $(M,g,I,J,K)$ be a Hyperk\"ahler manifold. Then we can 
consider $(M,g,I)$ as a K\"ahler manifold and decompose the 
complexified tangent bundle $TM$ into its holomorphic and 
anti-holomorphic parts 
$$
TM = T^+M \oplus T^-M\ ,\quad I\big|_{T^{\pm}M} = \pm i\ .
\eqno(3.64)$$
Let $P^{\pm}$ denote the projections $TM\lra T^{\pm}M$; then 
we have {}from (3.62) 
$$\eqalign{
&0 = P^{\pm}\{I,J\}P^{\pm}= \pm 2iP^{\pm}J P^{\pm}
\cr
&KP^{\pm}=  IJP^{\pm}= \mp i P^{\mp}J P^{\pm}
\cr}$$
which implies 
$$
J,K\,:\ T^{\pm}M \lra T^{\mp}M 
\eqno(3.65)$$
and 
$$
K\big|_{T^{\pm}M} = \mp iJ\big|_{T^{\pm}M}\ .
\eqno(3.66)$$
We define the {\sl holomorphic symplectic form} on $M$ by 
$$
\omega(\cdot,\cdot) = {1\over2} g\bigl( (J+iK)\cdot,\cdot\bigr)\ ;
\eqno(3.67)$$ 
if we denote by $\Omega_J$ and $\Omega_K$ the K\"ahler forms 
associated with the complex structures $J$ and $K$, resp.,  
then we have the relation 
$$
\omega = {1\over2} (\Omega_J +i\Omega_K)\ ,
\eqno(3.68)$$  
which shows that the form $\omega$ is indeed closed and -- 
with equations (3.65,66) -- also is a 
holomorphic 2-form (equivalently, $\overline{\omega}$ is an 
anti-holomorphic symplectic form). Moreover, (3.66) allows 
us to write  
$$
\omega(\cdot,\cdot) = g(J\cdot,\cdot)\quad {\rm on}\ T^+M
\eqno(3.69)$$
which shows that $\omega$ is non-degenerate and, therefore, 
that on a Hyperk\"ahler manifold there exists a {\sl 
holomorphic volume form} 
$$
\mu = \omega\wedge\ldots\wedge\omega 
$$ 
(${\rm dim}_{\R}M/4$ factors). 
\sn
Let $U$ be a coordinate neighborhood with holomorphic 
coordinates $z^1,\ldots,z^{n}$. {}From equations (3.65,66,69),
we can obtain the following local expressions which will 
be useful later on:
$$\eqalignno{
&J_{\mu}^{\phantom{\mu}\nu} = 
J_{\bar\mu}^{\phantom{\bar\mu}\bar\nu} =
K_{\mu}^{\phantom{\mu}\nu} = 
K_{\bar\mu}^{\phantom{\bar\mu}\bar\nu} = 0 
\phantom{xxxxxxxxxxx}&(3.70)\cr 
&K_{\mu}^{\phantom{\mu}\bar\nu} = 
-i J_{\mu}^{\phantom{\mu}\bar\nu} 
&(3.71)\cr
&K_{\bar\mu}^{\phantom{\bar\mu}\nu} = 
i J_{\bar\mu}^{\phantom{\bar\mu}\nu}
&(3.72)\cr
&\omega_{\mu\nu} = 
J_{\mu}^{\phantom{\mu}\bar\lambda}g_{\bar\lambda\nu}  =
J_{\mu\nu} 
&(3.73)\cr
&\omega = {1\over2} J_{\mu\nu} dz^{\mu}\wedge dz^{\nu}
&(3.74)\cr}$$
In local coordinates, the complex conjugate of the holomorphic 
symplectic form is given by $\overline{\omega} = 
{1\over2} J_{\bar\mu\bar\nu} dz^{\bar\mu}\wedge dz^{\bar\nu}\,$. 
\bn\bn
\vfil\eject
\leftline{\bf 3.5.2 The \Nfourfour data of a Hyperk\"ahler manifold}
\bn
In this section we prove that the canonical set of $N=(2,2)$
data on a Hyperk\"ahler manifold extends to what we will 
call $N=(4,4)$ spectral data. The canonical 
$N=(2,2)$ data on a K\"ahler manifold are given by the tuple
$(\cinfty, L^2\bigl(\bigwedge^{\bullet}M^{\C}\bigr), \partial,
\bar\partial, T)$, where 
$$
T\big\vert_{\bigwedge{}^{(p,q)}M} 
= p - {1\over2}\,{\rm dim}_{\C} M
\eqno(3.75)$$
counts the holomorphic degree of differential forms -- including 
a ``normalization term'' which makes the spectrum of $T$ symmetric
around zero. Let now $(M, g, I, J, K)$ be a 
Hyperk\"ahler manifold of complex dimension $n$, let 
and $(M, g, I)$ be the underlying K\"ahler 
manifold. We define operators 
$$\eqalignno{
&\square = \{\,\partial,\partial^* \} 
&(3.76)\cr
&G^{1+} = \partial
&(3.77)\cr
&G^{2+} = \lb\, \iota(\overline{\omega}), \partial\,\rb
&(3.78)\cr
&T^1 = {1\over2}\,\bigl(\iota(\overline{\omega})+\epsilon(\omega)\bigr) 
&(3.79)\cr
&T^2 = {i\over2}\,\bigl(\iota(\overline{\omega})-\epsilon(\omega)\bigr)
&(3.80)\cr
&T^3 = {1\over2}\,(p-{n\over2})\quad{\rm on}\ \bigwedge{}^{(p,q)}M
&(3.81)\cr}$$
where $\omega$ denotes the holomorphic symplectic form on $M$, 
$\epsilon(\omega)$ is the wedging operator by $\omega$ and 
$\iota(\overline{\omega})$ its adjoint, i.e.\ contraction by 
$\overline{\omega}$. In addition, for $a=1,2$ we introduce  
$$
G^{a-} = \bigl(G^{a+}\bigr)^* \ .
\eqno(3.82)$$
\mn
{\bf Theorem 3.40}\quad The operators defined by 
eqs.\ (3.76 - 82) satisfy the {(anti-)}\-commutation relations 
$$\eqalignno{
&\lb\, \square\,, G^{a+}\, \rb = 0\ , \quad a=1,2\ , 
&(3.83)\cr
&\lb\, \square\,, T^i\, \rb = 0\ , \quad i=1,2,3\ , 
&(3.84)\cr
&\{\, G^{a+}, G^{b+} \} = 0\ ,\quad a,b = 1,2\ ,  
&(3.85)\cr
&\{\, G^{a-}, G^{b+} \} = \delta^{ab}\,\square\,\ ,\quad a,b = 1,2\ ,  
&(3.86)\cr
&\lb \,T^i, G^{a+} \,\rb = {1\over2}\,\overline{\tau^i_{ab}}\, G^{b+}\ ,
\quad i=1,2,3\,,\ a=1,2\ ,  
&(3.87)\cr
&\lb \,T^i, T^j \,\rb = i \epsilon^{ijk}\, T^k\ ,\quad i,j = 1,2,3\ ;  
&(3.88)\cr}$$
$\tau^i$ are the Pauli matrices defined in the first section. 
In addition, the following Hermiticity conditions hold: 
$$
\square^* = \square\ ,\quad 
\bigl(G^{a\pm}\bigr)^*=  G^{a\mp}\ ,\quad  
\bigl(T^i\bigr)^* = T^i \ .
\eqno(3.89)$$
In particular, these operators generate a finite-dimensional
$\Z_2$-graded (or super) Lie algebra. 
\sn
{\kap Proof}: We begin with the SU(2) commutation relations, eq.\ (3.88). 
Since $\omega$ is a (2,0)-form, the definitions immediately give 
$$\eqalign{
&\lb\, T^3, T^1 \,\rb = {1\over2}\,\bigl(-\iota(\overline{\omega})
+\epsilon(\omega)\bigr) = i T^2 \ , 
\cr
&\lb\, T^3, T^2 \,\rb = {i\over2}\,\bigl(-\iota(\overline{\omega})
-\epsilon(\omega)\bigr) = -i T^1 \ .
\cr}$$
To obtain the last SU(2) commutator, we need the following 
\mn
{\bf Lemma 3.41}\quad $\lb\, \epsilon(\omega),\iota(\overline{\omega})
\,\rb = p - {n\over2}$
\sn
{\kap Proof}: Let $z^{\mu}$ be holomorphic local coordinates on $M$ 
with the properties of Theorem 3.31; set 
$a^{\mu\,*} = \epsilon(dz^{\mu})$ and 
$a^{\bar\mu} = \iota(d\bar z^{\mu})$. These operators satisfy the
anti-commutation relations 
$$
\{ a^{\mu\,*},a^{\nu\,*} \} = \{ a^{\bar\mu}, a^{\bar\nu} \}
=0\ ,\quad \{ a^{\mu\,*}, a^{\bar\nu} \} = g^{\mu\bar\nu}\ .
\eqno(3.90)$$
Then we can write $\epsilon(\omega)= {1\over2} \omega_{\mu\nu}
a^{\mu\,*}a^{\nu\,*}$ and $\iota(\bar\omega) = (\epsilon(\omega))^* 
= -{1\over2} \overline{\omega}_{\bar\rho\bar\sigma}
a^{\bar\rho}a^{\bar\sigma}$. Now the calculation is straightforward:
$$\eqalign{
\lb\, \epsilon&(\omega), \iota(\overline{\omega})\, \rb = 
-{1\over4}\,  \omega_{\mu\nu} \overline{\omega}_{\bar\rho\bar\sigma}
\lb\, a^{\mu\,*}a^{\nu\,*}, a^{\bar\rho}a^{\bar\sigma} \,\rb 
\cr
&= -{1\over4}\,  \omega_{\mu\nu} \overline{\omega}_{\bar\rho\bar\sigma}
\Bigl( g^{\nu\bar\rho}  a^{\mu\,*}a^{\bar\sigma} -
 g^{\nu\bar\sigma}  a^{\mu\,*}a^{\bar\rho} + 
g^{\mu\bar\rho} a^{\bar\sigma} a^{\nu\,*} -  
g^{\mu\bar\sigma} a^{\bar\rho} a^{\nu\,*}  \Bigr)
\cr
&= -{1\over4}\, J_{\mu\nu} \Bigl( 
J^{\nu}_{\phantom{\nu}\bar\sigma} a^{\mu\,*}a^{\bar\sigma} -
J_{\bar\rho}^{\phantom{\rho}\nu}  a^{\mu\,*}a^{\bar\rho} + 
J^{\mu}_{\phantom{\mu}\bar\sigma} a^{\bar\sigma}a^{\nu\,*} -  
J_{\bar\rho}^{\phantom{\rho}\mu} a^{\bar\rho} a^{\nu\,*}  \Bigr) 
\cr
&=-{1\over4}\, \Bigl( - g_{\mu\bar\sigma}  a^{\mu\,*}a^{\bar\sigma} -
 g_{\mu\bar\rho}  a^{\mu\,*}a^{\bar\rho} + 
g_{\nu\bar\sigma} a^{\bar\sigma} a^{\nu\,*} + 
g_{\nu\bar\rho} a^{\bar\rho} a^{\nu\,*}  \Bigr)
\cr
&= {1\over2}\, g_{\mu\bar\nu} \,
        ( a^{\mu\,*}a^{\bar\nu}-a^{\bar\nu}a^{\mu\,*} ) 
= g_{\mu\bar\nu}\,a^{\mu\,*}a^{\bar\nu}- {1\over2}\,g_{\mu\bar\nu} 
g^{\mu\bar\nu} 
\cr
&= p-{n\over2} 
\cr}$$
on $\bigwedge^{(p,q)}M^{\C}$, since $g_{\mu\bar\nu}a^{\mu\,*}a^{\bar\nu}$ 
is the ``number operator''.   \hfill\qed
\sn                                {}From
the preceding lemma, it follows easily that 
$$
\lb\, T^1, T^2\, \rb = {i\over2}\, (p-{n\over2}) = i T^3 
$$
on $\bigwedge^{(p,q)}M^{\C}$, which proves (3.88). 
\sn
To derive eq.\ (3.87), we recall that the symplectic form 
$\omega$ is closed, $\partial\omega=\bar\partial\omega=0$;  
since $\lb\, \epsilon(\omega), \partial\, \rb = 
- \epsilon(\partial\omega) = 0$, we obtain 
$$
\lb\, T^i, G^{1+} \,\rb = {1\over2} \overline{\tau^i_{1a}} G^{a+}\ .
$$
The commutation relations $\lb\, T^i, G^{2+} \,\rb$ will be 
computed with the help of another lemma:
\mn
{\bf Lemma 3.42}\quad $\bigl\lb\, \iota(\overline{\omega}),
\lb\, \iota(\overline{\omega}), \partial\, \rb \bigr\rb = 0$
\sn
{\kap Proof}: Using the same coordinates and notations 
as in the proof of Lemma 3.41, we have 
$$\eqalign{
\lb\, \iota(&\overline{\omega}), \partial\, \rb 
= -{1\over2}\, \overline{\omega}_{\bar\mu\bar\nu}\,
\lb\,a^{\bar\mu}a^{\bar\nu}, \partial\,\rb = 
-{1\over2}\, \overline{\omega}_{\bar\mu\bar\nu}\,
\Bigl( a^{\bar\mu}\,\{\,a^{\bar\nu},\partial\,\} 
- \{\, a^{\bar\mu},\partial\,\}\, a^{\bar\nu} \Bigr) 
\cr
&= -{1\over2}\, \overline{\omega}_{\bar\mu\bar\nu}\,
\Bigl( a^{\bar\mu} \delta^{\bar\nu\lambda} - 
a^{\bar\nu} \delta^{\bar\mu\lambda} \Bigr) \partial_{\lambda}
\cr}$$
at the center of the holomorphic geodesic coordinate system. 
This implies that
$$
\bigl\lb\, \iota(\overline{\omega}),
\lb\, \iota(\overline{\omega}), \partial\, \rb \bigr\rb =
-{1\over2}\, \overline{\omega}_{\bar\rho\bar\sigma}
\overline{\omega}_{\bar\mu\bar\nu} \lb\, a^{\bar\rho}a^{\bar\sigma},
a^{\bar\nu}\, \rb \delta^{\bar\mu\lambda} \partial_{\lambda} = 0\ .
\eqno{\qed}
$$
\sn
Together with the Jacobi identity, the last lemma yields
$$
\lb\, T^1, G^{2+} \,\rb = {1\over2}\,\bigl\lb\,\epsilon(\omega), \lb 
\,\iota(\overline{\omega}), \partial \,\rb \bigr\rb =
{1\over2}\,\bigl\lb \,\partial, \lb \,\iota(\overline{\omega}),
\epsilon(\omega) \,\rb \bigr\rb = {1\over2}\, \partial = 
{1\over2}\,\overline{\tau^1_{21}} G^{1+} 
$$
and analogously 
$$\eqalign{  
&\lb\, T^2, G^{2+} \,\rb = -{i\over2}\,\bigl\lb\,\epsilon(\omega), \lb 
\,\iota(\overline{\omega}), \partial \,\rb \bigr\rb = 
-{i\over2}\,\partial = {1\over2}\,\overline{\tau^2_{21}} G^{1+} \ ,
\cr
&\lb\, T^3, G^{2+} \,\rb = -{1\over2}\,G^{2+} 
= {1\over2}\,\overline{\tau^3_{22}}\, G^{2+} \ ,
\cr}$$
which proves (3.87). We proceed with eq.\ (3.86) in Theorem 3.40, 
$$
\{\, G^{1-}, G^{2+} \,\} 
= \{\, \partial^*, \lb \,\iota(\overline{\omega}),\partial \,\rb \} 
= \bigl\lb\,  \iota(\overline{\omega}),\{\,\partial, \partial^* \,\}\bigr\rb
+ \{\,\partial, \lb\, \partial^*, \iota(\overline{\omega})\,\rb \} 
= \lb\, \iota(\overline{\omega}), \square \,\rb 
$$
where we have used the Jacobi identity and 
$$
\lb\, \partial^*, \iota(\overline{\omega}) \,\rb = 
\lb\, \epsilon(\omega), \partial \,\rb^* = 0\ .
$$
On a K\"ahler manifold, we have $2\,\square = \triangle$, where 
$\triangle$ is the Laplace-Beltrami operator. One of the Hodge 
identities, see e.g.\ \q{Wel}, reads 
$$
\lb\, \triangle, \iota(\Omega)\, \rb = 0 
$$ 
where $\Omega$ denotes the K\"ahler form. Since $\overline{\omega}$ 
is a linear combination of the K\"ahler forms associated to the 
complex structures $J$ and $K$ on the Hyperk\"ahler manifold $M$, 
we have 
$$
\{\, G^{1-}, G^{2+} \} = \lb\, \iota(\overline{\omega}), \square \,\rb = 0\ .
\eqno(3.91)$$
The other commutation relations of the $G^{a\pm}$ are 
$$
\{\, G^{1-}, G^{1+} \} = \{\, \partial^*, \partial \,\} = \square
\eqno(3.92)$$
by definition, and using eqs.\ (3.87, 89, 91, 92) we obtain
$$\eqalign{
\{\, G^{2-}, G^{1+} \} &= \{\, G^{2+}, G^{1-} \} ^* = 0\ ,
\cr
\{\, G^{2-}, G^{2+} \} &= 2 \{\, G^{2-}, \lb\, T^1, G^{1+}\,\rb \} = 
2 \bigl\lb\, T^1, \{\, G^{1+}, G^{2-} \}\bigr\rb
+2 \{\, G^{1+}, \lb\, G^{2-}, T^1 \,\rb \} 
\cr 
&= \{\, G^{1+}, G^{1-} \} = \square\ ,
\cr}$$
which proves (3.86). The remaining equations are much simpler:
$$\eqalignno{
&\{\, G^{1+}, G^{1+} \} = \{\, \partial, \partial \,\} = 0 
&(3.93)\cr 
&\{\, G^{1+}, G^{2+} \} = 2 \{\, G^{1+}, \lb\, T^1, G^{1+}\,\rb \}=0 
&(3.94)\cr}$$
by (3.87) and the Jacobi identity; 
$$\eqalign{
\{\, G^{2+}, G^{2+} \} &= 2 \{\, G^{2+}, \lb\, T^1, G^{1+}\,\rb \}
=2 \bigl\lb\, T^1,  \{\, G^{1+}, G^{2+} \} \bigr\rb +
2 \{\, G^{1+}, \lb\, G^{2+}, T^1\, \rb \}
\cr &= - \{\,  G^{1+}, G^{1+} \} = 0 
\cr}$$
with the help of (3.87, 93, 94) and again the Jacobi identity; 
this proves (3.85). It only remains to show that the Laplace 
operator commutes with all other operators introduced at the
beginning of this section. The Hodge identity  
$$
\lb\, \square, \epsilon(\omega) \,\rb = 
\lb\, \square, \iota(\overline{\omega}) \,\rb = 0\ ,
$$
yields 
$$
\lb\, \square, T^1 \,\rb =  \lb\, \square, T^2 \,\rb = 0\ ,
$$
and $\lb\, \square, T^3 \,\rb = 0 $ follows {}from the fact 
that $\square\,$ has bi-degree $(0,0)$ -- proving (3.84). 
Finally, we have 
$$
\lb\, \square, G^{a+} \,\rb =
\bigl\lb \{\, G^{a+}, G^{a-} \}, G^{a+} \,\bigr\rb = 0 
$$
by (3.85,86) and the Jacobi identity. This completes the
proof of (3.83) and of Theorem 3.40.    \hfill\qed 
\mn
If we define the anti-holomorphic analogues $\overline{\square}\,$,
$\overline{G}{}^{a\pm}$ and $\overline{T}{}^i$ of the previous 
generators $\square, G^{a\pm}, T^i$, we get a second copy of 
the algebra established in Theorem 3.40. These two copies 
(anti-)commute, as is easily verified: It suffices to 
show that $\lb\, \partial, \iota(\omega) \,\rb = 0$ and
$\lb\, \partial, \epsilon(\overline{\omega}) \,\rb = 0$; 
the first equation follows {}from 
$$
\lb\, \partial, \iota(\omega) \,\rb = -{1\over2}\,
\lb\, a^{\mu\,*}\partial_{\mu},\, \omega_{\nu\rho}\, a^{\nu}
a^{\rho} \,\rb = 0
$$
at the center of a holomorphic geodesic system, and the second
equation simply means $\partial\overline{\omega} =0$. This leads 
us to the following 
\mn
{\bf Definition 3.43}\quad  A tuple $(\cinfty, L^2\bigl(
\bigwedge^{\bullet}M^{\C}\bigr), \partial,\bar\partial, T^i, 
\overline{T}{}^i, i=1,2,3)$ 
with operators subject to the relations of Theorem 3.40 and the 
corresponding relations for the  anti-holomorphic analogues -- which
(anti-)commute with the holomorphic operators --  
will be called a set of $N=(4,4)$ {\sl spectral data}. 
\bn\bn
{\bf 3.5.3 Characterization of Hyperk\"ahler manifolds}
\bn
In section 3.4, we have seen that K\"ahler manifolds can be 
algebraically characterized, i.e.\  that one can identify 
the operators $\partial = G^{1+}$ and $T^3 = {1\over2}\,(p-n/2)$ 
on $\bigwedge^{(p,q)}M^{\C}$ along with their anti-holomorphic 
partners, as soon as an $N=(2,2)$ structure on the bundle 
of differential forms is given. In this section, we will prove that 
if the $N=(2,2)$ extends to an $N=(4,4)$ structure, the underlying
manifold is Hyperk\"ahler. 
\mn
{\bf Theorem 3.44}\quad Let $M$ be a K\"ahler manifold and suppose 
that its canonical $N=(2,2)$ bundle carries a realization of two 
(anti-)commuting $N=4$ algebras, eqs.\ (3.83-89), where $G^{1+}, 
\overline{G}^{1+}$ and $T^3, \overline{T}{}^3$ have their usual 
meaning. Then $M$ is a Hyperk\"ahler manifold. 
\sn
{\kap Proof}: We have to show that there exists a holomorphic symplectic 
form $\omega$ on $M$, which by a theorem of A.\ Beauville \q{Beau,Bes} 
implies that $M$ is Hyperk\"ahler. The idea is to identify 
$T^+ := T^1 + i T^2$ with the wedging operator by $\omega$. {}From 
the $N=4$ commutation relations 
$$
\lb\, T^3, T^+ \,\rb = T^+\ ,\quad 
\lb\, \overline{T}^3, T^+ \,\rb = 0
$$
we see that $T^+$ has bi-degree (2,0), i.e.\ it maps 
$\Gamma\bigl(\bigwedge^{(p,q)}M^{\C}\bigr)$ to 
$\Gamma\bigl(\bigwedge^{(p+2,q)}M^{\C}\bigr)$. Furthermore, 
$$
\lb\, G^{1+}, T^+ \,\rb = 0\ ,\quad
\lb\, \overline{G}^{1+}, T^+ \,\rb = 0
$$
show that $\lb\, {\tt d}\,, T^+\,\rb = \lb\, G^{1+}+ 
\overline{G}^{1+}, T^+ \,\rb = 0$. In order to fully identify the 
operator in geometrical terms, we shall need the following 
\mn
{\bf Lemma 3.45}\quad Let $M$ be a compact Riemannian manifold and 
$$
W\,:\ \Gamma\bigl(\bigwedge{}^{\bullet}M^{\C}\bigr) \lra 
\Gamma\bigl(\bigwedge{}^{\bullet}M^{\C}\bigr)
$$
be a $\cinfty$-linear operator of degree $d$, i.e.\ $W$ maps 
$\Gamma\bigl(\bigwedge{}^{p}M^{\C}\bigr)$ to  
$\Gamma\bigl(\bigwedge{}^{p+d}M^{\C}\bigr)$, such that in addition 
$\lb\, W, {\tt d}\,\rb_g =0$. Then $W$ is a wedging operator 
by a closed form of degree $d$ uniquely determined by $W(1)$
where $1 \in \cinfty$ denotes the constant function equal to 1.  
\sn
{\kap Proof}: Fix a $p$-form $\eta$; then there is a finite set 
of functions $a_j^i \in \cinfty$ such that 
$$
\eta = \sum_i a_0^i\,\lb\,{\tt d},a_1^i\,\rb \ldots 
\lb\,{\tt d},a_p^i\,\rb \cdot 1\ .
$$
Using the $\cinfty$-linearity of $W$ and $\lb\, W, {\tt d}\,\rb_g =0\,$, 
we get 
$$\eqalign{
W \eta &= (-1)^{pd} \sum_i a_0^i\,\lb\,{\tt d},a_1^i\,\rb \ldots 
\lb\,{\tt d},a_p^i\,\rb \cdot W(1)
\cr
&= (-1)^{pd} \eta \wedge W(1) = (-1)^{pd} \eta \wedge 
\epsilon(W(1))\cdot 1 
\cr
&= W(1) \wedge \eta  
\cr}$$
which proves that $W$ is a wedging operator by a $d$-form. That 
$W(1)$ is closed follows {}from $\epsilon\bigl({\tt d}\,W(1)\bigr)
=\epsilon\bigl(\,\lb\,{\tt d},W\,\rb_g 
\cdot 1 \bigr) = 0$.     \hfill\qed
\mn
Since the operator $T^+$ satisfies the assumptions of Lemma 3.45, there 
exists a closed (2,0)-form $\omega$ such that $T^+ = \epsilon(\omega)$. 
We can identify this form with a holomorphic symplectic form on $M$ if
we can show that it is non-degenerate: {}From the $N=4$ commutation 
relations, we have 
$$
\lb\, T^+, T^- \,\rb = 2 T^3 = p - {n\over2} 
$$ 
on $\bigwedge^{(p,q)}M^{\C}$, where $n$ is the complex dimension of the 
manifold, and since $T^- = (T^+)^*$, we get $T^- = \iota(\overline{\omega})$. 
For any $f\in\cinfty$, we have 
$$
-{n\over2} f = 2 T^3 f = \lb\, T^+, T^- \,\rb\, f = 
- \iota(\overline{\omega})\epsilon(\omega)\, f 
= -{1\over2}\, \overline{\omega}^{\mu\nu}\omega_{\mu\nu}\, f\ , 
$$
and therefore 
$$
\overline{\omega}^{\mu\nu}\omega_{\mu\nu} = n\ .
$$
Using the last equation, we have for any (1,0)-form $\eta = 
\eta_{\lambda} dz^{\lambda}$
$$
(1-{n\over2})\, \eta = - \iota(\overline{\omega})\epsilon(\omega)\, \eta 
= -{1\over2}\, \overline{\omega}^{\mu\nu}\omega_{\mu\nu}\, \eta 
+ \overline{\omega}^{\mu\lambda}\omega_{\mu\nu}\, \eta_{\lambda} dz^{\nu}
= -{n\over2}\, \eta + \overline{\omega}^{\mu\lambda}
\omega_{\mu\nu}\, \eta_{\lambda} dz^{\nu}
$$
and it follows that $\eta = \overline{\omega}^{\mu\lambda}
\omega_{\mu\nu} \eta_{\lambda} dz^{\nu}$; in other words, the matrix 
$\omega_{\mu\nu}$ is invertible, independent of the coordinate system, 
and thus $\omega$ is non-degenerate. This concludes the proof of 
Theorem 3.44.    \hfill\qed
\bn\bn
\leftline{\bf 3.6 Symplectic geometry}
\bn
The remainder of this third section is devoted to the discussion 
of symplectic manifolds, which constitute an important 
class of manifolds where the central role of the 
Riemannian metric is taken over by the symplectic form. 
Again, there is a description of symplectic geometry in terms 
of a certain set of algebraic data, which turn out to be 
``close'' to the ones characterizing a K\"ahler manifold. 
More precisely, one can always choose a Riemannian metric 
$g$ on a manifold $M$ with given symplectic form $\omega$ 
such that $g$ and $\omega$ define a complex structure $J$ 
on the bundle of differential forms. The obstruction against 
$M$ being a complex K\"ahler manifold (with K\"ahler form 
$\omega$) can also be stated as an algebraic relation. 
\bn
Let $(M,\omega)$ be a compact symplectic manifold, i.e.,  
$M$ is a smooth manifold endowed with a non-degenerate 
closed two-form $\omega$. Using local coordinates $x^{\mu}$ 
in a neighborhood $U\subset M$, we define the following 
three operators on $\Gamma(\bigwedge^{p}M^{\C})$ for 
$0\leq p \leq n$, $n := {\rm dim}\,M$:
$$\eqalign{
L^3 &= p - {n\over2}\ ,  
\cr
L^+ &= {1\over2}\, \omega_{\mu\nu} a^{\mu\,*}a^{\nu\,*}\ ,
\cr
L^- &= {1\over2}\, \bigl(\omega^{-1}\bigr)^{\mu\nu} a_{\mu}a_{\nu}\ ,
\cr} \eqno(3.95)$$
where $a_{\mu} = \iota(\partial_{\mu})$ is contraction with the basis
vector field $\partial_{\mu}$ -- therefore, $\{\,a_{\mu},a^{\nu\,*}\,\}
= \delta_{\mu}^{\nu}$ --,  and $\bigl(\omega^{-1}\bigr)^{\mu\nu}$
is the inverse matrix of $\omega_{\mu\nu}$: $\omega_{\mu\nu}
\bigl(\omega^{-1}\bigr)^{\nu\lambda} = \delta^{\lambda}_{\mu}\,$. 
Obviously, $L^3, L^{\pm}$ are in fact globally defined. 
\mn
{\bf Proposition 3.46}\quad The operators $L^3$ and $L^{\pm}$ 
satisfy the su(2) commutation relations, i.e., 
$$ 
\lb\, L^3, L^{\pm}\,\rb = \pm 2 L^{\pm}\ ,\quad 
\lb\, L^+, L^-\,\rb = L^3\ .
$$
\sn
{\kap Proof}: The first commutator is clear since $L^+$ resp.\ 
$L^-$ increases resp.\ decreases the degree of a form by two. 
The second relation can be calculated in local coordinates:
$$\eqalignno{ 
\lb\, L^+, L^- \,\rb &= {1\over4}\,\omega_{\mu\nu}
\bigl(\omega^{-1}\bigr)^{\sigma\lambda}\,
\lb\, a^{\mu\,*}a^{\nu\,*}, a_{\sigma}a_{\lambda} \,\rb 
&\cr
&= {1\over4}\,\omega_{\mu\nu}\bigl(\omega^{-1}\bigr)^{\sigma\lambda}\,
\bigl( a^{\mu\,*} (\delta^{\nu}_{\sigma}a_{\lambda} - 
\delta^{\nu}_{\lambda} a_{\sigma}) + (\delta^{\mu}_{\sigma}a_{\lambda} - 
\delta^{\mu}_{\lambda} a_{\sigma}) a^{\nu\,*}\bigr) 
&\cr
&= {1\over2} \delta^{\lambda}_{\mu} ( a^{\mu\,*}a_{\lambda} - 
a_{\lambda}a^{\mu\,*}) = a^{\mu\,*}a_{\mu} - {n\over2} 
&\qed\cr}$$
\mn
The operators above allow us to introduce a second differential 
operator on $\Gamma(\bigwedge^{\bullet}M^{\C})$ in addition to 
the exterior differential {\tt d}, namely 
\def\tdst{\widetilde{{\tt d}}^*}\def\td{\widetilde{{\tt d}}}
$$
\tdst := \lb\,L^-, {\tt d}\,\rb\ ; 
\eqno(3.96)$$
since {\tt d} satisfies ${\tt d}^2 =0$, we conclude that $\tdst$ 
anti-commutes with {\tt d}.  
\mn
{\bf Proposition 3.47}\quad The operators {\tt d} and $\tdst$ 
form a two-dimensional representation of su(2) under the 
adjoint action of $L^3$ and $L^{\pm}$: 
$$\eqalign{
&\lb\, L^3, {\tt d} \,\rb = {\tt d}\ ,\quad\ 
\lb\, L^3, {\tdst} \,\rb = - \tdst\ ,
\cr
&\lb\, L^+, {\tt d} \,\rb = 0 \ ,\quad\  
\lb\, L^+, {\tdst} \,\rb = {\tt d}\ ,\quad
\cr 
&\lb\, L^-, {\tt d} \,\rb = \tdst\ ,\quad
\lb\, L^-, {\tdst} \,\rb = 0\ .
\cr}$$
\sn
{\kap Proof}: The commutators of $L$'s with {\tt d} are direct  
consequences of ${\tt d}\,\omega=0$ and of the definition of $\tdst$.  
Furthermore, we have 
$$
\lb\, L^+, {\tdst} \,\rb = \lb\, L^+,\lb\,L^-, {\tt d}\,\rb\rb = 
\lb\, {\tt d}, \lb\,L^-,L^+\,\rb\rb + 
\lb\, L^-, \lb\,L^+, {\tt d}\,\rb\rb = 
- \lb\,{\tt d}, L^3\,\rb = {\tt d}\ .
$$
In order to derive the last commutator, we choose a Darboux 
coordinate system, i.e., one in which the components $\omega_{\mu\nu}$
of the symplectic form are constant. Then the operator $\tdst$ has 
the explicit form 
$$
\tdst = \lb\, L^-, {\tt d}\,\rb = {1\over2}\, 
\bigl(\omega^{-1}\bigr)^{\mu\nu} \,
\lb\,a_{\mu}a_{\nu}, a^{\lambda\,*}\,\rb \partial_{\lambda} 
= \bigl(\omega^{-1}\bigr)^{\mu\lambda} a_{\mu}\partial_{\lambda}\ , 
$$
and it follows that 
$$
\lb\, L^-, {\tdst} \,\rb = {1\over2}\, \bigl(\omega^{-1}\bigr)^{\kappa\rho} 
\bigl(\omega^{-1}\bigr)^{\mu\lambda}\,  
\lb\, a_{\kappa} a_{\rho}, a_{\mu}\partial_{\lambda}\,\rb = 0 \ .
\eqno{\qed}$$
\mn 
{\bf Corollary 3.48}\quad The operator $\tdst$ is nilpotent:
$$
\{\, \tdst, \tdst \,\} = 0 
$$
\sn
{\kap Proof}: This is a direct consequence of the transformation
properties of {\tt d} and $\tdst$ under the su(2) generated by 
$L^3, L^{\pm}$, the Jacobi identity, and the fact that {\tt d} and 
$\tdst$ anti-commute. \hfill\qed 
\mn
In summary, we see that on a symplectic manifold $M$ the 
space $\Gamma(\bigwedge^{\bullet}M^{\C})$ carries a 
representation of the Lie algebra su(2) and that there are 
two anti-commuting and nilpotent operators spanning the 
spin ${1\over2}$ representation. However, as long as no 
Riemannian metric is given, the space $\Gamma(\bigwedge^{\bullet}M^{\C})$ 
has no scalar product, and therefore cannot be taken as the 
Hilbert space occurring in the set of algebraic data as introduced in the 
previous sections. 
As soon as we introduce a Riemannian metric, on the other hand, 
we have a notion of adjoint for operators on
$\Gamma(\bigwedge^{\bullet}M^{\C})$,
and we may ask whether the representation of su(2) is unitary, 
i.e., whether $L^+$ is the adjoint of $L^-$. 
\mn
{\bf Proposition 3.49}\quad Let $g$ be a Riemannian metric on the 
symplectic manifold $(M,\omega)$. Then the following statements 
are equivalent:
\smallskip
\item {1)} $L^+ = (L^-)^*\,$.
\smallskip
\item {2)} The (1,1) tensor field $J$ defined by 
$$
\omega(X,Y) = g(JX,Y) 
$$
for all $X,Y\in \Gamma(TM)$ is an almost complex structure. 
\sn
Furthermore, if the above conditions are satisfied then the 
manifold $(M,g,J)$ is almost K\"ahlerian with almost K\"ahler form
$\omega$. 
\sn
{\kap Proof}: Assume 1) holds; then we have 
$$
(L^+)^* = -{1\over2}\,\omega_{\mu\nu}a^{\mu}a^{\nu} = 
-{1\over2}\,\omega_{\mu\nu}g^{\mu\lambda}g^{\nu\sigma}a_{\lambda}
a_{\sigma} = L^- = {1\over2}\,\bigl(\omega^{-1}\bigr)^{\lambda\sigma}
a_{\lambda}a_{\sigma}\ ,
$$
which implies 
$$ 
\omega_{\mu\nu} \omega_{\kappa\lambda} g^{\nu\sigma} g^{\lambda\mu} 
= - \delta^{\sigma}_{\kappa}\ .
$$
By definition, the components of the tensor $J$ are given by 
$$
J_{\mu}^{\phantom{\mu}\kappa} = \omega_{\mu\nu}g^{\nu\kappa}\ ,
$$
and the previous equation implies that $J$ is indeed an almost
complex structure:
$$
J_{\mu}^{\phantom{\mu}\sigma}J_{\kappa}^{\phantom{\kappa}\mu}
= -\delta^{\sigma}_{\kappa}
$$
On the other hand, if 2) holds, then $J^2 = - {\rm id}$ and 
the above relation between the components of $J$ and $\omega$ 
yield 
$$
-\omega_{\mu\nu} g^{\mu\lambda} g^{\nu\sigma} =
\bigl(\omega^{-1}\bigr)^{\lambda\sigma}\ ,
$$
and thus $(L^+)^* = L^-$. \hfill\break
\noindent If either of the equivalent conditions 1) or 2) holds, 
we conclude that
$$
g(JX,JY) = \omega(X,JY) = -\omega(JY,X) = -g(JJY,X) = g(Y,X) = g(X,Y)
$$
for all $X,Y\in\Gamma(TM)$, and this means that $J$ is an almost 
Hermitian structure. The associated fundamental 2-form is clearly
$\omega$, and since it is closed, the manifold $(M,g,J)$ is in 
fact almost K\"ahlerian.    \hfill\qed
\mn
The next proposition, which is well-known, ensures that on a 
symplectic manifold there is always a Riemannian metric with 
the properties of the previous proposition. For a proof, we 
refer the reader e.g.\ to \q{LiM}.
\mn
{\bf Proposition 3.50}\quad If $(M,\omega)$ is a symplectic manifold, 
then there exists an almost complex structure $J$ on $M$ such 
that the tensor $g$ defined by 
$$
g(X,Y) = -\omega(JX,Y)
$$
for all $X,Y \in \Gamma(TM)$ is a Riemannian metric on $M$. 
\mn
Because of these facts, we will consider almost K\"ahler manifolds 
in the following. Since $g$ induces a scalar product on 
$\Gamma(\bigwedge^{\bullet}M^{\C})$ with respect to which the 
su(2) representation is unitary, we can introduce a second 
spin ${1\over2}$ doublet for the same su(2) simply by taking 
the adjoints of the operators {\tt d} and $\tdst$:
\mn
{\bf Proposition 3.51}\quad The operators ${\tt d}^*$ and 
$\td := (\tdst)^*$ satisfy the following relations:
$$\eqalign{
&\{\,{\tt d}^*, {\tt d}^*\,\} = 0\ ,\quad \{\,\td,\td\,\} = 0
\cr
&\{\, {\tt d}^*,\td\,\} = 0 
\cr
&\lb\, L^3, {\tt d}^* \,\rb = -{\tt d}^*\ ,\quad
\lb\, L^3, {\td} \,\rb = \td\ ,
\cr
&\lb\, L^+, {\tt d}^* \,\rb = -\td\ ,\quad 
\lb\, L^+, {\td} \,\rb = 0\ ,\quad
\cr 
&\lb\, L^-, {\tt d}^* \,\rb = 0\ ,\quad\ \ \,
\lb\, L^-, {\td} \,\rb = -{\tt d}^* \ .
\cr}$$
\sn
{\kap Proof}: All equations are obtained by taking the adjoints
of the corresponding relations in Propositions 3.47. \hfill\qed
\mn
Notice that the operators $L^3, L^{\pm}$ and ${\tt d}, \tdst$ together
with their adjoints, satisfy all the commutation relations of the 
$N=4$ supersymmetry algebra of Theorem 3.40, except for 
$\{\,{\tt d}, \td\,\}=0$ and all relations involving the Laplace operators. 
We shall see in section 4 that on a 
classical K\"ahler manifold there is always a realization of 
the full $N=4$ algebra -- which also explains the appearance of 
the SU(2) generators that may look somewhat surprising at the 
moment. The condition that {\tt d} anti-commutes with $\td$ is 
precisely the requirement for the almost complex structure to 
be integrable.
\mn
{\bf Proposition 3.52}\quad For an almost K\"ahler manifold $M$, 
the following two statements are equivalent:
\smallskip
\item {1)} The operators {\tt d} and $\td$ anti-commute, 
$$ 
\{\,{\tt d}, \td\,\}=0\ .
$$
\smallskip
\item {2)} The manifold $M$ is K\"ahler. 
\sn
In particular, if relation 1) holds, all the other equations in  
Theorem 3.40 are true as well. 
\sn
{\kap Proof}: The Jacobi identity and the su(2) transformation 
properties of {\tt d} imply 
$$
\{\,{\tt d}, \td\,\}= - \{\, {\tt d}, \lb\, L^+, {\tt d}^*\,\rb\} 
= - \{\, {\tt d}^*, \lb\, {\tt d}, L^+\,\rb\} - 
\lb\,L^+, \{\,{\tt d}^*, {\tt d}\,\}\rb = -\lb\,L^+,\triangle\,\rb
$$
where $\triangle$ denotes the Laplace-Beltrami operator on $M$. Thus
condition 1) holds if and only if $L^+$ commutes with $\triangle$. 
In section 4 we will show that this is true on any K\"ahler manifold, 
and therefore 2) implies 1).  \hfill\break
\noindent Let us now assume that 1) holds. Choose a point $p\in M$ 
and normal coordinates $x^{\mu}$ around $p$; then at the center of 
this system, we can write 
$$\eqalign{
\triangle &= - g^{\mu\nu} \partial_{\mu}\partial_{\nu} + \ldots 
\cr
L^+ &= {1\over2}\, \omega_{\rho\sigma} a^{\rho\,*} a^{\sigma\,*}
\cr}$$
where the dots substitute for terms that commute with functions, 
in other words, for zeroth order differential operators. We can now 
compute the commutator of $L^+$ with $\triangle$:
$$
\lb\,L^+,\triangle\,\rb = 
- {1\over2}\, g^{\mu\nu} a^{\rho\,*} a^{\sigma\,*}
\lb\, \omega_{\rho\sigma},\partial_{\mu}\partial_{\nu}\,\rb + \ldots 
= g^{\mu\nu} \partial_{\mu}\omega_{\rho\sigma} a^{\rho\,*} a^{\sigma\,*}
\partial_{\nu} + \ldots 
$$
If $\lb\,L^+,\triangle\,\rb =0$, then the zeroth and first order terms
should vanish separately, and we get the following series of implications, 
which taken together prove the claim:
$$\eqalign{
\lb\,L^+,\triangle\,\rb = 0\  &\Longrightarrow\  
\partial_{\mu}\omega_{\rho\sigma}=0\ \hbox{\rm in normal coordinates at}\ p 
\cr 
&\Longleftrightarrow\  \nabla\omega=0 
\cr 
&\Longleftrightarrow\  \nabla J = 0 
\cr 
&\Longleftrightarrow\  J\ \hbox{\rm is integrable}
\cr 
&\Longleftrightarrow\  M\ \hbox{\rm is K\"ahler}
\quad \Longrightarrow 
\lb\,L^+,\triangle\,\rb = 0  
\cr}$$
where $\nabla$ denotes the Levi-Civita connection; the last implication
is derived explicitly in section 4. If the symplectic manifold satisfies
the conditions of this proposition, we choose $\partial = {1\over2}\,
({\tt d} - i\td)$ and $\bar\partial = {1\over2}\,({\tt d} + i\td)$
as holomorphic resp.\ anti-holomorphic differential.   \hfill\qed 
\mn 
This concludes our derivation of the algebraic data  associated to 
a symplectic manifold. Because of Lemma 3.45 already used in 
subsection 3.5.3, we can also formulate a converse statement: 
Suppose we have an algebraic description of a classical manifold $M$ 
in terms of its canonical $N=(1,1)$ bundle with the usual operators 
{\tt d} and $L^3$ -- the exterior differential and the total 
degree of forms --, then $M$ is a symplectic manifold if we 
can identify further operators $L^{\pm}$ and $\tdst$, $\td$  acting 
on differential forms and satisfying the relations listed above. 
Moreover, the symplectic form $\omega$ of $M$ is determined by $L^+$. 
\eject 
\pageno=65
\leftline{\bf 4. Supersymmetry classification of geometries}
\bn 
In the preceding section, we have achieved an algebraic formulation of 
classical differential geometry. 
In each case, the basic data contain a commutative 
algebra of functions on the manifold $M$ and an appropriate Hilbert 
space of square-integrable sections of a spinor bundle $\S$ over
$M$. We always required a representation  
of the algebra of sections of the Clifford bundle on $\S$ in order 
to define a Dirac operator. While these building blocks 
of ``spectral data'', the algebra of functions and the 
Hilbert space, are basic for all geometries 
considered, it is the presence of additional operators and 
their algebraic relations with the Dirac operators 
that allows us to distinguish merely Riemannian {}from, say, 
K\"ahler geometry. 
\sn
In the next section, where we will describe natural non-commutative
generalizations of classical geometry, we will be able to take full 
advantage of the results of section 3. But first, we summarize and 
interpret our algebraic characterizations of classical differential 
geometry {}from the point of view of supersymmetry, which 
has been our -- up to now hidden -- guiding principle throughout 
the last section. In particular, notions like $N=(1,1)$ 
spectral data etc.\  will become transparent in the following. 
\bn
Supersymmetry was invented, at the beginning of the 1970's, in the
form of the ``spinning string'' \q{R,NS} and in the   
context of relativistic quantum field theory (QFT) on $3+1$-dimensional
Minkowski space-time \q{GL,WZ}. Within QFT, supersymmetry 
provided a surprising way to circumvent ``no-go theorems'' 
\q{CM} stating that in realistic relativistic QFTs -- i.e., disregarding 
1+1 dimensions, which at that time were thought to be irrelevant --
the space-time symmetries and the ``internal'' symmetries (gauge 
groups of the first kind) always show a direct product structure and 
cannot be non-trivially embedded into 
a common simple symmetry group. It turned out that,  
at least at the level of infinitesimal symmetry operations,  
one can install further symmetries by adding 
new generators, $Q$, to those of the Poincar\'e algebra, 
as long as they behave like fermions, i.e., they obey anti-commutation
relations among each other (and commutation relations with the old 
``bosonic'' generators).  The main consequence  of the new symmetry 
generated by the $Q$'s is to set up a correspondence between the 
bosonic and the fermionic fields occurring in a supersymmetric QFT -- 
with implications reaching {}from abstract concepts down
to computational details. Although nature obviously is not strictly
supersymmetric -- as this would e.g.\ imply that to each 
boson one can associate a fermion with the same mass --, physicists 
nevertheless hope that the fundamental theory of elementary particles
might display supersymmetry which is however ``broken'' at the 
low energy scales accessible to experiment. In a variety of physical 
contexts (critical phenomena, string theory) one also encounters 
supersymmetric quantum field theories on a $1+1$-dimensional space-time, 
in particular $1+1$-dimensional superconformal field theories. Many of 
our arguments are inspired by features of these theories. 
\bn\bn
\eject
\noindent The {\sl $1+1$-dimensional supersymmetry algebra 
with $N$ supersymmetry 
charges} is generated by fermionic self-adjoint operators 
$Q^{(i)}_L$, $Q^{(i)}_R$, $i=1,\ldots,N$, 
that are subject to the anti-commutation relations 
$$\eqalign{
&\{ Q^{(i)}_L,  Q^{(j)}_L \} = \delta^{ij}\,(H+P)\ ,
\cr
&\{ Q^{(i)}_R,  Q^{(j)}_R \} = \delta^{ij}\,(H-P)\ ,
\cr
&\{ Q^{(i)}_L,  Q^{(j)}_R \} = 0 \ .
\cr}\eqno(4.1)$$ 
Here, $H$ and $P$ denote the generators of time and 
space translations in the Poincar\'e group. On the rhs of (4.1), 
only the combinations $H+P$ and $H-P$ occur because we display 
the relations in light-cone coordinates:  
$Q^{(i)}_L$ behave like the left-moving, $Q^{(i)}_R$ 
like the right-moving chiral components of a fermion field in a 
$1+1$-dimensional QFT. To make this so-called ``chiral splitting'' 
explicit, we will often use notations like e.g.\ $N=(2,2)$ supersymmetry 
algebra for the relations (4.1) with, in this case, two left-moving and 
two right-moving supercharges.
\sn  
The operators $Q^{(i)}_{L,R}$ commute 
with $H$ and $P$, and, on a higher-dimensional space-time, we 
would additionally require that the supercharges transform 
as a spin $1/2$ spinor under rotations. \hfill\break
\noindent   The full algebra of (super) 
space-time transformations is then spanned by the 
fermionic {\sl supercharges} together with the {\sl Poincar\'e 
generators} -- and, for $N\geq2$, some additional 
{\sl Lie group generators} which will be introduced when necessary; 
they arise {}from chiral symmetries of the relations (4.1)  and 
act only on the supercharges, not on the Poincar\'e generators.
As was done in section 1.2, one could in fact always start {}from 
one supersymmetry charge and characterize the higher-$N$ superalgebras 
by the additional Lie group generators that act on this charge 
to create further fermionic operators.  
\sn 
Although the algebra introduced in (4.1) makes perfect sense 
for any number $N$ of supercharges, it turns out that within 
the area of QFT the special values $N=1,\ 2,\ 4$ and $8$ are the 
most interesting. Since, remarkably, for our classification of 
geometries we also primarily encounter those types of supersymmetry, 
other supersymmetry algebras will not be considered.
\bn
We first discuss the $N=(1,1)$ superalgebra: After renaming the 
generators as $Q := Q^{(1)}_L$, $\overline{Q}:=  Q^{(1)}_R$, 
$\triangle := {1\over2}(H+P)$ and 
$\overline{\triangle} := {1\over2}(H-P)$, the relations for 
the supercharges read
$$
\{ Q, \overline{Q} \} = 0\ ,\quad 
Q^2 = \triangle\ ,\quad 
\overline{Q}^2 = \overline{\triangle}\ . 
\eqno(4.2)$$ 
If we enforce the additional relation 
$$
 \triangle= \overline{\triangle}\ ,
\eqno(4.3)$$
i.e.\ $P=0$, we precisely find the relations of 
$N=(1,1)$ spectral data, see Definition 3.8: 
$Q$ and $\overline{Q}$ are to be identified with the 
anti-commuting Dirac operators $\d$ and $\bard$ on a 
Riemannian manifold $M$, and condition (4.3) expresses the 
fact that both these operators square to the Laplacian of $M$. 
Recall that the property $\d^2=\bard^2$ made it 
possible to define a nilpotent operator {\tt d} -- the exterior 
differential on $M$ -- by setting  
$$
{\tt d} = {1\over2}\,\bigl(\d - i\,\bard\bigr)\ .
$$
This enabled us to derive de Rham-Hodge theory directly {}from 
the $N=(1,1)$ spectral data of Riemannian geometry. 
\mn
In the $N=(1,1)$ data $(C^{\infty}(M), L^2({\cal S}), \d, \bard)$,   
{\sl two} chiral ``halves'' of the $N=1$ supersymmetry algebra 
(4.2,3) are present.  If we consider a theory with only one supercharge $Q$ 
(e.g.\ obtained {}from the $N=(1,1)$ operators as $Q= Q_L$ or $Q=Q_L+Q_R$),  
the remaining data may be identified with those of a {\sl spectral 
triple} $(C^{\infty}(M), L^2(S), D)$, as introduced by Connes.  
We have called such a triple a set of $N=1$ {\sl spectral data}. {}From 
the data $(C^{\infty}(M), L^2(S), D)$, one can again recover 
a complete geometric description of $M$, as was discovered by Connes; 
the natural algebra of forms arising here is $\Omega_D$, see Proposition 
3.6. Since $D$ acts on a spinor bundle $S$ over $M$, $N=1$ spectral 
data are the natural algebraic setting for {\sl spin geometry}. 
Note, however, that the constructions of section 3.1 show 
that a Dirac operator can be defined on any Riemannian manifold.  
Therefore $N=1$ spectral data in fact provide a description of 
general Riemannian geometry, not just of spin geometry.    
\sn
The algebra $\Omega_D$ is more difficult to handle  
than the usual algebra of forms defined with the help of the 
nilpotent differential ${\tt d} = {1\over2}(\d - i\bard)$. Therefore, 
it would be desirable to have a general construction enabling us 
to pass {}from an $N=1$ spectral triple to a set of $N=(1,1)$ 
spectral data. In the classical case, this is of course always 
possible; the concrete procedure follows the lines of section 3.2. 
Within the non-commutative context, however, the situation is   
more complicated -- see the end of this section and section 5.2.5 
for further remarks. 
\mn  
All the other special types of manifolds that were studied in section 3, 
Hermitian, K\"ahler, etc., are special cases of Riemannian manifolds; thus 
they will all be described by $N=(1,1)$ spectral data, but with additional 
operators and relations -- which, in fact, correspond to higher supersymmetry 
involving $n$ chiral supercharges and are denoted by $N=(n,n)$ with $n=1,2$ 
and 4.  \hfill\break 
\noindent It turns out that, in classical geometry, there is a kind of 
``staircase'' leading {}from $N=(1,1)$ to $N=8$ supersymmetric data. The 
supercharges of an $N=(n,n)$ set of data can always be re-interpreted 
as those of an $N=2n$ superalgebra. (More precisely, all the left- and 
right-moving $N=(n,n)$ charges together form an algebra with the relations 
of the left- {\sl or} the right-moving half of the $N=2n$ algebra, as long 
as the constraint $P=0$ is imposed in (4.1), cf.\ relation (4.3) on 
the two Laplace operators.) As was mentioned after eq.\ (4.1), for 
$N\geq2$ the full supersymmetry algebra contains also certain Lie group
generators; the dimension of these groups grows as $N$  increases, 
and thus the non-trivial task in enlarging $N=(n,n)$ data to 
$N=2n$ data is to uncover the additional symmetries.
In the classical situation, this is always possible, as will be shown 
in this section, case by case. Through this procedure, we can 
pass {}from $N=(n,n)$ to $N=2n$ data -- without gaining new structure in 
classical geometry. In a further step, upon ``doubling the algebra'',  
we arrive at a truly richer $N=(2n,2n)$ geometry.  
\mn
In the simplest, and most general, case of $N=(1,1)$ data, we have 
the operator ${\tt d} =  {1\over2}(\d - i\bard)$ and its adjoint 
${\tt d^*} =  {1\over2}(\d + i\bard)$; but in addition there is 
a $\Z$-grading given by the operator $\Tot$ counting the total degree 
of a differential form. Altogether, these  operators satisfy the 
algebra 
$$\eqalign{
&{\tt d}^2 = ({\tt d}^*)^2 = 0\ ,\quad 
\{ {\tt d}, {\tt d}^* \} = \triangle\ , 
\cr
&\lb\, \Tot, {\tt d}\,\rb = {\tt d}\ , \quad
\lb\, \Tot, {\tt d}^*\,\rb = - {\tt d}^*\ .
\cr}\eqno(4.4)$$
These are just the (anti-)commutation relations of the (non-chiral)  
$N=2$ supersymmetry algebra in $1+1$ dimensions.  
In the classical case we can always enlarge $N=(1,1)$  
spectral data to (non-chiral) $N=2$ data. 
\bn
In view of the relations (4.4), which  can be realized in terms of 
operators defined on any Riemannian manifold $M$, the natural question 
to ask is when one can find a ``geometrical representation'' of 
the full chiral $N=(2,2)$ supersymmetry algebra including left-
and right-moving supercharges. It turns out that this really 
requires additional structure on $M$. 
We discuss the $N=(2,2)$ supersymmetry algebra in terms of 
the supercharges 
$$
Q_{\pm} := {1\over2}\,\bigl(Q_L^{(1)}\pm i\,Q_L^{(2)}\bigr)\ ,
\quad\overline{Q}_{\pm} := {1\over2}\,\bigl(Q_R^{(1)}\pm 
i\,Q_R^{(2)}\bigr)\ ,
\eqno(4.5)$$
which satisfy $\bigl(Q_{\pm}\bigr)^*= Q_{\mp}$ and  
$\bigl(\overline{Q}_{\pm}\bigr)^* = \overline{Q}_{\mp}$ as well as 
the anti-commutation relations 
$$\eqalign{
&Q_{\pm}^2 = \overline{Q}_{\pm}^2=0\ ,\phantom{xxxxxxxx}\quad\ \  
\{Q_{\sharp},\overline{Q}_{\sharp} \} = 0\ ,
\cr
&\vphantom{\sum}\{Q_+,Q_-\} = {1\over2}\,(H+P)\ ,\quad\quad
\{ \overline{Q}_+, \overline{Q}_- \} = {1\over2}\,(H-P)\ . 
\cr}\eqno(4.6)$$ 
In addition, there are {\sl two} chiral self-adjoint operators 
$T$ and $\overline{T}$, which behave like bosons; $T$ and $\overline{T}$ 
commute with each other and with $H$ and $P$,  
and they act on the  $Q$'s as ``counting'' operators:
$$\eqalign{
&\lb\, T, Q_{\pm} \,\rb = \pm Q_{\pm}\ ,\quad 
\lb\, T, \overline{Q}_{\pm} \,\rb = 0 \ ,
\cr
&\lb\,\overline{T}, \overline{Q}_{\pm} \,\rb = \pm \overline{Q}_{\pm}\ ,\quad 
\lb\, \overline{T}, Q_{\pm} \,\rb = 0 \ . 
\cr}\eqno(4.7)$$
Note that $T$ and $\overline{T}$ generate the {\sl chiral  
U(1)$\,\times\,$U(1) symmetry} of the relations (4.6), which do 
not change if $Q_+$ and $\overline{Q}_+$ are  
separately multiplied by phases.  
\sn
Let us now show that the $N=(2,2)$ algebra (4.6,7) is realized 
by the operators included in {\sl K\"ahler complex data}, Definitions 
3.22, 3.27 and 3.32. We first use the (self-adjoint) operator $T$  
and the counting operator $\Tot$ as above to define another (self-adjoint) 
operator $\overline{T}$ by 
$$
\Tot = T + \overline{T}\ .
$$             The relation 
$\lb\, T, \overline{T} \,\rb = 0$ follows since \Tot\  and $T$ commute. 
The operators $G$, defined as $G= \lb\, T,{\tt d}\,\rb$, {\tt d} being 
the nilpotent differential as before, and $\overline{G} = {\tt d} - G$ 
are then easily seen to satisfy 
$$
\overline{G} = \lb\, \overline{T},{\tt d}\,\rb
$$
as well as the algebraic relations (4.7) -- if we identify $G$ with 
$Q_+$ and $\overline{G}$ with $\overline{Q}_+\,$. Taking adjoints 
we also obtain the supercharges  $Q_-$ and  $\overline{Q}_-\,$.
The remaining relations (4.6) of the $N=(2,2)$ supersymmetry algebra 
are proven in Lemma 3.23 and Theorem 3.32. Note that, via the 
identifications above, the charges $Q_+$ and $\overline{Q}_+$ 
simply correspond to the holomorphic and anti-holomorphic 
differentials $\partial$ and $\overline{\partial}$, 
respectively, on a Hermitian complex manifold if we work 
with the canonical Hermitian data of a Hermitian manifold 
(with untwisted spinor bundle). On a K\"ahler manifold, there is 
the additional requirement  that 
the holomorphic and anti-holomorphic Laplace operators coincide:
$$
\square = \overline{\square}
\eqno(4.8)$$
They are to be identified with operators occurring in (4.6) as 
$\,\square\, = {1\over2}(H+P)$ and $\,\overline{\square}\,= 
{1\over2}(H-P)$; relation (4.8) is analogous to condition (4.3) in the 
$N=(1,1)$ case. Taking everything together, we have identified  
$N=(2,2)$ {\sl spectral data} with {\sl K\"ahler geometry}. 
\sn
In the more general case of {\sl Hermitian complex geometry}, 
condition (4.8) and the basic relation 
$\{\,Q_+,\overline{Q}_-\,\} =0$ can be violated; see Theorem 3.32. 
Therefore, we cannot identify Hermitian geometry with 
a realization of an ordinary supersymmetry algebra. 
Note, however, that, in any case,  the combination 
$J_0 = T-\overline{T}$ of U(1) generators can be understood 
as the complex structure of the manifold; see eqs.\ (1.49) and 
(1.54) in section 1.2. 
\mn
Just as we could define an additional operator \Tot\ 
on a Riemannian manifold which enabled us to introduce a non-chiral 
$N=2$ structure on $N=(1,1)$ spectral data, we can construct a non-chiral 
$N=4$ supersymmetry algebra {}from the data of an $N=(2,2)$ 
classical K\"ahler manifold $M$. The main task is to find operators that 
generate the {\sl SU(2)-symmetry} of the $N=4$ 
anti-commutation relations -- whereas there is a simple choice for the 
representation of the supercharges on the algebra of forms over $M$: We 
can use the complex structure given on the K\"ahler manifold and put
$$\eqalign{
G_1 &= \partial\ ,\quad\   G_2 = i\overline{\partial}^*\ ,
\cr
G_1^* &= \partial^*\ ,\quad G_2^* = -i\overline{\partial}\ .
\cr}\eqno(4.9)$$
$M$ is also equipped with the K\"ahler form $\Omega\in\bigwedge^{(1,1)}M$, 
which is closed and real, i.e.\ $d\Omega=0$, $\overline{\Omega}= \Omega$. 
This permits us to introduce operators on $\Gamma(\bigwedge^{(p,q)}T^*M)$ by 
setting     $$
T^3\  = {1\over2}\,(p+q-n)\ ,\quad 
T^+ = \epsilon(\Omega)\ , \quad 
T^- = \iota(\Omega)\ ;
\eqno(4.10)$$
here, $n={\rm dim}_{\C}M$, $\epsilon(\Omega)$ is wedging by the K\"ahler form,
and $\iota(\Omega)= \epsilon(\Omega)^*$.   \hfill\break
\noindent  The anti-commutation relations of the operators $G_a$ follow 
directly {}from the K\"ahler conditions in Theorem 3.32 and the 
relations in Definition 3.22 and Lemma 3.23: 
$$
\{ G_a, G_b \} = \{ G_a^*, G_b^* \} =0\ , \quad
\{ G_a,  G_b^* \} = \delta_{ab}\,\square 
\eqno(4.11)$$
for $a,b=1,2$. The operators in (4.10), on the other hand, satisfy the 
commutation relations of the Lie algebra of SU(2), namely 
$$
\lb\, T^3, T^{\pm} \,\rb = \pm T^{\pm}\ ,\quad 
\lb\, T^+, T^- \,\rb = 2 T^3 \ . 
\eqno(4.12)$$
Only the last of these equations is not immediately obvious; to prove it, 
we locally choose holomorphic normal coordinates $z^{\mu}$, as in Theorem 
3.31, and introduce the operators $a^{\mu\,*} = \epsilon(dz^{\mu})$, 
$a^{\mu} = \iota(dz^{\mu})$. Then we can write 
$$
\epsilon(\Omega) = i g_{\mu\bar\nu}\, a^{\mu\,*} a^{\bar\nu\,*}\ ,\quad
\iota(\Omega) = - i g_{\rho\bar\sigma}\, a^{\rho} a^{\bar\sigma}\ ,
$$
since $\Omega =  i g_{\mu\bar\nu}\, dz^{\mu}\wedge d\bar z^{\mu}\,$. 
The simple anti-commutation rules (3.90) satisfied by the $a^{\mu\,*},\ 
a^{\bar\nu}$ allows for an easy verification of (4.12), much along the 
lines of Lemma 3.41.   \hfill\break
\noindent Similarly, we can show that the SU(2)-generators act on the 
supercharges as follows:
$$\eqalign{
&\lb\, T^3, G_1 \,\rb = {1\over2}\,G_1\ ,\quad 
\lb\, T^3, G_2 \,\rb = -{1\over2}\,G_2\ , 
\cr
&\lb\, T^+, G_1 \,\rb = 0\ ,\quad\quad\ \,  
\lb\, T^+, G_2 \,\rb = G_1\ , 
\cr
&\lb\, T^-, G_1 \,\rb = G_2\ ,\quad \ \ \,
\lb\, T^-, G_2 \,\rb = 0\ ;
\cr}\eqno(4.13)$$
to this end, we write $\partial = a^{\mu\,*}\partial_{\mu}$ and 
$\overline{\partial}^* = - a^{\mu}\partial_{\mu}$, which is true at the center of a  
complex geodesic coordinate system. Using $\bigl(T^3\bigr)^* = T^3$, 
$\bigl(T^{\pm}\bigr)^* = T^{\mp}$, 
we can also deduce the SU(2)-action on $G_a^*$ {}from (4.13).   \hfill\break
\noindent To complete the commutation relations of the $N=4$ 
superalgebra, we use 
(4.11) and (4.13) to show that the $T'$s commute with the operator $\square\,$,
e.g.\ 
$$
\lb\, T^+, \square \,\rb = \lb\, T^+, \{ G_1, G_1^* \}\,\rb = 
\{ G_1, \lb\, T^+, G_1^*\,\rb \} = - \{ G_1, G_2^* \} = 0\ .
$$
The relations (4.11-13) -- which we have derived for operators on a classical 
K\"ahler manifold originally associated to chiral $N=(2,2)$ spectral data -- 
are indeed the relations of an $N=4$ supersymmetry algebra. The non-chiral 
character of this $N=4$ algebra is not only obvious {}from the choice 
of $G_a$ in (4.9) -- note, that we could also have started with the de Rham 
differential {\tt d} as one of the supercharges --, but is also reflected 
in the fact that the SU(2)-generators are constructed in terms of the
``mixed'' (1,1)-form $\Omega$. 
\mn
We have seen in section 3.6 that there is a variant of the $N=4$ supersymmetry
algebra which provides an algebraic description of {\sl symplectic 
geometry}. Here the SU(2) generators are constructed with the symplectic 
form instead of the K\"ahler form used in (4.10), and the two SU(2) doublets
are generated {}from {\tt d} and ${\tt d}^*$, see Propositions 3.47 
and 3.51. The algebraic
relations satisfied by these operators i.g.\ do not coincide with the 
$N=4$ relations on a K\"ahler manifold, in particular the SU(2) generators 
need not commute with the Laplacian $\triangle$. Therefore, we cannot simply 
identify operators on a general symplectic manifold with generators of 
some supersymmetry algebra. As was shown at the end of section 3.6, the
symplectic relations deviate {}from the $N=4$ algebra if and only if 
the symplectic manifold has no integrable complex structure that is 
compatible with the symplectic structure.  
\bn
In analogy to the step {}from $N=2$ to $N=(2,2)$ data, the presence 
of $N=(4,4)$ spectral data on a K\"ahler manifold $M$ contains new 
information: In this case, $M$ is in fact a Hyperk\"ahler manifold 
equipped with three complex structures and a holomorphic symplectic form 
$\omega\in\bigwedge^{(2,0)}M$. We first present the complete  
commutation relations of the supersymmetry algebra 
with $N=4$ supercharges of each chirality  in abstract terms.  
The left-moving 
sector contains the fermionic operators  
$$
G_1 :=  {1\over\sqrt{2}}\,\bigl(Q_L^{(1)} + i\, Q_L^{(2)}\bigr)\ ,\quad
G_2 :=  {1\over\sqrt{2}}\,\bigl(Q_L^{(3)} - i\, Q_L^{(4)}\bigr) 
\eqno(4.14)$$
and their adjoints, and {}from the $Q_R^{(i)}$ we form $\overline{G}_a$, 
$\overline{G}_a^*$, $a=1,2$, analogously. The anti-commutators 
within chiral sectors read 
$$\eqalign{
&\{ G_a,G_b \} = \{ G_a^*,G_b^* \} = 0\ ,\quad 
\{ G_a,G_b^* \} = \delta_{ab}\, (H+P)\ , 
\cr
&\{\overline{G}_a,\overline{G}_b \} = 
\{\overline{G}_a^*,\overline{G}_b^* \} = 0\ ,\quad	
\{\overline{G}_a,\overline{G}_b^* \} =\delta_{ab}\, (H-P)\ ,
\cr}\eqno(4.15)$$
and left-moving charges $G_a, G_a^*$ anti-commute with right-movers
$\overline{G}_b$, $\overline{G}_b^*$. It is easy to see that 
(4.15) is left invariant by SU(2)$\,\times\,$SU(2) transformations
$$
\pmatrix{G_1\cr G_2} \longmapsto A \pmatrix{G_1\cr G_2}\ ,\quad
\pmatrix{\overline{G}_1\cr \overline{G}_2} \longmapsto 
\overline{A} \pmatrix{\overline{G}_1\cr\overline{G}_2}\ ,
\eqno(4.16)$$
where $A,\overline{A}\in\,$SU(2) are two independent matrices. \hfill\break
\noindent In addition, the relations within each of the two supercharge 
doublets are left invariant by rescaling with a phase, so the 
full symmetry Lie algebra consists of two commuting copies of 
su(2)$\,\oplus\,$u(1). For the moment, we ignore the u$(1)\,$, 
which can be associated with the generator $J_0$ of eq.\ (1.49) in 
section 1.2, but we will come to it at the end of this section. \hfill\break
\noindent As a consequence of (4.16), 
the complete $N=4$ supersymmetry algebra involves generators 
$T^{\alpha}, \overline{T}^{\alpha}$, $\alpha=1,2,3$, of two 
commuting SU(2) actions, with the $G_a$  transforming according to 
the fundamental representation: 
$$
\lb\, T^{\alpha}, T^{\beta} \,\rb = i \epsilon_{\alpha\beta\gamma}
   T^{\gamma}\ ,\quad\quad  
\lb\,  T^{\alpha}, G_a \,\rb  = {1\over2}\, 
   \bigl(\overline{\tau_{\alpha}}\bigr)_{ab} G_b \ .
\eqno(4.17)$$
The bar in the second equation simply means complex conjugation 
of the Pauli matrices. 
$T^{\alpha}$ commutes with all generators in the right-moving
sector; analogous statements hold for $\overline{T}^{\alpha}$ 
and  $\overline{G}_a$. \hfill\break
\noindent The operators and relations in the left-moving 
sector correspond to  
the operators on $(p,q)$-forms over a Hyperk\"ahler manifold
which were introduced in subsection 3.5.2 and satisfy the
relations of Theorem 3.40. The right-moving generators,
on the other hand, may be identified with their anti-holomorphic 
``partners''. In addition, since every Hyperk\"ahler manifold 
is automatically K\"ahler, the holomorphic and 
anti-holomorphic Laplacians $\square = H+P$ resp.\ 
$\overline{\square} = H-P$ have to coincide ($P=0$).  
All in all, we may say that $N=(4,4)$ {\sl supersymmetry} 
corresponds to {\sl Hyperk\"ahler geometry}. 
\mn
In view of the examples of Riemannian and K\"ahler manifolds, 
it is natural to ask whether the $N=(4,4)$ structure 
of Hyperk\"ahler data can be rewritten as an $N=8$ algebra 
involving eight supersymmetry charges as well as the generators 
of a larger Lie group which act on the fermions. This is indeed 
possible: \hfill\break
\noindent In order to simplify the presentation of the 
commutation relations later on, we introduce new notations
for the fermionic generators of section 3.5.2:
$$\eqalign{
G^{(1,0)} &= G^{1+} = \partial \ ,\phantom{xxxxxxxxxxxxx}
G^{(-1,0)} = G^{2+} 
= \lb\,\iota(\overline{\omega}),\partial\,\rb \ ,\phantom{MM}
\cr
G^{(0,1)} &= - \overline{G}^{2-} =\lb\, 
\epsilon(\overline{\omega}),\overline{\partial}^*\,\rb \ ,\quad\quad\quad
G^{(0,-1)} =- \overline{G}^{1-} = -\overline{\partial}^*\ ,\phantom{MM}
\cr}\eqno(4.18)$$
and 
$$\eqalign{
\overline{G}^{(1,0)} &= G^{2-} 
= - \lb\, \epsilon(\omega),\partial^*\,\rb  \ ,\quad\quad\phantom{N}
\overline{G}^{(-1,0)} = - G^{1-} = - \partial^*\ ,
\cr
\overline{G}^{(0,1)} &= \overline{G}^{1+} = \overline{\partial} \ ,
\phantom{xxxxxxxxxxx}\phantom{M}   \overline{G}^{(0,-1)} = - \overline{G}^{2+}
= - \lb\, \iota(\omega),\overline{\partial}\,\rb \ .
\cr}\eqno(4.19)$$ 
Note that the $G^{(\eta_1,\eta_2)}$ defined in (4.18) include 
both holomorphic and anti-holomorphic operators on $M$, and 
analogously for the $\overline{G}^{(\eta_1,\eta_2)}$ in (4.19); 
this ``mixing'' already occurred  in the 
passage {}from $N=(2,2)$ to $N=4$. We also introduce new  
normalizations and notations for the generators of the two 
commuting copies of su(2) acting on $\Gamma(\bigwedge^{(p,q)}T^*M)$
over a Hyperk\"ahler manifold of complex dimension $n\,$; $\omega$
and $\overline{\omega}$ are the holomorphic resp.\ anti-holomorphic
symplectic form: 
$$
H^1 = 2\,T^3 = p-{n\over2}\ ,\quad \ 
E^{(2,0)} = \sqrt{2}\,T^+ = \sqrt{2}\,\epsilon(\omega)\ ,\quad \,
E^{(-2,0)} = \sqrt{2}\,T^-= \sqrt{2}\,\iota(\overline{\omega})
\eqno(4.20)$$
and 
$$
H^2 = 2\,\overline{T}^3 = q-{n\over2}\ ,\quad \,
E^{(0,2)} = \sqrt{2}\, \overline{T}^+ = 
\sqrt{2}\,\epsilon(\overline{\omega})\ ,\quad \,
E^{(0,-2)} = \sqrt{2}\, \overline{T}^- = 
\sqrt{2}\, \iota(\omega)
\eqno(4.21)$$
Next, we want to extend the bosonic algebra su(2)$\,\oplus\,$su(2),
i.e., we want to define further generators (``currents'') of a larger
Lie algebra 
in terms of the geometric operations on $\Gamma(\bigwedge^{(p,q)}T^*M)$, 
which are the Riemannian metric $g$ and wedging 
and contraction operators built {}from $\omega$ and $\overline{\omega}$ --
or, equivalently, {}from the three complex structures $I, J$ and $K$. 
The new bosonic generators should act within the fibers of the 
complexified bundle of differential forms. Therefore it is sufficient 
to construct local expressions for the currents -- which must be 
bi-linears in the fermionic operators $a^{\mu}$ and $a^{\mu\,*}$, 
contracted with one of the rank two tensors listed above. 
We find four more bosonic operators:
$$\eqalign{
&E^{(1,1)} = \epsilon(-i\Omega_I) 
= g_{\mu\bar\nu} a^{\mu\,*} a^{\bar\nu\,*}\ ,\quad\quad\quad 
E^{(-1,-1)} = \iota(i\Omega_I) = g_{\mu\bar\nu} a^{\mu} a^{\bar\nu}\ ,
\cr
&E^{(1,-1)} = J_{\mu\nu} a^{\mu\,*} a^{\nu}\ , \phantom{MMMMMMM}
E^{(-1,1)} = - J_{\bar\mu\bar\nu} a^{\bar\mu\,*} a^{\bar\nu}\ , 
\cr}\eqno(4.22)$$
where $\Omega_I= i g_{\mu\bar\nu} dz^{\mu}\wedge d\bar z^{\nu}$
is the K\"ahler form associated to the complex structure $I$, and  
$J_{\mu\nu}$ are the components of the holomorphic 
symplectic form, see eqs.\ (3.70 - 74); again, $a^{\mu\,*}$ and
$a^{\nu}$ belong to some local coordinates, e.g.\ geodesic ones, 
as in Theorem 3.31. \hfill\break
\noindent Since the first derivatives of the metric and of the symplectic form 
vanish at the center of such a coordinate system, it is not very difficult to 
show that the operators introduced in (4.20-22) 
obey the following commutation relations: 
$$\eqalign{
&\lb\, H^i, E^{(n_1,n_2)} \,\rb = n_i\, E^{(n_1,n_2)}\ ,\phantom{MMMMMM}
\lb\, E^{(n_1,n_2)}, E^{(-n_1,-n_2)} \,\rb = n_1 H^1 + n_2 H^2 \ ,
\cr
&\lb\, E^{(n_1,n_2)}, E^{(m_1,m_2)} \,\rb = 0 \quad\quad {\rm if} \ \ 
\pmatrix{ n_1+m_1\cr n_2+m_2\cr} 
\notin \Bigl\{ \pmatrix{ \pm 2\cr0\cr}, \pmatrix{ 0\cr\pm 2\cr},
\pm \pmatrix{ 1\cr\pm 1\cr},\pmatrix{ 0\cr0\cr} \Bigr\} \ ,
\cr
&\lb\, E^{(2,0)}, E^{(-1,-1)} \,\rb = -\sqrt{2}\,  E^{(1,-1)}\ ,\phantom{MMMM}
\lb\, E^{(2,0)}, E^{(-1,1)} \,\rb = \sqrt{2}\,  E^{(1,1)}\ ,
\cr 
&\lb\, E^{(0,2)}, E^{(-1,-1)} \,\rb = -\sqrt{2}\,  E^{(-1,1)}\ ,\phantom{MMMM}
\lb\, E^{(0,2)}, E^{(1,-1)} \,\rb = \sqrt{2}\,  E^{(1,1)}\ ,
\cr
&\lb\, E^{(1,1)}, E^{(1,-1)} \,\rb = \sqrt{2}\,  E^{(2,0)}\ ,\phantom{MMNMMM}
\lb\, E^{(1,1)}, E^{(-1,1)} \,\rb = \sqrt{2}\,  E^{(0,2)}\ ;
\cr}\eqno(4.23)$$
further commutation relations follow {}from the Hermiticity properties
$$
H^{i\,*} = H^i\ ,\quad E^{(n_1,n_2)\,*}= E^{(-n_1,-n_2)}\ .
\eqno(4.24)$$
Equations (4.23,24) can be shown to be equivalent to the commutation relations
of the {\sl Lie algebra of Sp(4)}, e.g.\ by computing the Dynkin diagram. \hfill\break
\noindent With the help of the explicit realization of the currents in terms 
of fermionic operators $a^{\mu\,*}, a^{\mu}$, etc., it is also straightforward
to determine the transformation  properties of the fermionic charges
given in (4.18,19) under the Sp(4) generators. One finds that the 
$G^{(\eta_1,\eta_2)}$ form the fundamental representation 
of Sp(4): 
$$\eqalign{
&\lb\, H^i, G^{(\eta_1,\eta_2)} \,\rb = \eta_i\,  G^{(\eta_1,\eta_2)}\ ,
\cr
&\lb\, E^{(n_1,n_2)}, G^{(\eta_1,\eta_2)} \,\rb = 0 \quad\quad {\rm if} \ \ 
\pmatrix{ n_1+\eta_1\cr n_2+\eta_2\cr} 
\notin \Bigl\{\pmatrix{ 0\cr\pm 1\cr},\pmatrix{ \pm 1\cr0\cr} \Bigr\} \ ,
\cr
&\lb\, E^{(-2,0)}, G^{(1,0)} \,\rb = \sqrt{2}\, G^{(-1,0)}\ ,\phantom{MMIMMMM}
\lb\, E^{(-1,-1)}, G^{(1,0)} \,\rb = G^{(0,-1)}\ ,
\cr
&\lb\, E^{(-1,1)}, G^{(1,0)} \,\rb =  G^{(0,1)}\ ,
\cr
&\lb\, E^{(2,0)}, G^{(-1,0)} \,\rb = \sqrt{2}\, G^{(1,0)}\ ,\phantom{MMMMMMM'}
\lb\, E^{(1,1)}, G^{(-1,0)} \,\rb =  - G^{(0,1)}\ ,
\cr
&\lb\, E^{(1,-1)}, G^{(-1,0)} \,\rb =  G^{(0,-1)}\ ,
\cr
&\lb\, E^{(0,-2)}, G^{(0,1)} \,\rb = - \sqrt{2}\, G^{(0,-1)}\ ,\phantom{NMMMMM'}
\lb\, E^{(-1,-1)}, G^{(0,1)} \,\rb = -  G^{(-1,0)}\ ,
\cr
&\lb\, E^{(1,-1)}, G^{(0,1)} \,\rb = G^{(1,0)}\ ,
\cr
&\lb\, E^{(0,2)}, G^{(0,-1)} \,\rb = - \sqrt{2} G^{(0,1)}\ ,\phantom{MMMMNMM}
\lb\, E^{(1,1)}, G^{(0,-1)} \,\rb = G^{(1,0)}\ ,\quad
\cr
&\lb\, E^{(-1,1)}, G^{(0,-1)} \,\rb = G^{(-1,0)}\ .
\cr}\eqno(4.25)$$
The supercharges $\overline{G}^{(\eta_1,\eta_2)}$ span a second 
fundamental representation of Sp(4) with analogous relations. 
Note that $G$'s and $\overline{G}$'s are related by the 
following Hermiticity properties: 
$$\eqalign{
&\overline{G}^{(1,0)} = G^{(-1,0)\,*}\ ,\phantom{MMMM}
\overline{G}^{(-1,0)} = -G^{(1,0)\,*}\ ,
\cr
&\overline{G}^{(0,1)} = - G^{(0,-1)\,*}\ ,\phantom{MMMl}
\overline{G}^{(0,-1)} = G^{(0,1)\,*}\ .
\cr}\eqno(4.26)$$
Finally, the anti-commutators of the 
fermionic generators with each other are a direct consequence
of the Hyperk\"ahler relations of Theorem 3.40. They read 
$$\eqalign{
&\{\, G^{(\eta_1,\eta_2)}, G^{(\eta_1',\eta_2')} \,\} = 
\{\, \overline{G}^{(\eta_1,\eta_2)}, 
\overline{G}^{(\eta_1',\eta_2')} \,\} = 0\ ,
\cr
&\{\, G^{(\eta_1,\eta_2)}, \overline{G}^{(\eta_1',\eta_2')} \,\} = 0 
\quad\quad {\rm if}\ \ (\eta_1+\eta_1',\eta_2+\eta_2') \neq (0,0)\ ,
\cr
&\{\, G^{(1,0)}, \overline{G}^{(-1,0)} \,\} = 
\{\, G^{(0,-1)}, \overline{G}^{(0,1)} \,\} = -{1\over2}\,\triangle \ ,
\cr
&\{\, G^{(-1,0)}, \overline{G}^{(1,0)} \,\} = 
\{\, G^{(0,1)}, \overline{G}^{(0,-1)} \,\} = {1\over2}\,\triangle \ .
\cr}\eqno(4.27)$$
\mn
As before, it would be straightforward to proceed towards sets of $N=(8,8)$
supersymmetric data with $8+8$ supercharges and Sp(4)$\,\times\,$Sp(4) 
symmetry. Such spectral data describe a very rigid type of complex
geometry, related to manifolds discussed in \q{Joy}. We 
will not give details of the $N=(8,8)$ case here.  
\bn
Let us summarize the correspondence between different 
supersymmetry algebras and different types of classical differential 
geometry  in the following table: 
\bn
$$\vbox{\rm \halign {# & \quad #& \hfil # \hfil & \ # 
& \ #\hfill \cr
  & & $N=1$ & : & Riemannian (spin) geometry, Connes' complex 
   $\Omega_D$
\cr
& &\hfil $\vphantom{\displaystyle{\sum_s^s}}\futdarrow$\hfil & & \cr
 $N=2$ & $\frtlarrow\ \ \ $ & $N=(1,1)$  & : 
& Riemannian geometry, de Rham complex 
\cr
\hfil$\fuarrow$\hfil& &\hfil $\vphantom{\displaystyle{\sum_s^s}}\fuarrow$\hfil & & \cr
 $N=4^+$ & $\frtlarrow\ \ \ $ & $N=(2,2)$  & : 
& complex (K\"ahler) geometry, Dolbeault complex 
\cr
\hfil$\fuarrow$\hfil& &\hfil $\vphantom{\displaystyle{\sum_s^s}}\fuarrow$\hfil & & \cr
 $N=8$ & $\frtlarrow\ \ \ $ & $N=(4,4)$  & : 
& Hyperk\"ahler geometry 
\cr
}}$$
\mn
Additionally, there are symplectic geometry and complex Hermitian 
geometry, which we did not include in the above scheme since they 
correspond to relations that are only close to those of $N=4$ 
and $N=(2,2)$ supersymmetry, as discussed above. 
\sn 
In the table, fat arrows pointing e.g.\ {}from $N=(2n,2n)$ to $N=(n,n)$ data  
indicate that the algebraic relations of the latter are included within those
of the former. The necessary re-interpretations occur at a purely algebraic 
level. Therefore the corresponding ``transitions'' will be possible in 
the non-commutative case, too.  The same is true for the arrows {}from 
$N=2n$ to $N=n$ in the left row. Here, the notation $N=4^+$  means that 
the corresponding data possess $N=4$ supersymmetry with enlarged symmetry 
group SU(2)$\,\times\,$U(1). The extra U(1) factor was mentioned already
after eq.\ (4.16), as well as in section 1.2: It is always present 
if the $N=4$ data arise {}from a 
reduction of $N=8$ data (whose symmetry group has rank 2), and 
is in fact necessary if one wants to recover $N=(2,2)$ data (containing 
two commuting abelian Lie groups) {}from the $N=4$ data. Up to this 
subtlety, all the fat arrows pointing to the right merely involve 
re-interpretations of the $N=2n$ generators.      \hfill\break
In Connes' original context, $N=1$ data always correspond to a spin 
manifold, and thus the passage {}from $N=(1,1)$ data associated 
with an arbitrary Riemannian manifold to an 
$N=1$ description is i.g.\ obstructed. In the formulation of section 3.1, 
however, there is an $N=1$ set of data for every Riemannian 
manifold, which is why the table also includes a fat arrow {}from $N=(1,1)$
to $N=1$.     
\sn 
The thin arrows mark special extensions 
of a set of spectral data by constructing additional operators, either 
further symmetry generators for $N=(n,n) \lra N=2n$, or a second Dirac 
operator for $N=1 \lra N=(1,1)$. As we have seen, these transitions are always 
possible for spectral data describing a classical manifold. In the
non-commutative case, it seems that this is not generally true -- 
see the remarks below and in section 5.2.5.   \hfill\break
\noindent In both cases, steps against the upward arrows require 
true extensions of the spectral data with
passage to higher supersymmetry content and to a more rigid 
geometrical structure. Whether this is possible depends, of course, 
on the properties of the specific manifold or non-commutative space. 
One could also say it depends on the {\sl symmetries} ``hidden'' in the spectral 
data of the (non-commutative) space: In the introduction, we have seen 
that different types of spectral data associated with different types 
of classical manifolds can be written in the alternative form 
$(\a, \h, {\tt d}, {\tt d}^*, {\tt g})$ where {\tt d} is  the nilpotent 
de Rham differential on the manifold, and {\tt g} denotes some Lie algebra which 
acts on \h\ -- and on {\tt d}, ${\tt d}^*$, thereby generating further 
supersymmetry charges and determining the supersymmetry content of the 
spectral data. To be precise, ${\tt g}=$ u(1), su(2)$\,\oplus\,$u(1) or 
sp(4) for the $N=2$, $N=4^+$ or $N=8$ geometries, respectively. 
\sn
This concise classification of geometries according to their 
supersymmetry content -- or simply:\ according to their symmetry --  
opens up various new possibilities. First
of all, we gain a clear picture of what the natural 
non-commutative generalizations of Riemannian and complex 
geometry should look like: The algebra of functions $C^{\infty}(M)$ 
is to be replaced by a non-commutative $*$-algebra $\a$ now 
acting on some appropriate Hilbert space $\h$, but the algebraic 
relations for the collection of generalized Dirac operators 
given on $\h$ are just the same as in the classical cases. 
\sn
We remark that, within the non-commutative context, 
the transition {}from $N=(n,n)$ spectral data to an $N=2n$ 
structure will not always be possible, because the existence of the 
additional bosonic generators is not, in general, guaranteed. 
Likewise, it appears that one cannot always pass {}from $N=1$ 
to $N=(1,1)$ spectral data, i.e.\ introduce 
supersymmetry into the data. Comparing to the classical situation, 
where one essentially passes {}from the sections $\Gamma(S)$ of a 
spinor bundle $S$ to sections $\Gamma({\cal S}) = \Gamma(S) \otimes 
\Gamma(S)$ -- with the tensor product taken over 
the algebra of smooth functions --,  one might try to introduce 
two ``Dirac operators'' on a suitable tensor product space. 
We will discuss this question in more detail in section 5.2.5, and 
we will find that, given some extra structure on the $N=1$ 
spectral data, there is indeed a natural definition of the operators 
$\d$ and $\bard$ on the new Hilbert space, but in general they need not
obey the $N=(1,1)$ algebra. 
\sn
Another comment is in order, concerning the apparent identification 
of commutative algebras with classical geometry on the one hand, 
and non-commutative algebras with non--commutative geometry on the 
other hand. Recall the idea that was outlined in the introduction, namely 
that our spectral data describe how a quantum mechanical spinning  
particle (or a bound state of two such particles) would 
``see'' the manifold it propagates on. {}From the point 
of view of quantum mechanics, the natural object to study is 
quantized phase space rather than configuration space, with
the -- non-commutative -- algebra of pseudo-differential operators 
substituting for $C^{\infty}(M)$. More precisely, it is natural 
to require that the algebra $\a$ used in the spectral data is 
invariant under time evolution generated 
by the quantum mechanical Hamiltonian. 
{}From the physical point of view, there are reasons to believe that the 
non-commutative spectral data containing the algebra of ``functions 
over quantized phase space'' contains all essential information on 
the classical background manifold. However, at present we lack a 
complete understanding of how to extract it explicitly. We expect that
at least the cohomology of $M$ can be obtained {}from the differential 
forms over phase space in a rather straightforward procedure. These 
questions will be dealt with in a future publication.
\sn   
An important pay-off of our identification of the 
relations occurring in spectral data with the defining 
(anti-)commutators of supersymmetry algebras is that it  
immediately provides us with a large ``pool'' of physical 
examples for non-commutative geometries: In any supersymmetric 
QFT on a cylindrical space-time $S^1\times\R$, the full algebra 
formed by the charges $Q, \overline{Q}$ together with the 
Poincar\'e generators $H$ and $P$ and additional 
bosonic symmetry generators $T, \overline{T}$ is represented on 
the Hilbert space $\h$ of states, as well as on the algebra $\a$ of 
fields. Therefore we can simply start {}from the tuple $(\a,\h,Q, 
\overline{Q}, T, \overline{T})$ in order to produce admissible 
spectral data. Disregarding complex Hermitian and symplectic data, 
we encountered 
the additional requirement that the left- and right-moving ``Laplace'' 
operators coincide. In terms of the above identification, this 
simply means that we have to project onto the zero-momentum subspace, 
i.e.\ set $P=0$. At the level of differential forms, this 
restriction amounts to studying $S^1$-equivariant instead of ordinary 
cohomology. 
\sn     
In our view, the most important examples of supersymmetric QFTs are the 
$N=2$ superconformal field theories associated to perturbative vacua of 
string theory. General considerations suggest that the structure of these 
theories in principle also determines the features of space-time.
The methods used in the literature are, however, essentially limited to
uncover classical (and mainly topological) aspects of space-time {}from 
$N=2$ superconformal field theories, while our approach should make it 
possible to study the full non-commutative geometry of the underlying 
``quantum space-time''. 
\eject
\pageno=77
\leftline{\bf 5. Non-commutative geometry}
\bn
In the following, we  generalize the notions
of section 3 {}from classical differential geometry to the 
non-commutative setting. Again, a classification according 
to the ``supersymmetry content'' of the relevant spectral data
will be our guiding principle. In the first part, we 
review Connes' formulation of non-commutative geometry 
using just one generalized Dirac        
operator, whereas, in the following subsections, spectral data 
with genuine realizations of some supersymmetry algebras will 
be introduced, which allows us to define non-commutative 
generalizations of Riemannian, complex, K\"ahler and 
Hyperk\"ahler, as well as of symplectic geometry. 
\bn\bn 
{\bf 5.1 The \Neqone formulation of non-commutative geometry}
\bn
This section is devoted to the non-commutative generalization 
of the algebraic description of spin geometry -- and, according
to the results of section 3, of general Riemannian geometry --
following the ideas of Connes. The first two subsections contain 
the definition of abstract $N=1$ spectral data and of 
differential forms. In subsection 5.1.3, we describe a 
notion of integration which leads us to a definition of 
square integrable differential forms. After having introduced 
vector bundles and Hermitian structures in subsection 5.1.4, 
we show in subsection 5.1.5 that the module of square integrable 
forms always carries a generalized Hermitian structure. We then 
define connections, torsion, and Riemannian, Ricci 
and  scalar curvature in the next two subsections. Finally, in 
5.1.8 we derive non-commutative Cartan structure equations. 
Many of the notions introduced in section 5.1 will be 
basic for later sections. 
\bn\bn
{\bf 5.1.1 The \Neqone spectral data} 
\bn
{\bf Definition 5.1}\quad A quadruple $(\a,\h, D, \gamma)$ will 
be called a set of $N=1$ {\sl (even) spectral data} if 
\smallskip
\item {1)} \h\ is a separable Hilbert space; 
\smallskip
\item {2)} \a\ is a unital ${}^*$-algebra acting faithfully on \h\ 
by bounded operators;
\smallskip
\item {3)} $D$ is a self-adjoint operator on \h\  such that 
\itemitem {$i)$} for each $a\in\a$, the commutator $\lb\,D,a\,\rb$
defines a bounded operator on \h, 
\itemitem {$ii)$} the operator $\exp(-\varepsilon D^2)$ is trace class 
for all $\varepsilon >0\,$; 
\smallskip
\item {4)} $\gamma$ is a $\Z_2$-grading on \h, i.e., 
$\gamma = \gamma^* = \gamma^{-1}$, such that 
$$
\{\,\gamma,D\,\} = 0\ ,\quad 
\lb\, \gamma,a\,\rb = 0 \quad{\rm for\ all}\ \ a\in \a. 
$$ 
\mn
As mentioned before, in non-commutative geometry \a\  plays the role of 
the ``algebra of functions over a non-commutative space''. The 
existence of a unit in \a, together with property $3\,ii)$ above,  
reflects the fact that we are dealing with ``compact'' non-commutative 
spaces. Note that if the Hilbert space \h\ is infinite-dimensional, 
condition $3\,ii)$ implies that the operator $D$ is unbounded. By 
analogy with classical differential geometry, $D$ is interpreted as a 
(generalized) Dirac operator. 
\sn  
Also note that the fourth condition in Definition 5.1 does not impose 
any restriction on $N=1$ spectral data: In fact, given a triple 
$(\widetilde{\cal A}, \widetilde{\cal H}, \widetilde{D})$ satisfying 
the properties 1 - 3 {}from above, we can define a  set of $N=1$ 
{\sl even} spectral data $(\a,\h, D, \gamma)$ by setting   
$$\eqalign{
\h &= \widetilde{\cal H} \otimes \C^2\ ,\quad\quad 
\a = \widetilde{\cal A} \otimes \one_2\ ,
\cr
D &= \widetilde{D} \otimes \tau_1\ ,\quad\quad\  
\gamma\, = \one_{\tilde{\scriptscriptstyle \cal H}}\otimes \tau_3\ ,
\cr}$$
where $\tau_i$ are the Pauli matrices acting on $\C^2$. 
\bn\bn
{\bf 5.1.2 Differential forms}
\bn
The construction of differential forms follows the same lines
as in classical differential geometry: We define the unital, 
graded, differential ${}^*$-algebra of universal forms, 
$\Omega^{\bullet}(\a)$, as in section 3.1.4, see eqs.\ (3.18-23);
recall that, even in the classical case, no relations ensuring 
(graded) commutativity of $\Omega^{\bullet}(\a)$ were imposed. 
The complex conjugation in (3.22) is of course to be replaced 
by the ${}^*$-operation of \a, which, in turn, is inherited {}from
taking the adjoint on \h; in particular, we define  
$$
(\delta a)^* = - \delta(a^*)
\eqno(5.1)$$
for all $a\in\a$. With the help of the (self-adjoint) 
generalized Dirac operator $D$, we 
introduce a ${}^*$-representation $\pi$ of $\Omega^{\bullet}(\a)$ 
on \h\ as in eq.\ (3.24), namely
$$
\pi(a) = a\ ,\quad\quad \pi(\delta a) = \lb\,D,a\,\rb \ . 
$$
A graded ${}^*$-ideal $J$ of $\Omega^{\bullet}(\a)$ is defined by 
$$
J := \bigoplus_{k=0}^{\infty} J^k\ ,\quad\ \ 
J^k := {\rm ker}\,\pi\,|_{\Omega^k({\cal A})}\ .
\eqno(5.2)$$
Since $J$ is not a differential ideal, the graded quotient 
$\Omega^{\bullet}(\a)/J$ does not define a differential algebra 
and thus does not yield a satisfactory definition of the algebra 
of differential forms. This problem is solved as in the classical 
case:
\mn
{\bf Proposition 5.2}\ \q{Co1}\quad The graded sub-complex 
$$
J +\delta J=\bigoplus_{k=0}^{\infty}\,\bigl(J^k+\delta J^{k-1}\bigr)\ ,
$$
where $J^{-1} := 0$ and $\delta$ is the universal differential in 
$\Omega^{\bullet}(\a)$, is a two-sided graded differential ${}^*$-ideal 
of $\Omega^{\bullet}(\a)$. 
\mn
{\kap Proof}: Since $\delta^2=0$, it is clear that $J+\delta J$ is 
closed under $\delta$. We show that it is a two-sided ideal. 
By linearity, it is sufficient to consider homogeneous elements. 
Thus, let $\omega\in J^k$, $\eta\in J^{k-1}$, and 
$\phi\in\Omega^p(\a)$; then the Leibniz rule gives 
$$
\phi\,(\omega+\delta\eta) = \phi\,\omega 
- (-1)^p (\delta\phi)\eta + (-1)^p \delta(\phi\,\eta)\ .
$$
Since the first two terms are in $J^{p+k}$ and the last one is in 
$\delta J^{p+k-1}$, we have $\phi(\omega+\delta\eta) \in J + \delta J$. 
Analogously, we compute 
$$
(\omega+\delta\eta)\,\phi = \omega\,\phi - (-1)^{k-1} \eta\,\delta\phi 
+\delta(\eta\,\phi)
$$
which is again  an element of $J + \delta J$. \hfill\qed
\mn
We define the unital graded differential ${}^*$-algebra of 
differential forms, $\Omega_D^{\bullet}(\a)$, as the graded 
quotient $\Omega^{\bullet}(\a)/(J+\delta J)$, i.e., 
$$
\Omega_D^{\bullet}(\a) := \bigoplus_{k=0}^{\infty}\, \Omega_D^k(\a)\ ,
\quad\ \ 
\Omega_D^k(\a) := \Omega^k(\a)/(J^k+\delta J^{k-1})\ .
\eqno(5.3)$$
\sn    Since 
$\Omega_D^{\bullet}(\a)$ is a graded algebra, each $\Omega^k_D(\a)$ 
is, in particular, a bi-module over $\a=\Omega_D^0(\a)$. 
\sn 
On the other hand, note that $\pi$ does {\sl not} determine a 
representation of the algebra (or, for that matter, of the space) 
of differential forms $\Omega_D^{\bullet}(\a)$ on the Hilbert 
space \h: A differential $k$-form is an equivalence class 
$\lb\omega\rb\in\Omega_D^{k}(\a)$ with some representative 
$\omega\in\Omega^k(\a)$, and $\pi$ maps this class to a {\sl set} of 
bounded operators on \h, namely 
$$
\pi\bigl(\lb\omega\rb\bigr)=\pi(\omega)+\pi\bigl(\delta J^{k-1}\bigr)\ .
$$
In general, the only subspaces where we do not meet this complication 
are $\pi\bigl(\Omega^0_D(\a)\bigr) = \a$ and $\pi\bigl(
\Omega^1_D(\a)\bigr) \cong\pi\bigl(\Omega^1(\a)\bigr)$. However, the 
image of $\Omega_D^{\bullet}(\a)$ under $\pi$ is $\Z_2$-graded, 
$$
\pi\bigl(\Omega_D^{\bullet}(\a)\bigr) = 
\pi\Bigl(\,\bigoplus_{k=0}^{\infty} \Omega_D^{2k}(\a) \,\Bigr) \oplus
\pi\Bigl(\,\bigoplus_{k=0}^{\infty} \Omega_D^{2k+1}(\a) \,\Bigr)\ ,
$$
because of the (anti-)commutation properties of the 
$\Z_2$-grading $\gamma$ on \h, see Definition 5.1. 
\bn\bn
{\bf 5.1.3 Integration}
\bn
Property $3 ii)$ of the Dirac operator in Definition 5.1 allows
us to define the notion of integration over a non-commutative 
space in the same way as in the classical case. Note that, for 
certain sets of $N=1$ spectral data, we could use the Dixmier 
trace as Connes originally proposed; but the definition given 
below, first introduced in \q{CFF}, works in greater generality (cf.\ 
the remarks in section 3.1.5). Moreover, it is closer to notions 
coming up naturally in quantum field theory.    
\mn
{\bf Definition 5.3}\quad The {\sl integral} over the non-commutative
space described by the $N=1$ spectral data $(\a,\h,D,\gamma)$ is  
a state $\barint$ on $\pi\bigl(\Omega^{\bullet}(\a)\bigr)$
defined by 
$$
\Barint\,: \cases {&$\pi\bigl(\Omega^{\bullet}(\a)\bigr) \lra\, \C$\cr
&$\quad\quad \omega \phantom{xxx}\longmapsto\ \, 
{\displaystyle  \Barint\omega := \Lim_{\varepsilon\to 0^+} 
{ {\rm Tr}_{\scriptscriptstyle{\cal H}}\bigl( \omega e^{-\varepsilon
D^2}\bigr) 
\over {\rm Tr}_{\scriptscriptstyle{\cal H}}
\bigl( e^{-\varepsilon D^2}\bigr) } \ , }$\cr}
$$   
where $\Lim_{\varepsilon\to 0^+}$ denotes some limiting procedure 
making the functional $\barint$ linear and positive semi-definite;
existence of such a procedure can be shown analogously to \q{Co1,3}, 
where the Dixmier trace is discussed. 
\mn
For this integral $\barint$ to be a useful tool, we need an 
additional property that must be checked in each example: 
\mn
{\bf Assumption 5.4}\quad The state $\barint$ on 
$\pi\bigl(\Omega^{\bullet}(\a)\bigr)$ is {\sl cyclic}, i.e.,
$$
\Barint \omega\,\eta^* = \Barint \eta^*\,\omega 
$$
for all $\omega,\eta \in \pi\bigl(\Omega^{\bullet}(\a)\bigr)$.
\mn
The state $\barint$ determines a positive semi-definite 
sesqui-linear form on $\Omega^{\bullet}(\a)$ by setting
$$
(\omega,\eta) := \Barint \pi(\omega)\,\pi(\eta)^*
\eqno(5.4)$$
for all $\omega,\eta \in \Omega^{\bullet}(\a)$. In the formulas below, 
we will often drop the representation symbol $\pi$ under the integral, 
as there is no danger of confusion.
\sn
Note that the commutation relations of the grading $\gamma$ 
with the Dirac operator imply that forms of odd degree are 
orthogonal to those of even degree with respect to $(\cdot,\cdot)$. 
\sn  
By $K^k$ we denote the kernel of this sesqui-linear form restricted 
to $\Omega^k(\a)$. More precisely we set 
$$
K := \bigoplus_{k=0}^{\infty} K^k \ ,\quad\ 
K^k  := \{\, \omega\in\Omega^k(\a)\,|\, 
(\,\omega,\omega\,)= 0\,\}\ .
\eqno(5.5)$$ 
Obviously, $K^k$ contains the ideal $J^k$ defined in eq.\ (5.2); in 
the classical case they coincide. Assumption 5.4 is needed to 
show that $K$ is a two-sided ideal of the algebra of universal
forms, too, so that we can pass to the quotient algebra. 
\mn
{\bf Proposition 5.5}\quad The set $K$ is a two-sided, graded
${}^*$-ideal of $\Omega^{\bullet}(\a)$. 
\sn
{\kap Proof}: The Cauchy-Schwarz inequality for states implies
that $K$ is a vector space. If $\omega\in K^k$, then Assumption
5.4 gives  
$$
(\omega^*,\omega^*) = \Barint \pi(\omega)^* \pi(\omega)
= \Barint \pi(\omega) \pi(\omega)^* = 0\ ,
$$
i.e.\ that $K$ is closed under the involution. With $\omega$ as 
above and $\eta\in\Omega^p(\a)$, we have, on the other hand, that 
$$\eqalign{
(\eta\omega,\eta\omega) &= \Barint \pi(\eta)\pi(\omega)
\pi(\omega)^* \pi(\eta)^* 
= \Barint \pi(\omega)^* \pi(\eta)^* \pi(\eta)\pi(\omega)
\cr &\leq 
\Vert \pi(\eta)\Vert^2_{\scriptscriptstyle{\cal H}} 
\Barint \pi(\omega)^*\pi(\omega) = 0
\cr}$$
where $\Vert\cdot\Vert_{\scriptscriptstyle{\cal H}}$ is 
the operator norm on ${\cal B}(\h)$. 
On the other hand, we have that 
$$
(\omega\eta,\omega\eta) = 
 \Barint \pi(\omega) \pi(\eta) \pi(\eta)^*\pi(\omega)^* \leq 
\Vert \pi(\eta)\Vert^2_{\scriptscriptstyle{\cal H}} 
\Barint \pi(\omega)\pi(\omega)^* = 0\ ,
$$
and it follows that both $\omega\,\eta$ and $\eta\,\omega$ are 
elements of $K$, i.e., $K$ is a two-sided ideal.    \hfill\qed
\mn
We now define 
$$
\widetilde{\Omega}^{\bullet}(\a) 
:= \bigoplus_{k=0}^{\infty} \widetilde{\Omega}^k(\a)\ ,\quad
\widetilde{\Omega}^k(\a) := \Omega^k(\a)/K^k\ .
\eqno(5.6)$$
The sesqui-linear form $(\cdot,\cdot)$ descends to a positive 
definite scalar product on $\widetilde{\Omega}^k(\a)$, and we 
denote by $\widetilde{{\cal H}}^{k}$ the Hilbert space 
completion of this space with respect to the scalar product, 
$$
\widetilde{{\cal H}}^{\bullet} := 
\bigoplus_{k=0}^{\infty}\widetilde{{\cal H}}^k\ ,
\quad  \widetilde{{\cal H}}^k := 
\overline{\widetilde{\Omega}^k(\a)}^{\,{
\scriptscriptstyle (\cdot,\cdot)}}\ .
\eqno(5.7)$$
$\widetilde{{\cal H}}^k$ is to be interpreted as the {\sl space of 
square-integrable k-forms}. Note that $\widetilde{{\cal H}}^{\bullet}$ 
does not quite coincide with the Hilbert space that would arise {}from a 
GNS construction using the state $\barint$ on $\widetilde{\Omega}^{
\bullet}(\a)$: Whereas in $\widetilde{{\cal H}}^{\bullet}$,   
orthogonality of forms of different degree is installed by definition, 
there may occur mixings among forms of even degrees (or among the odd 
forms) in the GNS Hilbert space. 
\mn  
{\bf Corollary 5.6}\quad The space $\widetilde{\Omega}^{\bullet}(\a)$
is a unital graded ${}^*$-algebra. For any 
$\omega\in\widetilde{\Omega}^k(\a)$, the left and right actions of 
$\omega$ on $\widetilde{\Omega}^p(\a)$ with
values in $\widetilde{\Omega}^{p+k}(\a)$, 
$$
m_L(\omega)\eta := \omega\eta\ ,\quad 
m_R(\omega)\eta := \eta\omega\ ,
$$
are continuous in the norm given by $(\cdot,\cdot)$. In particular, 
the Hilbert space $\widetilde{{\cal H}}^{\bullet}$ is a bi-module over 
$\widetilde{\Omega}^{\bullet}(\a)$ with continuous actions. 
\sn
{\kap Proof}: The claim follows immediately {}from the two estimates 
given in the proof of the previous proposition, applied to 
$\omega\in\widetilde{\Omega}^k(\a)$ and 
$\eta\in\widetilde{\Omega}^p(\a)$.   \hfill\qed 
\sn 
This remark shows that $\widetilde{\Omega}^{\bullet}(\a)$ and 
$\widetilde{{\cal H}}^{\bullet}$ are ``well-behaved'' wrt.\ the 
$\widetilde{\Omega}^{\bullet}(\a)$-action. Furthermore, Corollary 5.6 
will be useful during our discussion of curvature and torsion in 
sections 5.1.7 and 5.1.8. 
\mn
Since the algebra $\widetilde{\Omega}^{\bullet}(\a)$ may fail to be 
differential, we introduce the unital graded differential ${}^*$-algebra 
of square-integrable differential forms 
$\widetilde{\Omega}_D^{\bullet}(\a)$ as the graded quotient of 
$\Omega^{\bullet}(\a)$ by $K+\delta K$, 
$$
\widetilde{\Omega}_D^{\bullet}(\a) := \bigoplus_{k=0}^{\infty} 
\widetilde{\Omega}_D^k(\a)\ ,\quad
\widetilde{\Omega}_D^{k}(\a) := \Omega^k(\a)/(K^k+\delta K^{k-1}) 
\cong \widetilde{\Omega}^{k}(\a)/ \delta K^{k-1} \ .
\eqno(5.8)$$
In order to show  that $\widetilde{\Omega}_D^{\bullet}(\a)$ has the 
stated properties one repeats the proof of Proposition 5.2. 
Note that we can regard the \a-bi-module 
$\widetilde{\Omega}_D^{\bullet}(\a)$ as a 
``smaller version'' of $\Omega_D^{\bullet}(\a)$ in the sense that 
there exists a projection {}from the latter onto the former; whenever 
one deals with a concrete set of $N=1$ spectral data that
satisfy Assumption 5.4, it will be advantageous to work with 
the ``smaller'' algebra of square-integrable differential forms. 
The algebra $\Omega_D^{\bullet}(\a)$, on the other hand, can be 
defined for arbitrary data. 
\sn
In the classical case, differential forms were identified 
with the orthogonal complement of $Cl^{(k-2)}$ within $Cl^{(k)}$, 
see the remark after eq.\ (3.32). Now, we use the 
scalar product $(\cdot,\cdot)$ on 
$\widetilde{{\cal H}}^{k}$ to introduce, for each 
$k\geq 1$,  the orthogonal projection 
$$
P_{\delta K^{k-1}}\,:\ 
\widetilde{{\cal H}}^k \lra\ \widetilde{{\cal H}}^k
\eqno(5.9)$$
onto the image of $\delta K^{k-1}$ in $\widetilde{{\cal H}}^k$, and 
we set 
$$
\omega^{\perp}:=(1-P_{\delta K^{k-1}})\,\omega\in\widetilde{{\cal H}}^k 
\eqno(5.10)$$
for each element $\lb\omega\rb\in\widetilde{\Omega}_D^{k}(\a)$. This 
allows us to define a positive definite scalar product on $\widetilde{
\Omega}_D^{k}(\a)$ via the representative $\omega^{\perp}\,$:
$$
(\,\lb\omega\rb, \lb\eta\rb\,) := (\,\omega^{\perp},\eta^{\perp}\,) 
\eqno(5.11)$$
for all $\lb\omega\rb, \lb\eta\rb \in \widetilde{\Omega}_D^{k}(\a)$. 
In the classical case, this is just 
the usual inner product on the space of square-integrable $k$-forms. 
\bn\bn
{\bf 5.1.4 Vector bundles and Hermitian structures}
\bn
Again, we simply follow the algebraic formulation of classical 
differential geometry in order to generalize the notion of a vector 
bundle to the non-commutative case:
\mn
{\bf Definition 5.7} \q{Co1}\quad A {\sl vector bundle} \e\ over the 
non-commutative space described by the $N=1$ spectral data 
$(\a,\h,D,\gamma)$ is a finitely generated projective left \a-module. 
\mn
Recall that a module \e\ is {\sl projective} if there exists another 
module ${\cal F}$ such that the direct sum ${\cal E} \oplus {\cal F}$
is {\sl free}, i.e., ${\cal E} \oplus {\cal F} \cong {\cal A}^n$ as 
left \a-modules,for some $n\in\N$. Since \a\ is an algebra, every 
\a-module is a vector space; therefore, left \a-modules are 
representations of the algebra \a, and \e\ is projective iff there 
exists a module ${\cal F}$ such that ${\cal E} \oplus {\cal F}$ is 
isomorphic to a multiple of the left-regular 
representation. 
\sn
By Swan's Lemma \q{Sw}, a finitely generated projective left module
corresponds, in  the commutative case, to the space of sections of 
a vector bundle. With this in mind, it is straightforward to define 
the notion of a Hermitian structure over a vector bundle \q{Co1}:
\mn
{\bf Definition 5.8}\quad A {\sl Hermitian structure} over a vector 
bundle \e\ is a sesqui-linear map (linear in the first argument) 
$$
\langle \cdot,\cdot \rangle\,:\ \e \times \e \lra\ \a 
$$
such that for all $a,b\in\a$ and all $s,t \in\e$ 
\smallskip
\item {1)}   $\langle\,as,bt\,\rangle = a\, \langle\,s,t\,\rangle\, 
   b^*\,$; 
\smallskip
\item {2)}   $\langle\,s,s\,\rangle \geq 0\,$;
\smallskip
\item {3)} the \a-linear map  
$$
g\,: \cases {&$\e\lra\, \ {\cal E}^*_R$\cr
&$\, s \longmapsto\ \,\vphantom{\sum^k}\langle\,s,\cdot\,\rangle  
   $\cr} \ ,
$$
where ${\cal E}^*_R := \{\, \phi\in{\rm Hom}(\e,\a)\,|\, 
\phi(as) = \phi(s)a^*\,\}$,\  
is an isomorphism of left \a-modules, i.e., $g$ can be regarded as 
a metric on \e. 
\sn 
In the second condition, the notion of positivity in \a\ 
is simply inherited {}from the algebra ${\cal B}(\h)$ of all 
bounded operators on the Hilbert space \h.
\bn\bn
\leftline{\bf 5.1.5 Generalized Hermitian structure on \Omti}
\bn
In this section we show that the \a-bi-modules 
$\widetilde{\Omega}^k(\a)$ carry Hermitian structures in a slightly 
generalized sense. Let
$\overline{{\cal A}}$ be the weak closure of the algebra \a\ acting on 
$\widetilde{{\cal H}}^0$, i.e., $\overline{{\cal A}}$ is the von 
Neumann algebra generated by $\widetilde{\Omega}^0(\a)$ acting on the 
Hilbert space  $\widetilde{{\cal H}}^0$. 
\mn
{\bf Theorem 5.9}\quad There is a canonically defined sesqui-linear 
map 
$$
\langle\cdot,\cdot\rangle_{\scriptscriptstyle D}\,:\
\widetilde{\Omega}^k(\a)\times\widetilde{\Omega}^k(\a)
\lra\ \overline{\a}
$$                             such that for all 
$a,b\in\a$ and all $\omega,\eta\in\widetilde{\Omega}^k(\a)$ 
\smallskip
\item {1)} $\langle\,a\,\omega,b\,\eta\,\rangle_{\scriptscriptstyle D} 
= a\,\langle\,\omega,\eta\,\rangle_{\scriptscriptstyle D}\, b^*\,$; 
\smallskip
\item {2)} $\langle\,\omega,\omega\,\rangle_{\scriptscriptstyle D}
     \geq 0\,$;   
\smallskip
\item {3)} $\langle\,\omega\,a,\eta\,\rangle_{\scriptscriptstyle D} = 
\langle\,\omega ,\eta\,a^*\,\rangle_{\scriptscriptstyle D}\,$.
\sn
We call $\langle\cdot,\cdot\rangle_{\scriptscriptstyle D}$ a 
{\sl generalized Hermitian structure on} $\widetilde{\Omega}^k(\a)$. 
It is the non-commutative analogue of the Riemannian metric on the 
bundle of differential forms. Note that $\langle\cdot,\cdot\rangle_{
\scriptscriptstyle D}$  takes values in $\overline{\a}$ 
and thus property 3) of Definition 5.8 is not directly applicable.
\sn
{\kap Proof}: Let $\omega,\eta\in\widetilde{\Omega}^k(\a)$ and define 
the $\C$-linear map 
$$
\varphi_{\omega,\eta}(a) = \Barint a\,\eta\,\omega^*
$$
for all $a\in\widetilde{\Omega}^0(\a)$. Note that $a$ on the rhs 
actually is a representative in \a\ of the class  $a\in
\widetilde{\Omega}^0(\a)$, and analogously for $\omega$ and $\eta$ 
(and we have omitted the representation symbol $\pi$).  
The value of the integral is, however, independent of the choice of 
these representatives, which is why we used the same letters.  
The map $\varphi$ satisfies 
$$
|\varphi_{\omega,\eta}(a)| 
\leq \Big|\,\Barint aa^*\,\Big|^{1\over2} \Big|\,\Barint
\omega\eta^*\eta\omega^*\,
\Big|^{1\over2} \leq (a,a)^{1\over2} 
\Big|\,\Barint \omega\eta^*\eta\omega^*\,\Big|^{1\over2} \ .
$$
Therefore, $\varphi_{\omega,\eta}$ extends to a bounded linear 
functional on $\widetilde{{\cal H}}^0$, and there exists an element
$\langle\,\omega,\eta\,\rangle_{\scriptscriptstyle D} 
\in\widetilde{\cal H}^0$ such that 
$$
\varphi_{\omega,\eta}(x) = 
(x, \langle\,\omega,\eta\,\rangle_{\scriptscriptstyle D}\,)
$$
for all $x\in\widetilde{{\cal H}}^0$; since $(\cdot,\cdot)$ is 
non-degenerate, $\langle\,\omega,\eta\,\rangle_{\scriptscriptstyle D}$ 
is a well-defined element; but it remains to
show that it also acts as a bounded operator on this Hilbert space. 
To this end, choose a net $\{a_{\iota}\}\subset
\widetilde{\Omega}^0(\a)$ which converges to
$\langle\,\omega,\eta\,\rangle_{\scriptscriptstyle D}$. Then,  
for all $b,c\in\widetilde{\Omega}^0(\a)$,  
$$\eqalign{
(\,\langle\,\omega,\eta\,\rangle_{\scriptscriptstyle D} b,c\,) 
&= \lim_{\iota\to\infty}(\,a_{\iota}b,c\,) 
= \lim_{\iota\to\infty} \Barint a_{\iota}bc^* = \lim_{\iota\to\infty}
\Barint a_{\iota}(cb^*)^* 
\cr
&= \lim_{\iota\to\infty}(\,a_{\iota},cb^*\,) =  
(\,\langle\,\omega,\eta\,\rangle_{\scriptscriptstyle D}, cb^*\,) \ ,
\cr}$$
and it follows that 
$$\eqalign{
|(\,\langle\,\omega,\eta\,\rangle_{\scriptscriptstyle D} b,c\,)| &= 
|(\,\langle\,\omega,\eta\,\rangle_{\scriptscriptstyle D}, cb^*\,)| 
= |(\,cb^*, \langle\,\omega,\eta\,\rangle_{\scriptscriptstyle D}\,)| 
\cr
&= \Big|\, \Barint cb^* \eta\,\omega^*\, \Big| 
= \Big| \,\Barint\omega^* cb^* \eta\, \Big| 
= \Big| \,\Barint b^* \eta\,\omega^* c\, \Big|
\cr 
&\leq \Big|\,\Barint b^*b\,\Big|^{1\over2} \,
\Big|\,\Barint c^*\omega\,\eta^*\eta\omega^*c\,\Big|^{1\over2}
\leq \Vert\, \omega\eta^*\Vert_{\scriptscriptstyle{\cal H}}\, \Big|
\,\Barint b^*b\,\Big|^{1\over2}\Big|\,\Barint c^*c\,\Big|^{1\over2}
\cr
&\leq \Vert\,\omega\eta^*\Vert_{\scriptscriptstyle{\cal H}}\, 
(b,b)^{1\over2}  (c,c)^{1\over2} \ .
\cr}$$
In the third line, we first use the Cauchy-Schwarz inequality for the 
positive state $\barint$, and then an estimate which is true for 
all positive operators on a Hilbert space; the upper bound 
$\Vert\,\omega\eta^*\Vert_{\scriptscriptstyle{\cal H}}$ again involves 
representatives $\omega,\eta\in\pi\bigl(\Omega^k(\a)\bigr)$, which was 
not explicitly indicated above, since any two will do. \hfill\break
\noindent As $\widetilde{\Omega}^0(\a)$ is dense in 
$\widetilde{{\cal H}}^0$, we see that 
$\langle\,\omega,\eta\,\rangle_{\scriptscriptstyle D}$ 
indeed defines a bounded operator in $\widetilde{{\cal H}}^0$, which, 
by definition, is the weak limit of elements in 
$\widetilde{\Omega}^0(\a)$, i.e., it belongs to $\overline{\a}$.
Properties 1-3 of $\langle\cdot,\cdot\rangle_{\scriptscriptstyle D}$ 
are easy to verify.\hfill\qed
\mn
Note that the definition of the metric $\langle\cdot,
\cdot\rangle_{\scriptscriptstyle D}$ given here differs slightly {}from 
the one of refs.\ \q{CFF,CFG}. One can, however, show that in the 
$N=1$ case both definitions agree; moreover, the present one is 
better suited for the $N=(1,1)$ formulation to be introduced later.
\bn\bn
\eject
\leftline{\bf 5.1.6 Connections}
\bn
{\bf Definition 5.10}\quad A {\sl connection} $\nabla$ on a vector 
bundle \e\ over a non-commutative space is a $\C$-linear map 
$$
\nabla\,:\ \e \lra\  \widetilde{\Omega}^1_D(\a) 
\otimes_{\scriptscriptstyle {\cal A}}\e 
$$
such that $$
\nabla(as) = \delta a\otimes s + a \nabla s 
$$
for all $a\in\a$ and all $s \in \e$. 
\mn
Given a vector bundle \e, we define a space of \e-valued 
differential forms by 
$$
\widetilde{\Omega}^{\bullet}_D(\e) := \widetilde{\Omega}^{
\bullet}_D(\a)\otimes_{\scriptscriptstyle {\cal A}}\e \ ;
$$
if $\nabla$ is a connection on \e, then it extends uniquely to a 
$\C$-linear map, again denoted $\nabla$, 
$$
\nabla\,:\ \widetilde{\Omega}^{\bullet}_D(\e)\lra
\widetilde{\Omega}^{\bullet+1}_D(\e)
\eqno(5.12)$$
such that 
$$
\nabla(\omega s) = \delta\omega\,s + (-1)^k \omega\,\nabla s 
\eqno(5.13)$$
for all $\omega\in\widetilde{\Omega}^{k}_D(\a)$ and all 
$s \in  \widetilde{\Omega}^{\bullet}_D(\e)$. 
\mn
{\bf Definition 5.11}\quad The {\sl curvature} of a connection 
$\nabla$ on a vector bundle \e\ is given by 
$$
{\ttR}\,(\nabla) = - \nabla^2\,:\ \e \lra\ \widetilde{\Omega}^2_D(\a) 
\otimes_{\scriptscriptstyle {\cal A}}\e \ .
$$
Note that the curvature extends to a map 
$$ 
{\ttR}\,(\nabla)\,:\ \widetilde{\Omega}^{\bullet}_D(\e)\lra
\widetilde{\Omega}^{\bullet+2}_D(\e)
$$                                                   which is left 
\a-linear, as follows easily {}from eq.\ (5.12) and Definition 5.10.
\mn
{\bf Definition 5.12}\quad A connection $\nabla$ on a Hermitian vector 
bundle $(\e,\langle\cdot,\cdot\rangle)$ is called {\sl unitary} if 
$$
\delta\,\langle\,s,t\,\rangle = \langle\,\nabla s,t\,\rangle -
\langle\,s,\nabla t\,\rangle 
$$
for all $s,t\in\e$, where the rhs of this equation is defined by 
$$
\langle\,\omega\otimes s,t\,\rangle = \omega\,\langle\,s,t\,\rangle\ ,
\quad\quad
\langle\,s,\eta\otimes t\,\rangle = \langle\,s,t\,\rangle\,\eta^*
\eqno(5.14)$$
for all $\omega,\eta\in \widetilde{\Omega}^1_D(\a)$ and  all $s,t\in\e$. 
\bn\bn
\vfil\eject
\leftline{\bf 5.1.7 Riemannian curvature and torsion}
\bn
Throughout this section, we  make three additional assumptions which 
limit the generality of our results, 
but turn out to be fulfilled in interesting examples. 
\mn
{\bf Assumption 5.13}\quad We assume that the $N=1$ spectral data under
consideration have the following additional properties:
\smallskip
\item {1)} $K^0= 0$. (This implies that $\widetilde{\Omega}_D^0(\a)=\a$ 
and $\widetilde{\Omega}_D^1(\a)=\widetilde{\Omega}^1(\a)$, 
thus $\widetilde{\Omega}_D^1(\a)$ carries a generalized Hermitian
structure.) 
\smallskip
\item {2)} $\widetilde{\Omega}_D^1(\a)$ is a vector bundle, called 
the {\sl cotangent bundle over} \a. ($\,\widetilde{\Omega}_D^1(\a)$ is 
always a left \a-module. Here, we assume, in addition, 
that it is {\sl finitely generated} and {\sl projective}.)
\smallskip
\item {3)} The generalized metric 
$\langle\cdot,\cdot\rangle_{\scriptscriptstyle D}$ on 
$\widetilde{\Omega}_D^1(\a)$ defines an isomorphism 
of left \a-modules between $\widetilde{\Omega}_D^1(\a)$ 
and the space of \a-anti-linear maps {}from $\widetilde{\Omega}_D^1(\a)$ 
to \a, i.e., for each \a-anti-linear map
$$
\phi\,:\ \widetilde{\Omega}^1_D(\a) \lra\ \a 
$$
satisfying $\phi(a\omega) = \phi(\omega)a^*$ for all
$\omega\in\widetilde{\Omega}^1_D(\a)$ and all $a\in\a$, there is a 
unique $\eta_{\phi}\in\widetilde{\Omega}^1_D(\a)$ with
$$
\phi(\omega) = \langle\,\eta_{\phi},\omega\,
\rangle_{\scriptscriptstyle D}\ .
$$
\mn
If $N=1$ spectral data $(\a,\h,D,\gamma)$ satisfy these assumptions, 
we are able to define non-commutative generalizations of  classical 
notions like torsion and curvature. Whereas torsion and Riemann 
curvature can be introduced whenever $\widetilde{\Omega}_D^1(\a)$ 
is a vector bundle, the last assumption in 5.13 will provide a 
substitute for the procedure of ``contracting indices'' leading to 
Ricci and scalar curvature. 
\mn
{\bf Definition 5.14}\quad Let $\nabla$ be a connection on the 
cotangent bundle $\widetilde{\Omega}^1_D(\a)$ over a non-commutative 
space $(\a,\h,D,\gamma)$ satisfying Assumption 5.13. The {\sl torsion} 
of $\nabla$ is the $\a$-linear map 
$$
{\ttT}(\nabla) := {\delta} - m \circ \nabla \,:\ 
\widetilde{\Omega}^1_D(\a)\lra\ \widetilde{\Omega}^2_D(\a)
$$ 
where $m\,:\ \widetilde{\Omega}^1_D(\a)\otimes_{\scriptscriptstyle 
{\cal A}}\widetilde{\Omega}^1_D(\a)
\lra\ \widetilde{\Omega}^2_D(\a)$ denotes the product of 1-forms in 
$\widetilde{\Omega}^{\bullet}_D(\a)$.  
\mn
Using the definition of a connection, \a-linearity of torsion is 
easy to verify. In analogy to the classical case, a unitary connection  
$\nabla$ with ${\ttT}(\nabla)=0$ is called a {\sl Levi-Civita connection}. 
Note, however, that, for a given set of non-commutative spectral data, 
there may be several Levi-Civita connections -- or none at all. 
\mn
Since we assume that $\widetilde{\Omega}^1_D(\a)$ is a vector bundle, 
we can define the {\sl Riemannian curvature} of a connection $\nabla$ 
on the cotangent bundle as a specialization of Definition 5.11. To 
proceed further, we make use of part 2) of Assumption 5.13, which 
implies that there exists a finite set of generators $\{\, E^A\,\}$ 
of  $\widetilde{\Omega}^1_D(\a)$ and an associated ``dual basis'' 
$\{\,\varepsilon_A\,\}\subset \widetilde{\Omega}^1_D(\a)^*$, 
$$
\widetilde{\Omega}^1_D(\a)^* := \bigl\{\,\phi\,:\ 
\widetilde{\Omega}^1_D(\a) \lra\ \a\,\big\vert\, 
\phi(a\omega) = a\phi(\omega)\quad {\rm  for\  all}\  
a\in\a, \omega\in\widetilde{\Omega}^1_D(\a)\,\bigr\}\ ,
$$ 
such that each $\omega\in\widetilde{\Omega}^1_D(\a)$ can be written 
as $\omega = \varepsilon_A(\omega) E^A\,$, see e.g.\ \q{Ja}. 
Because the curvature is \a-linear, there is 
a family of elements $\{\, {\ttR}^A_{\phantom{A}B}\,\} \subset
\widetilde{\Omega}^2_D(\a)$ with 
$$
{\ttR}\,(\nabla) = 
\varepsilon_A \otimes {\ttR}^A_{\phantom{A}B} \otimes E^B \ ;
\eqno(5.15)$$
here and in the following the summation convention is used. 
Put differently, we have applied the canonical isomorphism of 
vector spaces 
$$
{\rm Hom}_{\scriptscriptstyle {\cal A}}\bigl(\widetilde{\Omega}^1_D(\a),
\  \widetilde{\Omega}^2_D(\a) \otimes_{\scriptscriptstyle {\cal A}} 
\widetilde{\Omega}^1_D(\a)\bigr) \ \cong \  
\widetilde{\Omega}^1_D(\a)^* \otimes_{\scriptscriptstyle {\cal A}} 
\widetilde{\Omega}^2_D(\a)   
\otimes_{\scriptscriptstyle {\cal A}}  \widetilde{\Omega}^1_D(\a) 
$$
-- which exists because $\widetilde{\Omega}^1_D(\a)$ is projective 
-- and chosen explicit generators $E^A, \varepsilon_A$. Then we have  
that ${\ttR}\,(\nabla)\,\omega =
\varepsilon_A(\omega)\,{\ttR}^A_{\phantom{A}B} \otimes E^B$ for any 
1-form $\omega\in\widetilde{\Omega}^1_D(\a)$. 
\sn                         Note that although the components 
${\ttR}^A_{\phantom{A}B}$ need not be unique, the element on the rhs 
of eq.\ (5.15) is well-defined. Likewise, the Ricci and scalar 
curvature, to be introduced below, will be {\sl invariant} combinations  
of those components, as long as we make sure that all maps we use have 
the correct ``tensorial properties'' with respect to the \a-action. 
\sn 
The last part of Assumption 5.13 guarantees, furthermore, that to 
each $\varepsilon_A$ there exists a unique 1-form 
$e_A\in\widetilde{\Omega}^1_D(\a)$ such that 
$$
\varepsilon_A(\omega) = \langle\,\omega,e_A\,
\rangle_{\scriptscriptstyle D} 
$$
for all $\omega\in\widetilde{\Omega}^1_D(\a)$. By Corollary 5.6, every 
such $e_A$ determines a bounded operator $m_L(e_A)\,:\ 
\widetilde{{\cal H}}^1 \lra\widetilde{{\cal H}}^2$ acting on 
$\widetilde{{\cal H}}^1$ by left multiplication with $e_A$. The adjoint 
of this operator wrt.\ the scalar product $(\cdot,\cdot)$ on 
$\widetilde{\cal H}^{\bullet}$ is denoted by 
$$ 
e_A^{\rm ad}\,:\ \widetilde{{\cal H}}^2 \lra\ \widetilde{{\cal H}}^1 \ .
\eqno(5.16)$$
$e_A^{\rm ad}$ is a map of right \a-modules, and it is easy 
to see that also the correspondence $\varepsilon_A \mapsto 
e_A^{\rm ad}$ is right \a-linear: For all 
$b\in\a,\ \omega\in\widetilde{\Omega}^1_D(\a)$, we have 
$$
(\varepsilon_A\cdot b)(\omega) = \varepsilon_A(\omega)\cdot b = 
\langle\,\omega,e_A\,\rangle\,b=\langle\,\omega,b^* e_A\,\rangle\ ,
$$
and, furthermore, for all $\xi_1\in\widetilde{{\cal H}}^1,\ \xi_2\in
\widetilde{{\cal H}}^2$, 
$$
(\,b^*e_A(\xi_1),\xi_2\,) = (\,e_A(\xi_1),b \xi_2\,) = 
(\,\xi_1,e_A^{\rm ad}(b\xi_2)\,)\ ,
$$
where scalar products have to be taken in the appropriate spaces  
$\widetilde{{\cal H}}^k$. Altogether, the asserted right 
\a-linearity follows. Therefore, the map
$$
\varepsilon_A \otimes {\ttR}^A_{\phantom{A}B} \otimes E^B 
\longmapsto e_A^{\rm ad} \otimes {\ttR}^A_{\phantom{A}B} \otimes E^B
$$
is well-defined and has the desired tensorial properties. 
The definition of  Ricci curvature involves another operation which 
we require to be similarly well-behaved: 
\mn 
{\bf Lemma 5.15}\quad The orthogonal projections $P_{\delta K^{k-1}}$ 
on $\widetilde{{\cal H}}^k$, see eq.\ (5.9), satisfy 
$$
P_{\delta K^{k-1}}(axb) = a P_{\delta K^{k-1}}(x) b 
$$ 
for all $a,b\in\a$ and all $x\in\widetilde{{\cal H}}^k$. 
\sn
{\kap Proof}: Set $P := P_{\delta K^{k-1}}$, and let 
$y\in P\widetilde{{\cal H}}^k$. Then
$$
(\,P(axb), y\,) = (\,axb, P(y)\,) = (\,axb, y\,) = (\,x, a^*yb^*\,) 
= (\,x,P(a^*yb^*)\,) = (\,aP(x)b, y\,)\ , 
$$
where we have used that $P$ is self-adjoint wrt.\ $(\cdot,\cdot)$, that 
$Py=y$, and that the image of $P$ is an \a-bi-module.   \hfill\qed 
\mn
This lemma shows that projecting onto the ``2-form part'' of
${\ttR}^A_{\phantom{A}B}$ is an \a-bi-module map, i.e., we may apply  
$$
e_A^{\rm ad} \otimes {\ttR}^A_{\phantom{A}B} \otimes E^B
\longmapsto e_A^{\rm ad} 
\otimes \bigl({\ttR}^A_{\phantom{A}B}\bigr)^{\perp}\otimes E^B
$$ 
with $\bigl({\ttR}^A_{\phantom{A}B}\bigr)^{\perp}=(1-P_{\delta K^1})\,
{\ttR}^A_{\phantom{A}B}$ as in\ eq.\ (5.10).  Altogether, we arrive 
at the following definition of the {\sl Ricci curvature}, 
$$
{\ttRic}(\nabla) = e^{\rm ad}_A \Bigl(\bigl(
{\ttR}^A_{\phantom{A}B}\bigr)^{\perp}
\Bigr)\otimes E^B \in \widetilde{{\cal H}}^1 
\otimes_{{\cal A}} \widetilde{\Omega}^1_D(\a) \ ,
$$
which is in fact independent of any choices. In the following, 
we will also use the abbreviation 
$$
{\ttRic}_B := e^{\rm ad}_A \Bigl(\bigl({\ttR}^A_{\phantom{A}B}
\bigr)^{\perp}\Bigr) 
$$ 
for the components (which, again, are not uniquely defined).   
\sn     {}From  
the components ${\ttRic}_B$ we can pass to  scalar curvature. 
Again, we have to make sure that all maps occurring in this process are 
\a-covariant so as to obtain an invariant definition.  For any 1-form 
$\omega\in\widetilde{\Omega}^1_D(\a)$, right multiplication on 
$\widetilde{{\cal H}}^0$ with $\omega$ defines a bounded operator 
$m_R(\omega)\,:\ \widetilde{{\cal H}}^0 \lra\ \widetilde{{\cal H}}^1$, 
and we denote by  
$$
\omega_R^{\rm ad}\, :\ 
\widetilde{{\cal H}}^1 \lra\  \widetilde{{\cal H}}^0 
\eqno(5.17)$$
the adjoint of this operator. In a similar fashion as above, one 
establishes that 
$$
(\omega a)_R^{\rm ad}\,(x) = \omega_R^{\rm ad}\,(xa^*)
$$
for all $x\in\widetilde{{\cal H}}^1$ and $a\in\a$. This makes it 
possible to define the {\sl scalar curvature} ${\ttr}\,(\nabla)$ of 
a connection $\nabla$ as 
$$
{\ttr}\,(\nabla) = \bigl( E^{B*} \bigr)_R^{\rm ad} ({\ttRic}_B) 
\in \widetilde{{\cal H}}^0\ .
$$
As was the case for the Ricci tensor, acting with the adjoint of 
$m_R\bigl( E^{B*} \bigr)$ serves as a substitute for ``contraction of
indices''. We summarize our results in the following 
\mn
{\bf Definition 5.16}\quad Let  $\nabla$ be a connection on the 
cotangent bundle $\widetilde{\Omega}^1_D(\a)$ over a non-commutative 
space $(\a,\h,D,\gamma)$ satisfying Assumption 5.13. The {\sl 
Riemannian curvature} ${\ttR}\,(\nabla)$ is the left \a-linear map 
$$
{\ttR}\,(\nabla) = - \nabla^2\,:\ \widetilde{\Omega}^1_D(\a)
 \lra\ \widetilde{\Omega}^2_D(\a) 
\otimes_{\scriptscriptstyle {\cal A}} \widetilde{\Omega}^1_D(\a)\ .
$$
Choosing a set of generators $E^A$ of $\widetilde{\Omega}^1_D(\a)$  
and dual generators $\varepsilon_A$ of $\widetilde{\Omega}^1_D(\a)^*$, 
and writing  ${\ttR}\,(\nabla) = 
\varepsilon_A \otimes {\ttR}^A_{\phantom{A}B} \otimes E^B$
as above, the {\sl Ricci tensor} ${\ttRic}\,(\nabla)$ is given by 
$$
{\ttRic}(\nabla) =
{\ttRic}_B \otimes E^B \in \widetilde{{\cal H}}^1 
\otimes_{{\cal A}} \Omega^1_D(\a) \ ,
$$
where ${\ttRic}_B := e^{\rm ad}_A \Bigl(\bigl({\tt
R}^A_{\phantom{A}B}\bigr)^{\perp}\Bigr)$, see eqs.\ (5.10) and (5.16). 
Finally, the {\sl scalar curvature} ${\ttr}\,(\nabla)$ of the 
connection $\nabla$ is defined as 
$$
{\ttr}\,(\nabla) = \bigl( E^{B*} \bigr)_R^{\rm ad} ({\ttRic}_B) 
\in \widetilde{{\cal H}}^0\ , 
$$
with the notation of eq.\ (5.17). Both ${\ttRic}(\nabla)$ and
${\ttr}\,(\nabla)$ do not depend on the choice of generators.  
\bn\bn
{\bf 5.1.8 Non-commutative Cartan structure equations}
\bn
The classical Cartan structure equations are an important tool for
explicit calculations in differential geometry. Non-commutative 
analogues of those equations were obtained in \q{CFF,CFG}. Since in 
these references proofs were only sketched, we will give a rather 
detailed account of their results in the following. Throughout 
this section, we  assume that the space$\widetilde{\Omega}^1_D(\a)$ 
is a vector bundle over \a. In fact, no other properties of this 
space are used. Therefore all the statements on the non-commutative 
Cartan structure equations for the curvature will hold for any 
finitely generated projective module \e\ over \a; the torsion tensor, 
on the other hand, is defined only on the cotangent bundle over a 
non-commutative space. 
\mn                           Let $\nabla$ be a 
connection on the vector bundle $\widetilde{\Omega}^1_D(\a)$, 
then the curvature and the torsion of $\nabla$ are 
the left \a-linear maps given in Definitions 5.16 and 5.14, 
$$\eqalign{
\ttR\,(\nabla)\,:\ \ &\widetilde{\Omega}^1_D(\a) \lra\ 
\widetilde{\Omega}^2_D(\a) \otimes_{\scriptscriptstyle {\cal A}} 
\widetilde{\Omega}^1_D(\a)\ ,
\cr
\ttT\,(\nabla)\,:\ \ &\widetilde{\Omega}^1_D(\a) \lra\ 
\widetilde{\Omega}^2_D(\a)\ . \cr}$$
Since the left \a-module $\widetilde{\Omega}^1_D(\a)$ is finitely 
generated, we can choose a finite set of generators 
$\{\,E^A\,\}_{A=1,\ldots,N}\subset 
\widetilde{\Omega}^1_D(\a)$, and define the components
$\Omega^A_{\phantom{A}B}\in\widetilde{\Omega}^1_D(\a)$, 
${\ttR}^A_{\phantom{A}B}\in\widetilde{\Omega}^2_D(\a)$ and 
${\ttT}^A \in\widetilde{\Omega}^2_D(\a)$ of connection, curvature 
and torsion, resp., by setting
$$\eqalignno{
&\nabla\, E^A = - \,\Omega^A_{\phantom{A}B} \otimes E^B  \ ,
&(5.18)\cr
&\ttR\,(\nabla)\, E^A = {\ttR}^A_{\phantom{A}B} \otimes E^B\ , 
&(5.19)\cr
&\ttT\,(\nabla)\, E^A = {\ttT}^A \ .
&(5.20)\cr}$$
Note that the components $\Omega^A_{\phantom{A}B}$ and
${\ttR}^A_{\phantom{A}B}$ 
are {\sl not} uniquely defined if  $\widetilde{\Omega}^1_D(\a)$ is not 
a free module. Using Definitions 5.16 and 5.14, the components of the 
curvature and torsion tensors can be expressed in terms of the 
connection components: 
$$\eqalignno{
{\ttR}^A_{\phantom{A}B} &= \delta\,\Omega^A_{\phantom{A}B} + 
\Omega^A_{\phantom{A}C}\,\Omega^C_{\phantom{C}B}  \ ,
&(5.21)\cr
{\ttT}^A\   &= \delta E^A + \Omega^A_{\phantom{A}B} E^B\ .   
&(5.22)\cr}$$
As they stand, eqs.\ (5.21) and (5.22) cannot be applied for solving 
typical problems like finding a connection without torsion, because 
the connection components $\Omega^A_{\phantom{A}B}$ cannot be 
chosen at will unless $\widetilde{\Omega}^1_D(\a)$ is free.
We obtain more useful Cartan structure equations if we can 
relate the components $\Omega^A_{\phantom{A}B}$ to those of a 
connection $\widetilde{\nabla}$ on a free module $\a^N$. To this 
end, we employ some general constructions valid for any 
finitely generated projective left \a-module \e. 
\mn
Let $\{\,\widetilde{E}{}^A\,\}_{A=1,\ldots,N}$ be the canonical basis 
of the standard module $\a^N$, and define a left \a-module homomorphism 
$$
p\,:\ \cases{ &$\ \ \a^N\   \lra\ \widetilde{\Omega}^1_D(\a)$ \cr
&$a_A \widetilde{E}
\vphantom{M^{M^M}}{}^A \longmapsto\  a_A E^A$ \cr} 
\eqno(5.23)$$ 
for all $a_A\in\a$. Since $\widetilde{\Omega}^1_D(\a)$ is projective 
there exists a left \a-module ${\cal F}$ such that 
$$
\widetilde{\Omega}^1_D(\a) \oplus {\cal F} \cong \a^N \ .
\eqno(5.24)$$
Denote by  $i\,:\ \widetilde{\Omega}^1_D(\a) \lra\ \a^N$  the 
inclusion map determined by the isomorphism (5.24), which satisfies  
$p\circ i = {\rm id}$ on $\widetilde{\Omega}^1_D(\a)$. 
For each $A=1,\ldots,N$, we define a left \a-linear map 
$$
\widetilde{\varepsilon}_A\,:\ \cases{ 
&$\ \ \a^N\ \lra\ \a$\cr  
&$a_B\widetilde{E}\vphantom{M^{M^M}}{}^B \longmapsto\ a_A$\cr}\ .
\eqno(5.25)$$
It is clear that $\widetilde{\varepsilon}_A(\omega)\widetilde{E}{}^A  
= \omega$ for all $\omega\in\a^N$. With the help of the inclusion 
$i\,$, we can introduce the  left \a-linear maps  
$$
\varepsilon_A\,:\ \cases{ 
&$\widetilde{\Omega}^1_D(\a) \lra\ \a$\cr  
&$\quad \ \omega\ \ \ \longmapsto\ \widetilde{\varepsilon}_A
\vphantom{M^{M^M}}\bigl(i(\omega)\bigr) $\cr} 
\eqno(5.26)$$
for all $A=1,\ldots,N$. With these, 
$\omega\in\widetilde{\Omega}^1_D(\a)$ can be written as 
$$
\omega = p\bigl(i(\omega)\bigr) = 
p\bigl(\widetilde{\varepsilon}_A(i(\omega))\widetilde{E}{}^A \bigr)
= \varepsilon_A(\omega) E^A\ ,
\eqno(5.27)$$
and we see that $\{\,\varepsilon_A\,\}$ is the dual basis we already 
used in section 5.1.7. The first step towards the non-commutative 
Cartan structure equations is the following result; see also \q{Kar}. 
\mn
{\bf Proposition 5.17}\quad Every connection 
$\widetilde{\nabla}$ on $\a^N$
$$
\widetilde{\nabla}\,:\ \a^N \lra\ \widetilde{\Omega}^1_D(\a)
\otimes_{\scriptscriptstyle {\cal A}} \a^N
$$
determines a connection $\nabla$ on $\widetilde{\Omega}^1_D(\a)$ by 
$$
\nabla = ({\rm id}\,\otimes p) \circ \widetilde{\nabla} \circ i  ,
\eqno(5.28)$$
and every connection on $\widetilde{\Omega}^1_D(\a)$ is of this form. 
\sn
{\kap Proof}: Let $\widetilde{\nabla}$ be a connection on $\a^N$ -- 
which always exists (see the remarks after the proof).  
Clearly, $\nabla = ({\rm id}\,\otimes p) \circ \widetilde{\nabla} 
\circ i $ is a well-defined map, and it satisfies 
$$
\nabla (a\,\omega) =({\rm id}\,\otimes p)\bigl(\widetilde{\nabla}(a\,
i(\omega))\bigr)=({\rm id}\,\otimes p)\bigl(\delta a\otimes i(\omega) 
+a \widetilde{\nabla} i(\omega) \bigr) 
= \delta a \otimes \omega + a \nabla \omega 
$$
for all $a\in\a$ and all $\omega\in \widetilde{\Omega}^1_D(\a)$. 
This  proves that $\nabla$ is a connection on 
$\widetilde{\Omega}^1_D(\a)$.  \hfill\break 
\noindent If $\nabla'$ is any other connection on 
$\widetilde{\Omega}^1_D(\a)$, then 
$$
\nabla' - \nabla \in {\rm Hom}_{\scriptscriptstyle {\cal A}}
\bigl(\widetilde{\Omega}^1_D(\a),\ 
\widetilde{\Omega}^1_D(\a)\otimes_{\scriptscriptstyle {\cal A}} 
\widetilde{\Omega}^1_D(\a)\bigr)\ ,
$$
where Hom$_{\scriptscriptstyle {\cal A}}$ denotes the space of 
homomorphisms of left \a-modules. Since 
$$
{\rm id}\,\otimes p\,:\  \widetilde{\Omega}^1_D(\a) 
\otimes_{\scriptscriptstyle {\cal A}} \a^N \lra\ 
\widetilde{\Omega}^1_D(\a)\otimes_{\scriptscriptstyle {\cal A}} 
\widetilde{\Omega}^1_D(\a)
$$
is surjective and $\widetilde{\Omega}^1_D(\a)$ is a projective 
module, there exists a module map 
$$
\varphi\,:\ \widetilde{\Omega}^1_D(\a) \lra
\widetilde{\Omega}^1_D(\a) \otimes_{\scriptscriptstyle{\cal A}}\a^N 
$$
with 
$$\nabla' - \nabla = ({\rm id}\,\otimes p) \circ \varphi\ .
$$
Then $\widetilde{\varphi} := \varphi\circ p \in 
{\rm Hom}_{\scriptscriptstyle {\cal A}}\bigl( \a^N,\,\widetilde{
\Omega}^1_D(\a)\otimes_{\scriptscriptstyle {\cal A}}\a^N\,\bigr)$, 
and $\widetilde{\nabla} + \widetilde{\varphi}$ is a connection on 
$\a^N$ whose associated connection on $\widetilde{\Omega}^1_D(\a)$ is 
given by $\nabla'$:  
$$
({\rm id}\,\otimes p) \circ (\widetilde{\nabla} + \widetilde{\varphi}) 
\circ i = \nabla + ({\rm id}\,\otimes p) \circ \varphi = \nabla'\ .
$$
This proves that every connection on $\widetilde{\Omega}^1_D(\a)$  
comes {}from  a connection on $\a^N$.   \hfill\qed
\mn 
The importance of this proposition lies in the fact that an 
arbitrary collection of 1-forms 
$\bigl\{\,\widetilde{\Omega}^A_{\phantom{A}B}\,\bigr\}_{A,B=1,\ldots,N} 
\subset \widetilde{\Omega}^1_D(\a)$ defines a connection 
$\widetilde{\nabla}$ on $\a^N$ by the formula 
$$
\widetilde{\nabla}\,\bigl(a_A\widetilde{E}{}^A\bigr) 
= \delta a_A \otimes \widetilde{E}{}^A - 
a_A \widetilde{\Omega}^A_{\phantom{A}B} \otimes \widetilde{E}{}^B\ , 
$$ 
and conversely. Thus, not only the existence of connections on $\a^N$ 
and $\widetilde{\Omega}^1_D(\a)$ is guaranteed, but eq.\ (5.28) allows 
us to compute the components $\Omega^A_{\phantom{A}B}$ of the induced 
connection $\nabla$ on $\widetilde{\Omega}^1_D(\a)$. The action of 
$\nabla$ on the generators is 
$$\eqalign{
\nabla E^A&=({\rm id}\otimes p)\bigl(\widetilde{\nabla}\,i(E^A)\bigr) 
= ({\rm id}\otimes p)\bigl(\widetilde{\nabla}
\,\widetilde{\varepsilon}_B(i(E^A))\widetilde{E}{}^B\bigr) 
= ({\rm id}\otimes p) \bigl(
\widetilde{\nabla}\,\varepsilon_B(E^A) \widetilde{E}{}^B \bigr) 
\cr
&= ({\rm id}\otimes p)\bigl(\delta \varepsilon_B(E^A) \otimes
\widetilde{E}{}^B - \varepsilon_B(E^A)\, 
\widetilde{\Omega}^B_{\phantom{B}C} \otimes\widetilde{E}{}^C\bigr) 
\cr
&= \delta \varepsilon_B(E^A) \otimes E^B-  \varepsilon_C(E^A)\, 
\widetilde{\Omega}^C_{\phantom{C}B} \otimes E^B\ ,  
\cr}$$
where we have used some of the general properties listed before. 
In short, we get the relation 
$$
\Omega ^A_{\phantom{A}B} = \varepsilon_C(E^A)\, 
\widetilde{\Omega}^C_{\phantom{C}B} - \delta \varepsilon_B(E^A)
\eqno(5.29)$$
expressing the components of the connection $\nabla$ on 
$\widetilde{\Omega}^1_D(\a)$ in terms of the  components 
of the connection $\widetilde{\nabla}$ on $\a^N$. 
\sn
Upon inserting (5.29) into (5.21,22), one arrives at Cartan structure 
equations which express torsion and curvature in terms of these 
unrestricted components. We can, however, obtain equations of a 
simpler form if we exploit the fact that the map $\widetilde{\nabla}
\mapsto \nabla$ is many-to-one; this allows us to impose some 
extra symmetry relations on the components of $\widetilde{\nabla}$. 
\mn
{\bf Proposition 5.18}\quad Let $\widetilde{\Omega}^A_{\phantom{A}B}$ 
be the coefficients of a connection $\widetilde{\nabla}$ on $\a^N$, 
and denote by $\nadt$ the connection on $\a^N$ whose components 
are given by 
$\omdt{}^A_{\phantom{A}B} := \varepsilon_C(E^A)\,
\widetilde{\Omega}^C_{\phantom{C}D}\varepsilon_B(E^D)\,$.
Then, these components enjoy the symmetry relations 
$$
\varepsilon_C(E^A)\,\omdt{}^C_{\phantom{C}B}=\omdt{}^A_{\phantom{A}B}\ 
\quad\quad    \omdt{}^A_{\phantom{A}C}\,
\varepsilon_B(E^C)=\omdt{}^A_{\phantom{A}B}\ ,
\eqno(5.30)$$
and $\nadt$ and $\widetilde{\nabla}$ induce the same 
connection on $\widetilde{\Omega}_D^1(\a)$.  In particular, every 
connection on $\widetilde{\Omega}_D^1(\a)$ is induced by a connection 
on $\a^N$ that satisfies (5.30). 
\sn
{\kap Proof}: We explicitly compute the action of the connection $\nabla$
induced by $\nadt$ on a generator, using eqs.\ (5.27,28) and the fact 
that all maps and the tensor product are \a-linear: 
$$\eqalign{
\nabla\,E^A &= -\,\Omega^A_{\phantom{A}B} \otimes E^B = 
\delta \varepsilon_B(E^A) \otimes E^B
-  \varepsilon_C(E^A)\, \omdt{}^C_{\phantom{C}B} \otimes E^B
\cr
&=\delta\varepsilon_B(E^A)\otimes E^B-\varepsilon_C(E^A)
\varepsilon_D(E^C)\,\widetilde{\Omega}^D_{\phantom{D}F} 
\varepsilon_B(E^F) \otimes E^B 
\cr
&= \delta \varepsilon_B(E^A) \otimes E^B  - \varepsilon_D
\bigl(\varepsilon_C(E^A) E^C\bigr)\,\widetilde{\Omega}^D_{\phantom{D}F} 
\otimes \varepsilon_B(E^F) E^B 
\cr
&= \delta \varepsilon_B(E^A) \otimes E^B  - \varepsilon_D(E^A)
\,\widetilde{\Omega}^D_{\phantom{D}F} \otimes E^F 
\cr}$$
This shows that $\nabla$ is identical to the connection induced by 
$\widetilde{\nabla}$. The symmetry relations (5.30) follow directly 
{}from \a-linearity and (5.27).   \hfill\qed 
\mn
We are now in a position to state the {\sl Cartan structure equations} 
in a simple form.
\mn
{\bf Theorem 5.19}\quad Let $\widetilde{\Omega}^A_{\phantom{A}B}$ and 
$\omdt{}^A_{\phantom{A}B}$ be as in Proposition 5.18. Then the 
curvature and torsion components of the induced connection on  
$\widetilde{\Omega}_D^1(\a)$ are given by 
$$\eqalign{
&{\ttR}{}^A_{\phantom{A}B} = 
  \varepsilon_C(E^A)\,\delta\,\omdt{}^C_{\phantom{C}B} 
+ \omdt{}^A_{\phantom{A}C}\,\omdt{}^C_{\phantom{C}B}
+ \delta \varepsilon_C(E^A)\,\delta\,\varepsilon_B(E^C) \ ,
\cr
&{\ttT}{}^A\ \,= \varepsilon_B(E^A)\,\delta E^B 
  + \omdt{}^A_{\phantom{A}B} E^B\ .
\cr}$$
\sn
{\kap Proof}: With eqs.\ (5.21,29,30) and the Leibniz rule, we get 
$$\eqalign{
{\ttR}{}^A_{\phantom{A}B} &= \delta\,\omdt{}^A_{\phantom{A}B} 
+ \bigl(\,\omdt{}^A_{\phantom{A}C}-\delta\varepsilon_C(E^A) \bigr) 
\bigl(\,\omdt{}^C_{\phantom{C}B} - \delta\varepsilon_B(E^C) \bigr)
\cr
&= \delta\,\bigl(\varepsilon_C(E^A)\,\omdt{}^C_{\phantom{C}B}\bigr) 
+ \bigl(\,  \omdt{}^A_{\phantom{A}C} - \delta\varepsilon_C(E^A) \bigr) 
\bigl(\, \omdt{}^C_{\phantom{C}B} - \delta\varepsilon_B(E^C) \bigr)
\cr
&= \varepsilon_C(E^A)\delta\,\omdt{}^C_{\phantom{C}B} 
+ \omdt{}^A_{\phantom{A}C}\,\omdt{}^C_{\phantom{C}B}
+ \delta \varepsilon_C(E^A)\,\delta\,\varepsilon_B(E^C)
- \omdt{}^A_{\phantom{A}C} \delta\varepsilon_B(E^C)\ . 
\cr}$$
The last term does in fact not contribute to the curvature, as can be 
seen after tensoring with $E^B$:
$$
\omdt{}^A_{\phantom{A}C} \delta\varepsilon_B(E^C) \otimes E^B 
= - \delta\bigl(\,\omdt{}^A_{\phantom{A}B}\bigr) \otimes E^B 
+\bigl(\delta\,\omdt{}^A_{\phantom{A}C}\bigr)\otimes \varepsilon_B(E^C) 
E^B = 0\ ,
$$
where we have used the Leibniz rule, the relations (5.30) and 
\a-linearity of the tensor product.    \hfill\break
\noindent To compute the components of the torsion, we use 
eqs.\ (5.22,29) analogously, 
$$
{\ttT}{}^A = \delta E^A + \omdt{}^A_{\phantom{A}B}E^B - \delta 
\varepsilon_B(E^A) E^B 
=  \delta E^A + \omdt{}^A_{\phantom{A}B}E^B - \delta E^A 
+ \varepsilon_B(E^A)\delta E^B \ ,
$$ 
which gives the result.  \hfill\qed 
\mn
The Cartan structure equations of Theorem 5.19 are considerably simpler 
than those one would get directly {}from (5.29) and (5.21,22). The price 
to be paid is that the components $\omdt{}^A_{\phantom{A}B}$ are not 
quite independent {}from each other, but of course they can easily be 
expressed in terms of the arbitrary components 
$\widetilde{\Omega}^A_{\phantom{A}B}$ according to Proposition 5.18. 
Therefore, the equations of Theorem 5.19 are useful e.g.\ for 
determining connections on $\widetilde{\Omega}_D^1(\a)$ with special 
properties. We refer the reader to \q{CFG} for an explicit application 
of the Cartan structure equations. 
\bn\bn
{\bf 5.2 The \Noneone formulation of non-commutative geometry}
\bn
In this section, we introduce the non-commutative generalization 
of the description of Riemannian geometry by a set of $N=(1,1)$ 
spectral data, which was presented, in the classical case, in 
section 3.2. The advantage over the $N=1$ formulation is that now 
the algebra of differential forms is naturally represented on the 
Hilbert space \h. Therefore, calculations in concrete 
examples and also the study of cohomology rings will become much 
easier. There is, however, the drawback that the algebra of 
differential forms is no longer closed under the ${}^*$-operation 
on \h; but see section 6. \hfill\break
The $N=(1,1)$ framework explained in the following will  
also provide the basis for  the definition of various 
types of complex non-commutative geometries in sections 5.3 and 5.4. 
\bn\bn
{\bf 5.2.1 The \Noneone spectral data}
\bn
{\bf Definition 5.20}\quad A quintuple $(\a,\h, \dd, \gamma, *)$
is called a {\sl set of $N=(1,1)$ spectral data} if 
\smallskip
\item {1)} \h\ is a separable Hilbert space;
\smallskip
\item {2)} \a\ is a unital ${}^*$-algebra acting faithfully on \h\ 
by bounded operators;
\smallskip
\item {3)} \dd\ is a densely defined closed operator on \h\ such that 
\itemitem{$i)$} $\dd^2=0\,$,
\itemitem{$ii)$} for each $a\in\a$, the commutator $\lb\,\dd,a\,\rb$ 
extends uniquely to a bounded operator on \h,
\itemitem{$iii)$} the operator $\exp(-\varepsilon\triangle)$ with 
$\triangle=\dd\dd^*+\dd^*\dd$ is trace class for all $\varepsilon>0\,$; 
\smallskip
\item {4)} $\gamma$ is a $\Z_2$-grading on \h, i.e.,
$\gamma=\gamma^*=\gamma^{-1}$, such that 
\itemitem{$i)$} $\lb\,\gamma,a\,\rb=0$ for all $a\in\a\,$,
\itemitem{$ii)$} $\{\,\gamma,\dd\,\}=0\,$;
\smallskip
\item {5)} $*$ is a unitary operator on \h\ such that 
\itemitem{$i)$} $*\,\dd = \zeta\, \dd^*\,*\,$ for some $\zeta\in\C$ 
  with $|\zeta|=1\,$, 
\itemitem{$ii)$} $\lb\,*,a\,\rb=0$ for all $a\in\a\,$.
\mn
Several remarks are in order. First of all, note that we can 
introduce the two operators 
$$
\d = \dd + \dd^*\ ,\quad \bard = i\,(\dd- \dd^*\,) 
$$
on \h\ which satisfy the relations of Definition 3.8; thus, our 
notion of $N=(1,1)$ spectral data is an immediate generalization 
of a classical $N=(1,1)$ Dirac bundle -- except for the boundedness 
conditions to be required on infinite-dimensional Hilbert spaces, and 
the existence of the additional operator $*$ (see the comments below).   
\sn
As in the $N=1$ case, the $\Z_2$-grading $\gamma$ may always be 
introduced if not given {}from the start, simply by ``doubling'' the 
Hilbert space -- see the remarks following Definition 5.1.
\sn
Moreover, if $(\widetilde{\a}, \widetilde{{\cal H}}, \widetilde{\dd},
\tilde\gamma)$ is a quadruple satisfying conditions 1-4 of Definition
5.20, we obtain a full set of $N=(1,1)$ spectral data by setting 
$$\eqalign{
\h &= \widetilde{{\cal H}} \otimes \C^2\ ,\quad\quad\   
\a = \widetilde{{\cal A}} \otimes \one_2\ ,
\cr
\dd&= \widetilde{\dd}\otimes {1\over2}\,(\one_2+\tau_3)- 
\widetilde{\dd}^*  \otimes {1\over2}\,(\one_2-\tau_3)\ ,
\cr
\gamma &= \tilde\gamma\otimes\one_2\ , \quad\quad\quad 
*\, = \one_{\tilde{\scriptscriptstyle {\cal H}}} \otimes \tau_1 
\cr}$$
with the Pauli matrices $\tau_i$ as usual. Note that, in this example, 
$\zeta=-1$, and the $*$-operator additionally 
satisfies $*^2=1$ as well as  $\lb\,\gamma,*\,\rb=0\,$. 
\sn
The unitary operator $*$ was not present in our algebraic formulation 
of classical Riemannian geometry. But for a compact oriented manifold, 
the usual Hodge $*$-operator acting on differential forms satisfies 
all the properties listed above, after appropriate rescaling in each 
degree. (Moreover, one can always achieve $*^2=1$ {\sl or} $\zeta=-1$.)
For a non-orientable manifold, we can apply the construction of the 
previous paragraph to obtain a description of the differential forms 
in terms of $N=(1,1)$ spectral data including a Hodge operator. In 
our approach to the non-commutative case, we will make essential use  
of the existence of $*$, which we will also call {\sl Hodge operator},
in analogy to the classical case.   
\bn\bn
\leftline{\bf 5.2.2 Differential forms}
\bn
We first introduce an involution, $\natural$, called {\sl complex
conjugation}, on the algebra of universal forms: 
$$
\natural\,:\ \Omega^{\bullet}(\a) \lra\ \Omega^{\bullet}(\a) 
$$
is the unique $\C$-anti-linear anti-automorphism such that
$$
\natural(a) \equiv a^{\natural} := a^*\ ,\quad\quad
\natural(\delta a) \equiv (\delta a)^{\natural} := \delta(a^*\,) 
\eqno(5.31)$$
for all $a\in\a$. Here we choose a sign convention that 
differs {}from the $N=1$ case, eq.\ (5.1). 
If we write $\hat\gamma$ for the mod 2 reduction of the 
canonical $\Z$-grading on $\Omega^{\bullet}(\a)$, we have 
$$
\delta\natural\hat\gamma = \natural\delta\ .
\eqno(5.32) 
$$
We define a representation of $\Omega^{\bullet}(\a)$ on \h, 
again denoted by $\pi$, by
$$
\pi(a) := a\ ,\quad\quad 
\pi(\delta a) := \lb\,\dd,a\,\rb 
\eqno(5.33)$$
for all $a\in \a$. The map $\pi$ is a $\Z_2$-graded representation 
in the sense that 
$$
\pi(\hat\gamma \omega\hat\gamma) = \gamma\pi(\omega)\gamma
\eqno(5.34)$$
for all $\omega\in\Omega^{\bullet}(\a)$. 
\sn
Although the abstract algebra of universal forms is the 
same as in the $N=1$ setting, the interpretation of the universal
differential $\delta$ has changed: In the $N=(1,1)$ framework, it is 
represented on \h\ by the nilpotent operator \dd, instead of the 
self-adjoint Dirac operator $D$, as before. In particular, we now have 
$$
\pi(\delta\omega) = \lb\,\dd,\pi(\omega)\,\rb_g
\eqno(5.35)$$
for all $\omega\in\Omega^{\bullet}(\a)$, where $\lb\cdot,\cdot\rb_g$
denotes the graded commutator (defined with the  canonical 
$\Z_2$-grading on $\pi(\Omega^{\bullet}(\a))$, see (5.34)). 
The validity of eq.\ (5.35) is the main difference between the
$N=(1,1)$ and the $N=1$ formalism. It ensures that there do not exist 
any forms  $\omega\in\Omega^p(\a)$ with $\pi(\omega)=0$ but
$\pi(\delta\omega) \neq 0$, in other words:
\mn
{\bf Proposition 5.21}\quad The graded vector space 
$$
J=\bigoplus_{k=0}^{\infty} J^k\ ,\quad  
J^k := {\rm ker}\,\pi\,|_{\Omega^k({\cal A})} 
$$
with $\pi$ defined in (5.33) is a two-sided graded {\sl differential} 
${}^{\natural}$-ideal of $\Omega^{\bullet}(\a)$.
\sn
{\kap Proof}: The first two properties are obvious, the third one is 
the content of eq.\ (5.35). Using (5.31) and the relations satisfied 
by the Hodge $*$-operator according to part 5) of Definition 5.20, 
we find that  
$$\eqalign{
\pi\bigl((\delta a)^{\natural}\bigr) &= \pi(\delta( a^*\,)) = 
\lb\, \dd,a^*\,\rb = \lb\,a,\dd^*\,\rb^* 
= \zeta\, \lb\, a, *\,\dd\, *^{-1}\,\rb^* 
\cr
&= \zeta\, *\,\lb\,a,\dd\,\rb^*\,*^{-1}= 
-\zeta\,*\,\pi(\delta a)^*\,*^{-1}\ ,\cr}$$
which implies 
$$
\pi\bigl(\omega^{\natural}\bigr)=(-\zeta)^k\,*\,\pi(\omega)^*\,*^{-1}
\eqno(5.36)$$
for all  $\omega\in\Omega^k(\a)$. In particular, $J = {\rm ker}\,\pi$ 
is a ${}^{\natural}$-ideal.   \hfill\qed
\mn
As a consequence of this proposition, the algebra of differential forms
$$
\Omega_{\rm d}^{\bullet}(\a) := \bigoplus_{k=0}^{\infty}\, 
\Omega_{\rm d}^k(\a)\ ,\quad\ \ 
\Omega_{\rm d}^k(\a) := \Omega^k(\a)/J^k\ , 
\eqno(5.37)$$
is represented on the Hilbert space \h\ via $\pi$. For later purposes, 
we will also need an involution on $\Omega_{\rm d}^{\bullet}(\a)$, and 
according to Proposition 5.21, this is given by the anti-linear map
$\natural$ of (5.31). Note that the ``natural'' involution  
$\omega \mapsto \omega^*$, see eq.\ (5.1), which is inherited {}from \h\ 
and was used in the $N=1$ case, is no longer available here:  The 
space $\pi(\Omega^k(\a))$ is not closed under taking adjoints, simply
because \dd\ is not self-adjoint.   \hfill\break
In summary, the space $\Omega_{\rm d}^{\bullet}(\a)$ is a unital 
graded differential ${}^{\natural}$-algebra and the representation 
$\pi$ of $\Omega^{\bullet}(\a)$ determines a representation of 
$\Omega_{\rm d}^{\bullet}(\a)$ on \h\ as a unital differential algebra.
\bn\bn
\leftline{\bf 5.2.3 Integration}
\bn
The integration theory follows the same lines as in the $N=1$ case:
The state $\barint$ is given as in Definition 5.3 with $\d^2$
written as $\triangle= \dd\dd^*+\dd^*\dd$. Again, we make Assumption 
5.4 about the cyclicity of the integral. This yields a sesqui-linear
form on $\Omega_{\rm d}^{\bullet}(\a)$ as before:
$$
(\omega,\eta) = \Barint \omega\,\eta^* 
\eqno(5.38)$$
for all $\omega,\eta\in\Omega_{\rm d}^{\bullet}(\a)$, where we have 
dropped the representation symbols $\pi$ under the integral. 
\sn
Because of the presence of the Hodge $*$-operator, the form 
$(\cdot,\cdot)$ has an additional feature in the $N=(1,1)$ setting:
\mn
{\bf Proposition 5.22}\quad If the phase in part 5) of Definition 5.20 
is $\zeta=\pm 1$, then the  inner product defined in eq.\ (5.38)
behaves like a real functional with respect to the involution 
$\natural$, i.e., for $\omega,\eta\in\Omega_{\rm d}^{\bullet}(\a)$
we have 
$$
(\,\omega^{\natural},\eta^{\natural}\,) = \overline{(\omega,\eta)}
$$
where the bar denotes ordinary complex conjugation. 
\sn
{\kap Proof}: First, observe that the Hodge operator commutes with the 
Laplacian, which is verified e.g.\ by taking the adjoint of the relation  
$*\,\dd = \zeta\, \dd^*\,*\,$. Then the claim 
follows immediately using eq.\ (5.36), unitarity of 
the Hodge operator, and cyclicity of the trace on \h: Let
$\omega\in\Omega^p_{\rm d}(\a),\ \eta\in \Omega^q_{\rm d}(\a)$, then 
$$\eqalign{
(\,\omega^{\natural},\eta^{\natural}\,) 
&= \Barint \omega^{\natural} \bigl(\eta^{\natural}\bigr)^*
= (-\zeta)^p (-\bar\zeta)^q \Barint *\, \omega^* *^{-1}* \eta\, *^{-1} 
= (-\zeta)^{p-q} \Barint \omega^* \eta \cr
&= (-\zeta)^{p-q} \Barint \eta\, \omega^* = 
(-\zeta)^{p-q}\, \overline{(\omega,\eta)}\ ;\cr}
$$ 
again, we have suppressed the representation symbol $\pi$. The claim 
follows since the $\Z_2$-grading implies $(\omega,\eta)=0$ unless 
$p-q \equiv 0\ ({\rm mod}\,2)$.       \hfill\qed
\sn
Note that, in examples, $p$- and $q$-forms for $p\neq q$ are often
orthogonal wrt.\ the inner product $(\cdot,\cdot)$; then Proposition 
5.22 holds independently of the value of $\zeta$. 
\mn
Since $\Omega_{\rm d}^{\bullet}(\a)$ is a ${}^{\natural}$- and 
{\sl not} a ${}^*$-algebra, Proposition 5.5 is to be replaced by 
\mn 
{\bf Proposition 5.23}\quad The graded kernel $K$, see eq.\ (5.5),  
of the sesqui-linear form $(\cdot,\cdot)$ is a two-sided graded 
${}^{\natural}$-ideal of $\Omega_{\rm d}^{\bullet}(\a)$. 
\sn
{\kap Proof}: The proof that $K$ is a two-sided graded ideal is 
identical to the one of Proposition 5.5. That $K$ is closed under 
$\natural$ follows immediately {}from the proof of Proposition 
5.22.\hfill\qed
\mn
The remainder of section 5.1.3 carries over to the $N=(1,1)$ case, with 
the only differences that $\widetilde{\Omega}^{\bullet}(\a)$ is a 
${}^{\natural}$-algebra and that the quotients 
$\Omega^k(\a)/\bigl(K^k+\delta K^{k-1}\bigr) \cong
\widetilde{\Omega}^k(\a)/\delta K^{k-1}$ are denoted by 
$\widetilde{\Omega}_{\rm d}^k(\a)$. 
\sn
Upon passing {}from $\Omega_{\rm d}^{\bullet}(\a)$  to the algebra of 
{\sl square-integrable forms} 
$\widetilde{\Omega}^{\bullet}(\a)$, one might, however, lose the
advantage of working with differential ideals: Whereas $J$ has this 
property in the $N=(1,1)$ setting, there may exist $\omega\in K^{k-1}$ 
with $\delta\omega\notin K^k$. But it turns out that $K$ 
vanishes in many interesting examples, and, for these, we have a 
representation of the algebra $\widetilde{\Omega}^{\bullet}(\a)$ 
of square-integrable forms on $\widetilde{{\cal H}}^{\bullet}$. 
Apart {}from that, a sufficient condition ensuring that $K$ is a 
differential ideal is given by  
$$
\Barint \lb\,\dd,\omega\,\rb_g = 0 
$$     for all $\omega$ 
in the algebraic span of $\pi\bigl(\Omega_{\rm d}^{\bullet}
(\a)\bigr)+\pi\bigl(\Omega_{\rm d}^{\bullet}(\a)\bigr)^*$: If this 
is true, the Cauchy-Schwarz inequality leads to the estimate 
$$
\Big\vert\, \Barint \lb\,\dd,\omega\,\rb_g\eta^* \,\Big\vert 
= \Big\vert\, \Barint \omega\,\lb\,\dd,\eta^*\,\rb_g \Big\vert 
\leq \Bigl( \Barint \omega\,\omega^* \Bigr)^{1\over2} 
\Bigl( \Barint \lb\,\dd,\eta^*\,\rb_g\lb\,\dd,\eta^*\,\rb^*_g 
\Bigr)^{1\over2} \ .
$$
Applying this to $\eta= \lb\,\dd,\omega\,\rb_g$ for 
$\omega\in K^{k-1}$, we obtain that $\delta\omega\in K^k$. 
\bn\bn
{\bf 5.2.4 Unitary connections and scalar curvature}
\bn
Except for the notions of unitary connections and scalar curvature, 
all definitions and results of sections 5.1.4-8
literally apply to the $N=(1,1)$ case as well. The two exceptions 
explicitly involve the ${}^*$-involution on the algebra 
of differential forms, which is no longer available now. Therefore, 
we have to modify the definitions for $N=(1,1)$ non-commutative 
geometry as follows:
\mn
{\bf Definition 5.24}\quad A connection $\nabla$ on a Hermitian 
vector bundle $\bigl(\e,\langle\cdot,\cdot\rangle\bigr)$ over 
an \break \noindent \hbox{$N=(1,1)$} non-commutative space is called 
{\sl unitary} if 
$$
\dd\,\langle\,s,t\,\rangle = 
\langle\,\nabla s,t\,\rangle + \langle\,s,\nabla t\,\rangle
$$
for all $s,t\in\e$; the Hermitian structure on the rhs is extended 
to  
\e-valued differential forms by
$$
\langle\,\omega\otimes s,t\,\rangle = \omega\,\langle\,s,t\,\rangle\ ,
\quad\quad
\langle\, s,\eta \otimes t\,\rangle  
= \langle\,s,t\,\rangle\,\eta^{\natural}
$$
for all $\omega,\eta \in \widetilde{\Omega}^{\bullet}_{{\rm d}}(\a)$ 
and $s,t\in\e$. 
\mn
{\bf Definition 5.25}\quad The {\sl scalar curvature} of a connection
$\nabla$ on $\widetilde{\Omega}^{1}_{{\rm d}}(\a)$ is defined by 
$$
{\ttr}\,(\nabla) = \bigl( E^{B\,\natural}\bigr)^{\rm ad}_R ({\ttRic}_B)
\in \widetilde{{\cal H}}_0\ .
$$
\bn\bn
{\bf 5.2.5 Remarks on the relation of \Neqone and \Noneone 
spectral data}
\bn
The definitions of $N=1$ and $N=(1,1)$ non-commutative spectral
data provide two different generalizations of classical Riemannian 
differential geometry. In the latter context, one can always find 
an $N=(1,1)$ description of a manifold originally given by an 
$N=1$ set of data, whereas a non-commutative $N=(1,1)$ set of spectral 
data seems to require a different mathematical structure than a 
spectral triple, because of 
the additional generalized Dirac operator which must be given on the 
Hilbert space. Thus, it is a natural and important question under which 
conditions on an $N=1$ spectral triple $(\a, \h, D)$ there exists an 
associated $N=(1,1)$ set of data $(\a, \widetilde{\cal H}, {\tt d}, *)$ 
over the same non-commutative space \a. 
\sn
We have not been able yet to answer the question of how to pass {}from 
$N=1$ to $N=(1,1)$ data in a satisfactory way, but in the following  we 
present a procedure that probably will lead to a solution. Our 
guideline is the classical case, where the main step in passing {}from
$N=1$ to $N=(1,1)$ data is to replace the Hilbert space $\h=L^2(S)$ by 
$\widetilde{\cal H} = L^2(S)\otimes_{\cal A}L^2(S)$ carrying 
two actions of the Clifford algebra and therefore two anti-commuting 
Dirac operators $\d$ and $\bard$ -- which yield a description 
equivalent to the one involving the nilpotent differential {\tt d}, 
see the remark after Definition 5.20. \hfill\break 
\noindent It is plausible that there are other approaches 
to this question, in particular approaches of a more operator 
algebraic nature, e.g.\ using a ``Kasparov product of spectral 
triples'', but we will not enter these matters here. 
\sn                               The first problem 
one meets when trying to copy the classical step {}from $N=1$ to 
$N=(1,1)$ is that \h\ should be an \a-bi-module.  To ensure this, we 
require that the set of $N=1$ (even) spectral data $(\a,\h,D,\gamma)$
is endowed with a {\sl real structure} \q{Co4}, i.e.\ 
that there exists an anti-unitary operator $J$ on \h\ such that 
$$
J^2 = \epsilon\one\ ,\quad\quad J\gamma= \epsilon' \gamma J\ ,
\quad\quad      JD = DJ 
$$
for some (independent) signs $\epsilon, \epsilon' = \pm 1$, and 
such that, in addition,
$$
J a J^*\quad \hbox{commutes with $b$ and $\lb\,D,b\,\rb$ for all}\ 
a,b\in\a\ .
$$
\sn
This definition of a real structure was introduced by Connes in 
\q{Co4}; $J$ is of course a variant of Tomita's modular conjugation 
(cf.\ the next subsection) -- which is one reason why this notion 
is not only very useful for purely mathematical investigations, 
but should prove important also {}from the physical point of 
view.   \hfill\break
\noindent In the present context, $J$ simply provides a canonical 
right \a-module structure on \h\ by defining 
$$
\xi\cdot a := Ja^* J^* \xi 
$$
for all $a\in\a$, $\xi\in\h$, see \q{Co4}. We can extend this to a 
right action of $\Omega^1_D(\a)$ on \h\ if we set 
$$
\xi\cdot\omega := J\omega^* J^* \xi 
$$ 
for all $\omega\in\Omega^1_D(\a)$ and $\xi\in\h$; for simplicity, 
the representation symbol $\pi$ has been omitted. Note that by the 
assumptions on $J$, the right action commutes with the left action of 
\a. Thus \h\ is an \a-bi-module. Moreover, we can form tensor products 
of bi-modules {\sl over the algebra} \a\ just as in 
the classical case. 
\sn
The real structure $J$ in addition allows us to define the anti-linear
``flip'' operator 
\def\vps{\vphantom{m_{{\displaystyle \sum}}}}
$$
\Psi\,:\ \cases{&${\displaystyle \Omega^1_D(\a)\vps\otimes_{\cal A}\h
\lra \h\otimes_{\cal A}\Omega^1_D(\a)}$\cr
&$\phantom{xxxxxx}\omega\otimes\xi \longmapsto J\xi \otimes 
\omega^*$\cr} \ .
$$             It is straightforward 
to verify that $\Psi$ is well-defined and that it satisfies 
$$
\Psi(a\,s) = \Psi(s)\,a^*
$$
for all $a\in\a$, $s\in\Omega^1_D(\a)\otimes_{\cal A}\h$. 
\sn    {}From now on, we assume furthermore that \h\ is a 
{\sl projective} left \a-module. Then it admits connections 
$$
\nabla\,:\ \h \lra \Omega^1_D(\a)\otimes_{\cal A}\h\ ,
$$
i.e.\ $\C$-linear maps such that 
$$
\nabla(a\xi) = \delta a\otimes\xi + a \nabla\xi 
$$ 
for all $a\in\a$ and $\xi\in\h$. For each connection $\nabla$ on 
\h, there is an ``associated right-connection'' $\overline{\nabla}$ 
defined with the help of the flip $\Psi$: 
$$
\overline{\nabla}\,:\ \cases{&${\displaystyle \h\lra 
\h\vps\otimes_{\cal A}\Omega^1_D(\a)}$\cr
&$\phantom{l}\xi \longmapsto -\Psi(\nabla J^* \xi)$\cr}
$$
$\overline{\nabla}$ is again $\C$-linear and satisfies 
$$
\overline{\nabla}(\xi a)=\xi\otimes\delta a+(\overline{\nabla}\xi)a\ .
$$
A connection $\nabla$ on \h\ together with its associated right 
connection $\overline{\nabla}$ induces a $\C$-linear ``tensor product 
connection'' $\widetilde{\nabla}$ on $\h\otimes_{\cal A}\h$ of the form 
$$
\widetilde{\nabla}\,:\ \cases{&${\displaystyle \h\otimes_{\cal A}\h \lra 
\h\vps\otimes_{\cal A}\Omega^1_D(\a)\otimes_{\cal A}\h}$\cr
&\phantom{w}$\xi_1 \otimes \xi_2 \longmapsto \overline{\nabla}\xi_1 
\otimes \xi_2 + \xi_1 \otimes \nabla\xi_2$\cr}\ .
$$
Because of the position of the factor $\Omega^1_D(\a)$, 
$\widetilde{\nabla}$ is not quite a connection in the usual sense. 
In the classical case, the last ingredient needed for the definition of  
the two Dirac operators of an $N=(1,1)$ Dirac bundle were the two 
anti-commuting Clifford actions on $\widetilde{\h}$. Their obvious 
generalizations to the non-commutative case are the $\C$-linear maps 
$$
{\c}\,:\ \cases{&${\displaystyle \h\otimes_{\cal A}\Omega^1_D(\a)
\otimes_{\cal A}\h\lra \h\vps\otimes_{\cal A}\h}$\cr
&$\phantom{xxxxww}\xi_1\otimes\omega\otimes\xi_2 \longmapsto
\xi_1\otimes\omega\,\xi_2$\cr}
$$
and \def\c{\hbox{{\tpwrt c}}}
$$
\overline{{\c}}\,:\ \cases{&${\displaystyle \h\otimes_{\cal A}
\Omega^1_D(\a)\otimes_{\cal A}\h\lra \h\vps\otimes_{\cal A}\h}$\cr
&\phantom{xxxxxxw}$\xi_1\otimes\omega\otimes\xi_2 \longmapsto
\xi_1\,\omega\otimes\gamma\xi_2$\cr}\ .
$$
With these, we may introduce two operators $\d$ and $\bard$ on 
$\h\otimes_{\cal A}\h$ in analogy to the classical case:
$$
\d := \c \circ \widetilde{\nabla}\ ,\quad\quad
\bard := \overline{\c} \circ \widetilde{\nabla}\ .
$$
In order to obtain a set of $N=(1,1)$ spectral data, one has to find a 
connection $\nabla$ on \h\ which makes the operators $\d$ and $\bard$ 
self-adjoint and ensures that the anti-commutation relations of 
Definition 5.20 are satisfied. The $\Z_2$-grading on 
$\h\otimes_{\cal A}\h$ is simply the tensor product grading, and the 
Hodge operator can be taken to be $* = \gamma\otimes\one$. 
\sn      Although we 
are not able, in general, to prove the existence of a connection 
$\nabla$ on \h\ which supplies $\d$ and $\bard$ with the correct 
algebraic properties, the naturality of the construction presented 
above as well as the similarity with the procedure of section 3.2.2 
leads us to expect that this problem can be solved in many cases of 
interest.  
\bn\bn
{\bf 5.2.6 Riemannian and  Spin\CC\ ``manifolds'' in non-commutative 
geometry}
\bn
In this section, we want to address the following question: What are 
the additional structures that make an $N=(1,1)$ non-commutative 
space into a {\sl non-commutative ``manifold''}, into a {\sl Spin${}^c$ 
``manifold''}, or into a {\sl quantized phase space}? There exists a 
definition of non-commutative manifolds in terms of $K$-homology, see
e.g.\ \q{Co1}, but we feel that in the formalism introduced in the 
present work it is possible to find more direct criteria. In our 
search for the characteristic features of non-commutative manifolds
we will, as before, be guided by the classical case and by the 
principle that they should be natural {}from the physics point of view. 
\sn                                         Extrapolating {}from 
classical geometry, we are e.g.\ led to the following 
requirement an $N=(1,1)$ space $(\a, \h, {\tt d}, \gamma, *)$ should 
satisfy in order to become a ``manifold'': The data must extend to a 
set of $N=2$ {\sl spectral data} $(\a, \h, {\tt d}, T, *)$ where $T$ 
is a self-adjoint operator on \h\ such that 
\smallskip
\item{$i)$} $\lb\,T,a\,\rb=0$ for all $a\in\a\,$;
\smallskip
\item{$ii)$} $\lb\,T,{\tt d}\,\rb={\tt d}\,$; 
\smallskip
\item{$iii)$} $T$ has integral spectrum, and $\gamma$ is the mod 2 
reduction of $T$, i.e.\ $\gamma=\pm1$ on $\h_{\pm}$, where 
$$
\h_{\pm} = {\rm span}\,\bigl\{\, \xi\in\h\,|\, T\xi=n\,\xi 
\hbox{ for some } n\in\Z, (-1)^n = \pm 1 \,\bigr\}\ .
$$
\sn
Recall that $N=2$ data have been used in section 1.2 already, and 
have also been briefly discussed in section 4.  
\bn
Before we can formulate other properties that we suppose to 
characterize non-commu\-tative manifolds, we recall some  
basic facts about {\sl Tomita-Takesaki theory}. Let \m\  be a 
von Neumann algebra acting on a separable Hilbert space 
\h, and assume that $\xi_0\in\h$ is a cyclic and separating vector 
for \m, i.e.\ 
$$
\overline{\m\,\xi_0} = \h 
$$
and 
$$
a\,\xi_0 = 0 \quad \Longrightarrow\quad a=0 
$$
for any $a\in\m$, respectively. Then we may define an anti-linear 
operator $S_0$ on \h\ by setting 
$$
S_0\, a\, \xi_0 = a^* \xi_0 
$$
for all $a\in\m$. One can show that $S_0$ is closable, and we denote 
its closure by $S$. The polar decomposition of $S$ is written as 
$$
S= J \Delta^{1\over2}
$$
where $J$ is an anti-unitary involutive operator, referred  to as 
{\sl modular conjugation}, and the so-called {\sl modular operator} 
$\Delta$ is a positive self-adjoint operator on \h. The fundamental 
result of Tomita-Takesaki theory is the following theorem: 
$$
J \m J = \m'\ ,\quad \Delta^{it} \m \Delta^{-it} = \m
$$
for all $t\in\R$; here, $\m'$ denotes the commutant of \m\ on \h. 
Furthermore, the vector state $\omega_0(\cdot):=(\xi_0,\cdot\,\xi_0)$
is a {\sl KMS-state} for the automorphism $\sigma_t := 
{\rm Ad}_{\Delta^{it}}$ of \m, i.e. 
$$
\omega_0(\sigma_t(a)\,b) = \omega_0(b\,\sigma_{t-i}(a))
$$
for all $a,b\in\m$ and all real $t$. 
\bn
Let  $(\a, \h, {\tt d}, T, *)$ be a set of $N=2$ spectral data 
coming {}from an $N=(1,1)$ space as above. We define the analogue 
$Cl_{\cald}(\a)$ of the space of sections of the Clifford bundle, 
$$
Cl_{\cald}(\a)=\bigl\{\,a_0\,\lb\,\d,a_1\,\rb\ldots \lb\,\d,a_k\,\rb
   \,|\, k\in\Z_+,\ a_i\in\a\,\bigr\}\ ,
$$
where $\d = \dd + \dd^*$, and, corresponding to the second 
generalized Dirac operator $\bard = i(\dd - \dd^*)\,$, 
$$
Cl_{\calbd}(\a) = \bigl\{\, a_0\,\lb\,\bard,a_1\,\rb \ldots 
\lb\,\bard,a_k\,\rb   \,|\, k\in\Z_+,\ a_i\in\a\,\bigr\}\ .
$$
In the classical setting, the sections $Cl_{\cald}(\a)$ and 
$Cl_{\calbd}(\a)$ operate on \h\ by the two actions $c$ and 
$\overline{c}$, respectively, see Lemma 3.4 and section 3.2.4. In the 
general case, we notice that, in contrast to the objects 
$\Omega_{\rm d}(\a)$ and $\Omega_{\cald}(\a)$ introduced before,
$Cl_{\cald}(\a)$ and $Cl_{\calbd}(\a)$ form ${}^*$-algebras of 
operators on \h, but are neither $\Z$-graded nor differential. 
\sn
We want to apply Tomita-Takesaki theory to the von Neumann algebra 
$\m := \bigl(Cl_{\cald}(\a)\bigr)''\,$. Suppose there exists a vector 
$\xi_0\in\h$ which is cyclic and separating for $\m$, and let $J$ be 
the anti-unitary conjugation associated to $\m$ and $\xi_0$. Suppose, 
moreover, that for all $a\in {}^J\!\a := J\a J$ the operator 
$\lb\,\bard,a\,\rb$ uniquely extends to a bounded operator on \h. 
Then we can form the algebra of bounded operators $Cl_{\calbd}\,({}^J\!\a)$ 
on \h\ as above. The properties $J\a J \subset \a'$ and 
$\{\,\d,\bard\,\}=0$ imply that $Cl_{\cald}(\a)$ and 
$Cl_{\calbd}\,({}^J\!\a)$ commute in the graded sense; to arrive at 
truly commuting algebras, we first 
decompose $Cl_{\calbd}\,({}^J\!\a)$ into a direct sum 
$$
Cl_{\calbd}\,({}^J\!\a)=Cl_{\calbd}^+\,({}^J\!\a)\oplus 
Cl_{\calbd}^-\,({}^J\!\a)
$$                                      with 
$$
Cl_{\calbd}^{\pm}\,({}^J\!\a)=\bigl\{\,\omega\in Cl_{\calbd}\,({}^J\!\a)
\,|\, \gamma\,\omega = \pm\omega\,\gamma\bigr\}\ .
$$
Then we define the ``twisted algebra'' $\widetilde{Cl}_{\calbd}
\,({}^J\!\a) := Cl_{\calbd}^+\,({}^J\!\a) \oplus
\gamma\,Cl_{\calbd}^-\,({}^J\!\a)$. This algebra  
commutes with $Cl_{\cald}(\a)$.  
\sn
We propose the following definitions:  The $N=2$ spectral data 
$(\a, \h, {\tt d}, T, *)$ describe a {\sl non-commutative manifold} if 
$$
\widetilde{Cl}_{\calbd}\,({}^J\!\a) = J\, Cl_{\cald}(\a)\, J \ .
$$
Furthermore, inspired by classical geometry, we say that a 
non-commutative manifold $(\a, \h, {\tt d}, T, *, \xi_0)$ is 
{\sl spin${}^c$} if the Hilbert space factorizes as a 
$Cl_{\cald}(\a)\otimes \widetilde{Cl}_{\calbd}\,({}^J\!\a)$ module 
in the form 
$$
\h = \h_{\cald} \otimes_{\cal Z} \h_{\calbd} 
$$
where ${\cal Z}$ denotes the center of $\m$. 
\mn
Next, we introduce a notion of ``quantized phase space''. We consider a 
set of $N=(1,1)$ spectral data $(\a, \h, \dd, \gamma, *)$, where we now 
think of \a\ as the algebra of phase space ``functions'' (i.e.\ of
pseudo-differential operators, in the Schr\"odinger picture of quantum 
mechanics) rather than functions over configuration space. 
We are, therefore, not postulating the existence of a cyclic and 
separating  vector for the algebra $Cl_{\cald}(\a)$. 
Instead, we define for each $\beta>0$ the {\sl temperature} or 
{\sl KMS state} 
$$
\Barintbeta\,: \cases {&$Cl_{\cald}(\a) \lra\, \C$\cr
&$\quad\quad \omega \phantom{w}\longmapsto\ \, 
{\displaystyle  \Barintbeta\omega := { {\rm Tr}_{
\scriptscriptstyle{\cal H}}\bigl(\omega e^{-\beta\cald^2}\bigr) 
\over {\rm Tr}_{\scriptscriptstyle{\cal H}}
\bigl( e^{-\beta\cald^2}\bigr) } \ , }$\cr}
$$
with no limit $\beta\to 0$ taken, in contrast to Definition 5.3. The 
$\beta$-integral $\barintbeta$ clearly is a faithful state, and 
through the GNS-construction we obtain a faithful representation of 
$Cl_{\cald}(\a)$ on a Hilbert space $\h_{\beta}$ with a cyclic and 
separating vector $\xi_{\beta}\in \h_{\beta}$ for $\m$. Each bounded 
operator $A\in{\cal B}(\h)$ on \h\ induces a bounded operator 
$A_{\beta}$ on $\h_{\beta}$; this is 
easily seen by computing matrix elements of $A_{\beta}$,  
$$
\langle\,A_{\beta} x, y\,\rangle = \Barintbeta A x y^* 
$$ 
for all $x,y\in\m \subset \h_{\beta}$, and using the explicit 
form of the $\beta$-integral. We denote the modular conjugation 
and the modular operator on $\h_{\beta}$ by $J_{\beta}$ and
$\triangle_{\beta}$, respectively, and we assume that for 
each $a\in\m$ the commutator 
$$
\lb\,\bard, J_{\beta}aJ_{\beta}\,\rb = {1\over i}\,{d\over dt}\,
\left( \left( e^{it\calbd}\right)_{\beta} J_{\beta}a J_{\beta}
\left( e^{-it\calbd}\right)_{\beta} \right) \bigg\vert_{t=0} 
$$
defines a bounded operator on $\h_{\beta}$.   \hfill\break  
\noindent Then we can define an algebra of bounded operators 
$\widetilde{Cl}_{\calbd}\,({}^{J_{\beta}}\a)$ on $\h_{\beta}$, 
which is contained in the commutant of $Cl_{\cald}\,(\a)$, 
and we say that 
the $N=(1,1)$ spectral data $(\a, \h, \dd, \gamma, *)$ describe a  
{\sl quantized phase space} if the following equation holds:  
$$
J_{\beta}\,Cl_{\cald}\,(\a)\, J_{\beta} 
= \widetilde{Cl}_{\calbd}\,({}^{J_{\beta}}\a)
$$
\bn\bn
\leftline{\bf 5.3 Hermitian and K\"ahler non-commutative geometry}
\bn                     In this section, we  
introduce the spectral data describing complex non-commutative 
spaces, more specifically spaces that carry a Hermitian or a K\"ahler 
structure. Since these are more restrictive than the data of 
Riemannian non-commutative geometry, we will be able to derive some 
appealing properties of the space of differential forms. We also find a  
necessary condition for a set of $N=(1,1)$ spectral data to extend to 
Hermitian data. A different approach to complex non-commutative 
geometry has been proposed in \q{BC}. 
\bn\bn
{\bf 5.3.1 Hermitian and \Ntwotwo spectral data}
\bn
{\bf Definition 5.26}\quad A set of data $(\a,\h, \partial, \overline{\partial}, 
T, \overline{T}, \gamma, *)$ is called a set of {\sl Hermitian spectral data}
if 
\smallskip
\item {1)} the quintuple $(\a,\h, \partial+ \overline{\partial}, \gamma, *)$
forms a set of $N=(1,1)$ spectral data;
\smallskip
\item {2)} $T$ and $\overline{T}$ are self-adjoint bounded operators on \h, 
$\partial$ and $\overline{\partial}$ are densely defined, closed operators on \h\  
such that the following (anti-)commutation relations hold:
$$\eqalign{
&\partial^2 = \overline{\partial}{}^2 = 0\ , \quad\,
\{\,\partial, \overline{\partial}\,\} = 0\ , 
\cr
&\lb\, T, \partial\,\rb = \partial\ ,\quad\quad \lb\,
T,\overline{\partial}\,\rb = 0\ , 
\cr
&\lb\, \overline{T}, \partial\,\rb = 0\ ,\quad\quad 
\lb\, \overline{T},\overline{\partial}\,\rb = \overline{\partial}\ , 
\cr
&\lb\,T,\overline{T}\,\rb = 0 \ ;
\cr}$$ 
\smallskip
\item {3)} for any $a\in\a$, $\lb\,T,a\,\rb = \lb\,\overline{T},a\,\rb = 0\,$, 
and each of the operators $\lb\,\partial,a\,\rb$,  
$\lb\,\overline{\partial},a\,\rb$ and $\{\,\partial, \lb\,\overline{\partial},a\,\rb\,\}$
extends uniquely to a bounded operator on \h; 
\smallskip
\item {4)} the $Z_2$-grading $\gamma$ satisfies 
$$\eqalign{
\{\,\gamma,\partial\,\} &= \{\,\gamma,\overline{\partial}\,\} = 0\ ,
\cr
\lb\,\gamma, T\,\rb &=\, \lb\,\gamma, \overline{T}\,\rb\, = 0\ ;
\cr}$$
\smallskip
\item{ 5)} the Hodge $*$-operator satisfies 
$$
*\,\partial = \zeta\,\overline{\partial}{}^*\,*\ ,\quad\quad 
*\,\overline{\partial} = \zeta\,\partial^*\,*  
$$
for some phase $\zeta\in\C$. 
\mn
Some remarks on this definition may be useful: The Jacobi identity and the  
equation $\{\,\partial,\overline{\partial}\,\}=0$ show that condition 3) above 
is in fact symmetric in $\partial$ and $\overline{\partial}$. 
\sn
As in section 5.2.1, a set $(\a,\h,\partial,\overline{\partial},T,\overline{T})$ 
that satisfies all of the above conditions, but does not involve $\gamma$ or $*$, 
can be made into a complete set of Hermitian spectral data. 
\sn
In classical Hermitian geometry, the $*$-operator can always be taken to be 
the usual Hodge $*$-operator -- up to a multiplicative redefinition in each 
degree -- since complex manifolds are orientable. 
\mn
Next, we describe conditions sufficient to equip a set of 
$N=(1,1)$ spectral data with a Hermitian structure. In subsection 5.3.2, 
Corollary 5.34, a necessary criterion is given as well. 
\mn
{\bf Proposition 5.27}\quad Let $(\a,\h,{\tt d}, \gamma, *)$ be a set of 
$N=(1,1)$ spectral data with $\lb\,\gamma,*\,\rb=0$, and let $T$ be 
a self-adjoint bounded operator on \h\ such that 
\smallskip
\item {a)} the operator $\partial := \lb\,T,{\tt d}\,\rb$ is nilpotent: 
$\partial^2 =0$; 
\smallskip
\item {b)} $\lb\,T,\partial\,\rb = \partial\,$; 
\smallskip
\item {c)} $\lb\,T, a\,\rb = 0$ for all $a\in\a$; 
\smallskip
\item {d)} $\lb\,T,\omega\,\rb \in \pi(\Omega^1(\a))$ for all 
$\omega\in\pi(\Omega^1(\a))$;
\smallskip
\item {e)} the operator $\overline{\partial} := {\tt d} - \partial$ satisfies 
$*\,\partial = \zeta\,\overline{\partial}{}^*\,*\,$, where $\zeta$ is the
phase appearing in the relations of $*$ in the $N=(1,1)$ data; 
\smallskip
\item {f)} $\lb\, T,\gamma\,\rb = 0$ and $\lb\,T,\overline{T}\,\rb = 0$,  
where $\overline{T} := - *\,T\,*^{-1}\,$.  
\sn
Then $(\a,\h,\partial,\overline{\partial},T,\overline{T}, \gamma,*)$ forms a set 
of Hermitian spectral data. 
\sn
Notice that the conditions a - d) are identical to those in Definition 3.22 of
section 3.4.1. Requirement e) will turn out to correspond to part e) of that 
definition. The relations in f) ensure compatibility of the 
operators $T$, $\gamma$ and $*$ and were not needed in the classical setting. 
\sn
{\kap Proof}: We check each of the conditions in Definition 5.26: The first
one is satisfied by assumption, 
since ${\tt d} = \partial + \overline{\partial}$ is the differential 
of $N=(1,1)$ spectral data. \hfill\break
\noindent The equalities $\partial^2 = \overline{\partial}{}^2 =
\{\,\partial,\overline{\partial}\,\}= \lb\,T,\overline{\partial}\,\rb =0$
follow {}from a) and b), as in the proof of Lemma 3.23. With this, we compute 
$$
\lb\,\overline{T},\overline{\partial}\,\rb = - \lb\,*\,T\,*^{-1},
\overline{\partial}\,\rb
= - \zeta\,*\,\lb\,T,\partial^*\,\rb\,*^{-1} = \overline{\partial} \ , 
$$ 
and since
$$
\lb\,\overline{T},\dd\,\rb = \lb\,*\,T\,*^{-1},\dd^*\,\rb^* 
= \zeta\,*\,\lb\,T,\dd\,\rb^*\,*^{-1} =  \overline{\partial}\ ,
$$ 
we obtain $\lb\,\overline{T},\partial\,\rb =0$. 
The relation $\lb\,T,\overline{T}\,\rb=0$ and self-adjointness of $T$ 
were part of the assumptions, and $\overline{T}{}^*=\overline{T}$ is 
clear {}from the unitarity of the Hodge $*$-operator. \hfill\break
\noindent That $\lb\,\partial,a\,\rb$ and  $\lb\,\overline{\partial},a\,\rb$ 
are bounded for all $a\in\a$ follows {}from the corresponding property of {\tt d} 
and {}from the assumption that 
$T$ is bounded. As in the proof of Proposition 3.24, one shows 
that $\{\,\partial, \lb\,\overline{\partial},a\,\rb\}\in\pi(\Omega_{\rm d}^2(\a))$, 
and therefore $\{\,\partial, \lb\,\overline{\partial},a\,\rb\}$ 
is a bounded operator. $T$ and $*$ commute with all
$a\in\a$ by assumption, and thus the same is true for
$\overline{T}$. \hfill\break
\noindent Using f) and the Jacobi identity, we get 
$$
\{\,\gamma,\partial\,\} = \{\,\gamma, \lb\,T,{\tt d}\,\rb\,\} = 
\lb\,T,\{\,{\tt d},\gamma\,\}\,\rb + \{\,{\tt d},\lb\,\gamma,T\,\rb\,\} =0 
$$ 
and 
$$
\{\,\gamma,\overline{\partial}\,\} = \{\,\gamma, {\tt d}- \partial\,\} = 0 \ .
$$
By assumption, $\gamma$ commutes with $T$ and $*$, therefore also with
$\overline{T}$.    \hfill\break
\noindent Finally, the relations of condition 5) in Definition 5.26 between
the $*$-operator and $\partial$, $\overline{\partial}$ follow directly {}from e) and 
$*\,{\tt d} = \zeta\,{\tt d}^*\,*\,$. \hfill\qed 
\mn
As a special case of Hermitian geometry, K\"ahler spaces arise in the 
same way as in classical differential geometry. In particular, 
K\"ahler spectral data provide a realization of the $N=(2,2)$ 
supersymmetry algebra:
\mn
{\bf Definition 5.28}\quad Hermitian spectral data 
$(\a,\h, \partial, \overline{\partial}, T, \overline{T}, \gamma, *)$ are called
$N=(2,2)$ or {\sl K\"ahler spectral data} if 
$$\eqalign{
&\{\,\partial,\overline{\partial}{}^*\,\} = \{\,\overline{\partial},\partial^*\,\} = 0\ ,
\cr 
&\{\,\partial,\partial^*\,\} = \{\,\overline{\partial},\overline{\partial}{}^*\,\} \ .
\cr}$$
\sn
Note that the first line is a condition which is a consequence of  
the second one in classical complex geometry,  
but has to be imposed separately in the non-commutative setting. 
\bn
In our discussion of Hermitian and K\"ahler geometry, we have, {}from the 
start, worked with data whose $N=2$ supersymmetry algebra splits into two 
``chiral halves'' -- in the terminology of section 4. This is 
advantageous when we treat differential forms, which now form a bi-graded 
complex. Alternatively, one can define K\"ahler spectral data, as we did in 
section 1.2, as containing a nilpotent differential {\tt d} -- 
together with its adjoint ${\tt d}^*$ -- and two commuting
U(1) generators $L^3$ 
and $J_0$, say, which satisfy the relations (1.49-51). This approach has 
the virtue that the complex structure familiar {}from classical differential 
geometry is already present in the algebraic formulation; see eq.\ (1.54) 
for the precise relationship with $J_0$. Moreover, this way of 
introducing non-commutative complex geometry makes the role of Lie group 
symmetries of the spectral data explicit, which is somewhat hidden in 
the formulation of Definitions 5.26 and 5.28 and in Proposition 5.27: 
The presence of the U(1)$\,\times\,$U(1) symmetry, acting in an 
appropriate way on spectral data, ensures that a set of  $N=(1,1)$ acquires 
an $N=(2,2)$ structure.  \hfill\break
\noindent               Because of the 
advantages in the treatment of differential forms, we will 
stick to the setting using $\partial$ and $\overline{\partial}$ 
for the time being, but the data with generators $L^3$ and $J_0$ will 
appear naturally in the context of symplectic geometry in section 5.5. 
\bn\bn 
\eject
\leftline{\bf 5.3.2 Differential forms}
\bn
In the context of Hermitian non-commutative geometry, we have two 
differential operators $\partial$ and $\overline{\partial}$ at our disposal. 
We begin this section with the definition of an abstract
algebra of universal forms which is appropriate for this situation. 
\mn
{\bf Definition 5.29}\quad A {\sl bi-differential algebra} \b\ is 
a unital algebra together with two {\sl anti-commuting} nilpotent 
derivations $\delta, \overline{\delta}\,:\ \b \lra\ \b\,$.   \hfill\break
\noindent A {\sl homomorphism of bi-differential algebras} 
$\varphi : \b \lra\ \b'$ is a unital algebra homomorphism  
which intertwines the derivations. 
\mn
{\bf Definition 5.30}\quad The {\sl algebra of complex universal forms}
$\Omega^{\bullet,\bullet}(\a)$ over a unital algebra \a\ is the 
(up to isomorphism) unique pair $(\iota, \Omega^{\bullet,\bullet}(\a))$
consisting of a unital bi-differential algebra $\Omega^{\bullet,\bullet}(\a)$ 
and an injective unital algebra homomorphism 
$\iota : \a \lra\ \Omega^{\bullet,\bullet}(\a)$
such that the following universal property holds: For any bi-differential 
algebra \b\ and any unital algebra homomorphism $\varphi : \a \lra\ \b\,$,  
there is a unique homomorphism $\widetilde{\varphi} : 
\Omega^{\bullet,\bullet}(\a) \lra\ \b$ of bi-differential algebras such 
that $\varphi = \widetilde{\varphi}\circ\iota$. 
\mn
The description of $\Omega^{\bullet,\bullet}(\a)$ in terms of generators 
and relations is analogous to the case of $\Omega^{\bullet}(\a)$, 
and it shows that $\Omega^{\bullet,\bullet}(\a)$ is a {\sl bi-graded
bi-differential algebra}  
$$
\Omega^{\bullet,\bullet}(\a) = \bigoplus_{r,s=0}^{\infty} 
\Omega^{r,s}(\a)
\eqno(5.39)$$
by declaring the generators $a, \delta a, \overline{\delta} a$ and 
$\delta\overline{\delta} a$, $a\in\a$, to have bi-degrees 
(0,0), (1,0), (0,1) and (1,1), respectively. 
\sn
As in the $N=(1,1)$ framework, we introduce an involution $\natural$, 
called {\sl complex conjugation}, on the algebra of complex universal 
forms, provided \a\ is a ${}^*$-algebra: 
$$
\natural\,:\ \Omega^{\bullet,\bullet}(\a) \lra\ \Omega^{\bullet,\bullet}(\a)
$$
is the unique anti-linear anti-automorphism acting on generators by 
$$\eqalignno{
&\natural(a) \equiv a^{\natural} :=  a^*\ , 
&\cr
&\natural(\delta a) \equiv (\delta a)^{\natural} := \overline{\delta} ( a^*\,)\ ,
\quad\quad\quad\quad  
\natural(\overline{\delta} a) \equiv (\overline{\delta} a)^{\natural} := \delta ( a^*\,)\ , 
\phantom{XXX}&(5.40)\cr
&\natural(\delta\overline{\delta} a) \equiv (\delta\overline{\delta} a)^{\natural} 
:=  \delta\overline{\delta}(a^*\,)\ . 
&\cr}$$
Let $\tilde\gamma$ be the $\Z_2$-reduction of the total 
grading on $\Omega^{\bullet,\bullet}(\a)$, i.e., $\tilde\gamma = (-1)^{r+s}$ 
on $\Omega^{r,s}(\a)$. Then it is easy to verify that 
$$
\overline{\delta} \natural \tilde\gamma = \natural \delta \ .
\eqno(5.41)$$
This makes $\Omega^{\bullet,\bullet}(\a)$ into a unital bi-graded
bi-differential ${}^{\natural}$-algebra.
\mn
Let $(\a,\h,\partial,\overline{\partial},T,\overline{T},\gamma,*)$ be a set of
Hermitian spectral data. Then we define a $\Z_2$-graded representation
$\pi$ of $\Omega^{\bullet,\bullet}(\a)$ as a unital bi-differential 
algebra on \h\ by setting
$$\eqalignno{
&\pi(a) = a\ , 
&\cr
&\pi(\delta a) = \lb\,\partial,a\,\rb\ ,\quad\quad\quad 
\pi(\overline{\delta} a) = \lb\,\overline{\partial},a\,\rb\ ,\phantom{XXXXX}
&(5.42)\cr
&\pi(\delta\overline{\delta} a) = \{\,\partial,\lb\,\overline{\partial},a\,\rb\,\}\ .
&\cr}$$
Note that, by the Jacobi identity, the last equation is compatible with 
the anti-commutati\-vi\-ty of $\delta$ and $\overline{\delta}$.  \hfill\break
\noindent As in the case of $N=(1,1)$ geometry, we have that
$$
\pi(\delta\omega) = \lb\,\partial, \pi(\omega)\,\rb_g\ ,\quad\quad\quad\quad  
\pi(\overline{\delta}\omega) = \lb\,\overline{\partial}, \pi(\omega)\,\rb_g\ ,
\eqno(5.43)$$
for any $\omega\in\Omega^{\bullet,\bullet}(\a)$,  
and therefore the graded kernel of the representation $\pi$ has good 
properties: We define 
$$
J^{\bullet,\bullet} := \bigoplus_{r,s=0}^{\infty} J^{r,s}\ ,\quad\quad
J^{r,s} := \{\, \omega\in\Omega^{r,s}(\a)\,|\, \pi(\omega) = 0\,\}\ , 
\eqno(5.44)$$
and we prove the following statement in the same way as 
Proposition 5.21: 
\mn
{\bf Proposition 5.31}\quad The set $J$ is a two-sided, bi-graded, 
bi-differential ${}^{\natural}$-ideal of $\Omega^{\bullet,\bullet}(\a)$. 
\mn
We introduce the space of complex differential forms as 
$$
\Omega_{\partial,\bar\partial}^{\bullet,\bullet}(\a) := \bigoplus_{r,s=0}^{\infty}
\Omega_{\partial,\bar\partial}^{r,s}(\a)\ , \quad\quad
\Omega_{\partial,\bar\partial}^{r,s}(\a):= \Omega^{r,s}(\a)/ J^{r,s}\ . 
\eqno(5.45)$$     The algebra 
$\Omega_{\partial,\bar\partial}^{\bullet,\bullet}(\a)$ is a unital bi-graded
bi-differential ${}^{\natural}$-algebra, too, and the representation $\pi$ 
determines a representation, still denoted $\pi$,  of this algebra on \h. 
\mn
Due to the presence of the operators $T$ and $\overline{T}$ among the Hermitian 
spectral data, the image of $\Omega_{\partial,\bar\partial}^{\bullet,\bullet}(\a)$  
under $\pi$ enjoys a property not  present in the $N=(1,1)$ case:
\mn
{\bf Proposition 5.32}\quad The representation of the algebra of complex 
differential forms satisfies 
$$
\pi\bigl(\Omega_{\partial,\bar\partial}^{\bullet,\bullet}(\a) \bigr)
= \bigoplus_{r,s=0}^{\infty}\pi\bigl(\Omega_{\partial,\bar\partial}^{r,s}(\a)\bigr)\ .
\eqno(5.46)$$
In particular, $\pi$ is a representation of
$\Omega_{\partial,\bar\partial}^{\bullet,\bullet}(\a)$ as a unital, bi-graded, 
bi-differential 
${}^{\natural}$-algebra. The ${}^\natural$-operation is implemented on 
$\pi\bigl(\Omega_{\partial,\bar\partial}^{\bullet,\bullet}(\a) \bigr)$ with
the help of the Hodge $*$-operator and the ${}^*$-operation on ${\cal B}(\h)$:
$$
\natural\,:\ \cases{
&${\displaystyle \pi\bigl(\Omega_{\partial,\bar\partial}^{r,s}(\a)\bigr) 
\lra\ \pi\bigl(\Omega_{\partial,\bar\partial}^{r,s}(\a)\bigr) }$ 
\cr
&$\phantom{MMM}\vphantom{\sum^{M}}\omega\quad \longmapsto \omega^{\natural} 
:= (-\zeta)^{r+s}\,*\,\omega^*\,*^{-1} $\cr}\ . 
$$
\sn
{\kap Proof}: Let
$\omega\in\pi\bigl(\Omega_{\partial,\bar\partial}^{r,s}(\a)\bigr)$. 
Then part 2) of Definition 5.26 implies that 
$$
\lb\,T, \omega\,\rb = r\,\omega\ ,\quad\quad 
\lb\,\overline{T}, \omega\,\rb = s\,\omega\ ,
$$
which gives the direct sum decomposition (5.46). It remains to 
show that the ${}^\natural$-operation is implemented on the space 
$\pi\bigl(\Omega_{\partial,\bar\partial}^{\bullet,\bullet}(\a) \bigr)$: 
For $a\in\a$, we have that  
$$\eqalign{
\pi\bigl((\delta a)^{\natural}\bigr) 
&=\pi\bigl(\overline{\delta}(a^*\,)\bigr) = \lb\,\overline{\partial},a^*\,\rb 
= - \lb\,\overline{\partial}{}^*, a\,\rb^* 
= - \lb\,\bar\zeta\,*\,\partial\,*^{-1}, a\,\rb^* 
= -\zeta\,*\,\lb\,\partial, a\,\rb^*\,*^{-1} 
\cr
&= -\zeta\,*\,\pi(\delta a)^*\,*^{-1}\ ,
\cr}$$ 
and, similarly, using (5.40) and the properties of the Hodge
$*$-operator,   
$$
\pi\bigl((\overline{\delta} a)^{\natural}\bigr) 
= -\zeta\,*\,\pi(\overline{\delta} a)^*\,*^{-1}\ ,
\quad\quad\ \ \pi\bigl((\delta\overline{\delta} a)^{\natural}\bigr) 
= \zeta^2\,*\,\pi(\delta\overline{\delta} a)^*\,*^{-1}\ .
$$ 
This proves that $\pi(\omega^{\natural}) = \pi(\omega)^{\natural}$. \hfill\qed
\sn
As an aside, we mention that the implementation of $\natural$ on 
$\pi\bigl(\Omega_{\partial,\bar\partial}^{\bullet,\bullet}(\a) \bigr)$ 
via the Hodge $*$-operator shows that the conditions e) of the 
``classical'' Definition 3.22 and of Proposition 5.27 are related; 
more precisely, the former is a consequence of the latter.   
\mn
Hermitian spectral data carry, in particular, an $N=(1,1)$ structure, and thus we 
have two notions of differential forms available. Their relation is described 
in our next proposition.  
\mn
{\bf Proposition 5.33}\quad The space of $N=(1,1)$ differential forms is 
included in the space of Hermitian forms, i.e., 
$$
\pi\bigl(\Omega_{\rm d}^{p}(\a) \bigr) \subset \bigoplus_{r+s=p} 
\pi\bigl(\Omega_{\partial,\bar\partial}^{r,s}(\a) \bigr) \ ,
\eqno(5.47)$$ 
and the spaces coincide if and only if 
$$
\lb\,T, \omega\,\rb \in \pi\bigl(\Omega_{\rm d}^{1}(\a) \bigr)
\quad\ {\rm for\ all}\quad \omega\in\Omega_{\rm d}^{1}(\a)\ .
\eqno(5.48)$$
\sn
{\kap Proof}: The inclusion (5.47) follows simply {}from ${\tt d} = 
\partial + \overline{\partial}$. If the spaces are equal then the equation   
$$
\lb\,T, \omega\,\rb = r\,\omega\ , 
$$ 
for all $\omega\in \pi\bigl(\Omega_{\partial,\bar\partial}^{r,s}(\a)
\bigr)\,$,   
implies (5.48). The converse is shown as in the proof of Proposition 
3.24, section 3.4.1, concerning classical Hermitian geometry.    \hfill\qed
\mn
Note that even if the spaces of differential forms do not coincide, 
the algebra of complex forms contains a graded differential algebra 
$\bigl( \Omega_{\partial,\bar\partial}^{\ \bullet}(\a), {\tt d}\bigr)$
with ${\tt d} = \partial + \overline{\partial}$ and
$$
\Omega_{\partial,\bar\partial}^{\ \bullet}(\a) 
   := \bigoplus_p \Omega_{\partial,\bar\partial}^{\ p}(\a)\ ,  \quad\quad
\Omega_{\partial,\bar\partial}^{\ p}(\a) := \bigoplus_{r+s=p} 
\Omega_{\partial,\bar\partial}^{r,s}(\a)\ .
\eqno(5.49)$$
By Proposition 5.32, we know that 
$$
\pi\bigl( \Omega_{\partial,\bar\partial}^{\bullet,\bullet}(\a)\bigr)
= \bigoplus_p \pi\bigl(\Omega_{\partial,\bar\partial}^{\ p}(\a)\bigr) \ ,
$$
and hence we obtain a necessary condition for $N=(1,1)$ spectral data 
to extend to Hermitian spectral data:
\mn
{\bf Corollary 5.34}\quad If a set of $N=(1,1)$ spectral data extends to  
a set of Hermitian spectral data then
$$
\pi\bigl( \Omega_{\rm d}^{\bullet}(\a)\bigr)
= \bigoplus_p \pi\bigl(\Omega_{\rm d}^{p}(\a)\bigr) \ .
$$
\mn
This condition is clearly not sufficient since it is always satisfied 
in classical differential geometry. 
\mn
Beyond the complexes (5.45) and (5.49), one can of course also consider 
the analogue of the {\sl Dolbeault complex} using only the differential 
$\overline{\partial}$ acting on
$\Omega^{\bullet,\bullet}_{\partial,\bar\partial}(\a)$. 
The details are straightforward. 
\bn           We conclude this  
subsection with some remarks concerning possible variations of our 
Definition 5.26 of Hermitian spectral data. E.g., one may wish to drop 
the boundedness condition on the operators $T$ and $\overline{T}$, in 
order to include infinite-dimensional spaces into the theory.   
This is possible, but then one has to make some stronger assumptions 
in Proposition 5.27. 
\sn
Another relaxation of the requirements in Hermitian spectral data  
is to avoid introducing $T$ and $\overline{T}$ altogether, and to 
replace them by a decomposition of the $\Z_2$-grading 
$$
\gamma = \gamma_{\partial} + \gamma_{\bar\partial}
$$
such that 
$$\eqalign{
&\{\,\gamma_{\partial}, \partial\,\} = 0\ ,\quad\quad       
\lb\,\gamma_{\partial}, \overline{\partial}\,\rb = 0\ , 
\cr
&\{\,\gamma_{\bar\partial}, \overline{\partial}\,\} = 0\ ,\quad\quad 
\lb\,\gamma_{\bar\partial}, \partial\,\rb = 0\ . 
\cr}$$
Then the space of differential forms may be defined as above, but 
Propositions 5.32 and 5.33, as well as the good properties of the 
integral established in the next subsection, will not hold in general.
\bn\bn
\leftline{\bf 5.3.3 Integration in complex non-commutative geometry}
\bn
The definition of the integral is completely analogous to the $N=(1,1)$ 
setting: Again we use the operator $\triangle = {\tt d}\,{\tt d}^* + 
{\tt d}^*\,{\tt d}$, where now ${\tt d} = \partial + \overline{\partial}$. 
Due to the larger set of data, the space of square-integrable,  
complex differential forms, now obtained after quotienting by the
two-sided {\sl bi-graded} ${}^{\natural}$-ideal $K$, has better properties 
than the corresponding space of forms in Riemannian non-commutative 
geometry. There, two elements 
$\omega\in\Omega^p_{\rm d}(\a)$ and $\eta\in\Omega^q_{\rm d}(\a)$
with $p\neq q$ were not necessarily orthogonal wrt.\ the sesqui-linear form 
$(\cdot,\cdot)$ induced by the integral. For Hermitian and K\"ahler
non-commutative geometry, however, we can prove the following orthogonality 
statements: 
\mn
{\bf Proposition 5.35}\quad Let $\omega_i \in
\pi\bigl(\Omega_{\partial,\bar\partial}^{r_i,s_i}(\a)\bigr)$, $i=1,2$. 
Then 
$$
(\omega_1, \omega_2) = 0  
\eqno(5.50)$$
if $r_1+s_1 \neq r_2+s_2$ in the Hermitian case; if the spectral data also 
carry an $N=(2,2)$ structure, then eq.\  (5.50) holds as soon as $r_1\neq
r_2$ or $s_1\neq s_2$. 
\sn
{\kap Proof}: In the case of Hermitian spectral data, the assertion follows
immediately {}from cyclicity of the trace, {}from the commutation relations 
$$
\lb\, T, \omega_i\,\rb = r_i\,\omega_i\ , \quad\quad 
\lb\, \overline{T}, \omega_i\,\rb = s_i\,\omega_i\ , 
$$
which means that $T+\overline{T}$ counts the total degree of a differential 
form, and {}from the equation  
$$
\lb\, T+\overline{T}, \triangle\,\rb = 0\ .
$$
In the K\"ahler case, Definition 5.28 implies the stronger relations 
$$
\lb\, T, \triangle\,\rb = \lb\, \overline{T}, \triangle\,\rb = 0\ .
\eqno{\qed} 
$$
\bn\bn
{\bf 5.3.4 Generalized metric on \Omtipar} 
\bn
The notions of vector bundles, Hermitian structure, torsion, etc.\ 
are defined just as for $N=(1,1)$ spectral data in section 5.2.  
The definitions of holomorphic vector bundles and connections 
can be carried over {}from the classical case; see section 3.4.4.
Again, we pass {}from $\Omega^{\ 1}_{\partial,\bar\partial}$, 
see eq.\ (5.49), to the space of all square-integrable 1-forms  
$\widetilde{\Omega}^{\ 1}_{\partial,\bar\partial}$, which is equipped with 
a generalized Hermitian structure $\langle\cdot,\cdot\rangle_{\scpbp}\,$ 
according to the construction in Theorem 5.9. Starting {}from here, we 
can define an analogue 
$$
\langle\!\langle\cdot,\cdot\rangle\!\rangle\,:\ 
\widetilde{\Omega}^{\ 1}_{\partial,\bar\partial}(\a)
\times \widetilde{\Omega}^{\ 1}_{\partial,\bar\partial}(\a)  \lra\ \C 
$$
of the $\C$-bi-linear metric in classical complex geometry by 
$$
\langle\!\langle\,\omega,\eta\,\rangle\!\rangle := 
\langle\,\omega,\eta^{\natural}\,\rangle_{\scpbp}
\ .
$$
\mn
{\bf Proposition 5.36}\quad  The generalized metric 
$\langle\!\langle\cdot,\cdot\rangle\!\rangle$ on 
$\widetilde{\Omega}^{\ 1}_{\partial,\bar\partial}(\a)$
has the following properties: 
\smallskip
\item {1)} $\langle\!\langle\,a\omega,\eta\,b\,\rangle\!\rangle
= a\,\langle\!\langle\,\omega,\eta\,\rangle\!\rangle\,b\,$;
\smallskip
\item {2)} $\langle\!\langle\,\omega\,a,\eta\,\rangle\!\rangle
= \langle\!\langle\,\omega,a\ \eta\,\rangle\!\rangle\,$;
\smallskip
\item {3)} $\langle\!\langle\,\omega,\omega^{\natural}\,\rangle\!\rangle
\geq 0\ $; 
\sn
here $\omega,\,\eta\in\widetilde{\Omega}^{\ 1}_{\partial,\bar\partial}(\a)$ 
and $a,b\in\a$. If the underlying spectral data are K\"ahlerian, one has that 
$$
\langle\!\langle\,\omega,\eta\,\rangle\!\rangle= 0
$$ 
if $\,\omega,\eta\in\widetilde{\Omega}^{0,1}_{\partial,\bar\partial}(\a)\,$
or $\,\omega,\eta\in\widetilde{\Omega}^{1,0}_{\partial,\bar\partial}(\a)\,$.
\sn
{\kap Proof}: The first three statements follow directly {}from the 
definition of $\langle\!\langle\cdot,\cdot\rangle\!\rangle$ and the 
corresponding properties of $\langle\cdot,\cdot\rangle_{\scpbp}$ 
listed in Theorem 5.9. The last 
assertion is a consequence of Proposition 5.35, using the 
fact that the spaces $\widetilde{\Omega}^{r,s}_{\partial,\bar\partial}(\a)$
are \a-bi-modules. Note that this property of the metric $\langle\!\langle
\cdot,\cdot\rangle\!\rangle$ corresponds to the property 
$g_{\mu\nu}=g_{\bar\mu\bar\nu}=0$ (in complex coordinates) in the 
classical case.  \hfill\qed
\bn\bn
{\bf 5.4 The \Nfourfour spectral data}
\bn
We just present the definition of spectral data 
describing non-commutative Hyperk\"ahler spaces.
Obviously, it is chosen in analogy to the discussion of the classical 
case in section 3.5.
\mn
{\bf Definition 5.37}\quad A set of data $(\a,\h,G^{a\pm}, 
\overline{G}{}^{a\pm}, T^i, \overline{T}{}^i, \gamma,*)$ with 
$a=1,2$, $i=1,2,3$, is called a set of $N=(4,4)$ or {\sl 
Hyperk\"ahler spectral data} if 
\smallskip
\item {1)} the subset $(\a,\h,G^{1+}, \overline{G}{}^{1+}, 
T^3, \overline{T}{}^3, \gamma,*)$ forms a set of $N=(2,2)$ 
spectral data; 
\smallskip
\item {2)} $G^{a\pm}$, $a=1,2$  are closed, densely defined operators   
on \h, and $T^i$, $i=1,2,3$,  are bounded operators on \h\ such that
the (anti-)commutation relations and Hermiticity conditions given
in eqs.\ (3.83-89) hold;
\smallskip
\item {3)} the operators $\overline{G}{}^{a\pm}$, $a=1,2$, and   
$\overline{T}{}^i$, $i=1,2,3$, also satisfy the conditions in 
2) and (anti-)commute with $G^{a\pm}$ and  $T^i$. 
\mn
The construction of non-commutative differential forms and the 
integration theory 
is precisely the same as for $N=(2,2)$ spectral data. We therefore  
refrain {}from giving more details. It might, however, be interesting to see  
whether the additional information encoded in $N=(4,4)$ 
spectral data gives rise to special properties,  beyond 
the ones found for K\"ahler data in subsection 5.3.3. 
\bn\bn
{\bf 5.5 Symplectic non-commutative geometry}
\bn             
Again, our description in the non-commutative context 
follows the  algebraic characterization 
of classical symplectic manifolds given in section 3.6. 
The difference between our approaches to the classical
and to the non-commutative case is that, in the former, we 
could derive most of the algebraic relations -- including  
the SU(2) structure showing up on symplectic manifolds -- {}from 
the specific properties of the symplectic 2-form, whereas now 
we will instead include those relations into the defining data, as 
a ``substitute'' for the symplectic form. 
\def\tdst{\widetilde{{\tt d}}^*}\def\td{\widetilde{{\tt d}}}
\mn
{\bf Definition 5.38}\quad The set of data $(\a,\h, {\tt d}, L^3, 
L^+, L^-, \gamma,*)$ is called a set of {\sl symplectic  
spectral data} if 
\smallskip
\item {1)} $(\a,\h, {\tt d}, \gamma,*)$ is a set of $N=(1,1)$ 
spectral data; 
\smallskip
\item {2)} $L^3$, $L^+$ and $L^-$ are bounded operators on \h\ which  
commute with all $a\in\a$ and satisfy the sl$_2$ commutation relations 
$$ 
\lb\, L^3, L^{\pm}\,\rb = \pm 2 L^{\pm}\ ,\quad 
\lb\, L^+, L^-\,\rb = L^3\ 
$$                   as well as 
the Hermiticity properties $(L^3)^*=L^3$, $(L^{\pm})^*= L^{\mp}$;
furthermore, they commute with the grading $\gamma$ on \h; 
\item {3)} the operator $\tdst := \lb\,L^-, {\tt d}\,\rb$ 
is densely defined and closed, and, together with {\tt d}, it forms 
an SU(2) doublet, i.e.\ the following commutation relations hold:  
$$\eqalign{
&\lb\, L^3, {\tt d} \,\rb = {\tt d}\ ,\quad\ 
\lb\, L^3, {\tdst} \,\rb = - \tdst\ ,
\cr
&\lb\, L^+, {\tt d} \,\rb = 0 \ ,\quad\  
\lb\, L^+, {\tdst} \,\rb = {\tt d}\ ,\quad
\cr 
&\lb\, L^-, {\tt d} \,\rb = \tdst\ ,\quad
\lb\, L^-, {\tdst} \,\rb = 0\ .
\cr}$$
\mn
As in the classical case, there is a second SU(2) doublet spanned 
by the adjoints ${\tt d}^*$ and $\td$. The Jacobi identity 
shows (see section 3.6) that $\tdst$ is nilpotent and that it 
anti-commutes with {\tt d}. 
\sn
Differential forms and integration theory are formulated just as for 
$N=(1,1)$ spectral data, but the presence of SU(2) generators among 
the symplectic spectral data leads to additional interesting features, 
such as the following: Let $\omega\in\Omega_{\rm d}^k(\a)$ 
and $\eta\in\Omega_{\rm d}^l(\a)$ be two differential forms. 
Then their scalar product, see eq.\ (5.38), vanishes unless $k=l$:
$$
(\omega,\eta) = 0 \quad\hbox{if $k\neq l$}\ .
\eqno(5.51)$$ 
This is true because, by the SU(2) commutation relations listed above, 
the operator $L^3$ induces a $\Z$-grading on differential forms, and 
because $L^3$ commutes with the Laplacian $\triangle = {\tt d}^*{\tt d}
+{\tt d}{\tt d}^*$. One consequence of (5.51) is that 
the reality property of $(\cdot,\cdot)$ stated in   
Proposition 5.22 is valid independently of the phase occurring in the 
Hodge relations. 
\mn
The following proposition shows that we can introduce an $N=(2,2)$ 
structure on a set of symplectic spectral data if certain 
additional properties are satisfied. As was the case for 
Definition 5.38, the extra requirements are slightly stronger 
than in the classical situation, where some structural elements like 
the almost-complex structure are given automatically. In the 
K\"ahler case, the latter allows for a separate counting holomorphic 
resp.\ anti-holomorphic degrees of differential forms, which in turn 
ensures that the symmetry group of the symplectic data associated 
to a classical K\"ahler manifold  
is in fact SU(2)$\,\times\,$U(1) -- i.e., in the terminology used at 
the end of section 4, that we are dealing with $N=4^+$ data. Without 
this enlarged symmetry group, it is impossible to re-interpret the 
$N=4$ as an $N=(2,2)$ supersymmetry algebra. Therefore, we explicitly 
postulate the existence of an additional U(1) generator  in the 
non-commutative context --  which coincides with the U(1) generator $J_0$ in  
eq.\ (1.49) of section 1.2 and is intimately related to the complex structure.  
\sn
{\bf Proposition 5.39}\quad Suppose that the SU(2) generators of a  
set of symplectic spectral data satisfy the following relations with 
the Hodge operator:
$$
*\,L^3  = - L^3\,*\ ,\quad\quad *\,L^+ = - \zeta^2\,L^-\,*\ ,
$$
where $\zeta$ is the phase appearing in the Hodge relations of the $N=(1,1)$ 
subset of the symplectic data. Assume, furthermore, that there exists a 
bounded self-adjoint operator $J_0$ on \h\ which commutes with all $a\in\a$, 
with the grading $\gamma$, and with $L^3$, whereas it acts like  
$$
\lb\,J_0, {\tt d}\,\rb = -i\,\td\ ,\quad\quad
\lb\,J_0, \td\,\rb = i\,{\tt d}
$$
between the SU(2) doublets. If, finally, the operators {\tt d} 
and $\td$ anti-commute, 
$$
\{\,{\tt d},\td\,\}=0\ ,
\eqno(5.52)$$
then the set of symplectic data carries an $N=(2,2)$ K\"ahler structure 
with 
$$\eqalign{
\partial &= {1\over2}\,({\tt d} - i\,\td)\ ,\quad\quad\quad 
\overline{\partial} =  {1\over2}\,({\tt d} + i\,\td)\ ,
\cr
T &= {1\over2}\,(L^3 + J_0)\ ,\quad\quad\, 
\overline{T} = {1\over2}\,(L^3 - J_0)\ .
\cr}$$
\sn
{\kap Proof}: All the conditions listed in Definition 5.26 of Hermitian
spectral data can be verified easily: Nilpotency of $\partial$ and 
$\overline{\partial}$ is equivalent to the anti-commutator (5.52)   
in the assumptions, and the action of the Hodge operator on the SU(2) generators 
ensures that $*$ intertwines $\partial$ and $\overline{\partial}$ in the 
right way. As for the extra conditions in Definition 5.28 of K\"ahler 
Hermitian data, one sees that the first one is always true for 
symplectic spectral data, whereas the second one, namely the equality of 
the ``holomorphic'' and ``anti-holomorphic'' Laplacians, is again a 
consequence of relation (5.52).      \hfill\qed 
\bn\bn
\vfill\eject 
\leftline{\bf 6. Directions for future work}
\bn
In this work, we have presented an approach to geometry rooted 
in supersymmetric quantum theory. We have systematized the various 
types of classical and of non-commutative geometries by ordering them 
according to the symmetries, or the ``supersymmetry content'', of their  
associated spectral data. Obviously, many natural 
and important questions remain to be studied. In this 
concluding section, we describe a few of these open problems and sketch 
once more some of the physical motivations underlying our work. 
\sn
{\bf(1)}\quad An obvious fundamental question is whether one can give a 
complete classification of the possible types of spectral data in terms of 
graded Lie algebras (and perhaps $q$-deformed graded Lie algebras). As an 
example, recall the structure of $N=4^+$ spectral data, which is an extension 
of K\"ahler geometry (see section 1.2). The spectral data involve the 
operators $\ttd,\ \ttdst,\ \td,\ \tdst,\ L^3,\ L^+,\ L^-,\ J_0$ and 
$\triangle$, which close under taking (anti-)commutators: They generate 
a graded Lie algebra defined by 
$$\eqalign{
&\lb\,L^3,L^{\pm}\,\rb=\pm2L^{\pm}\ ,\quad
\lb\,L^+,L^{-}\,\rb=L^{3}\ ,\quad 
\lb\,J_0,L^{3}\,\rb=\lb\,J_0,L^{+}\,\rb=0\ ,
\cr
&\lb\,L^3,\ttd\,\rb=\ttd\ ,\quad
\lb\,L^+,\ttd\,\rb=0\ ,\quad
\lb\,L^-,\ttd\,\rb=\tdst\ ,\phantom{NN}
\lb\,J_0,\ttd\,\rb=-i\,\td\ ,    
\cr
&\lb\,L^3,\td\,\rb=\td\ ,\quad
\lb\,L^+,\td\,\rb=0\ ,\quad
\lb\,L^-,\td\,\rb=-\ttdst\ ,\quad
\lb\,J_0,\td\,\rb= i\,\ttd\ ,    
\cr
&\{\,\ttd,\ttd\,\}=\{\,\td,\td\,\}= \{\,\ttd,\td\,\}=
\{\ttd,\tdst\,\}=0\ ,
\cr
&\{\,\ttd,\ttdst\,\}=  \{\,\td,\tdst\,\}=   \triangle \ ,  
\cr}$$
and $\triangle$ is central in the graded Lie algebra. The remaining 
(anti-)commutation relations follow by taking adjoints, with the rules 
that $\triangle$, $J_0$ and $L^3$ are self-adjoint, 
and $(L^-)^*=L^+$. \hfill\break
\noindent It would be interesting to determine all graded Lie algebras 
(and their representations) that can arise as the spectral data of a
(non-commutative) space. In the case of classical geometry, we have given 
a classification up to $N=(4,4)$ spectral data, and we expect that 
there is enough information available to settle the problem completely. 
In the non-commutative setting, however, further algebraic structures 
might occur, including $q$-deformations of graded Lie algebras.\hfill\break 
\noindent To give a list of all graded Lie algebras that are possible {}from 
an abstract point of view is most certainly possible with the help of the 
results available in the literature. However, in view of the classical 
case, where we only found the groups U(1), SU(2), Sp(4) and direct 
products thereof, we expect that not all of the abstractly possible 
Lie group symmetries are realized in (non-commutative) geometry. 
\sn
Determining the graded Lie algebras that actually occur in the spectral data 
of geometric spaces is clearly just the first step towards a classification 
of non-commutative spaces. It will be harder to determine the class of all 
${}^*$-algebras \a\ that admit a given type of spectral data, i.e.\ such 
that the algebra possesses a $K$-cycle $(\h, \ttd_i)$, with a collection of 
differentials $\ttd_i$ generating a given graded Lie
algebra, and such that the ordinary Lie group generators $X_j$ contained 
in the graded Lie algebra commute with the elements of \a. 
\mn
{\bf(2)}\quad Given some set of spectral data, it is natural to 
investigate symmetries such as diffeomorphisms. 
For definiteness, let us start {}from a set of data $(\a, \h,
{\tt d}, {\tt d}^*, T, *)$  with an $N=2$ structure, cf.\ section 5.2.6. The
study of this question involves the algebra $\Phi_{\rm d}^{\bullet}(\a)$
defined as the smallest ${}^*$-algebra of (unbounded) operators containing 
${\cal B}:= \pi(\Omega^{\bullet}(\a))\lor \pi(\Omega^{\bullet}(\a))^*$ 
and arbitrary graded commutators of \ttd\ and \ttdst\ with elements of 
$\cal B$. With 
the help of the $\Z$-grading $T$, $\Phi_{\rm d}^{\bullet}(\a)$ decomposes into 
a direct sum 
$$
\Phi_{\rm d}^{\bullet}(\a) := \bigoplus_{n\in\Z} \Phi_{\rm d}^{n}(\a)\ ,
\quad\quad \Phi_{\rm d}^{n}(\a) := \bigl\{\,\phi\in\Phi_{\rm d}^{\bullet}
(\a)\,\big\vert\,\lb\,T,\phi\,\rb_g=n\,\phi\,\bigr\}\ ;
$$
note that both positive and negative degrees occur. 
Thus, $\Phi_{\rm d}^{\bullet}(\a)$ is a graded ${}^*$-algebra, and is in fact 
a natural object to introduce when dealing with $N=2$ spectral data, as 
the algebra $\Omega_{\rm d}^{\bullet}(\a)$ of differential forms 
does not have a ${}^*$-representation on \h\ simply because 
{\tt d} is not self-adjoint.    \hfill\break
\noindent Ignoring certain domain problems which arise since the 
(anti-)commutator 
of \ttd\ with the adjoint of a differential form is unbounded in general, 
we observe that $\Phi_{\rm d}^{\bullet}(\a)$ has the interesting property that  
it forms a complex with respect to the action of \ttd\ by the graded
commutator, and in view of examples {}from quantum field theory, we will 
call it the {\sl field complex} in the following. \hfill\break
\noindent 
If we start with $N=(2,2)$ non-commutative K\"ahler data containing 
the holomorphic and anti-holomorphic gradings $T$ and $\overline{T}$, see 
Definition 5.26, we can introduce a {\sl bi-graded complex} 
$\Phi_{\partial,\overline{\partial}}^{\bullet,\bullet}(\a)$ in an analogous 
way. Such bi-graded field complexes occur e.g.\ in $N=(2,2)$ superconformal 
field theory, but there we also meet a more general type of 
``$(n,m)$-forms'' $\phi\in\Phi_{\partial,\overline{\partial}}^{n,m}(\a)$
which have $n+m \in\Z$ but {\sl arbitrary real} holomorphic and 
anti-holomorphic degrees $n$ and $m$.
\sn
Let us now describe the role of the field complex in connection with 
{\sl diffeomorphisms} of a geometric space presented by a set of $N=2$ 
spectral data. One possible generalization of the notion of diffeomorphisms 
to non-commutative geometry would be to identify them with
${}^*$-automorphisms of the algebra \a\ of ``smooth functions over the 
manifold''. It may be better, though, to follow concepts of classical 
geometry more closely: There, an infinitesimal diffeomorphism is given by a 
derivation $\delta(\cdot) := \lb\,L,\cdot\,\rb$ of \a\ with an element $L$ 
of $\Phi_{\rm d}^0$ such that $\delta$ commutes with {\tt d}, i.e.\   
$$
\lb\,{\tt d},L\,\rb =0 \ .
$$
The derivation can then be extended to all of $\pi(\Omega_{\rm
d}^{\bullet}(\a))$, and 
we see that $\delta$ leaves the degree of a differential form invariant 
iff $L$ commutes with $T$, i.e.\ iff $L\in \Phi_{\rm d}^0$.  \hfill\break 
\noindent For a classical manifold $M$, it turns out that each $L$ with 
the above properties can be written as 
$$
L= \{\,{\tt d}, X\,\}
$$ 
for some vector field $X\in\Phi_{\rm d}^{-1}$, i.e.\ $L$ is the Lie 
derivative in the direction of this vector field on the manifold. 
In the general non-commutative case, however, it might happen  that 
the cohomology of the field complex at the zeroth position is non-trivial. 
In this case, the study of diffeomorphisms of the non-commutative space 
necessitates studying the cohomology of the field complex 
$\Phi_{\rm d}^{\bullet}(\a)$.   
\sn
Besides general diffeomorphisms, it is interesting to investigate 
{\sl special} diffeomorphisms, i.e.\ ones that respect additional 
structures of the spectral data. E.g., derivations  $\delta(\cdot) = 
\lb\,L,\cdot\,\rb$ such that $L$ commutes with \ttd\ {\sl and} \ttdst\ 
correspond to {\sl isometries} of the non-commutative space. If we 
consider complex spectral data, the diffeomorphisms of interest 
are {\sl bi-holomorphic} maps, which can be studied with the help of 
the bi-graded field complex 
$\Phi_{\partial,\overline{\partial}}^{\bullet,\bullet}(\a)$. In the 
symplectic case, the relevant diffeomorphisms are those which leave 
the symplectic form invariant; in algebraic terms, we define 
infinitesimal {\sl symplectomorphisms} as derivations which 
annihilate the SU(2)-generator $L^+$ of the symplectic spectral data. 
\mn
{\bf(3)}\quad {\sl Deformation theory} is another important subject, also 
relevant for the problem of classifying non-commutative geometries. Given a 
set of spectral data with generators $\{\,X_j,\ \ttd,\ \ttd_{\alpha},\ 
\triangle\,\}$ of a graded Lie algebra as in paragraph (1), we may study 
one-parameter families 
$\{\,X_j^{(t)},\ \ttd,\ \ttd_{\alpha}^{(t)},\ \triangle^{(t)}\,\}_{t\in\R}$ 
of deformations. Here, we chose to keep one generator $\ttd$ fixed, and we 
require that the algebraic relations of the graded Lie algebra are the same 
for all $t$ -- which means that we study deformations of the (non-commutative) 
complex or symplectic structure of a given space \a\ while preserving the 
differential and the de Rham complex. In view of paragraph (2), we regard as 
non-trivial only those deformations which do not arise {}from diffeomorphisms of 
the (non-commutative) space. In the classical case, a number of important facts 
about such deformations are known (Kodaira-Spencer theory, Moser's Theorem).  
But even in classical geometry, many questions remain to be answered. 
\sn
Alternatively, we can ``move'' all the differentials, including $\ttd$, 
and study deformations 
$$
{\tt d}' := {\tt d} + \omega \ ,
$$
where, e.g., \ttd\ is the nilpotent differential in a
set of $N=2$ data $(\a, \h, {\tt d}, {\tt d}^*, T, *)$, and $\omega\in
\Phi^{\bullet}_{\rm d}(\a)$.  We require that ${\tt d}'$ again squares 
to zero. Therefore, $\omega$ has to satisfy a kind of ``zero curvature 
condition'' 
$$
\omega^2 + \{\,{\tt d}, \omega\,\} = 0\ .
\eqno(6.1)$$
We distinguish between several possibilities: First, let us require 
that the deformed data still carry an $N=2$ structure with the same 
$\Z$-grading $T$ as before. Then, $\omega$ must be an element of 
$\Phi_{\rm d}^{1}(\a)$ satisfying (6.1), and we can identify it with 
the connection 1-form of a flat connection on some vector bundle, 
similar to the discussion of the structure of classical $N=(1,1)$ Dirac 
bundles in section 3.2.3.    \hfill\break
\noindent  More generally, we might require the deformed data to be merely of 
$N=(1,1)$ type, but with a $\Z_2$-grading $\gamma$ inherited as the mod$\,2$
reduction of $T$. Then, $\omega$ can be the sum of different homogeneous parts, 
each taken {}from some $\Phi_{\rm d}^{2n+1}$, and condition (6.1) decouples into 
homogeneous equations which relate the different components of $\omega$. 
The simplest situation occurs if $\omega\in{\Phi}_{\rm d}^{2n+1}$ is homogeneous 
of degree $2n+1$ with $n\neq0$; in this case, the zero curvature condition 
implies that $\omega$ itself is nilpotent and closed with respect to 
{\tt d}. Thus, the cohomology of the complex  ${\Phi}_{\rm d}^{\bullet}$ 
will again play an interesting role.  \hfill\break
\noindent In the most general case, when we do not even require that 
the mod$\,2$ reduction of the original $\Z$-grading $T$ is preserved 
by the deformation, eq.\ (6.1) is the only restriction on the 
deformation operator $\omega$. Again, the splitting into homogeneous degrees 
leads to a number of independent relations. 
\mn
{\bf (4)}\quad In the introduction (and in section 4), 
we have stressed  
that {}from the point of view of supersymmetric quantum theory it is not 
natural to exclusively focus on the algebra of ``functions'' over a 
(non-commutative) space, but that ``functions'' over {\sl phase 
space } are at least as important, physically. Therefore, we have 
to study the relations between spectral data formed with an 
algebra \a\ of configuration space functions and those built 
with a (deformed) algebra ${\cal F}_{\hbar}$ of functions over phase 
space. One question to be asked is, e.g., whether there is a simple 
connection between the cohomology groups derived {}from both sets of 
data by the methods presented in \q{Co1} and in this paper.
Furthermore, it is important whether one can determine \a, i.e., 
in the classical case, the manifold itself, when given only the 
spectral data of phase space. {}From physics, we expect that 
ambiguities arise here, known under the name of {\sl T-duality};  
see \q{FG,KS} for some discussion of this feature in the context 
of non-commutative geometry and quantum field theory. T-duality is the 
phenomenon that 
two $\sigma$-models are equivalent as quantum theories although 
they have different classical target manifolds $M_1$ and $M_2$, 
which are then said to be dual to each other. In algebraic 
terms, this implies that the phase space algebras provided by the 
two models can be identified, while the algebras $\a_1$ and 
$\a_2$ of functions over the two manifolds are different 
(maximal) commuting sub-algebras of this phase space ${\cal F}_{\hbar}$.  
Thus, the classification of T-dual targets should essentially 
be the classification of ``maximal tori'' within a given 
phase space algebra.   \hfill\break
\noindent These remarks apply to the classical case. In view of 
the definitions suggested in section 5.2.6, one might want to 
discuss the relation between ``phase space'' and ``configuration 
space'' also in the truly non-commutative setting, when the sub-algebra
\a\ of ${\cal F}_{\hbar}$ is non-commutative, too. In this situation, 
group symmetries of a set of spectral data should prove useful 
to find a suitable replacement for the notion of ``maximal tori''. 
Recall that we have imposed the condition $\lb\,X_j,a\,\rb=0$ 
for all Lie group generators $X_j$ and all $a\in\a$. This requirement 
cannot be extended to $a\in{\cal F}_{\hbar}$, not even in the classical 
case. It thus appears natural to regard fixed point sub-algebras 
of ${\cal F}_{\hbar}$ under a given Lie algebra action 
as candidates for a non-commutative algebra of ``functions over 
configuration space''. In addition, we use the properties described 
in section 5.2.6 characteristic of  non-commutative manifolds, 
namely we require that there exists a vector $\xi_0\in\h$ which is 
cyclic and separating for the algebra $\bigl(Cl_{\cald}(\a)\bigr)''$
associated to a ``candidate maximal torus'' \a\ in ${\cal F}_{\hbar}$. 
\sn 
Applications of our general formalism to physical examples will lead us to the
study of {\sl loop spaces} $M^{S^1}$, or more generally of spaces $M^L$ of
maps {}from some parameter manifold $L$ into a (non-commutative) target $M$.  As
was briefly discussed in sections 1 and 4, every supersymmetric quantum field
theory over a $d+1$-dimensional parameter space-time with $d\geq1$ provides
such a loop- or mapping-space.  Again we have to ask how the geometry of the
underlying target $M$ can be recovered {}from the spectral data of such a loop
space. This question is not only of purely mathematical interest, but becomes
important in connection with string vacua, see the comments below.
\mn
{\bf (5)}\quad A subject bearing some similarity to T-duality 
is {\sl mirror symmetry}. At the end of section 3.4.3, we have made 
some remarks on the definition of mirrors of classical Calabi-Yau manifolds, 
and it is natural to look for a non-commutative generalization of this 
notion. Guided by $N=(2,2)$ superconformal field theory, we suggest the 
following definition (see also \q{FG}): Given two sets of $N=(2,2)$ 
spectral data $(\a_i,\h, \partial_i, \overline{\partial}_i, T_i, 
\overline{T}_i, *_i)$, $i=1,2$, where the algebras $\a_i$ act on the same 
Hilbert space \h, we say that the space $\a_2$ 
is the {\sl mirror} of $\a_1$ if 
$$\partial_2=\partial_1\ ,\quad \overline{\partial}_2=
\overline{\partial}{}_1^*\ ,\quad 
T_2=T_1\ ,\quad \overline{T}_2=-\overline{T}_1\ ,
$$ 
and if the dimensions 
$b_i^{p,q}$ of the cohomology of the Dolbeault complexes (5.45) 
satisfy  $b_2^{p,q}=b_{1}^{n-p,q}$, where $n$ is the top dimension 
of differential forms (recall that in Definition 5.26 we required $T$ 
and $\overline{T}$ to be bounded operators).  \hfill\break 
\noindent Let \a\ be a non-commutative K\"ahler space with mirror
$\widetilde{\a}$. Within superconformal field theory, there is the 
following additional relation between the two algebras: Viewing \a\ as the 
algebra of functions over a (non-commutative) target $M$, and analogously 
for $\widetilde{\a}$ and $\widetilde{M}$, then the loop spaces over the
phase spaces over $M$ and $\widetilde{M}$ coincide. Casting this relation 
between \a\  and $\widetilde{\a}$ into a precise 
operator algebraic form should prove useful for a classification 
of mirror pairs in non-commutative geometry. 
\mn
{\bf(6)}\quad The success of the theory presented in this paper will
ultimately be measured in terms of the applications it has to interesting
concrete problems of geometry. In particular, it is important to apply the
notions introduced here to examples of truly non-commutative spaces like
quantum groups, or the non-commutative (``fuzzy'') sphere, see e.g.\
\q{Ber,Ho,Ma,GKP}, the non-commutative torus, see \q{Co1} and references
therein, non-commutative Riemann surfaces \q{KL} and non-commutative symmetric
spaces $\lb {\sl BLU,}\break\noindent{\sl BLR,GP,BBEW}\rb$.  In these cases, 
it is natural to ask whether
the ``deformed'' spaces carry a complex or K\"ahler structure in the sense of
section 5.3.
\sn   {}From our point of view, however, the most interesting examples
for the general theory and the strongest motivation to study spectral data
with supersymmetric structure come {}from string theory: The ``ground states''
of string theory are described by certain $N=(2,2)$ superconformal quantum
field theories.  They provide the spectral data of the loop space over a
target which is a ``quantization'' of classical space -- or rather of
3-dimensional physical space times an internal compact manifold.  We have
already mentioned in the introduction -- in the example of the WZW model --
that starting {}from a classical $\sigma$-model of maps into a classical target,
quantizing it and then analyzing the geometry encoded in the associated set of
spectral data, one discovers that the original target is deformed upon
quantization of the field theory. The same phenomenon is expected to occur in
string theory, when the $N=(2,2)$ supersymmetric conformal field theory one
starts {}from is a $\sigma$-model with a complex 3-dimensional Calabi-Yau
manifold (times Minkowski space-time) as a target. But instead, we may as well
apply the methods developed in this paper to a supersymmetric CFT with just
the right algebraic properties and without an a priori connection to a
classical manifold, and re-interpret it as a $\sigma$-model over some
non-commutative target space. The details of reconstructing models of
non-commutative target spaces {}from the spectral data of models of
supersymmetric conformal field theory are not altogether straightforward and
require some amount of technical work. We plan to come back to these problems
in future work.
\sn 
\vfill\eject 
\parindent=35pt \vsize=23.5truecm
\halign{#\hfil&\vtop{\parindent=0pt \hsize=35.5em #\strut}\cr
\noalign{\leftline{{\bf References }}} \noalign{\vskip.4cm}
\q{AG} &L.\ Alvarez-Gaum\'e, {\sl Supersymmetry and the Atiyah-Singer 
   index theorem},  Commun.\ Math.\ Phys. {\bf90} (1983) 161-173\cr 
\q{AGF}&L.\ Alvarez-Gaum\'e, D.Z.\ Freedman, {\sl Geometrical structure and 
   ultraviolet finiteness in the supersymmetric $\sigma$-model}, 
   Commun.\ Math.\ Phys. {\bf80} (1981) 443-451\cr 
\q{Au}&O.\ Augenstein, {\sl Supersymmetry, non-commutative geometry 
   and model building}, Diploma Thesis ETH Z\"urich, March 1996\cr
\q{BC} &P.\ Beazley Cohen, {\sl Structure complexe non commutative et
   superconnexions}, pre\-print MPI f\"ur Mathematik, Bonn, MPI/92-19\cr
\q{Beau} &A.\ Beauville, 
   {\sl Variet\'es K\"ahleriennes dont la 1\`ere classe de 
   Chern est nulle}, J.\ Diff.\ Geom.\ {\bf18} (1983) 755-782\cr
\q{Ber} &F.A.\ Berezin, {\sl General concept of quantization},  
   Commun.\ Math.\ Phys.\ {\bf40} (1975) 153-174\cr
\q{Bes} &A.L.\ Besse, {\sl Einstein Manifolds}, Springer Verlag 1987\cr
\q{BGV} &N.\ Berline, E.\ Getzler, M.\ Vergne, {\sl Heat Kernels 
   and Dirac Operators}, Springer Verlag 1992\cr 
\q{BLR} &D.\ Borthwick, A.\ Lesniewski, M.\ Rinaldi, {\sl Hermitian 
   symmetric superspaces of type IV}, J.\ Math.\ Phys.\ {\bf343} (1993) 
   4817-4833\cr
\q{BLU} &D.\ Borthwick, A.\ Lesniewski, H.\ Upmeier, {\sl Non-perturbative
   quantization of Cartan domains}, J.\ Funct.\ Anal.\ {\bf113} (1993)
   153-176\cr
\q{BBEW} &M.\ Bordemann, M.\ Brischle, C.\ Emmrich, S.\ Waldmann, 
   {\sl Phase space reduction for star-products: An explicit 
   construction for $\C{\rm P}^n$}, Lett.\ Math.\ Phys.\ {\bf36} 
   (1996) 357-371\cr
\q{BSN} &E.\ Bergshoeff, E.\ Szegin, H.\ Nishino, {\sl (8,0) Locally 
   supersymmetric $\sigma$-models with conformal invariance in two 
   dimensions}, Phys.\ Lett.\ B {\bf186} (1987) 167-172\cr
\q{Cal} &E.\ Calabi, {\sl On K\"ahler manifolds with vanishing 
   canonical class}, in ``Algebraic geometry and topology, a 
   symposium in honor of S.\ Lefschetz'', Princeton University 
   Press 1955, p.\ 78-89\cr
\q{Can} &P.\ Candelas, {\sl Lectures on complex manifolds}, in 
   ``Superstrings '87'', Proceedings of the Trieste Summer School, 
   Eds.\ L.\ Alvarez-Gaum\'e et al., p.\ 1-88\cr
\q{CE} &C.\ Chevalley, S.\ Eilenberg, {\sl Cohomology of Lie groups 
   and Lie algebras}, Trans.\ Amer.\ Math.\ Soc.\ {\bf63} (1948) 
   85-124\cr 
\q{CFF} &A.H.\ Chamseddine, G.\ Felder, J.\ Fr\"ohlich, {\sl Gravity
   in Non-commutative Geometry}, Commun.\ Math.\ Phys.\ {\bf155}
   (1993), 205-217\cr 
\q{CFG} &A.H.\ Chamseddine, J.\ Fr\"ohlich, O.\ Grandjean, {\sl The
   gravitational sector in the Connes-Lott formulation of the 
   standard model}, J.\ Math.\ Phys.\ {\bf36} (1995) 6255-6275\cr
\q{CM} &S.\ Coleman, J.\ Mandula, {\sl All possible symmetries of 
   the S-matrix}, Phys.\ Rev.\ {\bf159} (1967) 1251-1256\cr  
\q{Co1} &A.\ Connes, {\sl Noncommutative Geometry}, Academic Press
   1994\cr
\q{Co2} &A. Connes, {\sl Noncommutative differential geometry}, 
   Inst.\ Hautes \'Etudes Sci.\ Publ.\ Math.\ {\bf62} (1985) 257-360\cr
\q{Co3} &A.\ Connes, {\sl The action functional in noncommutative 
   geometry}, Commun.\ Math.\ Phys.\ {\bf117} (1988) 673-683\cr 
\q{Co4} &A.\ Connes, {\sl Reality and noncommutative geometry}, 
   J.\ Math.\ Phys.\ {\bf36} (1995) 6194-6231\cr
\q{CoK} &A.\ Connes, M.\ Karoubi, {\sl Caract\`ere multiplicatif 
   d'un module de Fredholm}, $K$-Theory {\bf2} (1988)\cr
\q{DFR} &S.\ Doplicher, K.\ Fredenhagen, J.E.\ Roberts, {\sl The quantum 
   structure of space-time at the Planck scale and quantum fields}, 
   Commun.\ Math.\ Phys.\ {\bf172} (1995) 187-220\cr
\q{F} &J.\ Fr\"ohlich, {\sl The non-commutative geometry of 
   two-dimensional supersymmetric conformal field theory}, in: PASCOS, 
   Proc.\ of the Fourth Intl. Symp.\ on Particles, Strings and Cosmology, 
   K.C.\ Wali (ed.), World Scientific 1995\cr   
\q{FG} &J.\ Fr\"ohlich, K.\ \gaw, {\sl Conformal Field Theory 
   and the Geometry of Strings}, CRM Proceedings and Lecture Notes 
   Vol.\ {\bf 7} (1994), 57-97\cr
\q{FGK} &G.\ Felder, K.\ \gaw, A.\ Kupiainen,  {\sl Spectral of 
   Wess-Zumino-Witten models with arbitrary simple groups}, 
   Commun.\ Math.\ Phys.\ {\bf117} (1988) 127-158\cr
\q{FLL} &J.\ Fr\"ohlich, E.H.\ Lieb, M.\ Loss, {\sl Stability of 
   Coulomb systems with magnetic fields I}, Commun.\ Math.\ Phys.\ 
   {\bf104} (1986) 251-270\cr
\q{FW} &D.\ Friedan, P.\ Windey, {\sl Supersymmetric derivation of the 
   Atiyah-Singer index theorem and the chiral anomaly}, Nucl.\ Phys.\ B 
   {\bf235} (1984) 395-416\cr
\q{Ge} &E.\ Getzler, {\sl Pseudo-differential operators on supermanifolds 
   and the Atiyah-Singer index theorem}, Commun.\ Math.\ Phys.\ 
   {\bf92} (1983) 163-178; {\sl A short proof of the Atiyah-Singer index 
   theorem}, Topology {\bf25} (1986) 111-117\cr
\q{GeW} &D.\ Gepner, E.\ Witten, {\sl String theory on group manifolds},
   Nucl.\ Phys.\ B {\bf278} (1986) 493-549\cr  
\q{GKP} &H.\ Grosse, C.\ Klim\v cik, P.\ Pre\v snajder, {\sl Towards 
   finite quantum field theory in non-commutative geometry}, CERN preprint,
   hep-th/9505175\cr
\q{GL} &Y.A.\ Gol'fand, E.P.\ Likhtman, {\sl Extension of the algebra of 
   Poincar\'e group generators and violation of P-invariance}, 
   JETP Lett.\ {\bf 13} (1971) 323-326\cr
\q{GP} &H.\ Grosse, P.\ Pre\v snajder, {\sl The construction of
   non-commutative manifolds using coherent states}, Lett.\ Math.\ Phys.\ 
   {\bf28} (1993) 239-250\cr
\q{GSW} &M.B.\ Green, J.H.\ Schwarz, E.\ Witten, {\sl Superstring Theory I,II},
   Cambridge University Press 1987\cr
\q{HKLR} &N.J.\ Hitchin, A.\ Karlhede, U.\ Lindstrom, M.\ Rocek, {\sl
   Hyperk\"ahler metrics and supersymmetry}, 
   Commun.\ Math.\ Phys.\ {\bf108} (1987) 535-589\cr
\q{Ho} &J.\ Hoppe, {\sl Quantum theory of a massless relativistic 
   surface and a two-dimensional boundstate problem}, PhD Thesis, MIT 1982;
   \quad {\sl Quantum theory of a relativistic surface}, in: Constraint's 
   theory and relativistic dynamics, G.\ Longhi, L.\ Lusanna (eds.), 
   Proceedings Florence 1986, World Scientific\cr
\q{Ja} &N.\ Jacobson, {\sl Basic Algebra II}, W.H.\ Freeman and Company, 
   1985\cr
\q{Joy} &D.D.\ Joyce, {\sl Compact hypercomplex and quaternionic manifolds}, 
   J.\ Differ.\ Geom.\ {\bf35} (1992) 743-762; 
   {\sl Manifolds with many complex structures}, Q.\ J.\ Math.\ 
   Oxf.\ II.\ Ser.\ {\bf46} (1995) 169-184\cr
\q{Kar} &M.\ Karoubi, {\sl Homologie cyclique et K-th\'eorie}, 
   Soci\'et\'e Math\'ematique de France, Ast\'erisque {\bf 149} (1987)\cr 
\q{KL} &S.\ Klimek, A.\ Lesniewski, {\sl Quantum Riemann surfaces I. The 
   unit disc}, Commun.\ Math.\ Phys.\ {\bf146} (1992) 103-122; \quad
   {\sl Quantum Riemann surfaces II. The discrete series}, Lett.\ Math.\ 
   Phys.\ {\bf24} (1992) 125-139\cr 
\q{KN} &S.\ Kobayashi, K.\ Nomizu, {\sl Foundations of differential 
   geometry I, II}, Interscience Publishers 1963\cr
\q{Ko} &S.\ Kobayashi, {\sl Differential geometry of complex vector 
   bundles}, Iwanami Shoten Publishers and Princeton University 
   Press 1987\cr
\q{KS} &C.\ Klim\v cik, P.\ \v Severa, {\sl Dual non-abelian duality 
   and the Drinfeld double}, Phys.\ Lett.\ B {\bf351} (1995) 455-462\cr
\q{LaM} &H.B.\ Lawson, M.-L.\ Michelsohn, {\sl Spin Geometry}, 
   Princeton University Press 1989\cr
\q{LiM} &P.\ Libermann, C.-M.\ Marle, {\sl Symplectic geometry and 
   analytical mechanics}, D.\ Reidel Publishing Company 1987\cr
\q{LL} &E.H.\ Lieb, M.\ Loss, {\sl Stability of Coulomb systems with 
   magnetic fields II}, Commun.\ Math.\ Phys.\ {\bf104} (1986) 271-282\cr
\q{LY} &M.\ Loss, H.-T.\ Yau, {\sl Stability of Coulomb systems with 
   magnetic fields III}, Commun.\ Math.\ Phys.\ {\bf104} (1986) 283-290\cr
\q{Ma} &J.\ Madore, {\sl The commutative limit of a matrix geometry},
   J.\ Math.\ Phys {\bf32} (1991) 332-335\cr
\q{Ni} &H.\ Nicolai, {\sl The integrability of $N=16$ supergravity}, 
   Phys.\ Lett.\ B {\bf194} (1987) 402-407\cr
\q{NS} &A.\ Neveu, J.H.\ Schwarz, {\sl Quark model of dual pions}, 
   Phys.\ Rev.\ D {\bf4} (1971) 1109-1111\cr
\q{PS} &A.\ Pressley, G.\ Segal, {\sl Loop groups}, Clarendon Press 1986\cr  
\q{R} &R.\ Ramond, {\sl Dual theory for free fermions}, Phys.\ Rev.\ D 
   {\bf3} (1971) 2415-2418\cr
\q{Sa} &D.\ Salamon, {\sl Spin geometry and Seiberg-Witten invariants},
   preprint 1995, to be published by Birkh\"auser Verlag\cr
\q{Su} &D.\ Sullivan, {\sl Exterior {\tt d}, the local degree, and 
   smoothability}, IHES preprint 1995\cr
\q{Sw} &R.G.\ Swan, {\sl Vector bundles and projective modules}, 
   Trans.\ Amer.\ Math.\ Soc.\ {\bf105} 
   (1962) 264-277\cr
\q{VGB} &J.C.\ V\'arilly, J.M.\ Gracia-Bondia, {\sl Connes' 
   non-commutative differential geometry and the standard model}, 
   J.\ Geom.\ Phys.\ {\bf12} (1993) 223-301\cr
\q{WeB} &J.\ Wess, J.\ Bagger, {\sl Supersymmetry and supergravity}, 
   Princeton University Press 1983\cr
\q{Wel} &R.O.\ Wells, {\sl Differential analysis, on complex manifolds},
   Springer Verlag 1979\cr
\q{Wes} &P.\ West, {\sl Introduction to supersymmetry and supergravity}, 
   World Scientific 1986\cr
\q{Wi1} &E.\ Witten, {\sl Constraints on supersymmetry breaking}, 
   Nucl.\ Phys.\ B {\bf202} (1982) 253-316\cr
\q{Wi2} &E.\ Witten, {\sl Supersymmetry and Morse theory}, 
   J.\ Diff.\ Geom.\ {\bf17} (1982) 661-692\cr
\q{Wi3} &E.\ Witten, {\sl Non-abelian bosonization in two dimensions}
   Commun.\ Math.\ Phys.\ {\bf92} (1984) 455-472\cr
\q{WN} &B.\ de Wit, P.\ van Nieuwenhuizen, {\sl Rigidly and locally 
   supersymmetric two-dimen\-sio\-nal nonlinear $\sigma$-models with 
   torsion}, Nucl.\ Phys.\ B {\bf 312} (1989) 58-94\cr
\q{WTN} &B.\ de Wit, A.K.\ Tollst\'en, H.\ Nicolai, {\sl Locally 
   supersymmetric $D=3$ non-linear sigma models}, 
   Nucl.\ Phys.\ B {\bf 392} (1993) 3-38\cr
\q{WZ} &J.\ Wess, B.\ Zumino, {\sl Supergauge transformations in 
   four dimensions}, Nucl.\ Phys.\ B {\bf70} (1974) 39-50\cr
\q{Y} &S.T.\ Yau, {\sl On the Ricci curvature of a compact K\"ahler 
   manifold and the complex Monge-Amp\`ere equation I}, Comm.\ Pure 
   and Appl.\ Math.\ {\bf31} (1978) 339-411\cr
}\bye